\documentclass[11pt,a4paper]{article}   	
\pdfoutput=1
\usepackage{geometry}

\usepackage{graphicx}		
\usepackage{xcolor}

\usepackage{xspace}

\usepackage{amsmath} 
\usepackage{ifthen}  
\usepackage{amssymb}
\usepackage{amsfonts}
\usepackage{upgreek} 
\usepackage{bigints}

\usepackage{lineno}  % for line numbering during review

\usepackage{hyperref}
\usepackage{cancel}
\newboolean{uprightparticles}
\setboolean{uprightparticles}{false}

\def\ux85 {\mbox{UX85}\xspace}

\ifthenelse{\boolean{uprightparticles}}%
{

 \def\Pmu         {\ensuremath{\upmu}\xspace}

 \def\Ppsi        {\ensuremath{\uppsi}\xspace}

 \def\PDelta      {\ensuremath{\Delta}\xspace}                 
 \def\PXi      {\ensuremath{\Xi}\xspace}                 
 \def\PLambda      {\ensuremath{\Lambda}\xspace}                 
 \def\PSigma      {\ensuremath{\Sigma}\xspace}                 
 \def\POmega      {\ensuremath{\Omega}\xspace}                 
 \def\PUpsilon      {\ensuremath{\Upsilon}\xspace}                 
 
 \def\PB      {\ensuremath{\mathrm{B}}\xspace}                 
                  
 \def\PD      {\ensuremath{\mathrm{D}}\xspace}

 \def\PJ      {\ensuremath{\mathrm{J}}\xspace}                 
 \def\PK      {\ensuremath{\mathrm{K}}\xspace}

 \def\Pb      {\ensuremath{\mathrm{b}}\xspace}

 \def\Pe      {\ensuremath{\mathrm{e}}\xspace}

 \def\Pi      {\ensuremath{\mathrm{i}}\xspace}

}
{

 \def\Pmu         {\ensuremath{\mu}\xspace}

 \def\Ppsi        {\ensuremath{\psi}\xspace}                 
                  
 \mathchardef\PDelta="7101
 \mathchardef\PXi="7104
 \mathchardef\PLambda="7103
 \mathchardef\PSigma="7106
 \mathchardef\POmega="710A
 \mathchardef\PUpsilon="7107
                  
 \def\PB      {\ensuremath{B}\xspace}                 
                  
 \def\PD      {\ensuremath{D}\xspace}

 \def\PJ      {\ensuremath{J}\xspace}                 
 \def\PK      {\ensuremath{K}\xspace}

 \def\Pb      {\ensuremath{b}\xspace}

 \def\Pe      {\ensuremath{e}\xspace}

 \def\Pi      {\ensuremath{i}\xspace}

}

\def\en         {\ensuremath{\Pe^-}\xspace}   % electron negative (\em is taken)
\def\ep         {\ensuremath{\Pe^+}\xspace}

\def\mumu       {\ensuremath{\Pmu^+\Pmu^-}\xspace}

\def\ellell     {\ensuremath{\ell^+ \ell^-}\xspace}

\def\bquark    {\ensuremath{\Pb}\xspace}
\def\bquarkbar {\ensuremath{\overline \bquark}\xspace}
\def\bbbar     {\ensuremath{\bquark\bquarkbar}\xspace}

\def\kaon  {\ensuremath{\PK}\xspace}
  \def\Kbar  {\kern 0.2em\overline{\kern -0.2em \PK}{}\xspace}

\def\Kz    {\ensuremath{\kaon^0}\xspace}
\def\Kzb   {\ensuremath{\Kbar^0}\xspace}
\def\KzKzb {\ensuremath{\Kz \kern -0.16em \Kzb}\xspace}
\def\Kp    {\ensuremath{\kaon^+}\xspace}
\def\Km    {\ensuremath{\kaon^-}\xspace}

\def\KpKm  {\ensuremath{\Kp \kern -0.16em \Km}\xspace}

\def\Kstarz  {\ensuremath{\kaon^{*0}}\xspace}

  \def\Dbar    {\kern 0.2em\overline{\kern -0.2em \PD}{}\xspace}
\def\D       {\ensuremath{\PD}\xspace}

\def\Dz      {\ensuremath{\D^0}\xspace}
\def\Dzb     {\ensuremath{\Dbar^0}\xspace}
\def\DzDzb   {\ensuremath{\Dz {\kern -0.16em \Dzb}}\xspace}
\def\Dp      {\ensuremath{\D^+}\xspace}
\def\Dm      {\ensuremath{\D^-}\xspace}

\def\DpDm    {\ensuremath{\Dp {\kern -0.16em \Dm}}\xspace}

\def\B       {\ensuremath{\PB}\xspace}
  \def\Bbar    {\kern 0.18em\overline{\kern -0.18em \PB}{}\xspace}

\def\Bz      {\ensuremath{\B^0}\xspace}

\def\jpsi     {\ensuremath{{\PJ\mskip -3mu/\mskip -2mu\Ppsi\mskip 2mu}}\xspace}

  \def\Y#1S{\ensuremath{\PUpsilon{(#1S)}}\xspace}% no space before {...}!

\def\Lz {\ensuremath{\PLambda}\xspace}
\def\Lbar {\ensuremath{\kern 0.1em\overline{\kern -0.1em\Lambda\kern -0.05em}\kern 0.05em{}}\xspace}

\def\Lb      {\ensuremath{\Lz_\bquark}\xspace}

\newcommand{\decay}[2]{\ensuremath{#1\!\to #2}\xspace}         % {\Pa}{\Pb \Pc}

\def\to                 {\ensuremath{\rightarrow}\xspace}

\def\qsq       {\ensuremath{q^2}\xspace}

\def\AT#1     {\ensuremath{A_{\mathrm{T}}^{#1}}\xspace}           % 2

\def\C#1      {\ensuremath{\mathcal{C}_{#1}}\xspace}                       % 9
\def\Cp#1     {\ensuremath{\mathcal{C}_{#1}^{'}}\xspace}                    % 7
\def\Ceff#1   {\ensuremath{\mathcal{C}_{#1}^{\mathrm{(eff)}}}\xspace}        % 9  
\def\Cpeff#1  {\ensuremath{\mathcal{C}_{#1}^{'\mathrm{(eff)}}}\xspace}       % 7
\def\Ope#1    {\ensuremath{\mathcal{O}_{#1}}\xspace}                       % 2
\def\Opep#1   {\ensuremath{\mathcal{O}_{#1}^{'}}\xspace}                    % 7

\newcommand{\bra}[1]{\ensuremath{\langle #1|}}             % {a}
\newcommand{\ket}[1]{\ensuremath{|#1\rangle}}              % {b}
\newcommand{\tev}{\ensuremath{\mathrm{\,Te\kern -0.1em V}}\xspace}
\newcommand{\gev}{\ensuremath{\mathrm{\,Ge\kern -0.1em V}}\xspace}
\newcommand{\mev}{\ensuremath{\mathrm{\,Me\kern -0.1em V}}\xspace}
\newcommand{\kev}{\ensuremath{\mathrm{\,ke\kern -0.1em V}}\xspace}
\newcommand{\ev}{\ensuremath{\mathrm{\,e\kern -0.1em V}}\xspace}
\newcommand{\gevc}{\ensuremath{{\mathrm{\,Ge\kern -0.1em V\!/}c}}\xspace}
\newcommand{\mevc}{\ensuremath{{\mathrm{\,Me\kern -0.1em V\!/}c}}\xspace}
\newcommand{\gevcc}{\ensuremath{{\mathrm{\,Ge\kern -0.1em V\!/}c^2}}\xspace}
\newcommand{\gevgevcccc}{\ensuremath{{\mathrm{\,Ge\kern -0.1em V^2\!/}c^4}}\xspace}
\newcommand{\mevcc}{\ensuremath{{\mathrm{\,Me\kern -0.1em V\!/}c^2}}\xspace}

\def\invfb   {\ensuremath{\mbox{\,fb}^{-1}}\xspace}

\def\deriv {\ensuremath{\mathrm{d}}}

\def\gsim{{~\raise.15em\hbox{$>$}\kern-.85em
          \lower.35em\hbox{$\sim$}~}\xspace}
\def\lsim{{~\raise.15em\hbox{$<$}\kern-.85em
          \lower.35em\hbox{$\sim$}~}\xspace}

\def\tell1  {TELL1\xspace}
\def\ukl1   {UKL1\xspace}

\newcommand{\eg}{\mbox{\itshape e.g.}\xspace}
\newcommand{\ie}{\mbox{\itshape i.e.}}

\newboolean{articletitles}
\setboolean{articletitles}{true}

\usepackage{cite}
\usepackage{mciteplus}

\newboolean{inbibliography}

\newcommand{\asq}[3]{\ensuremath{|A_{#1 #2}^{\rm #3}|^{2}}\xspace} 
\newcommand{\aprod}[6]{\ensuremath{A_{#1 #2}^{\rm #3} A_{#4 #5}^{*{\rm #6}}}\xspace}

\begin{document}

\begin{flushright} EOS-2017-02 \end{flushright}

\vspace*{1.0cm}
\begin{center}
{\huge\bfseries 
Angular distribution of polarised $\Lb$ baryons decaying to \boldmath{$\Lz \ell^+\ell^-$}
}\\[1.0 cm]
{\Large
Thomas Blake$^{a}$ and Michal~Kreps$^{a}$, 
%T.\,Blake$^{a}$, G.\,Chatzikonstantinidis$^{b}$, M.\,Kreps$^{a}$ and N.\,Watson$^{b}$
}\\[0.4 cm] 
{\small
$^a$ Department of Physics, University of Warwick, Coventry CV4\,7AL, UK 
%$^a$ Department of Physics, University of Warwick, Coventry CV4\,7AL, UK \\
%$^b$ School of Physics and Astronomy, University of Birmingham, Birmingham B15\,2TT, UK
} \\[0.5 cm]
\small
E-Mail:
\texttt{\href{mailto:thomas.blake@cern.ch}{thomas.blake@cern.ch}},
\texttt{\href{mailto:Michal.Kreps@warwick.ac.uk}{Michal.Kreps@warwick.ac.uk}}.
%\texttt{\href{mailto:thomas.blake@cern.ch}{thomas.blake@cern.ch}},
%\texttt{\href{mailto:gc.hep.ph.bham.ac.uk}{gc.hep.ph.bham.ac.uk}}, \\
%\texttt{\href{mailto:Michal.Kreps@warwick.ac.uk}{Michal.Kreps@warwick.ac.uk}},
%\texttt{\href{mailto:nigel.watson@cern.ch}{nigel.watson@cern.ch}}.
\end{center}

\vspace*{2.0cm}

\begin{abstract} 
  \noindent 
  Rare $b \to s\ellell$ flavour-changing-neutral-current processes provide important tests of the Standard Model of particle physics. 
  Angular observables in exclusive $b \to s\ellell$ processes can be particularly powerful as they allow hadronic uncertainties to be controlled. 
  Amongst the exclusive processes that have been studied by experiments, the decay  \decay{\Lb}{\Lz\ellell} is unique in that the \Lb baryon can be produced polarised. 
  In this paper, we derive an expression for the angular distribution of the \decay{\Lb}{\Lz\ellell} decay for the case where the \Lb baryon is produced polarised. 
  This extends the number of angular observables in this decay from 10 to 34.
  Standard Model expectations for the new observables are provided and the sensitivity of the observables is explored under a variety of new physics models.
  At low-hadronic recoil, four of the new observables have a new short distance dependence that is absent in the unpolarised case. 
  The remaining observables depend on the same short distance contributions as the unpolarised observables, but with different dependence on hadronic form-factors.
  These relations provide possibilities for novel tests of the SM that could be carried out with the data that will become available at the LHC or a future $\ep\en$ collider.
\end{abstract} 

\clearpage

\section{Introduction} 
\label{sec:introduction}

Rare $b\to s\ellell$ have been studied extensively by experiments at the B-factories as well as experiments at the Tevatron and Large Hadron Collider (LHC). 
Amongst the $b\to s\ellell$ processes that have been studied, the decay \decay{\Lb}{\Lz\mumu} is unique for two reasons: 
it is the only baryonic decay that has been studied; and the \Lz baryon decays weakly leading to new hadron-side observables.   
The angular distribution of \decay{\Lb}{\Lz\mumu} decays has been studied in Refs.~\cite{Gutsche:2013pp,Boer:2014kda} for the case of unpolarised \Lb baryons. 
The resulting angular distribution is described by 10 angular observables.
The decay rate and lepton side angular distribution has also been studied in the SM and in several extensions of the SM (NP models) in Refs.~\cite{Aslam:2008hp,Wang:2008sm,Huang:1998ek,Chen:2001ki,Chen:2001zc,Chen:2001sj,Mott:2011cx,Aliev:2010uy,Mohanta:2010eb,Sahoo:2009zz}.
If the \Lb is produced polarised, a much larger number of observables are measurable. 
These observables are explored in this paper. 
The exploitation of production polarisation in radiative \decay{\Lb}{\Lz^{(*)}\gamma} decays has previously been studied in Refs.~\cite{Gremm:1995nx,Hiller:2001zj,Legger:2006cq,Hiller:2007ur}.

In $\ep\en$ collisions, \Lb baryons can be produced with large longitudinal polarisations. 
The longitudinal polarisation of \Lb baryons and \bquark-quarks produced via $\ep\en \to Z^0 (\to \bbbar)$ decays has been studied by the LEP experiments in Refs.~\cite{Buskulic:1995mf,Abbiendi:1998uz,Abreu:1999gf}. 
The production of \Lb baryons with longitudinal polarisation is forbidden in strong interactions, due to parity conservation. 
The \Lb can, however, be produced with transverse polarisation in $pp$ collisions.
In this paper, we focus on the transverse polarisation of the \Lb baryon.
The transverse  polarisation of \Lb baryons produced in $pp$ collisions at $\sqrt{s}=7$ and 8\tev has been studied by the LHCb and CMS experiments  in Refs.~\cite{LHCb-PAPER-2012-057} and \cite{CMS:2016iaf}, respectively.  
The LHCb experiment measures $P_{\Lb} = 0.06\pm 0.07 \pm 0.02$  at  $\sqrt{s}=7\tev$.
The CMS experiment measures  $P_{\Lb} = 0.00 \pm 0.06 \pm 0.02$ combining data from $\sqrt{s}=7$ and 8\tev.
In both cases, the production polarisation is determined from the observed angular distribution of \decay{\Lb}{\jpsi\Lz} decays.
Whilst the measured transverse production polarisation is small, polarisations of $\mathcal{O}(10\%)$ cannot be excluded. 
Polarised \Lb baryons can also be obtained from decays of heavier \bquark-baryons, for example in decays of the $\Sigma_b^{(*)}$~\cite{Galanti:2015pqa}. 

The only existing measurements of the angular distribution of the \decay{\Lb}{\Lz\ellell} decay come from the LHCb experiment~\cite{LHCb-PAPER-2015-009}.
Due to the limited size of their dataset, LHCb only studied a subset of the angular distribution that could be accessed from single angle projections on the lepton- and hadron-side. 
With the much larger data sets that will be available at the LHC experiments after run\,2 of the LHC, the experiments will be able to probe the full angular distribution.
However, the sheer number of observables involved will most likely require an analysis of the moments of the angular distribution (see for example Ref.~\cite{Beaujean:2015xea}) rather than the conventional approach of fitting for the angular observables. 
This approach is discussed in Sec.~\ref{sec:weighting}, where we provide the weighting functions needed to extract the observables.

\section{Angular distribution} 
\label{sec:Angular}

\begin{figure}
\centering
\includegraphics[width=\linewidth]{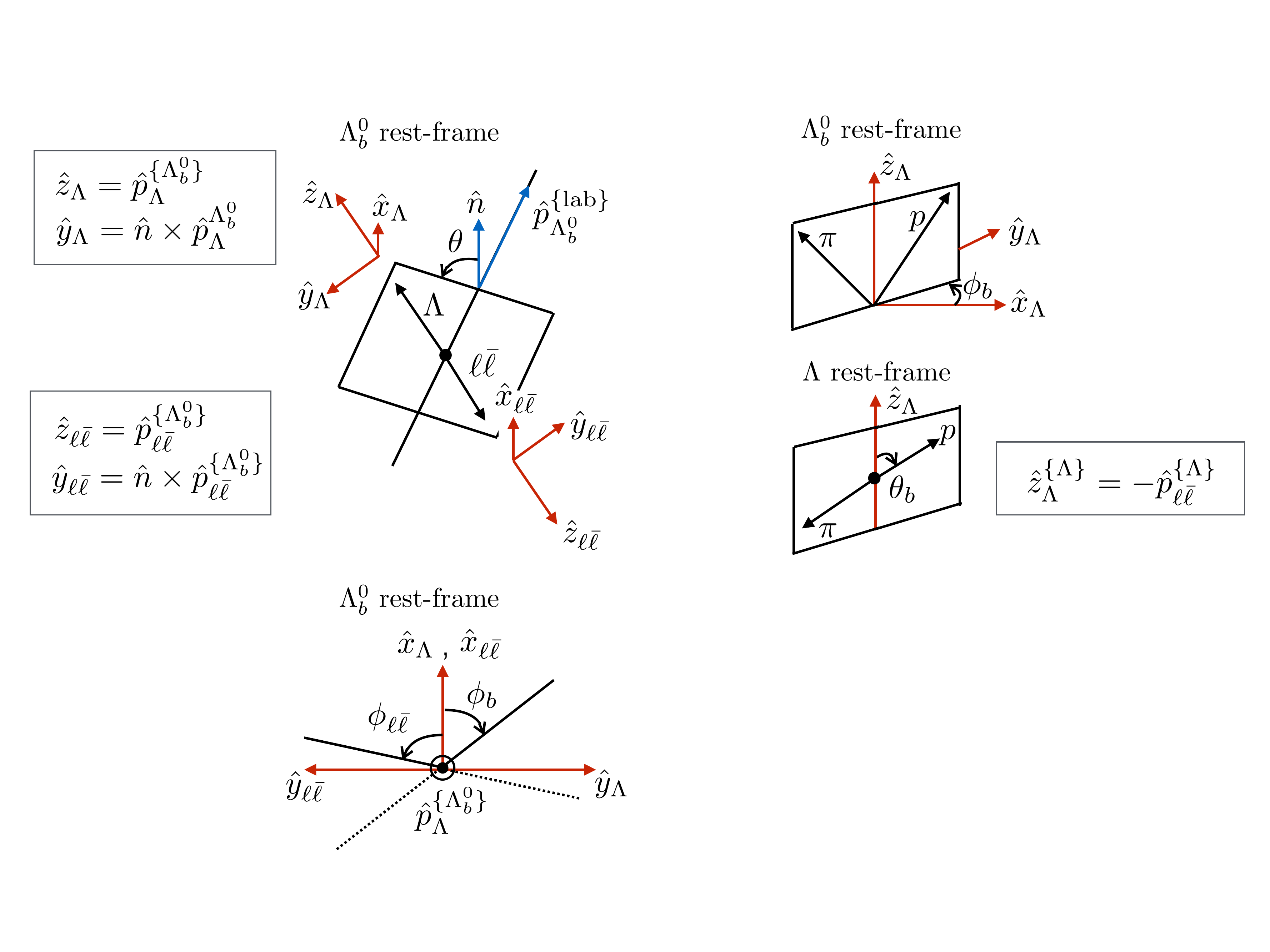}
\caption{
The \decay{\Lb}{\Lz\ell^+\ell^-} decay is described by five angles: the angle, $\theta$, between the direction of the \Lz baryon and the normal vector $\hat{n}$ in the \Lb rest-frame; and two sets of helicity angles, describing the decays of the \Lz baryon ($\theta_b,\phi_b$) and the dilepton system ($\theta_l,\phi_l$). 
For transverse production polarisation $\hat{n}$ is chosen to be $\hat{p}_{\Lb} \times \hat{p}_{\rm beam}$. 
The helicity angles are then defined with respect to this normal vector through the coordinate systems $(\hat{x}_{\Lz},\hat{y}_\Lz,\hat{z}_\Lz)$ and $(\hat{x}_{\ell\bar{\ell}},\hat{y}_{\ell\bar{\ell}},\hat{z}_{\ell\bar{\ell}})$. 
The $\hat{z}$ axis points in the direction of the \Lz/dilepton system in the \Lb rest-frame. 
The angle between the two decay planes in the \Lb rest frame is $\chi = \phi_l + \phi_b$. 
The angles $\theta_l$, $\theta_b$ and $\chi$ are sufficient to parameterise the angular distribution of the decay in the case of zero production polarisation 
\label{fig:angles}
}
\end{figure}

The angular distribution of the \decay{\Lb}{\Lz\ell^+\ell^-} decay has been previously studied in Refs.~\cite{Gutsche:2013pp,Boer:2014kda}. 
In this paper we extend those studies to include the case where the $\Lb$ baryon is produced with a transverse polarisation. 
We start by expanding the differential decay rate for the \decay{\Lb}{\Lz\ell^+\ell^-} decay in terms of generalised helicity amplitudes
\begin{equation}
\begin{split}
\frac{\deriv^{6}\Gamma}{\deriv\qsq\,\deriv\vec{\Omega}} \propto
\sum_{\substack{ \lambda_1,\lambda_{2}, \lambda_{p}, \lambda_{\ell\ell},\lambda'_{\ell\ell}, \\ J,J',m,m',\lambda_{\Lambda}, \lambda'_{\Lambda},}}  \Big(  
& (-1)^{J + J'} \times \\[-25pt]
& \rho_{\lambda_{\Lz} - \lambda_{\ell\ell},\lambda'_{\Lz}-\lambda'_{\ell\ell}}(\theta) \times \\
& H^{m,J}_{\lambda_{\Lz},\lambda_{\ell\ell}}(\qsq) H^{\dagger\, m',J'}_{\lambda'_{\Lz},\lambda'_{\ell\ell}}(\qsq) \times \\ 
& h^{m,J}_{\lambda_1,\lambda_2}(\qsq) h^{\dagger\, m',J'}_{\lambda_1,\lambda_2}(\qsq)  \times \\
& D^{J\,*}_{\lambda_{\ell\ell},\lambda_1-\lambda_{2}}(\phi_l,\theta_l,-\phi_l)  D^{J'}_{\lambda'_{\ell\ell},\lambda_1-\lambda_{2}}(\phi_l,\theta_l,-\phi_l) \times \\ 
& h^{\Lz}_{\lambda_{p},0}h^{\dagger\,\Lz}_{\lambda_{p}0}  \times \\ 
& D^{1/2\,*}_{\lambda_{\Lz},\lambda_{p}}(\phi_b,\theta_{b},-\phi_b)  D^{1/2}_{\lambda'_{\Lz},\lambda_{p}}(\phi_b,\theta_{b},-\phi_b) \Big) \,,
\end{split}
\label{eq:base}
\end{equation}
which depends on five angles, $\vec{\Omega} = (\theta_l, \phi_l,\theta_b,\phi_b,\theta)$, and the dilepton invariant mass squared, $q^2$. 
The angular basis is illustrated in Fig.~\ref{fig:angles}. 
The helicity basis is defined starting from the normal vector between the direction of the \Lb baryon in the lab-frame and the beam-axis of the experiment ($\hat{n} = \hat{p}_{\Lz_b} \times \hat{p}_{\rm beam}$). 
This is an appropriate choice when considering transverse production polarisation of the \Lb baryon. 

Equation~\ref{eq:base} involves three sets of helicity amplitudes: 
$H^{m,J}_{\lambda_{\Lz},\lambda_{\ell\ell}}(\qsq)$ describing the decay of the \Lb baryon into a \Lz baryon with helicity $\lambda_\Lz$ and a dilepton pair with helicity $\lambda_{\ell\ell}$; 
$h^{m,J}_{\lambda_{1},\lambda_{2}}$ describing the decay of the dilepton system to leptons with helicities $\lambda_1$ and $\lambda_2$; 
and $h^{\Lz}_{\lambda_p, 0}$ describing the decay \decay{\Lz}{p\pi} to a proton with helicity $\lambda_p$. 
The index $J$ refers to the spin of the dilepton system, which can either be zero or one. 
When $J = 0$, $\lambda_{\ell\ell} = 0$, and when $J=1$, $\lambda_{\ell\ell}$ can take the values $-1, 0, +1$. 
The helicity labels $\lambda_p$, $\lambda_\Lambda$, $\lambda_1$ and $\lambda_2$ can take the values $\pm 1/2$. 
Angular momentum conservation in the \Lb decay requires $|\lambda_{\Lz} - \lambda_{\ell\ell}| = 1/2$. 
The factor $(-1)^{J+J'}$ originates from the structure of the Minkowski metric tensor, see Ref.~\cite{Kadeer:2005aq} for details. 
The remaining index, $m=V,\,A$, denotes the decay of the dilepton system by either a vector or an axial-vector current. 
The term $\rho_{\lambda_{\Lz}-\lambda_{\ell\ell},\lambda'_{\Lz}-\lambda'_{\ell\ell}}$ is the polarisation density matrix for the transverse polarisation of the \Lb. The matrix is a two-by-two matrix (with ${\rm Tr}(\rho) = 1$) given by
\begin{equation}
\begin{split}
\rho_{+1/2,+1/2}(\theta) &=  \tfrac{1}{2}(1 + P_{\Lb}) \cos\theta~, \\
\rho_{+1/2,-1/2}(\theta) &=  \tfrac{1}{2}P_{\Lb} \sin\theta~, \\
\rho_{-1/2,-1/2}(\theta) &= \tfrac{1}{2}(1- P_{\Lb}) \cos\theta~, \\
\rho_{-1/2,+1/2}(\theta) &=  \tfrac{1}{2} P_{\Lb} \sin\theta~.
\end{split}
\end{equation} 
Finally, the $D^{j}_{m,m'}(\phi,\theta,-\phi)$ are Wigner-$D$ functions.
An explicit form of the Wigner-$D$  functions is given in Appendix~\ref{appendix:Dfunctions}.

\subsection{Lepton system amplitudes}
\label{sec:leptonside}

There are two sets of amplitudes for the dilepon system, with either a vector or an axial-vector current, 
\begin{equation}
\begin{split}
h^{V,J}_{\lambda_1,\lambda_2} &= \bar{\ell}(\lambda_2) \gamma^\mu \ell(\lambda_1) \varepsilon_\mu^{*}( \lambda_1 - \lambda_2 ) \\
h^{A,J}_{\lambda_1,\lambda_2} &= \bar{\ell}(\lambda_2) \gamma^\mu \gamma^5 \ell(\lambda_1) \varepsilon_\mu^{*}( \lambda_1 - \lambda_2 ) ~,\\ 
\end{split}
\end{equation} 
where $\gamma^{\mu}$ is a Dirac $\gamma$-matrix and $\varepsilon_{\mu}$ is a polarisation vector.
These amplitudes evaluate to~\cite{Gutsche:2013pp}
\begin{alignat}{3}
& h^{V,0}_{+1/2, +1/2} && = 0\,,  &&  h^{A,0}_{+1/2, +1/2} = 2 m_l  = \sqrt{ q^{2} ( 1- \beta_l^2 ) }\,,  \nonumber \\ 
& h^{V,0}_{+1/2, -1/2}  && =  0\,, && h^{A,0}_{+1/2, -1/2}  =  0\,,  \\
& h^{V,1}_{+1/2,+1/2}  && = 2 m_l  = \sqrt{ q^{2} ( 1- \beta_l^2 ) }\,, \quad && h^{A,1}_{+1/2, +1/2}  =  0\,,   \nonumber \\
& h^{V,1}_{+1/2, -1/2}  && = -\sqrt{2 q^2}\,,  && h^{A,1}_{+1/2, -1/2}  = \sqrt{2 q^2} \beta_{l}~, \nonumber  
\end{alignat}
where $m_l$ is the lepton mass and  $\beta_l$ is the lepton velocity in the dilepton rest frame ($|\vec{p}_{l}|/E_{l}$), \ie
\begin{equation}
\beta_{l} = \sqrt{ 1 - \frac{4 m_{l}^{2}}{q^2} }\,.
\end{equation}
The amplitudes with $J=0$ vanish in the case that the lepton mass is zero (when $\beta_l = 1$).
Under the Parity transformation
\begin{equation}
\begin{split}
h^{V,J}_{-\lambda_1,-\lambda_2} & = h^{V,J}_{\lambda_1,\lambda_2} \\
h^{A,J}_{-\lambda_1,-\lambda_2} & = -h^{A,J}_{\lambda_1,\lambda_2} ~.
\end{split}
\end{equation}

\subsection{Hadron system amplitudes} 
\label{sec:hadronside}

On the hadron side, the \Lz decay amplitudes can be expressed in terms of the well known \Lz asymmetry parameter~\cite{PDG2016}
\begin{equation}
\alpha_{\Lambda}\,=\,\frac{|h^{b}_{{\frac{1}{2}},0}|^{2}-
|h^{b}_{-{\frac{1}{2}},0}|^{2}}
{|h^{b}_{{\frac{1}{2}},0}|^{2}
+|h^{b}_{-{\frac{1}{2}},0}|^{2}} =  0.642 \pm 0.013\,.
\end{equation}
The hadron side amplitudes are normalised such that
\begin{equation}
|h^{b}_{{\frac{1}{2}},0}|^{2} +|h^{b}_{-{\frac{1}{2}},0}|^{2} = 1\,.
\end{equation}

\subsection{Helicity and transversity amplitudes}

After replacing the lepton and hadron-side amplitudes with the expressions given in Secs.~\ref{sec:hadronside} and \ref{sec:leptonside}, the angular distribution can be expanded in terms of 10 helicity amplitudes,
\begin{displaymath}
 H^{m,1}_{+1/2,+1},~H^{m,1}_{-1/2,-1},~H^{m,1}_{+1/2,0},~H^{m,1}_{-1/2,0},~H^{A, 0}_{+1/2,0}~\text{and}~H^{A, 0}_{-1/2,0}\,,
 \end{displaymath}
$P_{\Lb}$, $\alpha_{\Lz}$ and a set of kinematic factors that come from the lepton-side amplitudes. 
For the remainder of this paper it is convenient to absorb a common factor of $\sqrt{q^2}$ from the lepton-side amplitudes into these helicity amplitudes, \ie
\begin{align}
\sqrt{q^2} H^{m,J}_{\lambda_\Lz,\lambda_{\ell\ell}} = H'\,^{m,J}_{\lambda_\Lz,\lambda_{\ell\ell}}~.
\end{align}
By absorbing this factor, the only kinematic dependence outside of $H'\,^{m,J}_{\lambda_\Lz,\lambda_{\ell\ell}}(q^2)$ comes from factors of $\beta_l$.

The helicity amplitudes can be replaced by a corresponding set of transversity amplitudes for the decay that separate the vector and axial-vector contributions on the hadron-side: 
the amplitudes $A_{\parallel 1}^{\rm R,\,L}$ and $A_{\parallel 0}^{\rm R,\,L}$ depend only on the vector contribution to $H'_{\lambda_\Lz, \lambda_{\ell\ell}}$ (\ie~on $\bra{\Lz}{\bar{s} \gamma^{\mu} b}\ket{\Lb}$); 
and the amplitudes $A_{\perp 1}^{\rm R,\,L}$ and $A_{\perp 0}^{\rm R,\,L}$ depend only on the axial-vector contribution to $H'_{\lambda_{\Lz}, \lambda_{\ell\ell}}$ (\ie~on $\bra{\Lz}{\bar{s} \gamma^{\mu} \gamma^5 b}\ket{\Lb}$). 
To do this, we start by re-writing the original helicity amplitudes as
\begin{align}
H'\,^{{\rm \{R,L\}},J}_{\lambda_\Lz,\lambda_{\ell\ell}} = \frac{1}{\sqrt{2}}\left( H'\,^{V,J}_{\lambda_\Lz,\lambda_{\ell\ell}} \pm H'\,^{A,J}_{\lambda_\Lz,\lambda_{\ell\ell}} \right)~,
\end{align} 
where the indices L and R refer to left- and right-handed chiralities of the dilepton system, respectively. 
This is followed by the replacements
\begin{equation}
\begin{split}
A_{\perp 1}^{\rm \{R,L\}} &= \frac{1}{\sqrt{2}} \left(H'\,^{\rm \{R,L\}, 1}_{+1/2,+1} - H'\,^{\rm \{R,L\}, 1}_{-1/2,-1} \right) \,,\\ 
A_{\parallel 1}^{\rm \{R,L\}} &= \frac{1}{\sqrt{2}} \left(H'\,^{\rm \{R,L\},1}_{+1/2,+1} + H'\,^{\rm \{R,L\},1}_{-1/2,-1} \right) \,,\\ 
A_{\perp 0}^{\rm \{R,L\}} &= \frac{1}{\sqrt{2}} \left(H'\,^{\rm \{R,L\},1}_{+1/2,0} - H'\,^{\rm \{R,L\}, 1}_{-1/2,0} \right) \,,\\ 
A_{\parallel 0}^{\rm \{R,L\}} &= \frac{1}{\sqrt{2}} \left(H'\,^{\rm \{R,L\}, 1}_{+1/2,0} + H'\,^{\rm \{R,L\}, 1}_{-1/2,0} \right) \,,\\ 
A_{\perp t} &= \frac{1}{\sqrt{2}} \left(H'\,^{A,0}_{+1/2,0} - H'\,^{A,0}_{-1/2,0} \right)  \,, \\ 
A_{\parallel t} &= \frac{1}{\sqrt{2}} \left(H'\,^{A,0}_{+1/2,0} + H'\,^{A,0}_{-1/2,0} \right)~.
\end{split}
\end{equation} 
Here, the subscript $t$ refers to the time-like polarisation vector of the dilepton system.

\section{Observables}
\label{observables}

Expanding out the sum in Eq.~\ref{eq:base}, gives 34 different angular terms
\begin{equation}
\begin{split}
\frac{\deriv^{6}\Gamma}{\deriv\qsq\,\deriv\vec{\Omega}} = \frac{3}{32\pi^{2}}  \Big( & \sum\limits_{i=0}^{34} K_i(\qsq) f_{i}(\vec{\Omega}) \Big) \\
\frac{\deriv^{6}\Gamma}{\deriv\qsq\,\deriv\vec{\Omega}} = \frac{3}{32\pi^{2}} \Big(
& \left(K_1\sin^2\theta_l+K_2\cos^2\theta_l+K_3\cos\theta_l\right)  +  \\[-5pt]
& \left(K_4\sin^2\theta_l+K_5\cos^2\theta_l+K_6\cos\theta_l\right)\cos\theta_b +  \\
& \left(K_7\sin\theta_l\cos\theta_l+K_8\sin\theta_l\right)\sin\theta_b\cos\left(\phi_b+\phi_l\right) +  \\
&\left(K_9\sin\theta_l\cos\theta_l+K_{10}\sin\theta_l\right)\sin\theta_b\sin\left(\phi_b+\phi_l\right) +  \\
& \left(K_{11}\sin^2\theta_l+K_{12}\cos^2\theta_l+K_{13}\cos\theta_l\right)\cos\theta +  \\
& \left( K_{14}\sin^2\theta_l+K_{15}\cos^2\theta_l+K_{16}\cos\theta_l\right)\cos\theta_b \cos\theta +  \\
& \left(K_{17}\sin\theta_l\cos\theta_l+K_{18}\sin\theta_l\right)\sin\theta_b\cos\left(\phi_b+\phi_l\right)\cos\theta  +  \\
& \left(K_{19}\sin\theta_l\cos\theta_l+K_{20}\sin\theta_l\right)\sin\theta_b\sin\left(\phi_b+\phi_l\right) \cos\theta +  \\
& \left(K_{21}\cos\theta_l\sin\theta_l+K_{22}\sin\theta_l\right)\sin\phi_l \sin\theta +  \\
& \left(K_{23}\cos\theta_l\sin\theta_l+K_{24}\sin\theta_l\right)\cos\phi_l  \sin\theta +  \\
& \left(K_{25}\cos\theta_l\sin\theta_l+K_{26}\sin\theta_l\right)\sin\phi_l\cos\theta_b  \sin\theta +  \\
& \left(K_{27}\cos\theta_l\sin\theta_l+K_{28}\sin\theta_l\right)\cos\phi_l\cos\theta_b  \sin\theta  +  \\
& \left(K_{29}\cos^2\theta_l+K_{30}\sin^2\theta_l\right)\sin\theta_b\sin\phi_b  \sin\theta +  \\
& \left(K_{31}\cos^2\theta_l+K_{32}\sin^2\theta_l\right)\sin\theta_b\cos\phi_b  \sin\theta +  \\
& \left(K_{33}\sin^2\theta_l \right) \sin\theta_b\cos\left(2\phi_l+\phi_b\right) \sin\theta  +  \\
& \left(K_{34}\sin^2\theta_l \right) \sin\theta_b\sin\left(2\phi_l+\phi_b\right)  \sin\theta  \Big)~.
\end{split}
\label{eq:34terms}
\end{equation} 
Integrating this expression over $\vec{\Omega}$ yields the differential decay rate as a function of \qsq, 
\begin{equation}
\frac{\deriv\Gamma}{\deriv\qsq} = 2 K_{1} + K_{2}~.
\end{equation} 
This can be used to define a set of normalised angular observables
\begin{equation}
M_{i} = \frac{K_i}{2 K_{1} + K_{2} }~.
\end{equation}

\section{Angular terms}
\label{sec:terms}

The first ten angular terms are
\begin{equation}
\begin{split}
K_{1} &= \tfrac{1}{4}\left(\asq{\parallel}{1}{L} +  \asq{\perp}{1}{L} +  \asq{\parallel}{1}{R}  +  \asq{\perp}{1}{R}\right) \\ 
& + \tfrac{1}{4}( 1 + \beta_l^2 ) \left( 
\asq{\parallel}{0}{L} +  \asq{\perp}{0}{L} +  \asq{\parallel}{0}{R}  +  \asq{\perp}{0}{R} 
\right )  \\
& + \tfrac{1}{2}(1-\beta_l^{2}) {\rm Re}\left( 
\aprod{\parallel}{1}{R}{\parallel}{1}{L} +  \aprod{\perp}{1}{R}{\perp}{1}{L} + 
\aprod{\parallel}{0}{R}{\parallel}{0}{L} +  \aprod{\perp}{0}{R}{\perp}{0}{L} 
\right) \\
& + \tfrac{1}{2}(1-\beta_l^2)\left( \asq{\parallel}{t}{} + \asq{\perp}{t}{} \right)  \,, \\ 
K_{2} & =   \tfrac{1}{4}(1 + \beta^{2}_l) \left( 
  \asq{\parallel}{1}{R}  + \asq{\perp}{1}{R} + \asq{\parallel}{1}{R}  + \asq{\perp}{1}{L} 
  \right)  \\
& + \tfrac{1}{4}(1 - \beta^2_l) \left( 
\asq{\parallel}{0}{R}  + \asq{\perp}{0}{R} +  \asq{\parallel}{0}{L} +  \asq{\perp}{0}{L} 
\right) \\ 
& + \tfrac{1}{2}( 1- \beta^2_l) {\rm Re}\left( 
\aprod{\parallel}{1}{R}{\parallel}{1}{L}  + \aprod{\perp}{1}{R}{\perp}{1}{L} +  
\aprod{\parallel}{0}{R}{\parallel}{0}{L}  + \aprod{\perp}{0}{R}{\perp}{0}{L}
\right) \\ 
& + \tfrac{1}{2}(1-\beta^2_l) \left( \asq{\parallel}{t}{} + \asq{\perp}{t}{} \right) \,, \\ 
K_{3} & = - \beta_l {\rm Re}\left( 
A_{\perp 1}^{\rm R} A_{\parallel 1}^{*{\rm R}} -  A_{\perp 1}^{\rm L} A_{\parallel 1}^{*{\rm L}} 
\right) \\
K_{4} &= \tfrac{1}{2} \alpha_{\Lz} {\rm Re} \left( 
\aprod{\perp}{1}{R}{\parallel}{1}{R} + \aprod{\perp}{1}{L}{\parallel}{1}{L} \right) \\
& + \tfrac{1}{2} \alpha_{\Lz} (1+\beta^{2}_l) {\rm Re} \left( \aprod{\perp}{0}{R}{\parallel}{0}{R} + \aprod{\perp}{0}{L}{\parallel}{0}{L} \right) \\
& + \tfrac{1}{2} \alpha_{\Lz} (1-\beta^2_l){\rm Re}\left( 
\aprod{\perp}{1}{R}{\parallel}{1}{L} +  \aprod{\parallel}{1}{R}{\perp}{1}{L} + 
\aprod{\perp}{0}{R}{\parallel}{0}{L} + \aprod{\parallel}{0}{R}{\perp}{0}{L} 
\right) \\
& + \alpha_{\Lz} (1-\beta_l^2){\rm Re}\left( \aprod{\perp}{t}{}{\parallel}{t}{} \right) \,, \\
K_{5} &= \tfrac{1}{2} \alpha_{\Lz} (1+\beta_l^2) {\rm Re}\left( 
\aprod{\perp}{1}{R}{\parallel}{1}{R} + \aprod{\perp}{1}{L}{\parallel}{1}{L} 
\right) \\
& +  \tfrac{1}{2} \alpha_{\Lz} (1-\beta_l^2) {\rm Re} \left( 
\aprod{\parallel}{0}{R}{\perp}{0}{R} +  \aprod{\parallel}{0}{L}{\perp}{0}{L}
\right) \\
& +  \tfrac{1}{2} \alpha_{\Lz} (1-\beta_l^2) {\rm Re} \left(
 \aprod{\perp}{1}{R}{\parallel}{1}{L}  + \aprod{\parallel}{1}{R}{\perp}{1}{L} +
\aprod{\perp}{0}{R}{\parallel}{0}{L} + \aprod{\parallel}{0}{R}{\perp}{0}{L}
\right) \\  
& + \alpha_{\Lz} (1-\beta_l^2){\rm Re}\left( \aprod{\perp}{t}{}{\parallel}{t}{} \right)  \,, \\
K_{6} &= - \tfrac{1}{2} \alpha_{\Lz}  \beta_l  \left( 
 | A_{\parallel 1}^{\rm R}|^{2} + |A_{\perp 1}^{\rm R}|^{2} - | A_{\parallel 1}^{\rm L}|^{2} - |A_{\perp 1}^{\rm L}|^{2}  \right) \,, \\
K_{7} &= \tfrac{1}{\sqrt{2}}  \alpha_{\Lz} \beta_l^2 {\rm Re} \left( 
A_{\perp 1}^{\rm R} A_{\parallel 0}^{*{\rm R}}  -
A_{\parallel 1}^{\rm R} A_{\perp 0}^{*{\rm R}}  +
A_{\perp 1}^{\rm L}  A_{\parallel 0}^{*{\rm L}}  - 
A_{\parallel 1}^{\rm L} A_{\perp 0}^{*{\rm L}}  
\right) \,, \\ 
K_{8} &= \tfrac{1}{\sqrt{2}}  \alpha_{\Lz} \beta_l {\rm Re} \left(
A_{\perp  1}^{\rm R} A_{\perp 0}^{*{\rm R}} - A_{\parallel 1}^{\rm R} A_{\parallel 0}^{*{\rm R}}  - 
A_{\perp  1}^{\rm L} A_{\perp 0}^{*{\rm L}} + A_{\parallel 1}^{\rm L} A_{\parallel 0}^{*{\rm L}}   
\right) \,, \\
K_{9} &= \tfrac{1}{\sqrt{2}} \alpha_{\Lz} \beta_l^2  {\rm Im} \left( 
A_{\perp 1}^{R} A_{\perp 0}^{*{\rm R}} - A_{\parallel 1}^{\rm R}  A_{\parallel 0}^{*{\rm R}} +  
A_{\perp 1}^{\rm L} A_{\perp 0}^{*{\rm L}} - A_{\parallel 1}^{\rm L}  A_{\parallel 0}^{*{\rm L}}  
\right) \,,  \\
K_{10} &= \tfrac{1}{\sqrt{2}} \alpha_{\Lz} \beta_l {\rm Im} \left( 
\aprod{\perp}{1}{R}{\parallel}{0}{R}  - \aprod{\parallel}{1}{R}{\perp}{0}{R} -  
\aprod{\perp}{1}{L}{\parallel}{0}{L} + \aprod{\parallel}{1}{L}{\perp}{0}{L} 
\right) \,.
\end{split}
\end{equation}

\noindent These terms are accessible even if the $\Lb$ baryon is unpolarised and have been previously studied in Refs.~\cite{Boer:2014kda,Detmold:2016pkz}. 
There is a straightforward relationship between our observables and those of Ref.~\cite{Boer:2014kda}, with 
$K_{1ss} = K_{1}$, 
$K_{1cc} = K_{2}$, 
$K_{1c} = K_{3}$, 
$K_{2ss} = K_{4}$, 
$K_{2cc} = K_{5}$, 
$K_{2c} = K_{6}$, 
$K_{4sc} = K_{7}$, 
$K_{4s} = K_{8}$, 
$K_{3sc} = K_{9}$ and
$K_{3s} = K_{10}$. 

The remaining 24 terms are only non-vanishing if $P_{\Lb}$ is non-zero. 
Terms $K_{11}$ through  $K_{16}$ have a similar dependence to $K_{1}$ through $K_{6}$. 
These are
\begin{equation}
\begin{split}
K_{11} &= -\tfrac{1}{2} P_{\Lb} {\rm Re}\left( \aprod{\parallel}{1}{R}{\perp}{1}{R} + \aprod{\parallel}{1}{L}{\perp}{1}{L} \right) \\
& + \tfrac{1}{2} P_{\Lb} (1 +  \beta^2_l) {\rm Re} \left( \aprod{\parallel}{0}{R}{\perp}{0}{R} + \aprod{\parallel}{0}{L}{\perp}{0}{L} \right) \\ 
& - \tfrac{1}{2} P_{\Lb} (1 - \beta^2_l) {\rm Re} \left(  
\aprod{\parallel}{1}{R}{\perp}{1}{L} + \aprod{\perp}{1}{R}{\parallel}{1}{L}  -  \aprod{\parallel}{0}{R}{\perp}{0}{L} - \aprod{\perp}{0}{R}{\parallel}{0}{L} 
\right) \\ 
& +  P_{\Lb} (1 - \beta^2_l) {\rm Re}\left( \aprod{\parallel}{t}{}{\perp}{t}{} \right) \,, \\
K_{12} &= -\tfrac{1}{2} P_{\Lb} (1 + \beta^2_l) {\rm Re}\left( \aprod{\parallel}{1}{R}{\perp}{1}{R} + \aprod{\parallel}{1}{L}{\perp}{1}{L} \right) \\
& + \tfrac{1}{2} P_{\Lb} (1- \beta^2_l) {\rm Re} \left( \aprod{\parallel}{0}{R}{\perp}{0}{R} +  \aprod{\parallel}{0}{L}{\perp}{0}{L} \right)  \\ 
& - \tfrac{1}{2} P_{\Lb} (1 - \beta^2_l) {\rm Re} \left( 
\aprod{\parallel}{1}{R}{\perp}{1}{L} + \aprod{\perp}{1}{R}{\parallel}{1}{L} -  \aprod{\parallel}{0}{R}{\perp}{0}{L} - \aprod{\perp}{0}{R}{\parallel}{0}{L} 
\right) \\ 
& + P_{\Lb} (1 - \beta^2_l ) {\rm Re} \left( \aprod{\parallel}{t}{}{\perp}{t}{} \right) \,,  \\
K_{13} &= \tfrac{1}{2} P_{\Lb} \beta_l \left( 
\asq{\parallel}{1}{R} + \asq{\perp}{1}{R} - \asq{\parallel}{1}{L} - \asq{\perp}{1}{L} 
\right) \,, \\
K_{14} &= -\tfrac{1}{4} \alpha_{\Lz} P_{\Lb}  \left(\asq{\parallel}{1}{R} + \asq{\perp}{1}{R} + \asq{\parallel}{1}{L} + \asq{\perp}{1}{L} \right) \\
& +  \tfrac{1}{4}  \alpha_{\Lz} P_{\Lb} ( 1 + \beta^2_l ) \left( \asq{\parallel}{0}{R} + \asq{\perp}{0}{R} + \asq{\parallel}{0}{L} + \asq{\perp}{0}{L} \right) \\ 
& + \tfrac{1}{2}  \alpha_{\Lz} P_{\Lb} ( 1 - \beta^2_l) \left( \asq{\parallel}{t}{} + \asq{\perp}{t}{} \right) \\
& - \tfrac{1}{2}  \alpha_{\Lz} P_{\Lb} ( 1 - \beta^2_l) {\rm Re} \left( 
\aprod{\parallel}{1}{R}{\parallel}{1}{L} + 
\aprod{\perp}{1}{R}{\perp}{1}{L} - 
\aprod{\parallel}{0}{R}{\parallel}{0}{L} - 
\aprod{\perp}{0}{R}{\perp}{0}{L}
\right) \,, \\
K_{15} &=   -\tfrac{1}{4}  \alpha_{\Lz} P_{\Lb} ( 1 + \beta^2_l ) \left( \asq{\parallel}{1}{R} + \asq{\perp}{1}{R} + \asq{\parallel}{1}{L} + \asq{\perp}{1}{L} \right) \\
& +   \tfrac{1}{4} \alpha_{\Lz} P_{\Lb} ( 1 - \beta^2_l ) \left(  \asq{\parallel}{0}{R} + \asq{\perp}{0}{R} + \asq{\parallel}{0}{L} + \asq{\perp}{0}{L} \right) \\ 
& -  \tfrac{1}{2}  \alpha_{\Lz} P_{\Lb} (1 - \beta^2_l) {\rm Re} \left( 
\aprod{\parallel}{1}{R}{\parallel}{1}{L} +  \aprod{\perp}{1}{R}{\perp}{1}{L} - \aprod{\parallel}{0}{R}{\parallel}{0}{L} -  \aprod{\perp}{0}{R}{\perp}{0}{L} 
\right) \\
& + \tfrac{1}{2} \alpha_{\Lz} P_{\Lb} (1 - \beta^2_l) \left( \asq{\parallel}{t}{}  + \asq{\perp}{t}{} \right) \,, \\
K_{16} & =  \alpha_\Lz P_{\Lb} \beta_l {\rm Re} \left( 
\aprod{\perp}{1}{R}{\parallel}{1}{R} - \aprod{\perp}{1}{L}{\parallel}{1}{L} \right) \,.\\
\end{split}
\end{equation}
The observables $K_{13}$ and $K_{16}$ are trivially related to $K_{6}$ and $K_{3}$ through $K_{13} = - P_{\Lb} K_{6}$ and $K_{16} = - P_{\Lb} K_{3}$ and can therefore be used as an experimental consistency check or to determine $P_{\Lb}$.
The observables $K_{11}$, $K_{12}$, $K_{14}$ and $K_{15}$ have a similar structure to $K_{1}$, $K_{2}$, $K_{4}$ and $K_{5}$ but, unlike in those observables, the  amplitudes with $\lambda_{\ell\ell} = 0$ enter with a different relative sign to those with $\lambda_{\ell\ell} = \pm 1$.

The observables $K_{17}$ through $K_{34}$ also involve new combinations of amplitudes that are not accessible if the \Lb baryon is unpolarised. They are
~
\begin{equation}
\begin{split}
K_{17} & = -\tfrac{1}{\sqrt{2}} \alpha_\Lz P_{\Lb} \beta^2_l {\rm Re} \left( 
\aprod{\parallel}{1}{R}{\parallel}{0}{R} - \aprod{\perp}{1}{R}{\perp}{0}{R} + 
\aprod{\parallel}{1}{L}{\parallel}{0}{L} - \aprod{\perp}{1}{L}{\perp}{0}{L}  \right) \,, \\ 
K_{18} & = -\tfrac{1}{\sqrt{2}} \alpha_\Lz P_{\Lb} \beta_l {\rm Re}\left(
\aprod{\parallel}{1}{R}{\perp}{0}{R} - \aprod{\perp}{1}{R}{\parallel}{0}{R}  - 
\aprod{\parallel}{1}{L}{\perp}{0}{L} +\aprod{\perp}{1}{L}{\parallel}{0}{L} \right)  \,, \\
K_{19} &= -\tfrac{1}{\sqrt{2}} \alpha_\Lz P_{\Lb} \beta^2_l {\rm Im}\left( 
\aprod{\parallel}{1}{R}{\perp}{0}{R} - \aprod{\perp}{1}{R}{\parallel}{0}{R} + 
\aprod{\parallel}{1}{L}{\perp}{0}{L} - \aprod{\perp}{1}{L}{\parallel}{0}{L}   \right) \,, \\
K_{20} &= -\tfrac{1}{\sqrt{2}}\alpha_\Lz P_{\Lb} \beta_l {\rm Im} \left( 
\aprod{\parallel}{1}{R}{\parallel}{0}{R} - \aprod{\perp}{1}{R}{\perp}{0}{R} -
\aprod{\parallel}{1}{L}{\parallel}{0}{L} + \aprod{\perp}{1}{L}{\perp}{0}{L}  \right) \,, \\
K_{21} &= \tfrac{1}{\sqrt{2}} P_{\Lb} \beta^2_l {\rm Im} \left( 
\aprod{\parallel}{1}{R}{\parallel}{0}{R} + \aprod{\perp}{1}{R}{\perp}{0}{R}  + 
\aprod{\parallel}{1}{L}{\parallel}{0}{L} + \aprod{\perp}{1}{L}{\perp}{0}{L}  
\right) \,, \\
K_{22} &= -\tfrac{1}{\sqrt{2}} P_{\Lb} \beta_l {\rm Im} \left( 
\aprod{\parallel}{1}{R}{\perp}{0}{R} + \aprod{\perp}{1}{R}{\parallel}{0}{R} - 
\aprod{\parallel}{1}{L}{\perp}{0}{L} - \aprod{\perp}{1}{L}{\parallel}{0}{L} 
\right) \,, \\
K_{23} &= -\tfrac{1}{\sqrt{2}} P_{\Lb} \beta^2_l {\rm Re} \left( 
\aprod{\parallel}{1}{R}{\perp}{0}{R} + \aprod{\perp}{1}{R}{\parallel}{0}{R} + 
\aprod{\parallel}{1}{L}{\perp}{0}{L} + \aprod{\perp}{1}{L}{\parallel}{0}{L} 
\right) \,, \\ 
K_{24} &= \tfrac{1}{\sqrt{2}} P_{\Lb} \beta_l {\rm Re}  \left( 
\aprod{\parallel}{1}{R}{\parallel}{0}{R} + \aprod{\perp}{1}{R}{\perp}{0}{R} - 
\aprod{\parallel}{1}{L}{\parallel}{0}{L} - \aprod{\perp}{1}{L}{\perp}{0}{L} 
\right) \,, \\
K_{25} &= \tfrac{1}{\sqrt{2}} \alpha_{\Lz} P_{\Lb} \beta^2_l {\rm Im}  \left( 
\aprod{\parallel}{1}{R}{\perp}{0}{R} + \aprod{\perp}{1}{R}{\parallel}{0}{R} + 
\aprod{\parallel}{1}{L}{\perp}{0}{L} + \aprod{\perp}{1}{L}{\parallel}{0}{L} 
\right) \,, \\
K_{26} &= -\tfrac{1}{\sqrt{2}} \alpha_{\Lz} P_{\Lb} \beta_l {\rm Im} \left(
\aprod{\parallel}{1}{R}{\parallel}{0}{R} + \aprod{\perp}{1}{R}{\perp}{0}{R}  - 
\aprod{\parallel}{1}{L}{\parallel}{0}{L} - \aprod{\perp}{1}{L}{\perp}{0}{L}  
\right) \,, \\
K_{27} &= -\tfrac{1}{\sqrt{2}} \alpha_{\Lz} P_{\Lb} \beta^2_l {\rm Re} \left( 
\aprod{\parallel}{1}{R}{\parallel}{0}{R} + \aprod{\perp}{1}{R}{\perp}{0}{R}  + 
\aprod{\parallel}{1}{L}{\parallel}{0}{L} + \aprod{\perp}{1}{L}{\perp}{0}{L}  
\right) \,, \\
K_{28} & = \tfrac{1}{\sqrt{2}} \alpha_\Lz P_{\Lb} \beta_l {\rm Re} \left( 
\aprod{\parallel}{1}{R}{\perp}{0}{R} + \aprod{\perp}{1}{R}{\parallel}{0}{R} - 
\aprod{\parallel}{1}{L}{\perp}{0}{L} - \aprod{\perp}{1}{L}{\parallel}{0}{L}  
\right) \,, \\
K_{29} &=   \tfrac{1}{2} \alpha_\Lz P_{\Lb} (1 - \beta^2_l) {\rm Im} \left(  
\aprod{\perp}{0}{R}{\parallel}{0}{R} + \aprod{\perp}{0}{L}{\parallel}{0}{L} + 
\aprod{\perp}{0}{R}{\parallel}{0}{L} - \aprod{\parallel}{0}{R}{\perp}{0}{L} 
 \right) \\
& + \alpha_\Lz P_{\Lb} (1 - \beta^2_l) {\rm Im} \left(  \aprod{\perp}{t}{}{\parallel}{t}{}  \right) \,, \\
 K_{30} &= \tfrac{1}{2} \alpha_\Lz P_{\Lb} (1 + \beta^2_l ){\rm Im} \left( 
 \aprod{\perp}{0}{R}{\parallel}{0}{R} +  \aprod{\perp}{0}{L}{\parallel}{0}{L} \right) \\ 
 & + \tfrac{1}{2} \alpha_\Lz P_{\Lb} ( 1- \beta^2_l ) {\rm Im} \left( 
 \aprod{\perp}{0}{R}{\parallel}{0}{L} - \aprod{\parallel}{0}{R}{\perp}{0}{L}
 \right) \\
 & + \alpha_{\Lz} P_{\Lb} (1-\beta^2_l) {\rm Im} \left( \aprod{\perp}{t}{}{\parallel}{t}{} \right)  \,, \\
 K_{31} &= \tfrac{1}{4} \alpha_{\Lz} P_{\Lb} (1-\beta^2_l) \left( 
 \asq{\perp}{0}{R} - \asq{\parallel}{0}{R} + \asq{\perp}{0}{L} - \asq{\parallel}{0}{L} 
 \right) \\
 & + \tfrac{1}{2} \alpha_\Lz P_{\Lb} (1-\beta^{2}_l) {\rm Re} \left( \aprod{\perp}{0}{R}{\perp}{0}{L} - \aprod{\parallel}{0}{R}{\parallel}{0}{L} \right) \\ 
 & + \tfrac{1}{2} \alpha_\Lz P_{\Lb} (1-\beta^{2}_l) \left( \asq{\perp}{t}{} -  \asq{\parallel}{t}{} \right)  \,, \\
 K_{32} &= \tfrac{1}{4} \alpha_\Lz P_{\Lb} (1+\beta^{2}_l)\left( \asq{\perp}{0}{R} + \asq{\perp}{0}{L} - \asq{\parallel}{0}{R} - \asq{\parallel}{0}{L} \right) \\
 & + \tfrac{1}{2} \alpha_\Lz P_{\Lb} (1-\beta^{2}_l) {\rm Re}\left(
 \aprod{\perp}{0}{R}{\perp}{0}{L} - \aprod{\parallel}{0}{R}{\parallel}{0}{L}  \right) \\
 & + \tfrac{1}{2} \alpha_\Lz P_{\Lb} (1 - \beta^2_l)\left( \asq{\perp}{t}{} - \asq{\parallel}{t}{} \right) \,, \\ 
 K_{33} &= \tfrac{1}{4} \alpha_{\Lz} P_{\Lb} \beta_l^{2} \left(
 \asq{\perp}{1}{R}  - \asq{\parallel}{1}{R} + \asq{\perp}{1}{L} - \asq{\parallel}{1}{L} \right) \,, \\ 
 K_{34} &= \tfrac{1}{2} \alpha_\Lz P_{\Lb} \beta^{2}_l {\rm Im}\left( 
 \aprod{\perp}{1}{R}{\parallel}{1}{R} + \aprod{\perp}{1}{L}{\parallel}{1}{L}  \right) \,. \\
\end{split}
\end{equation}

\noindent The angular terms $K_{29}$ and $K_{31}$ are zero in the massless lepton limit. 

\section[Angular distribution of charmonium decays]{Angular distribution of \decay{\Lb}{\jpsi\Lz}}

The angular distribution of the \decay{\Lb}{\jpsi\Lz} decay is a limiting case of Eq.~\ref{eq:base}, with a pure vector current in the dilepton system. 
In this limit, the expression collapses to the one given in Refs.~\cite{Lednicky:1985zx,Hrivnac:1994jx} with $\beta_{l} \sim 1$. 
The amplitudes $a_{\pm}$ and $b_{\pm}$ in Ref.~\cite{Lednicky:1985zx,Hrivnac:1994jx} are related to the ones in this paper by 
\begin{align}
\begin{split}
a_{-} = H'\,^{V,1}_{-1/2, \,0~}~,~ &  
a_{+} = H'\,^{V,1}_{+1/2, \,0}~, \\
b_{-} = H'\,^{V,1}_{+1/2, +1}~,~&
b_{+} = H'\,^{V,1}_{-1/2, -1}~. 
\end{split}
\end{align}  

\section{Weighting functions}
\label{sec:weighting}

The values of the normalised angular observables can be determined experimentally from an analysis of the moments of the angular distribution, 
\begin{align}
M_i =  \frac{3}{32\pi^{2}} \bigintss\limits_{}^{}   \left( \sum\limits_{j=0}^{34} M_j f_{j}(\vec{\Omega}) \right) g_i (\vec\Omega) \deriv\vec{\Omega}
\end{align} 
if the weighting functions $g_i(\vec{\Omega})$ are chosen such that they satisfy
\begin{align}
\int f_{j}(\vec{\Omega}) g_{i}(\vec{\Omega}) \deriv\vec{\Omega} = \left(\frac{32\pi^{2}}{3} \right) \delta_{ij}~.
\end{align} 
In this case, the moments can be extracted from data using Monte Carlo integration. 
The statistical uncertainty and correlation between the moments can be determined from the single sample covariance or by bootstrapping the measurement (see for example Ref.~\cite{Efron:1979}). 

The weighting functions for $M_1$--$M_{10}$ are
\begin{align}
g_{1}(\vec{\Omega}) = & \tfrac{1}{4}(3-5\cos^2\theta_l)~,  
&&  g_{6}(\vec{\Omega}) ~= 3\cos\theta_l\cos\theta_b ~,    \\ 
g_{2}(\vec{\Omega}) = & \tfrac{1}{2}(5\cos^2\theta_l-1)~,  
&& g_{7}(\vec{\Omega}) ~= \tfrac{15}{2}\cos\theta_l\sin\theta_l\sin\theta_b\cos(\phi_l + \phi_b)~,   \nonumber  \\
g_{3}(\vec{\Omega}) = & \cos\theta_l ~,  
&& g_{8}(\vec{\Omega}) ~= \tfrac{3}{2}\sin\theta_l\sin\theta_b\cos(\phi_l + \phi_b), \nonumber \\ 
g_{4}(\vec{\Omega}) = & \tfrac{3}{4}(3 - 5 \cos^{2}\theta_l) \cos\theta_b ~, 
&& g_{9}(\vec{\Omega}) ~= \tfrac{15}{2} \cos\theta_l\sin\theta_l\sin\theta_b\sin(\phi_l + \phi_b)~,  \nonumber  \\
g_{5}(\vec{\Omega}) = & \tfrac{3}{2}( 5\cos^2\theta_l - 1) \cos\theta_b~,
&& g_{10}(\vec{\Omega}) = \tfrac{3}{2} \sin\theta_l\sin\theta_b\sin(\phi_l + \phi_b)~.  \nonumber
\end{align}
These weighting functions have been previously derived in Ref.~\cite{Beaujean:2015xea}.
The weighting functions for the polarisation-dependent terms can be derived in a similar manner,  
they are
\begin{align}
g_{11}(\vec{\Omega}) =  & \tfrac{3}{4}(3-5\cos^2\theta_l)\cos\theta ~,
&& g_{23}(\vec{\Omega}) =  \tfrac{15}{2} \cos\theta_l\sin\theta_l\sin\theta\cos\phi_l~,  \nonumber \\ 
g_{12}(\vec{\Omega}) =  & \tfrac{3}{2}(5\cos^2\theta_l-1)\cos\theta ~,
&& g_{24}(\vec{\Omega}) = \tfrac{3}{2} \sin\theta\sin\theta_l\cos\phi_l ~,  \\
g_{13}(\vec{\Omega}) =  & 3\cos\theta_l\cos\theta~, 
&& g_{25}(\vec{\Omega}) =  \tfrac{45}{2}\cos\theta_l\sin\theta_l\cos\theta_b\sin\theta\sin\phi_l ~, \nonumber \\ 
g_{14}(\vec{\Omega}) =  & \tfrac{9}{4}(3 - 5\cos^{2}\theta_l) \cos\theta_b \cos\theta~, 
&& g_{26}(\vec{\Omega}) = \tfrac{9}{2}\sin\theta\sin\theta_l\cos\theta_b\sin\phi_l~, \nonumber \\
g_{15}(\vec{\Omega}) =  & \tfrac{9}{2} ( 5\cos^2\theta_l - 1 ) \cos\theta_b\cos\theta~,
&& g_{27}(\vec{\Omega}) =  \tfrac{45}{2}\cos\theta_l\sin\theta_l\cos\theta_b\sin\theta\cos\phi_l~, \nonumber \\
g_{16}(\vec{\Omega}) =  & 9\cos\theta\cos\theta_l\cos\theta_b ~,
&& g_{28}(\vec{\Omega}) = \tfrac{9}{2} \sin\theta\sin\theta_l\cos\theta_b\cos\phi_l ~, \nonumber \\
g_{17}(\vec{\Omega}) =  & \tfrac{45}{2} \cos\theta_l\sin\theta_l\sin\theta_b\cos\theta\cos(\phi_l + \phi_b)~, 
&& g_{29}(\vec{\Omega}) = \tfrac{9}{4}(5\cos^2\theta_l - 1)\sin\theta_b\sin\theta\sin\phi_b~, \nonumber \\
g_{18}(\vec{\Omega}) =  & \tfrac{9}{2}\sin\theta_l\sin\theta_b\cos\theta\cos(\phi_l + \phi_b)~, 
&& g_{30}(\vec{\Omega}) =  \tfrac{9}{8}(3 - 5\cos^2\theta_l)\sin\theta_b\sin\theta\sin\phi_b~,  \nonumber \\
g_{19}(\vec{\Omega}) =  & \tfrac{45}{2}\cos\theta_l\sin\theta_l\sin\theta_b\cos\theta\sin(\phi_l+\phi_b)~, 
&&  g_{31}(\vec{\Omega}) =  \tfrac{9}{4}(5 \cos^2\theta_l - 1)\sin\theta_b\sin\theta\cos\phi_b~, \nonumber \\
g_{20}(\vec{\Omega}) =  & \tfrac{9}{2}\sin\theta_l\sin\theta_b\cos\theta\sin(\phi_l + \phi_b)~, 
&& g_{32}(\vec{\Omega}) =  \tfrac{9}{8}(3 - 5\cos^2\theta_l)\sin\theta_b\sin\theta\cos\phi_b~, \nonumber \\
g_{21}(\vec{\Omega}) =  & \tfrac{15}{2}\cos\theta_l\sin\theta_l\sin\theta\sin\phi_l ~,
&& g_{33}(\vec{\Omega}) =  \tfrac{9}{4}\sin\theta_b\sin\theta\cos(2\phi_l + \phi_b), \nonumber \\
g_{22}(\vec{\Omega}) =  & \tfrac{3}{2} \sin\theta\sin\theta_l\sin\phi_l~,  
&& g_{34}(\vec{\Omega}) =  \tfrac{9}{4}\sin\theta_b\sin\theta\sin(2\phi_l + \phi_b)~. \nonumber 
\end{align}
The weighting functions are not unique and a more compact set can be formed by exploiting the fact that the integral of $\sin\theta_b$ over $\deriv\!\cos\theta_b$ is $\pi/2$ \eg the weighting functions for $M_{33}$ and $M_{34}$ can be written in a shorter form as
\begin{equation}
\begin{split}
g_{33}(\vec{\Omega}) &=  \tfrac{6}{\pi}\sin\theta\cos\left(\phi_b+2\phi_l\right)~,  \\
g_{34}(\vec{\Omega}) & = \tfrac{6}{\pi} \sin\theta\sin\left(\phi_b+2\phi_l\right)~.  
\end{split}
\end{equation}
More compact expressions can also be found for many of the other observables. 
Note, the different sets of weighting functions can lead to different experimental precision on the normalised moments. 
In general,  the longer form of the weighting functions provides the best precision. 

\section{Standard Model predictions}

In order to describe the SM contribution to the decay amplitudes, an effective field theory approach is used. 
The Hamiltonian for the decay is factorised into local four-fermion operators and Wilson coefficients (see for example Ref.~\cite{Bobeth:1999mk}).
The Wilson coefficients describe the short-distance contributions from the heavy SM particles. 

Numerical values for the SM predictions, in the case that $P_{\Lb} = 1$, are provided in Appendix~\ref{appendix:numerical} in two \qsq ranges: 
at large hadronic recoil, in the range $1 < \qsq < 6\gev^2/c^4$, and at low hadronic recoil, in the range $15 < \qsq < 20\gev^2/c^4$.
To evaluate SM predictions for the different angular observables we use the \texttt{EOS} flavour tool~\cite{OurEOSVersion}. 
At low hadronic recoil, the SM calculations employ an operator product expansion of the four-quark contributions to the matrix element in powers of $\Lz_{\rm QCD}/\sqrt{\qsq}$~\cite{Grinstein:2004vb}. 
%At large recoil we make use of some of the known $\alpha_s$ corrections to the perturbative charm loop. 
At large recoil, \texttt{EOS} uses some of the known $\alpha_s$ corrections to charm loop processes.
However, potentially large contributions from hard spectator scattering~\cite{Beneke:2004dp} and soft gluon emission~\cite{Khodjamirian:2010vf} are neglected. 
The form-factors for the $\Lb \to \Lz$ transition are taken from a recent Lattice QCD calculation in Ref.~\cite{Detmold:2016pkz}.
These form-factors enable the observables to be computed with high-precision. 
The form-factors at large hadronic recoil have also been calculated in the framework of light-cone-sum-rules, see for example Refs.~\cite{Mannel:2011xg} and \cite{Wang:2015ndk}.
The SM Wilson coefficients are computed in \texttt{EOS} to NNLO in QCD. 
The \Lb lifetime and CKM matrix elements are taken from the latest experimental values~\cite{PDG2016}.
The quark masses are taken in the $\overline{\rm MS}$ scheme.

Tables~\ref{tab:eos:predictions:largerecoil} and \ref{tab:eos:predictions:lowrecoil}  in Appendix~\ref{appendix:numerical} also provide 68\% confidence level intervals for the SM predictions. 
To evaluate these intervals: 
the form-factors from Ref.~\cite{Detmold:2016pkz} have been varied within their full covariance matrix;  
the \Lb lifetime, the \Lz asymmetry parameter and CKM matrix elements are varied within their experimental precision~\cite{Amhis:2016xyh,PDG2016}; 
the scale dependence of Wilson coefficients $C_i(\mu)$ is explored by varying the scale, $\mu$, in the range $m_b/2 < \mu < 2 m_b$; 
and in keeping with Ref.~\cite{Meinel:2016grj} a 3\% correction to the amplitudes from hadronic matrix elements is considered (see also Ref.~\cite{Beylich:2011aq}). 

\subsection{Low-hadronic recoil}

At low hadronic recoil the observables are precisely predicted in the SM. 
The uncertainties on the predictions are worse at large recoil, where a large extrapolation in $q^2$ of the form-factors is needed.
Figures~\ref{fig:scan:c9:c10:largerecoil}--\ref{fig:scan:c9:c9p:lowrecoil:pol}  in Appendix~\ref{appendix:NP} demonstrate how the observables depend on NP contributions to the Wilson coefficients. 
In the large-recoil region there is sensitivity  to $C_9^{\rm NP}$ from both the polarised and unpolarised observables. 
Interestingly, the observables $M_{23}$ and $M_{27}$ can also distinguish between two of the possibilities that are favoured by global fits to $b\to s\ell^+\ell^-$ processes: where $C_9^{\rm NP} \simeq -1$ with $C_{10}^{\rm NP} = 0$ and where $C_9^{\rm NP} = -C_{10}^{\rm NP} \simeq -1$~\cite{Altmannshofer:2014rta,Descotes-Genon:2015uva,Hurth:2016fbr}.
In the low-recoil range the sensitivity to $C_9^{\rm NP}$ is reduced. 

In Ref.~\cite{Boer:2014kda}, the authors point out that the observables at low hadronic recoil place constraints on six combinations of Wilson coefficients
\begin{equation}
\begin{split}
\rho^{\pm}_{1} &= | C_{\rm V} \pm C'_{\rm V} |^{2} + |C_{10} \pm C'_{10}|^{2} \\ 
\rho_{2} &= {\rm Re}\left( C_{\rm V} C^{*}_{10} - C'_{\rm V} C'^{*}_{10} \right) - i {\rm Im}\left( C_{\rm V}C'^{*}_{\rm V} + C_{10} C'^{*}_{10} \right)\\
\rho^{\pm}_{3} &= 2 {\rm Re}\left((C_{V} \pm C'_{\rm V})(C_{10} \pm C'_{10})^{*}\right) \\ 
\rho_{4} &= |C_{\rm V}|^{2} - |C'_{\rm V}|^{2} + |C_{10}|^{2} - |C'_{10}|^{2} - i {\rm Im}\left( C_{\rm V} C^{*}_{10} - C'_{\rm V} C'^{*}_{10} \right)~,
\end{split}
\end{equation} 
where $C_{\rm V}$ contains contributions from $C_{7}$ and $C_9$. 
The primed coefficients correspond to right-handed currents whose contribution is vanishingly small in the SM.
The short-distance dependence of $K_{1}$--$K_{34}$ on $\rho_{1}^{\pm}$, $\rho_{3}^{\pm}$, $\rho_{2}$ and $\rho_{4}$ is provided for completeness in Appendix~\ref{appendix:rhodependence}. 

If the \Lb is unpolarised, the decay rate is insensitive to the short-distance contribution ${\rm Im}(\rho_{2})$ but provides sensitivity to $\rho^{\pm}_{1}$, ${\rm Re}(\rho_{2})$, $\rho^{\pm}_{3}$, ${\rm Re}(\rho_{4})$ and ${\rm Im}(\rho_{4})$.
The polarised observables also depend on these short-distance contributions but have different form-factor dependencies. 
This permits a new set of checks of the OPE and the form-factors.
The short-distance combination ${\rm Im}(\rho_{2})$ can also be determined from $M_{19}$, $M_{25}$, $M_{30}$ and $M_{34}$.
Furthermore, in $K_1$--$K_{10}$ the short-distance contributions $\rho_{1}^{+}$ and $\rho_{1}^{-}$ always appear together as a sum.
Using the polarised observables , $\rho_{1}^{+}$ and $\rho_{1}^{-}$ can be separated, \eg by using
\begin{align}
\begin{split}
K_{2} + \frac{2}{\alpha_{\Lz} P_{\Lb}} K_{33} &= 16 s_{-} |f_{\perp}^{V}|^{2} \rho_{1}^{+} ~, \\ 
K_{2} - \frac{2}{\alpha_{\Lz} P_{\Lb}} K_{33}  &= 16 s_{+} |f_{\perp}^{A}|^{2} \rho_{1}^{-} ~,
\end{split}
\end{align} 
where $f_{\perp}^{V}$ and $f_{\perp}^{A}$ are helicity form-factors (see for example Ref.~\cite{Feldmann:2011xf}). 
A similar trick can be used to separate $\rho_{3}^{+}$ and $\rho_{3}^{-}$ using $K_{24}$ and $K_{8}$.
It is also possible to form new short-distance relationships, in which the form-factors cancel by taking ratios of the $K_{i}$, 
\begin{align}
\frac{K_{16}}{K_{34}} = 2 \frac{ {\rm Re}(\rho_{2})}{{\rm Im}(\rho_{2})}~,\quad 
\frac{K_{25}}{K_{22}} = - \frac{ {\rm Im}(\rho_{2})}{{\rm Im}(\rho_{4})}~,\quad
\frac{K_{23}}{K_{10}} = -\frac{ {\rm Re}(\rho_{4})}{{\rm Im}(\rho_{4})} P_{\Lb}~.
\label{eq:Ki:ratio}
\end{align} 
The short-distance combinations $\rho_{2}$ and $\rho_{4}$ can then be determined up-to their overall normalisation, independent of the hadronic form-factors, using Eq.~\ref{eq:Ki:ratio} and  the relationship
\begin{align}
\frac{K_{3}}{K_{5}} = -\frac{1}{\alpha_{\Lz}} \frac{{\rm Re}(\rho_{2})}{{\rm Re}(\rho_{4})}
\end{align} 
from Ref.~\cite{Boer:2014kda}.
Similarly, one can form short-distance relationships that depend only on $\rho_{1}^{\pm}$ and $\rho_{3}^{\pm}$
\begin{align}
\frac{P_{\Lb} K_{8} + \alpha_{\Lz}K_{24}}{K_{27}-K_{17}} = -\frac{\rho_{3}^{-}}{\rho_{1}^{-}}\quad,\quad
\frac{P_{\Lb} K_{8} - \alpha_{\Lz}K_{24}}{K_{27}+K_{17}} = \frac{\rho_{3}^{+}}{\rho_{1}^{+}}~.
\end{align}
Alternatively, it is possible to form ratios that depend only on the form-factors and not on the short-distance physics. 
For example, 
\begin{align}
\begin{split}
\frac{K_{7}}{K_{5}} &= \frac{1}{2} \left( 
\frac{(m_{\Lb} + m_{\Lz})}{\sqrt{q^2}} \frac{f_{0}^{V}}{f_{\perp}^{V}} - 
\frac{(m_{\Lb} - m_{\Lz})}{\sqrt{q^2}} \frac{f_{0}^{A}}{f_{\perp}^{A}} 
\right) ~, \\ 
\frac{K_{23}}{K_{5}} &= \frac{1}{2} \left( 
\frac{(m_{\Lb} + m_{\Lz})}{\sqrt{q^2}} \frac{f_{0}^{V}}{f_{\perp}^{V}} +
\frac{(m_{\Lb} - m_{\Lz})}{\sqrt{q^2}} \frac{f_{0}^{A}}{f_{\perp}^{A}} 
\right) P_{\Lb}~
\end{split}
\end{align}
allow the ratios $f_{0}^{V}/f_{\perp}^{V}$ and $f_{0}^{A}/f_{\perp}^{A}$ to be determined independent of the $\rho_{i}$.

\subsection{Photon-polarisation at large hadronic-recoil}

At very large hadronic recoil ($q^2 \ll 1\gev^{2}/c^{4}$), the angular distribution of the \decay{\Lb}{\Lz\mumu} decay is sensitive primarily to the Wilson coefficients $C_7$ and $C'_7$ due to a pole-like enhancement of the amplitudes. 
The observable $K_{33}$ is proportional to ${\rm Re}(C_7 C'_7)$ and can therefore provide a null test of the size of $C'_7$ (in the same way as  the $S_3$ observable in the \decay{\Bz}{\Kstarz\mumu} decay).
In this case, however, the observable is suppressed by the size of $P_{\Lb}$.

\section{Expected experimental precision} 
\label{sec:experiment}

Table~\ref{tab:experiment} indicates the typical precision on the angular moments that could be achieved at the LHCb experiment. 
The experimental precision has been estimated using pseudo-experiments corresponding approximately to the expected signal yield in the current and in a future LHCb dataset.
Experimental backgrounds and non-uniform angular acceptance have been neglected in this estimate.
However,  these are expected to have only a small impact on the experiments sensitivity.
The sensitivity that can be achieved with the large datasets that will be available at an upgraded LHCb experiment is interesting event for modest values of $P_{\Lb}$.

\begin{table}[!htb]
\caption{
Expected experimental precision on the angular moments of the \decay{\Lb}{\Lz\mumu} decay at the LHCb experiment. 
The four columns correspond to: the observed yield of 300 \decay{\Lb}{\Lz\mumu} candidates with $15 < \qsq < 20\gev^2/c^4$ in the LHC run\,1 dataset~\cite{LHCb-PAPER-2015-009}; an expected yield of $\sim$1000 candidates at the end of  run\,2 of the LHC;  an expected  yield of $\sim$8\,000 candidates in 50\invfb of integrated luminosity with an upgraded LHCb experiment; and an expected yield of $\sim$50\,000 candidates in 300\invfb with the proposed LHCb phase II upgrade. 
}
\centering
\begin{tabular}{lcccc|lcccc}
\hline
Obs. & Run\,1  & Run\,2   &  Upgrade  &Phase\,II  & Obs. &  Run\,1  & Run\,2   &  Upgrade  & Phase\,II  \\
\hline
$M_{1}$     &  0.021  & 0.011   &  0.004    & 0.002  & $M_{18}$    &  0.071  & 0.038   &  0.014    & 0.006  \\
$M_{2}$     &  0.042  & 0.023   &  0.008    & 0.003  & $M_{19}$    &  0.156  & 0.084   &  0.030    & 0.012  \\
$M_{3}$     &  0.030  & 0.016   &  0.006    & 0.002 & $M_{20}$    &  0.071  & 0.038   &  0.014    & 0.006  \\
$M_{4}$     &  0.050  & 0.026   &  0.010    & 0.004  & $M_{21}$    &  0.090  & 0.048   &  0.017    & 0.007  \\
$M_{5}$     &  0.078  & 0.042   &  0.015    & 0.006  & $M_{22}$    &  0.041  & 0.022   &  0.008    & 0.003  \\
$M_{6}$     &  0.055  & 0.030   &  0.011    & 0.004  & $M_{23}$    &  0.089  & 0.047   &  0.017    & 0.007  \\
$M_{7}$     &  0.090  & 0.048   &  0.017    & 0.007  & $M_{24}$    &  0.036  & 0.019   &  0.007    & 0.003  \\
$M_{8}$     &  0.041  & 0.022   &  0.008    & 0.003  & $M_{25}$    &  0.156  & 0.083   &  0.030    & 0.012  \\
$M_{9}$     &  0.090  & 0.048   &  0.017    & 0.007  & $M_{26}$    &  0.071  & 0.038   &  0.014    & 0.006  \\
$M_{10}$    &  0.041  & 0.022   &  0.008    & 0.003  & $M_{27}$    &  0.156  & 0.083   &  0.030    & 0.012  \\
$M_{11}$    &  0.051  & 0.027   &  0.010    & 0.004  & $M_{28}$    &  0.071  & 0.038   &  0.014    & 0.005  \\
$M_{12}$    &  0.078  & 0.041   &  0.015    & 0.006  & $M_{29}$    &  0.097  & 0.052   &  0.019    & 0.008  \\
$M_{13}$    &  0.054  & 0.029   &  0.010    & 0.004  & $M_{30}$    &  0.062  & 0.033   &  0.012    & 0.005  \\
$M_{14}$    &  0.088  & 0.047   &  0.017    & 0.007  & $M_{31}$    &  0.097  & 0.052   &  0.019    & 0.008  \\
$M_{15}$    &  0.136  & 0.073   &  0.026    & 0.011  & $M_{32}$    &  0.062  & 0.033   &  0.012    & 0.005  \\
$M_{16}$    &  0.097  & 0.052   &  0.019    & 0.008  & $M_{33}$    &  0.061  & 0.033   &  0.012    & 0.005  \\
$M_{17}$    &  0.156  & 0.084   &  0.030    & 0.012  & $M_{34}$    &  0.061  & 0.033   &  0.012    & 0.005  \\
\hline
\end{tabular}
\label{tab:experiment}
\end{table}

\section{Conclusion} 
\label{sec:conclusion} 

In this paper we have derived an expression for the angular distribution of the \decay{\Lb}{\Lz\mumu} in the case of non-zero production polarisation. 
This extends the number of observables in the decay from 10 to 34. 
These observables can be determined from moments of the \decay{\Lb}{\Lz\mumu} angular distribution. 
Explicit expressions have been provided for the observables in terms of the angular moments to enable an experiment to determine the new observables from their dataset.
A phenomenological analysis has also been performed to illustrate how these observables might vary in extensions of the Standard Model. 
The analysis shows that there is interesting new sensitivity that can be gained if the \Lb baryon is produced polarised.

\section{Acknowledgements} 

We would like to thank Danny Van Dyk, Georgios Chatzikonstantinidis, Tim Gershon and Nigel Watson for their useful feedback on the manuscript. 
We would also like to thank Danny Van Dyk for his help in implementing the observables in the \texttt{EOS} flavour tool. 
T.\,B.\ acknowledges support from the Royal Society (United Kingdom). 
M.\,K.\ acknowledges support from the Science \& Technology Facilities Council (United Kingdom).

\clearpage

{\noindent\bf\Large Appendices}

\appendix

\section{Wigner $D$-functions} 
\label{appendix:Dfunctions}

The Wigner $D$-functions are 
\begin{align}
D^{J}_{m',m}(\alpha,\beta,\gamma) = e^{-i m' \alpha} d^{J}_{m',m}(\beta) e^{-i m \gamma}   
\end{align} 
where the $\alpha$, $\beta$ and $\gamma$ correspond to the Euler rotation angles needed to rotate between the reference frame of the mother particle and the helicity frame of its daughters. 
The relevant small $d$-functions are
\begin{align}
\begin{split} 
d^{1/2}_{1/2,1/2}(\beta) & = \cos (\beta/2)~,  \\ 
d^{1/2}_{1/2,-1/2}(\beta) &= -\sin (\beta/2)~, \\
d^{1}_{1,1}(\beta) & = \cos^{2} (\beta/2)~, \\
d^{1}_{1,-1}(\beta) &= \sin^{2} (\beta/2)~, \\
d^{1}_{1,0}(\beta) & = \cos (\beta/2) \sin(\beta/2)~, \\
d^{1}_{0,0}(\beta) & = \cos (\beta)~, \\
\end{split}
\end{align}
with 
\begin{align}
d^{J}_{m',m}(\beta) = d^{J}_{-m,-m'}(\beta) = (-1)^{m-m'}d^{J}_{m,m'}(\beta)~.
\end{align} 

\section{Numerical results}
\label{appendix:numerical}

Standard Model predictions for the angular observables with $P_{\Lb} =1$ are provided in Tables~\ref{tab:eos:predictions:largerecoil} and \ref{tab:eos:predictions:lowrecoil}. 
Predictions are provided in two \qsq ranges:
at large hadronic recoil, in the range $1 < \qsq < 6\gev^2/c^4$, and at low hadronic recoil, in the range $15 < \qsq < 20\gev^2/c^4$.
The SM predictions are evaluated using the \texttt{EOS} flavour-tool.
For any other choice of $P_{\Lb}$, predictions for $M_{11}$--$M_{34}$ can be achieved by multiplying the values in Tables~\ref{tab:eos:predictions:largerecoil} and \ref{tab:eos:predictions:lowrecoil} by the new value of $P_{\Lb}$.

\begin{table}[!htb]
\caption{
Predictions from \texttt{EOS} for the angular observables of the \decay{\Lb}{\Lz\mumu} decay with $P_{\Lb}=1$ in the range $1 < \qsq < 6\gev^2/c^4$. 
The SM calculation is described in the text.
The observables $M_{31}$ and $M_{34}$ vanish due to the small size of the muon mass. 
Observables that depend on the imaginary part of the product of two transversity amplitudes also tend to be vanishingly small, due to the small strong phase difference between pairs of amplitudes in the SM.
\label{tab:eos:predictions:largerecoil}
} 
\begin{center}
\begin{tabular}{lrc|lrc}
\hline
Obs. & Value & 68\% interval & Obs. & Value & 68\% interval \\
\hline
 $M_{1}$ & $0.459$ & $[ 0.453 , 0.465 ]$ & $M_{6}$ & $0.000$ & $[ -0.005 , 0.006 ]$ \\                                                                                                                       
 $M_{2}$ & $0.081$ & $[ 0.071 , 0.094 ]$ &  $M_{7}$ & $-0.025$ & $[-0.034,  -0.014 ]$ \\                                                                                                                                                                                                                                       
 $M_{3}$ & $-0.005$ & $[ -0.014 , -0.001 ]$ &   $M_{8}$ & $-0.003$ & $[ -0.016, 0.012]$ \\                                                                                                                                                                                                                              
 $M_{4}$ & $-0.280$ & $[ -0.290 , -0.262 ]$ &   $M_{9}$ & $0.002$ & $[ 0.001 , 0.002 ]$ \\                                                                                                                                                                                                                                
 $M_{5}$ & $-0.045$ & $[ -0.053 , -0.037 ]$ &  $M_{10}$ & $0.002$ & $[ 0.001 , 0.002 ]$ \\                                                                                                                   
\hline
$M_{11}$ & $-0.366$ & $[ -0.383 , -0.338 ]$ &  $M_{23}$ & $-0.147$ & $[ -0.162 , -0.133 ]$ \\                                                                                                                        
 $M_{12}$ & $0.071$ & $[ 0.058 , 0.081 ]$ &   $M_{24}$ & $0.132$ & $[ 0.120 , 0.150 ]$ \\                                                                                                         
 $M_{13}$ & $0.001$ & $[ -0.010 , 0.007 ]$  & $M_{25}$ & $-0.001$ & $[ -0.001 , -0.000 ]$ \\                                                                                                                 
 $M_{14}$ & $0.243$ & $[ 0.230 , 0.254 ]$ &   $M_{26}$ & $0.004$ & $[ 0.003 , 0.005 ]$ \\                                                                                                                                                                                                                                 
 $M_{15}$ & $-0.052$ & $[ -0.060 , -0.045 ]$ &  $M_{27}$ & $0.089$ & $[ 0.081 , 0.099 ]$ \\                                                                                                                                                            
 $M_{16}$ & $0.003$ & $[ 0.001 , 0.009 ]$ &  $M_{28}$ & $-0.089$ & $[ -0.100 , -0.080 ]$ \\                                                                                                               
 $M_{17}$ & $0.004$ & $[ -0.012 , 0.018 ]$ &  $M_{29}$ & $0.000$ & $[ 0.000 , 0.000 ]$ \\                                                                                                                   
 $M_{18}$ & $0.029$ & $[ 0.018 , 0.037 ]$ &  $M_{30}$ & $0.000$ & $[ 0.000 , 0.000 ]$ \\                                                                                                                  
 $M_{19}$ & $-0.001$ & $[ -0.002 , -0.001 ]$ &  $M_{31}$ & $0.000$ & $[ 0.000 , 0.000 ]$ \\                                                                                                                    
 $M_{20}$ & $-0.003$ & $[ -0.003 , 0.002 ]$ & $M_{32}$ & $0.075$ & $[ 0.035 , 0.118 ]$ \\                                                                                                                     
 $M_{21}$ & $0.002$ & $[ 0.001 , 0.003 ]$ &  $M_{33}$ & $0.007$ & $[ 0.001 , 0.012 ]$ \\                                                                                                                    
 $M_{22}$ & $-0.005$ & $[ -0.006 , -0.003 ]$ & $M_{34}$ & $0.000$ & $[ -0.000 , 0.000 ]$ \\    
\hline
\end{tabular} 
\end{center}
\end{table}

\begin{table}[!htb]
\caption{
Predictions from \texttt{EOS} for the angular observables  of the \decay{\Lb}{\Lz\mumu} decay with $P_{\Lb}=1$ in the range $15 < \qsq < 20\gev^2/c^4$. 
The SM calculation is described in the text.
The observables $M_{31}$ and $M_{34}$ vanish due to the small size of the muon mass. 
Observables that depend on the imaginary part of the product of two transversity amplitudes also tend to be vanishingly small, due to the small strong phase difference between pairs of amplitudes in the SM.
\label{tab:eos:predictions:lowrecoil}} 
\begin{center}
\begin{tabular}{lrc|lrc}
\hline
Obs. & Value & 68\% interval & Obs. & Value & 68\% interval \\
\hline
 $M_{1}$ & $0.351$ & $[ 0.349 , 0.353 ]$ & $M_{6}$ & $0.187$ & $[ 0.183 , 0.192 ]$ \\
 $M_{2}$ & $0.298$ & $[ 0.294 , 0.301 ]$ & $M_{7}$ & $-0.022$ & $[-0.025,  -0.019 ]$ \\
 $M_{3}$ & $-0.236$ & $[ -0.240 , -0.230 ]$ &  $M_{8}$ & $-0.100$ & $[-0.105,  -0.095 ]$ \\
 $M_{4}$ & $-0.195$ & $[ -0.200 , -0.190 ]$ &  $M_{9}$ & $0.000$ & $[ 0.000 , 0.001 ]$ \\
 $M_{5}$ & $-0.154$ & $[ -0.159 , -0.149 ]$ & $M_{10}$ & $-0.001$ & $[ -0.001 , -0.000 ]$ \\
\hline
 $M_{11}$ & $-0.064$ & $[ -0.069 , -0.058 ]$ & $M_{23}$ & $-0.299$ & $[ -0.303 , -0.295 ]$ \\
 $M_{12}$ & $0.240$ & $[ 0.235 , 0.245 ]$ & $M_{24}$ & $0.337$ & $[ 0.335 , 0.338 ]$ \\
 $M_{13}$ & $-0.292$ & $[ -0.295 , -0.288 ]$ & $M_{25}$ & $-0.001$ & $[ -0.001 , -0.000 ]$ \\
 $M_{14}$ & $0.034$ & $[ 0.031 , 0.038 ]$ & $M_{26}$ & $0.001$ & $[ 0.000 , 0.001 ]$ \\
 $M_{15}$ & $-0.191$ & $[ -0.196 , -0.186 ]$ & $M_{27}$ & $0.221$ & $[ 0.216 , 0.226 ]$ \\
 $M_{16}$ & $0.151$ & $[ 0.146 , 0.156 ]$ & $M_{28}$ & $-0.187$ & $[ -0.191 , -0.183 ]$ \\
 $M_{17}$ & $0.102$ & $[ 0.096 , 0.107 ]$ & $M_{29}$ & $0.000$ & $[ 0.000 , 0.000 ]$ \\
 $M_{18}$ & $0.021$ & $[ 0.018 , 0.024 ]$ & $M_{30}$ & $-0.001$ & $[ -0.001 , -0.000 ]$ \\
 $M_{19}$ & $0.000$ & $[ 0.000 , 0.000 ]$ & $M_{31}$ & $0.000$ & $[ 0.000 , 0.000 ]$ \\
 $M_{20}$ & $-0.001$ & $[ -0.001 , -0.001 ]$ & $M_{32}$ & $-0.046$ & $[ -0.050 , -0.043 ]$ \\
 $M_{21}$ & $0.000$ & $[ 0.000 , 0.001 ]$ & $M_{33}$ & $-0.053$ & $[ -0.056 , -0.050 ]$ \\
 $M_{22}$ & $-0.002$ & $[ -0.002 , -0.001 ]$ &  $M_{34}$ & $0.000$ & $[ 0.000 , 0.000 ]$ \\
\hline
\end{tabular} 
\end{center}
\end{table}

\clearpage

\section{Variation of observables with NP contributions}
\label{appendix:NP}

\begin{figure}[!htb]
\centering
\includegraphics[width=0.24\linewidth]{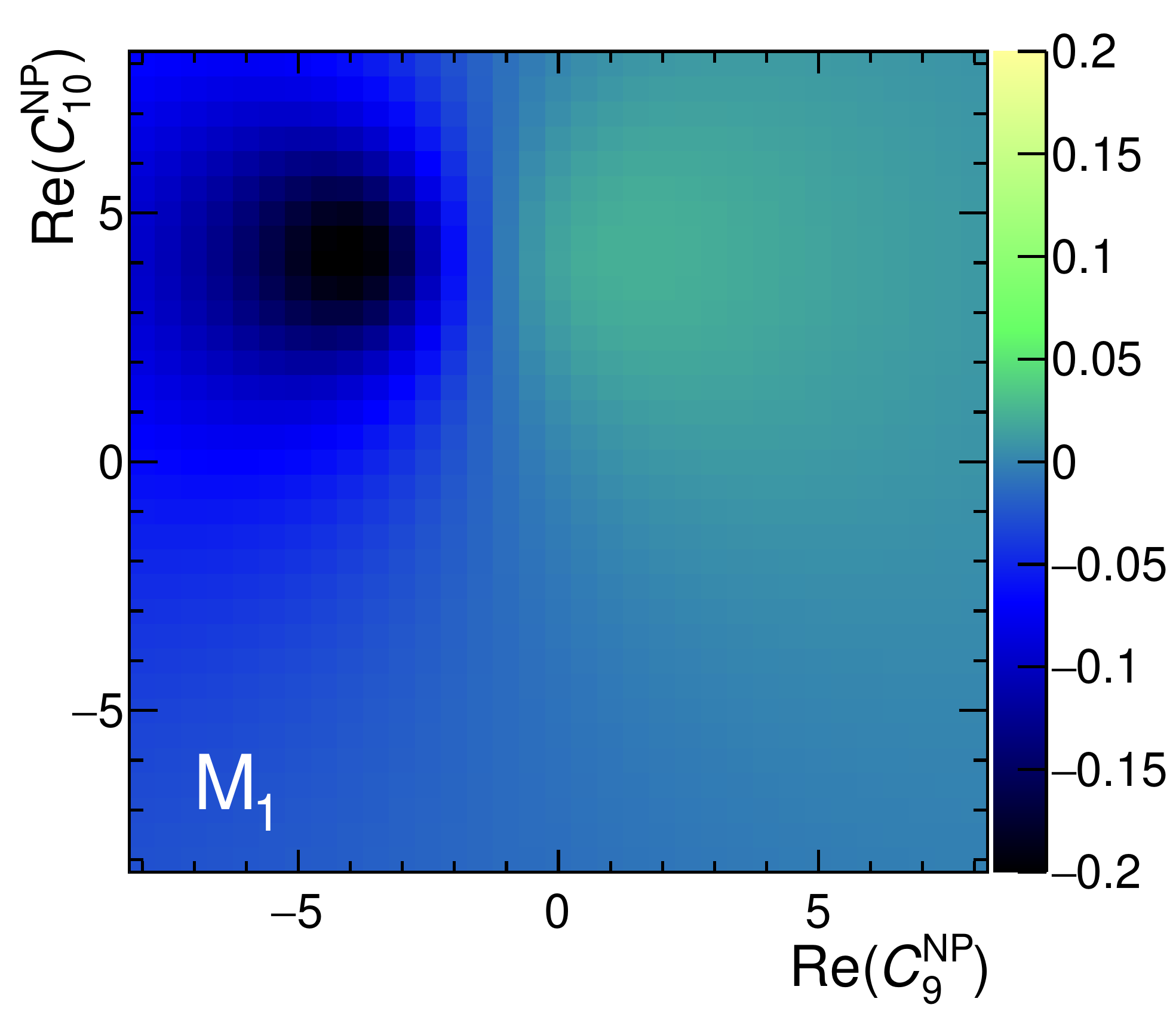}  
\includegraphics[width=0.24\linewidth]{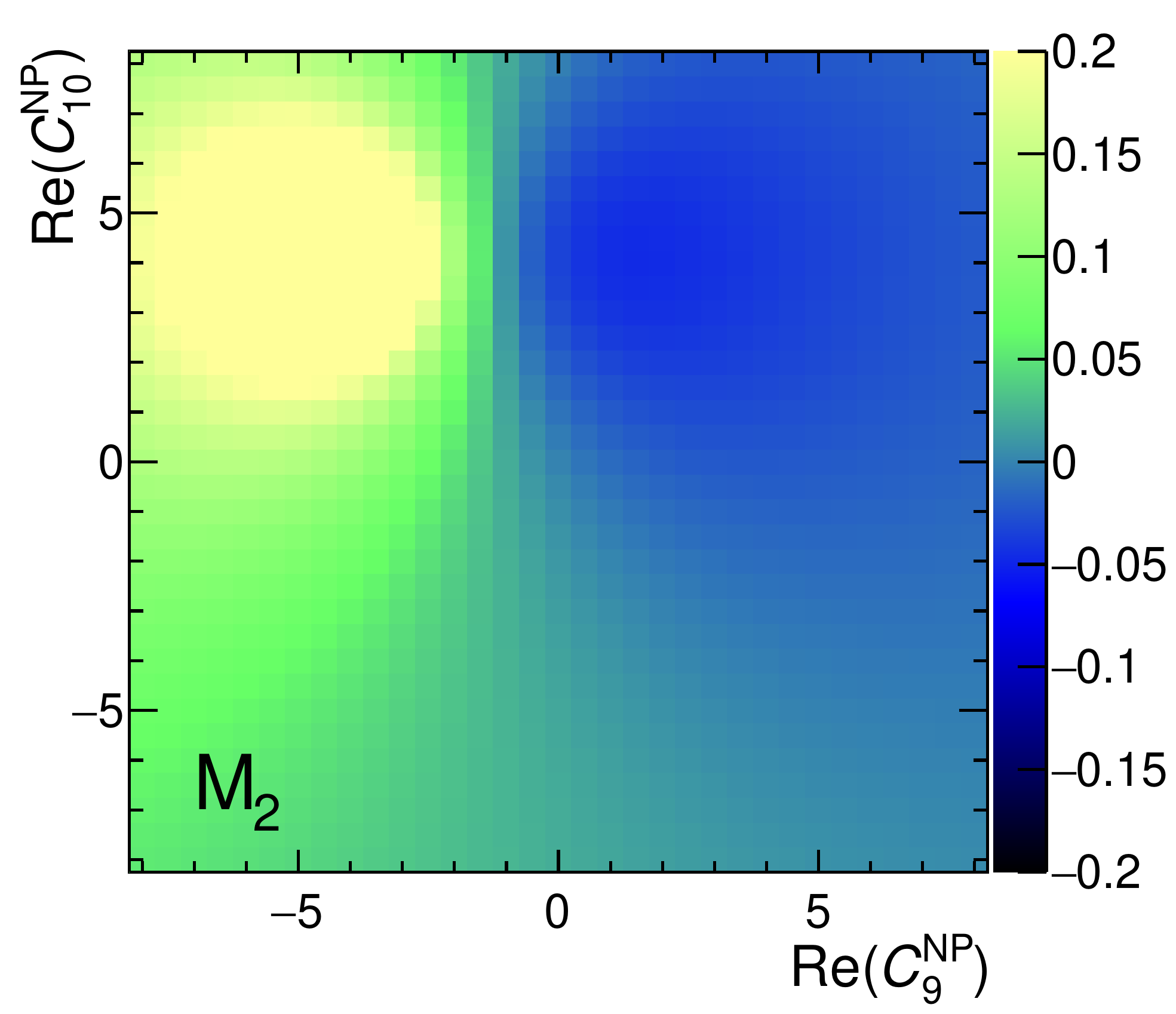} 
\includegraphics[width=0.24\linewidth]{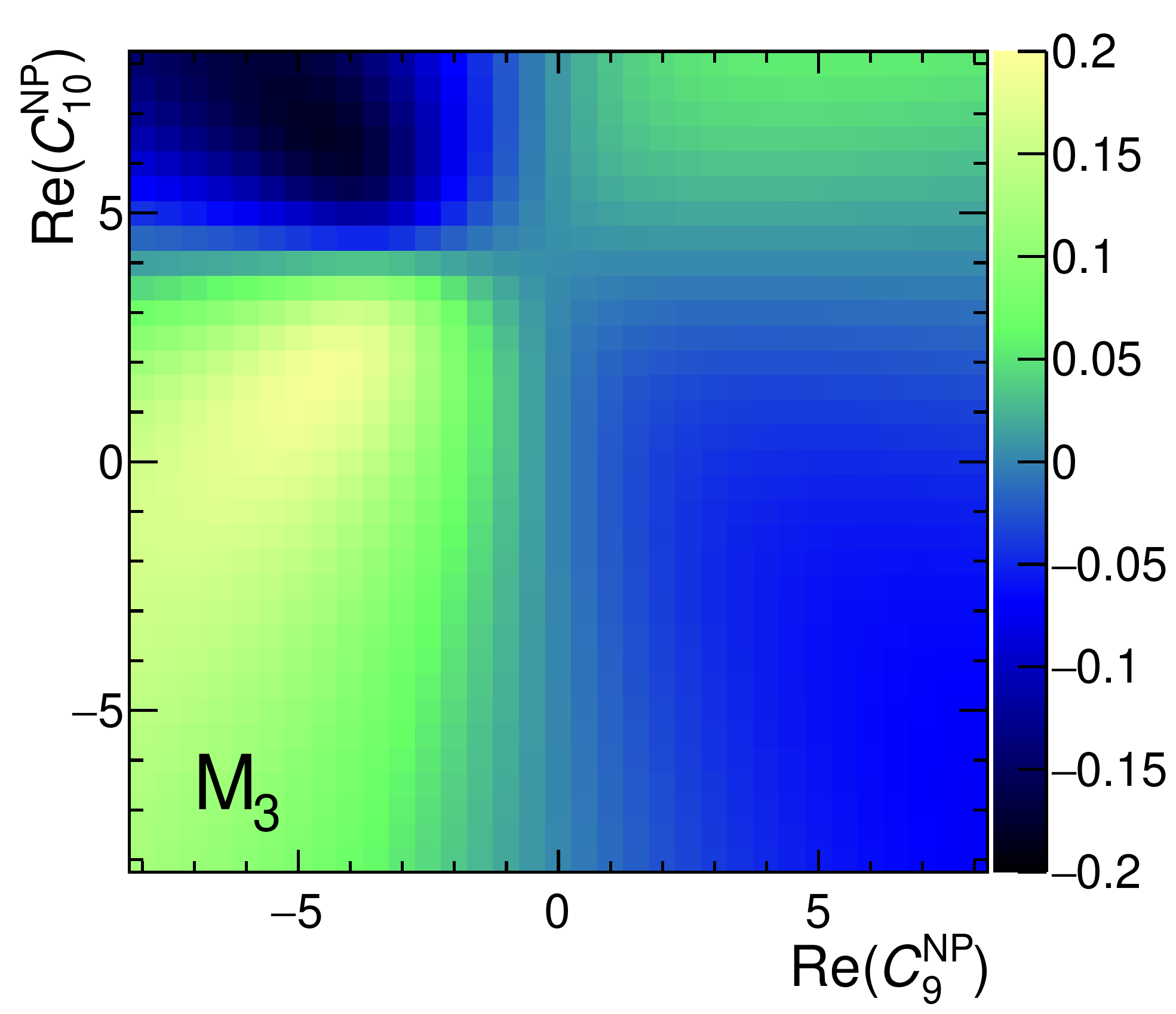} 
\includegraphics[width=0.24\linewidth]{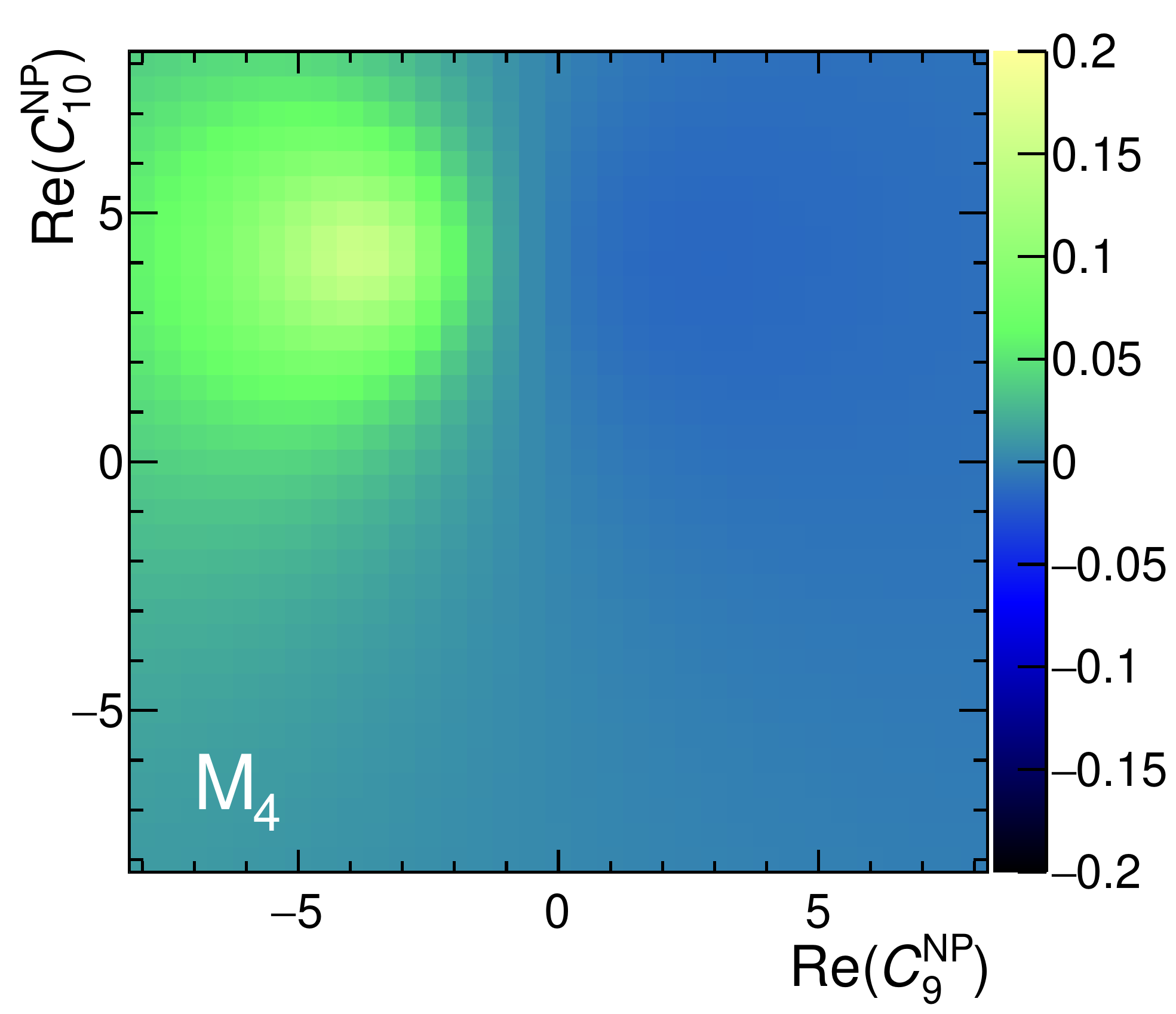} \\ 
\includegraphics[width=0.24\linewidth]{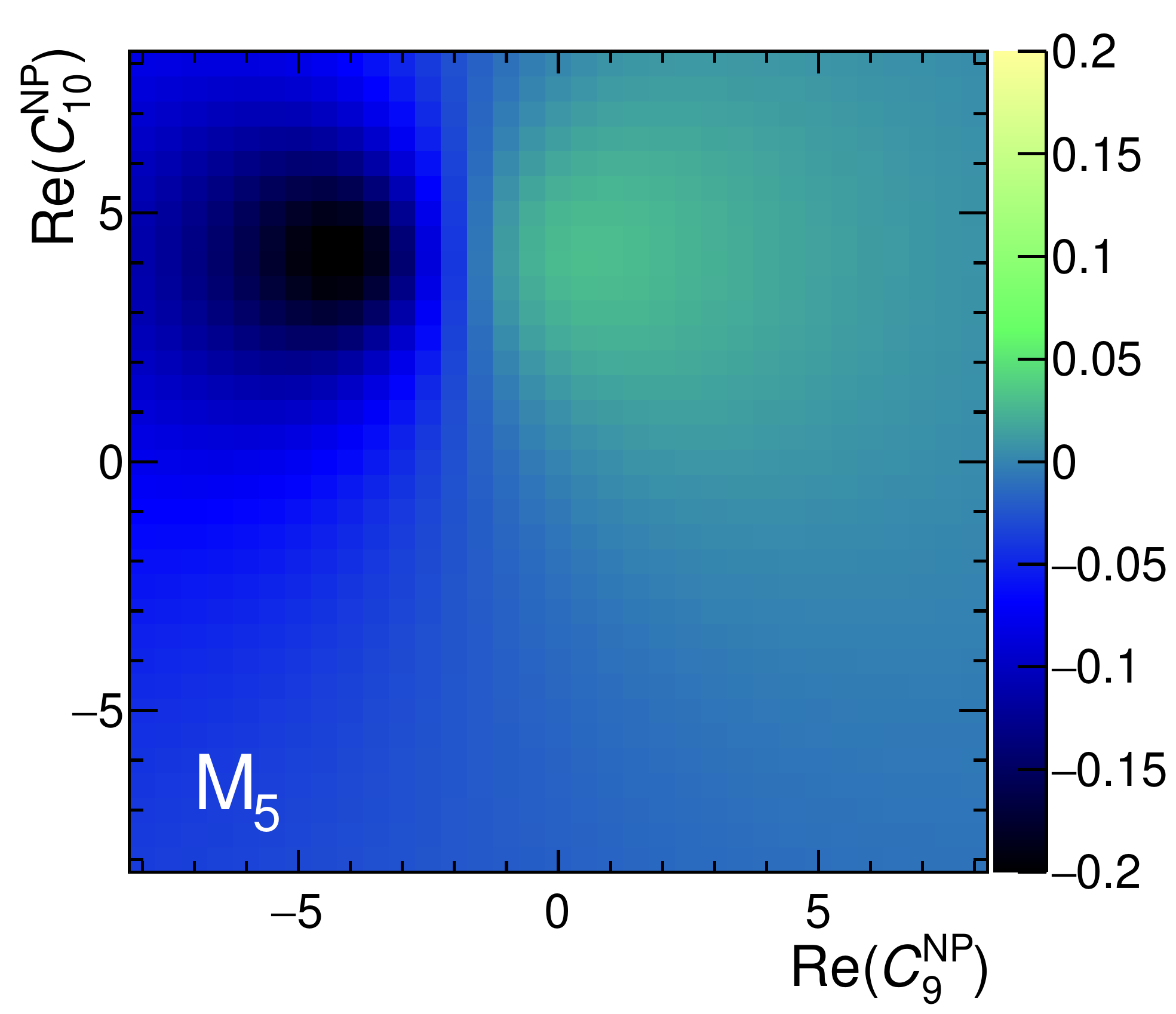}  
\includegraphics[width=0.24\linewidth]{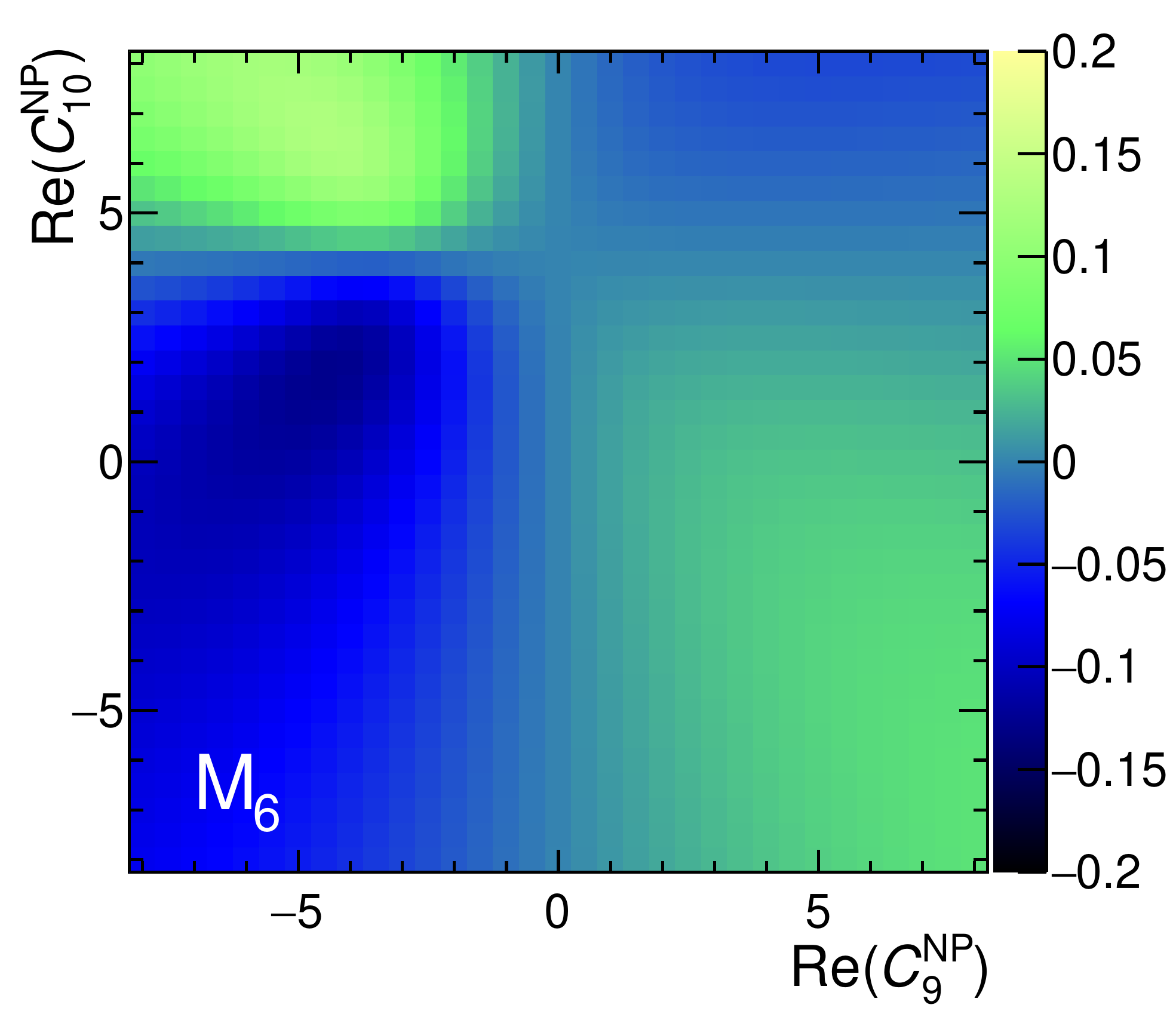} 
\includegraphics[width=0.24\linewidth]{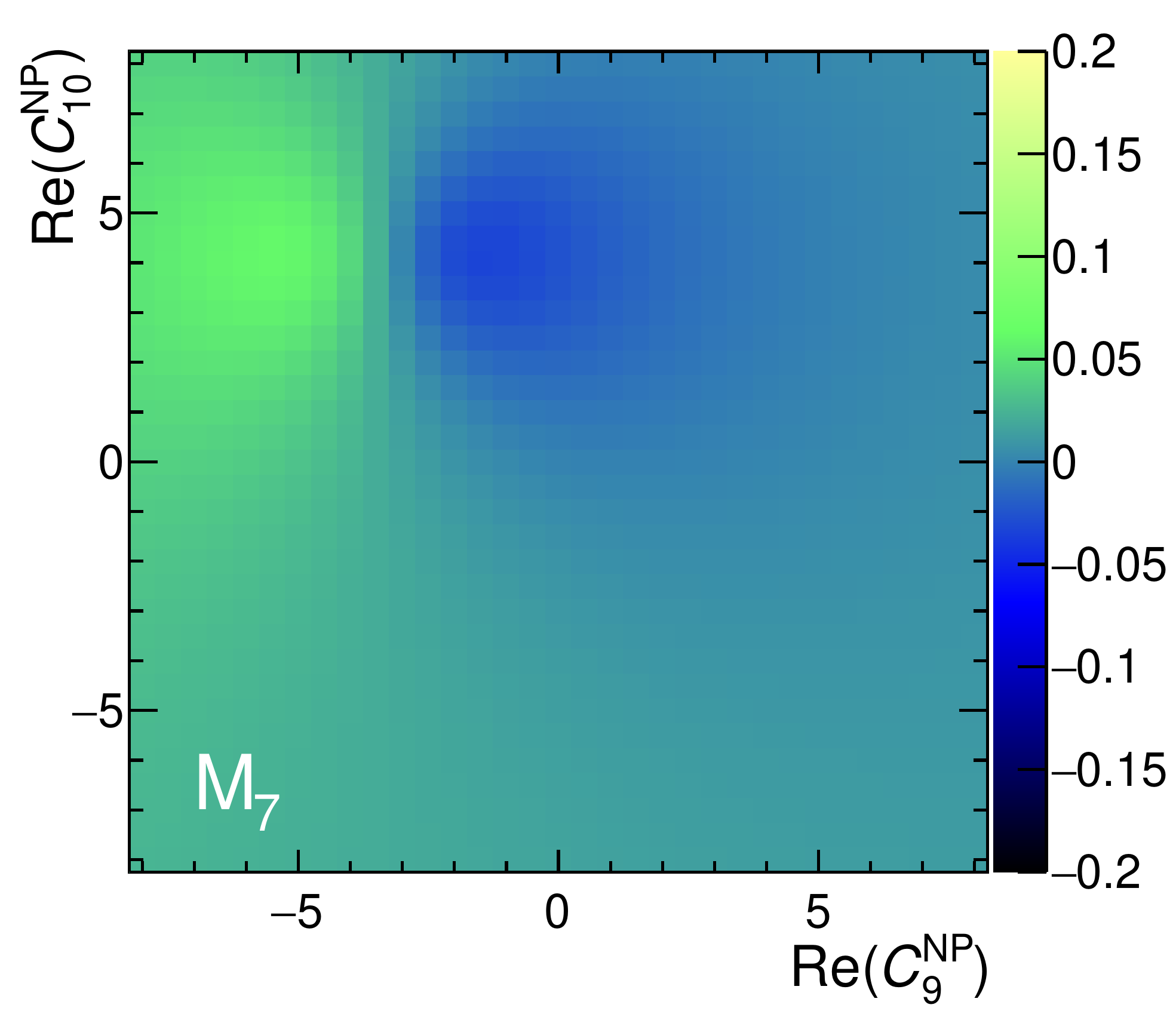} 
\includegraphics[width=0.24\linewidth]{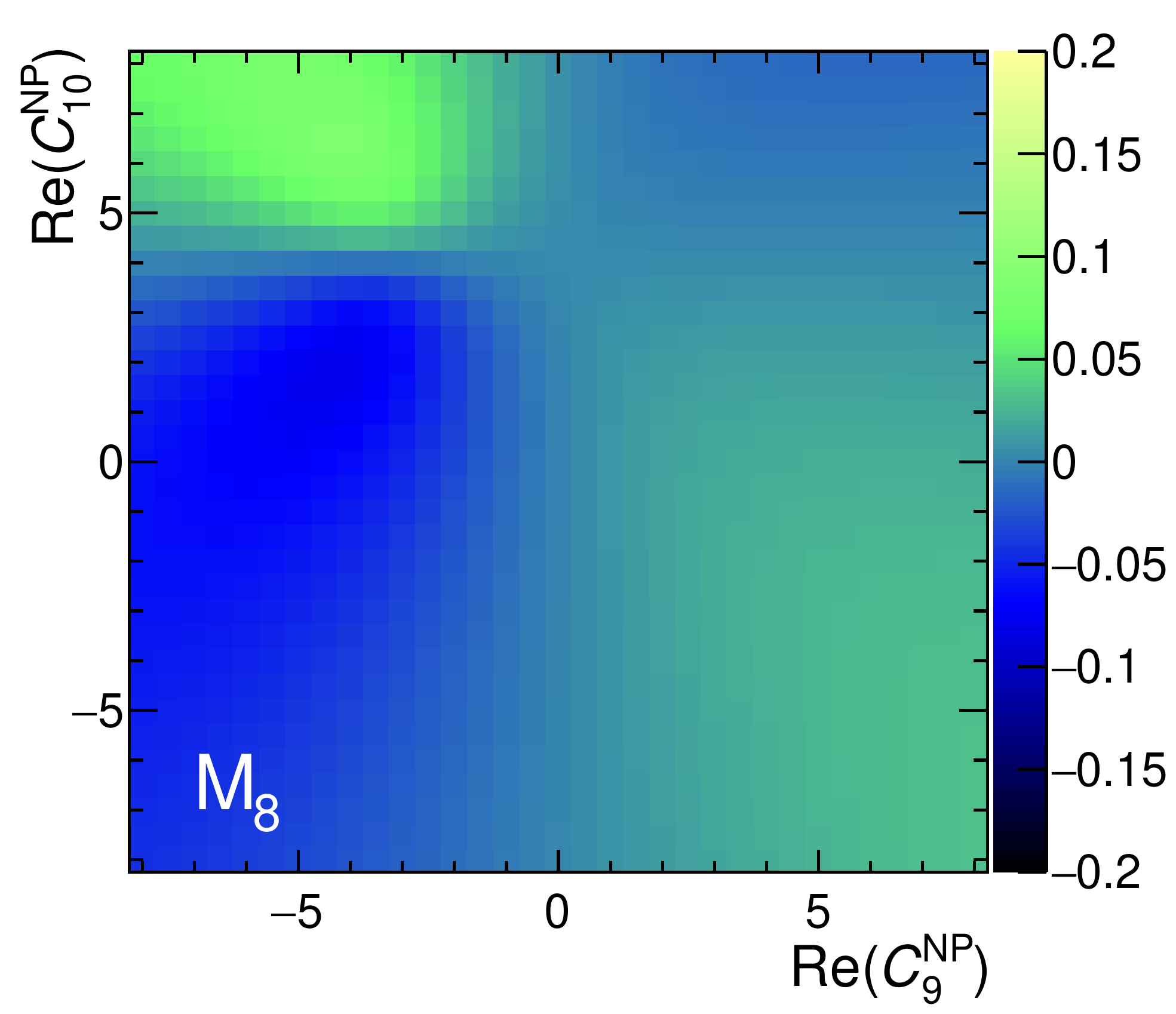} \\ 
\includegraphics[width=0.24\linewidth]{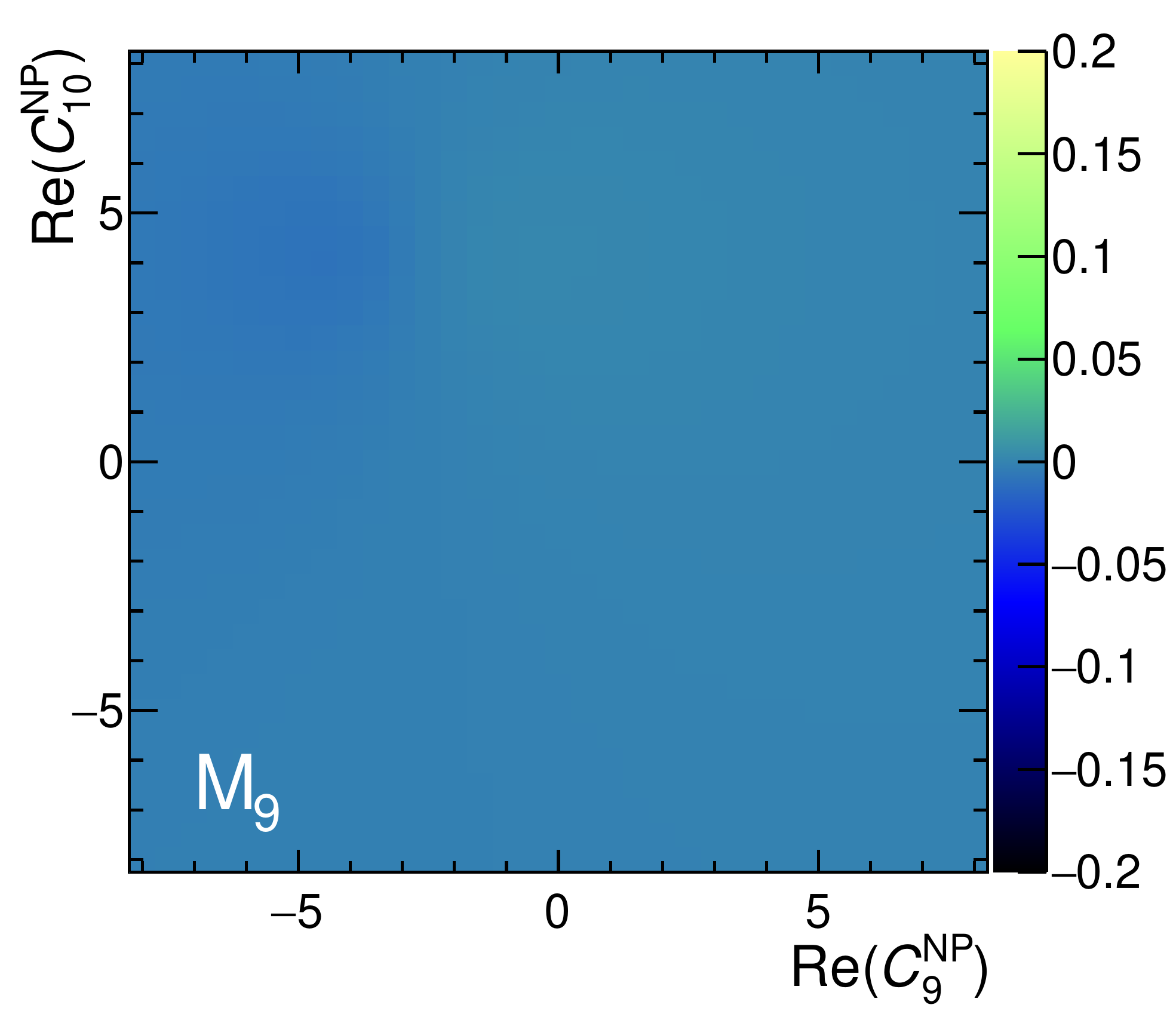}  
\includegraphics[width=0.24\linewidth]{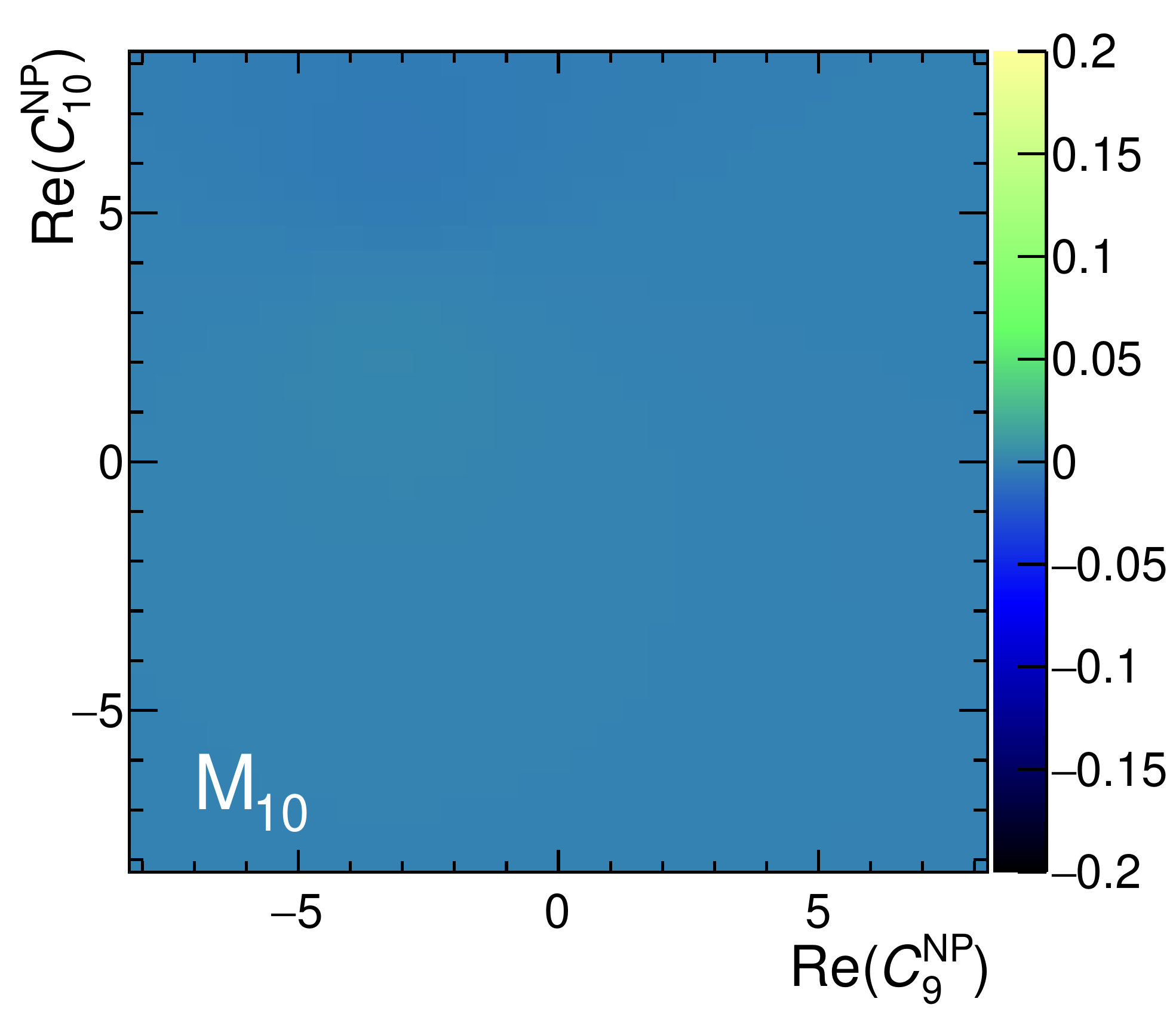} 
\caption{
Variation of the observables $M_{1}$--$M_{10}$ of the \decay{\Lb}{\Lz\mumu} decay from their SM central values in the large-recoil region ($1 < \qsq < 6\gev^{2}/c^{4}$) with a NP contribution to ${\rm Re}(C_9)$ or ${\rm Re}(C_{10})$. 
The SM point is at $(0,0)$.
\label{fig:scan:c9:c10:largerecoil} 
}
\end{figure}

\begin{figure}[!htb]
\centering
\includegraphics[width=0.24\linewidth]{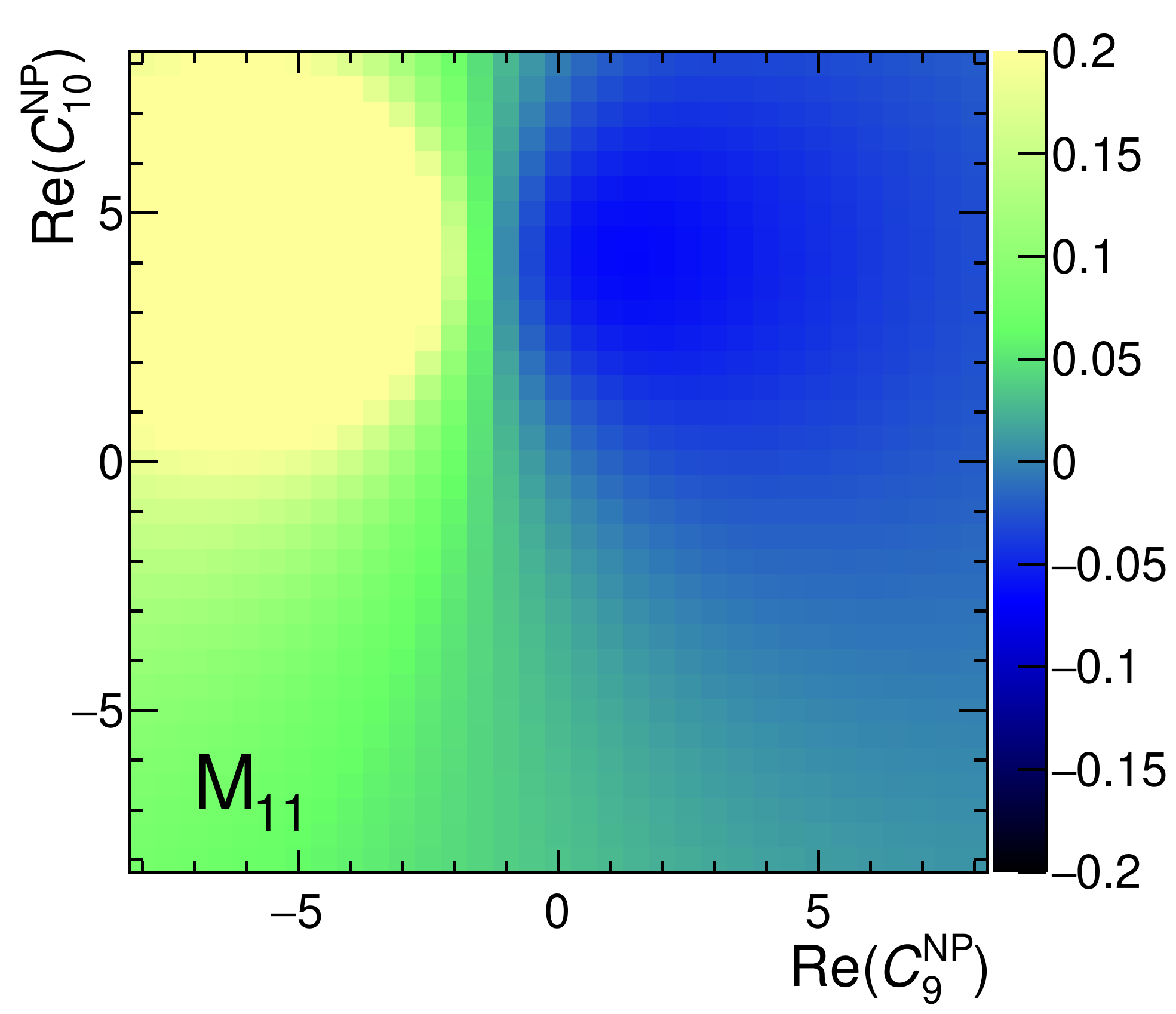} 
\includegraphics[width=0.24\linewidth]{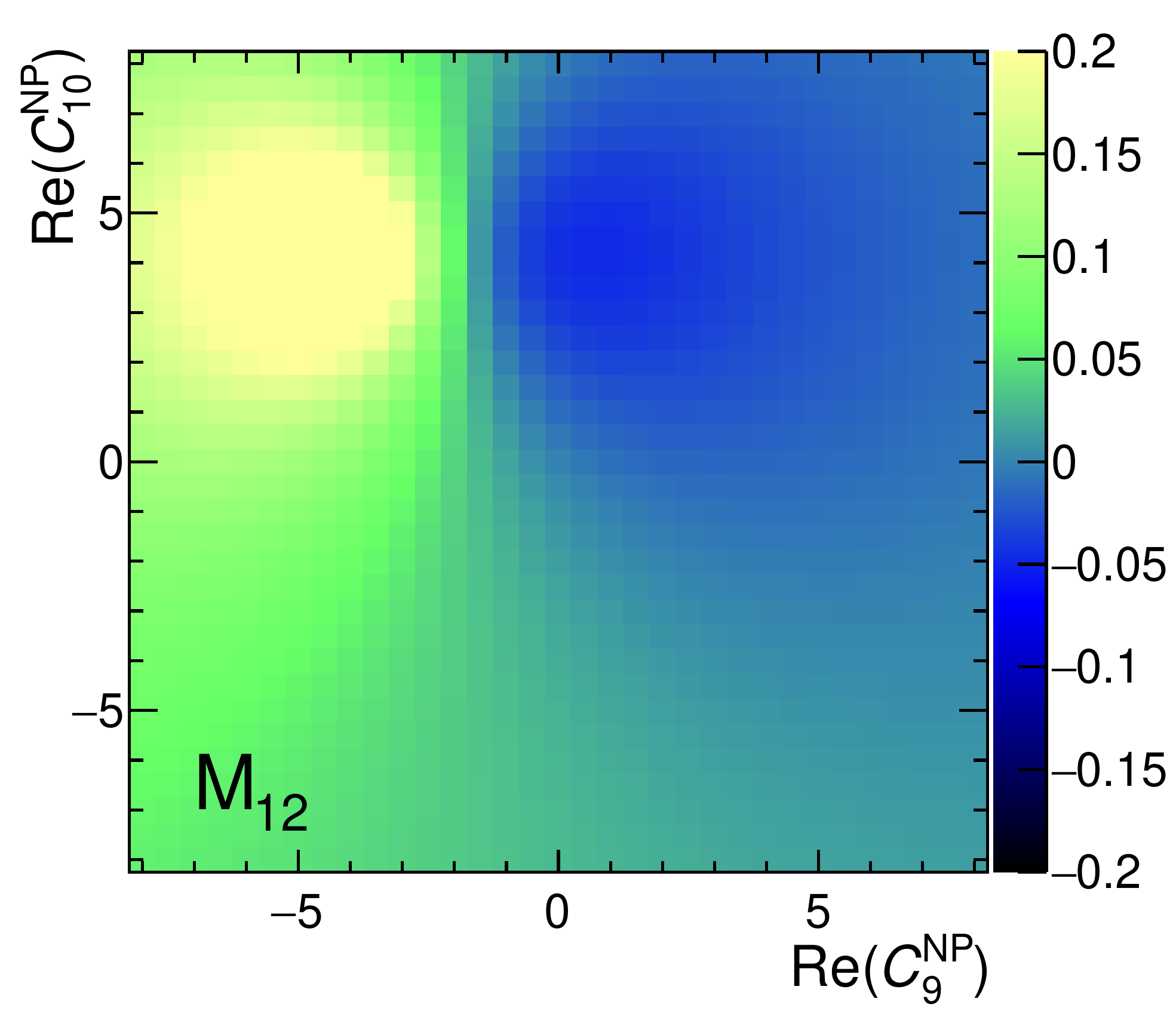} 
\includegraphics[width=0.24\linewidth]{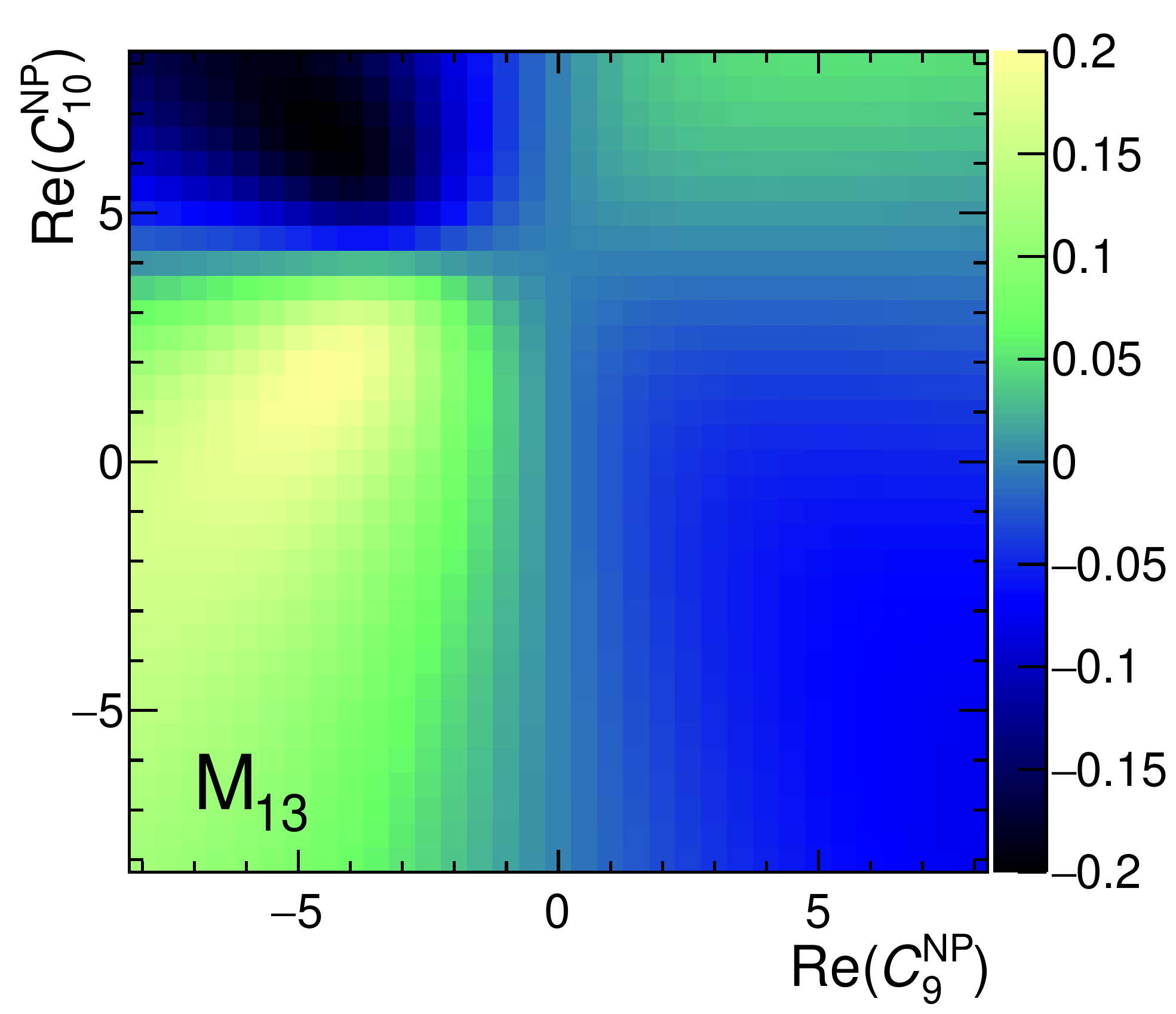}  
\includegraphics[width=0.24\linewidth]{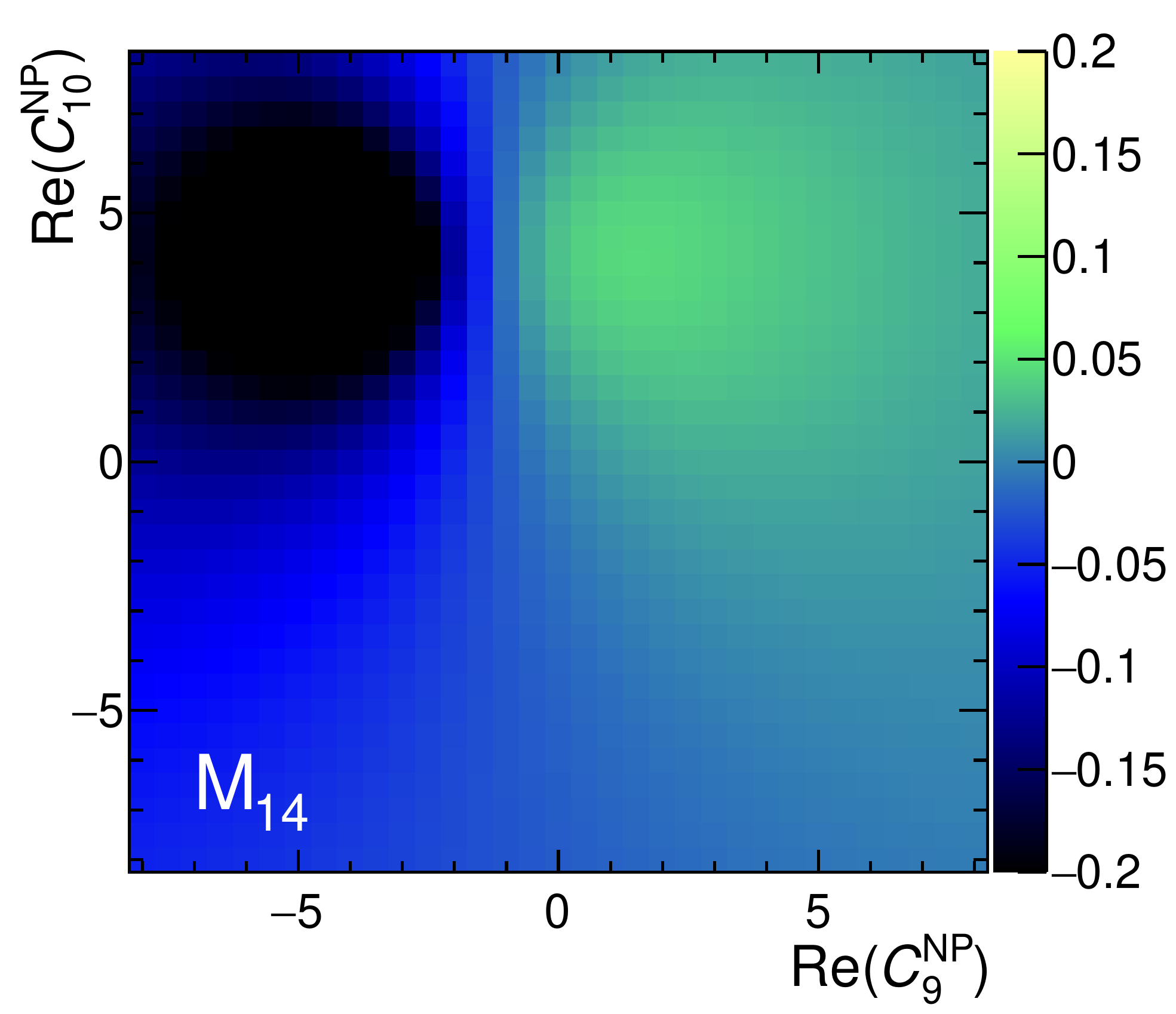}  \\ 
\includegraphics[width=0.24\linewidth]{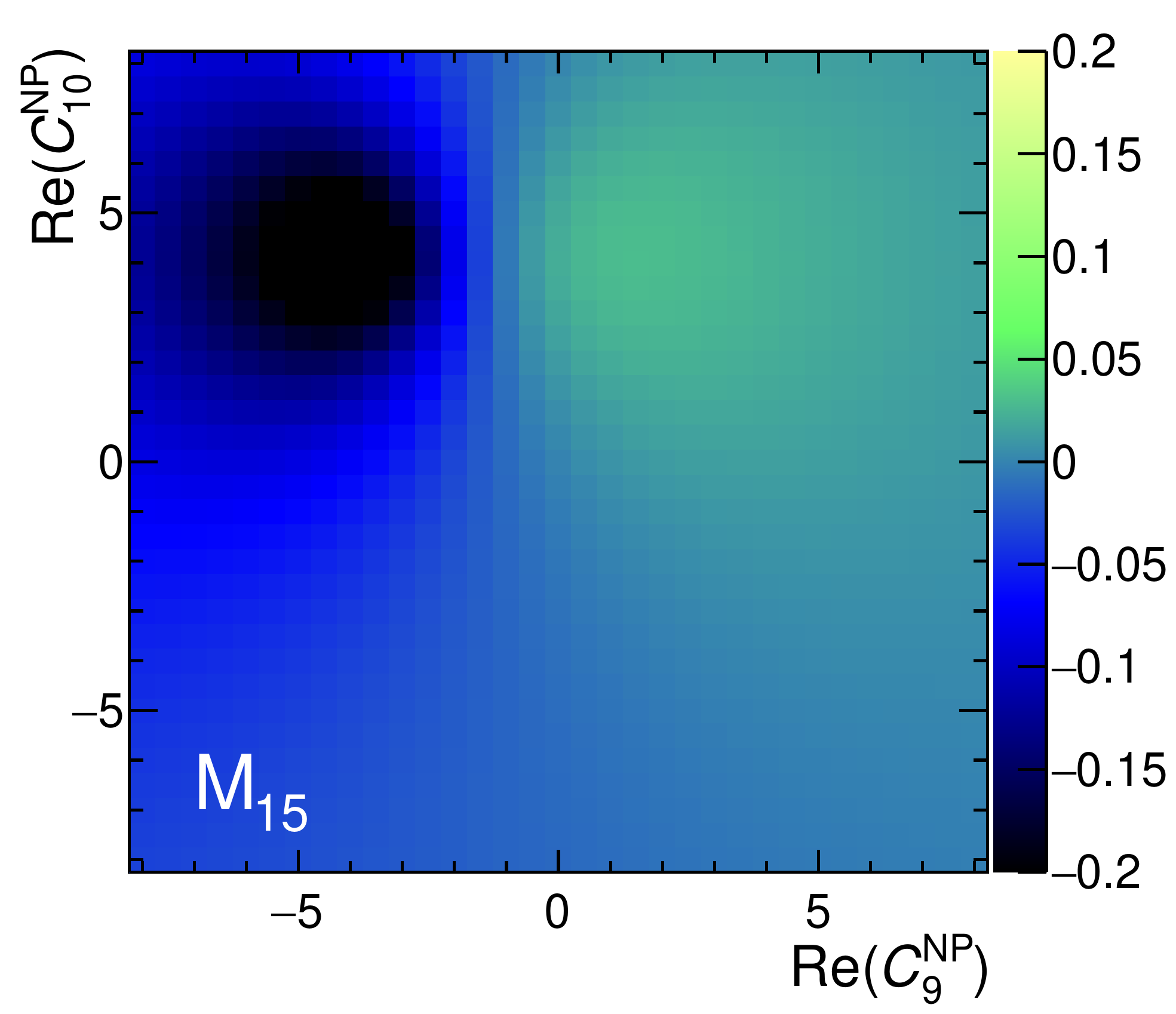} 
\includegraphics[width=0.24\linewidth]{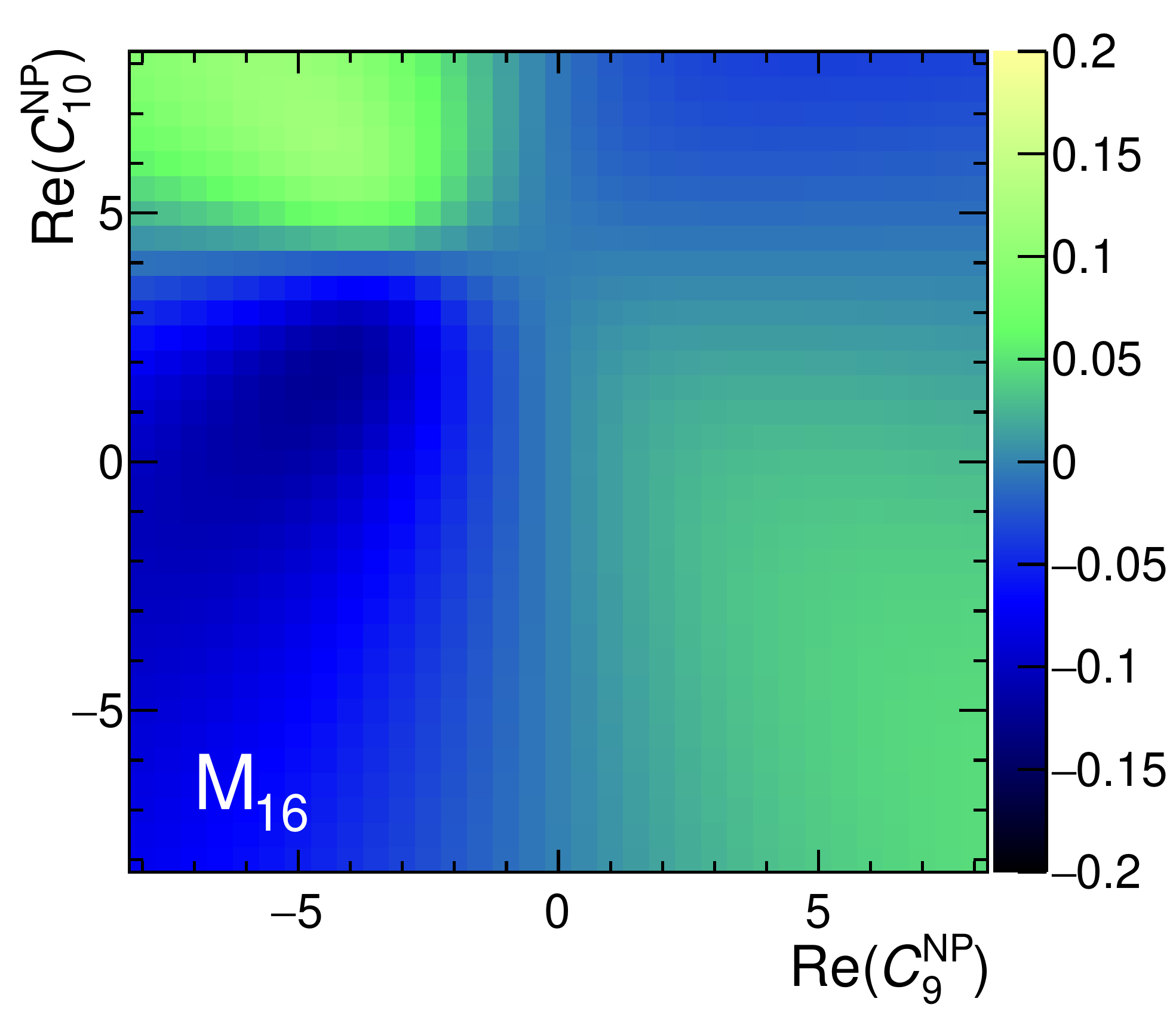} 
\includegraphics[width=0.24\linewidth]{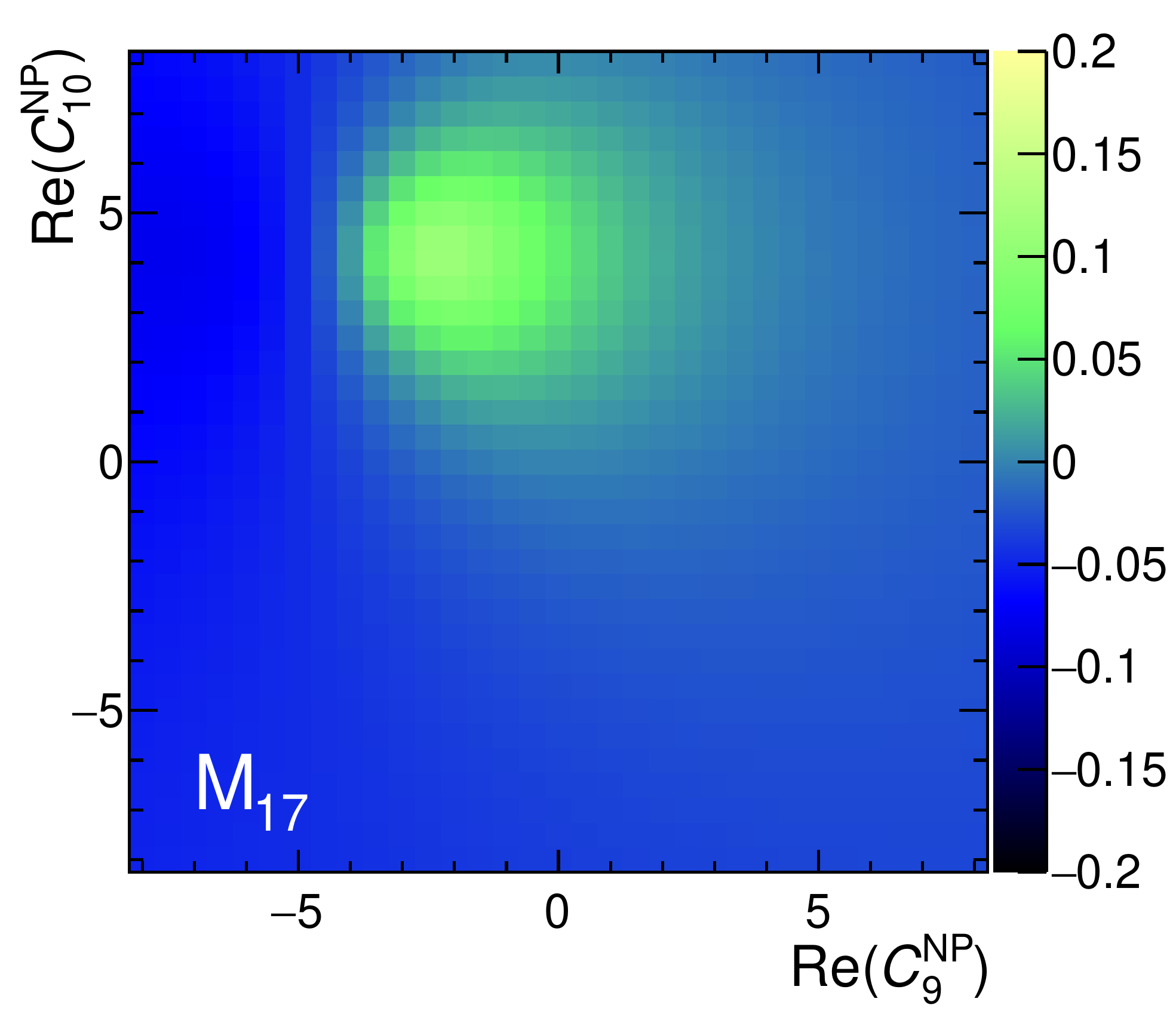}  
\includegraphics[width=0.24\linewidth]{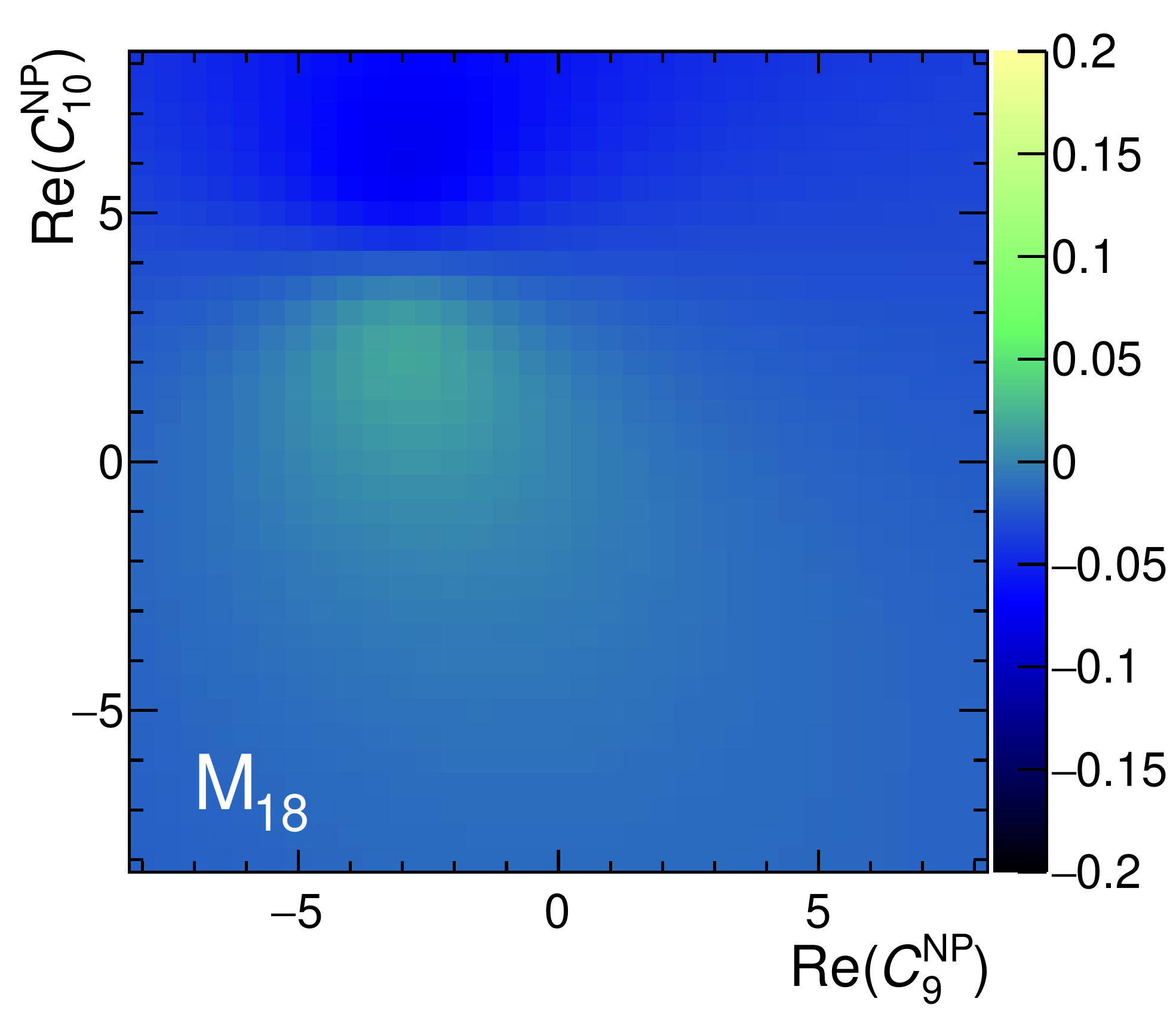} \\
\includegraphics[width=0.24\linewidth]{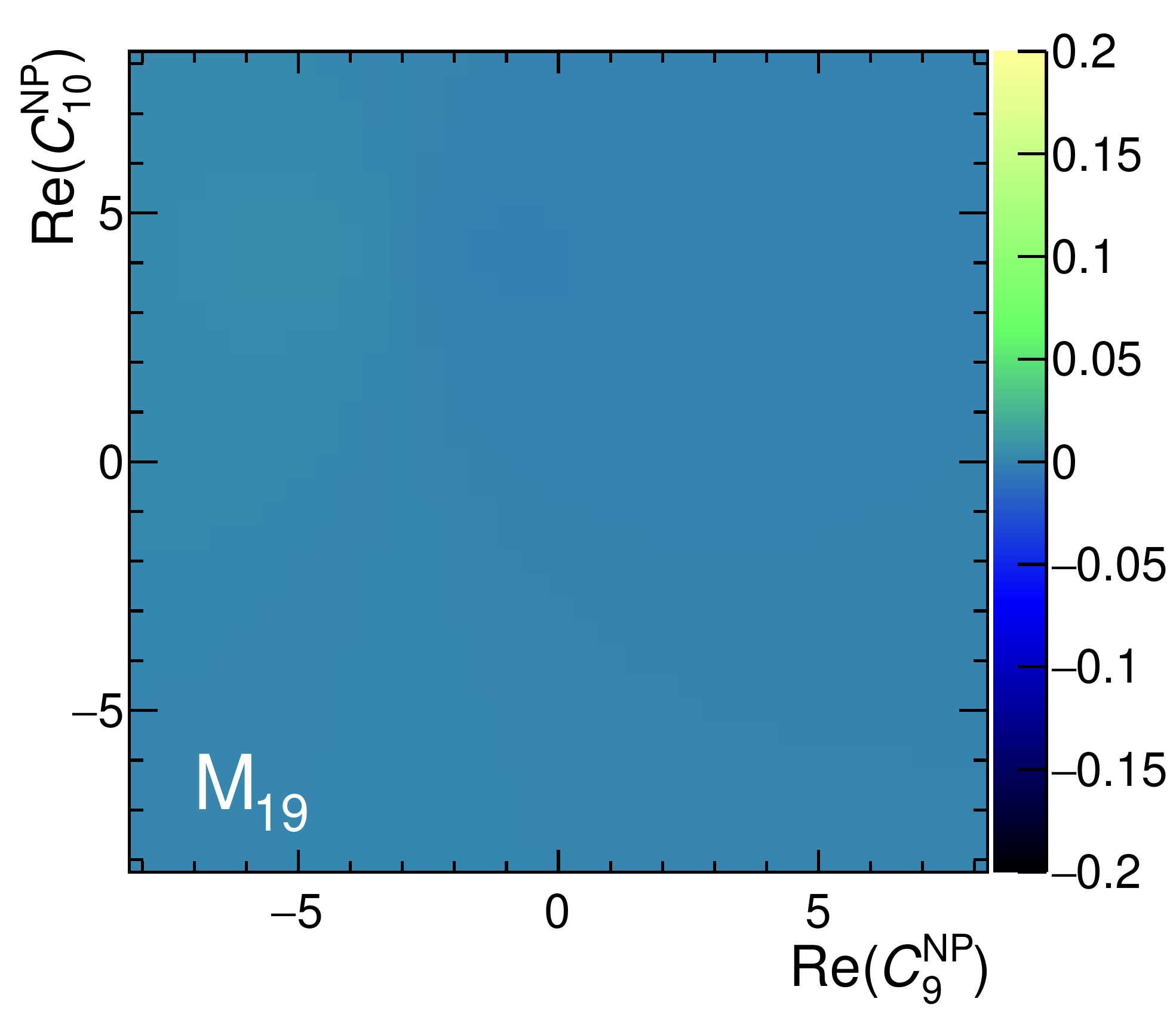} 
\includegraphics[width=0.24\linewidth]{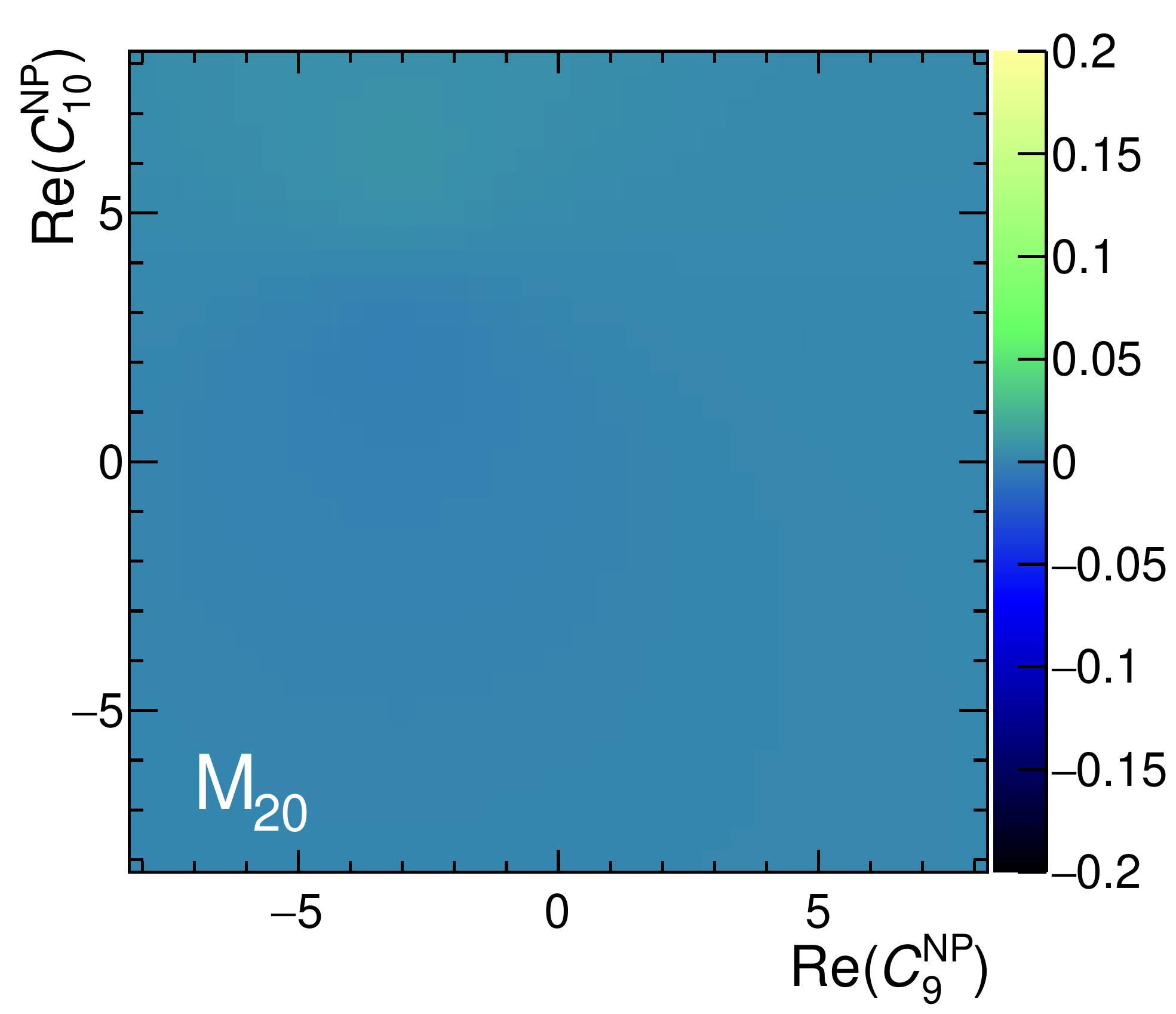}  
\includegraphics[width=0.24\linewidth]{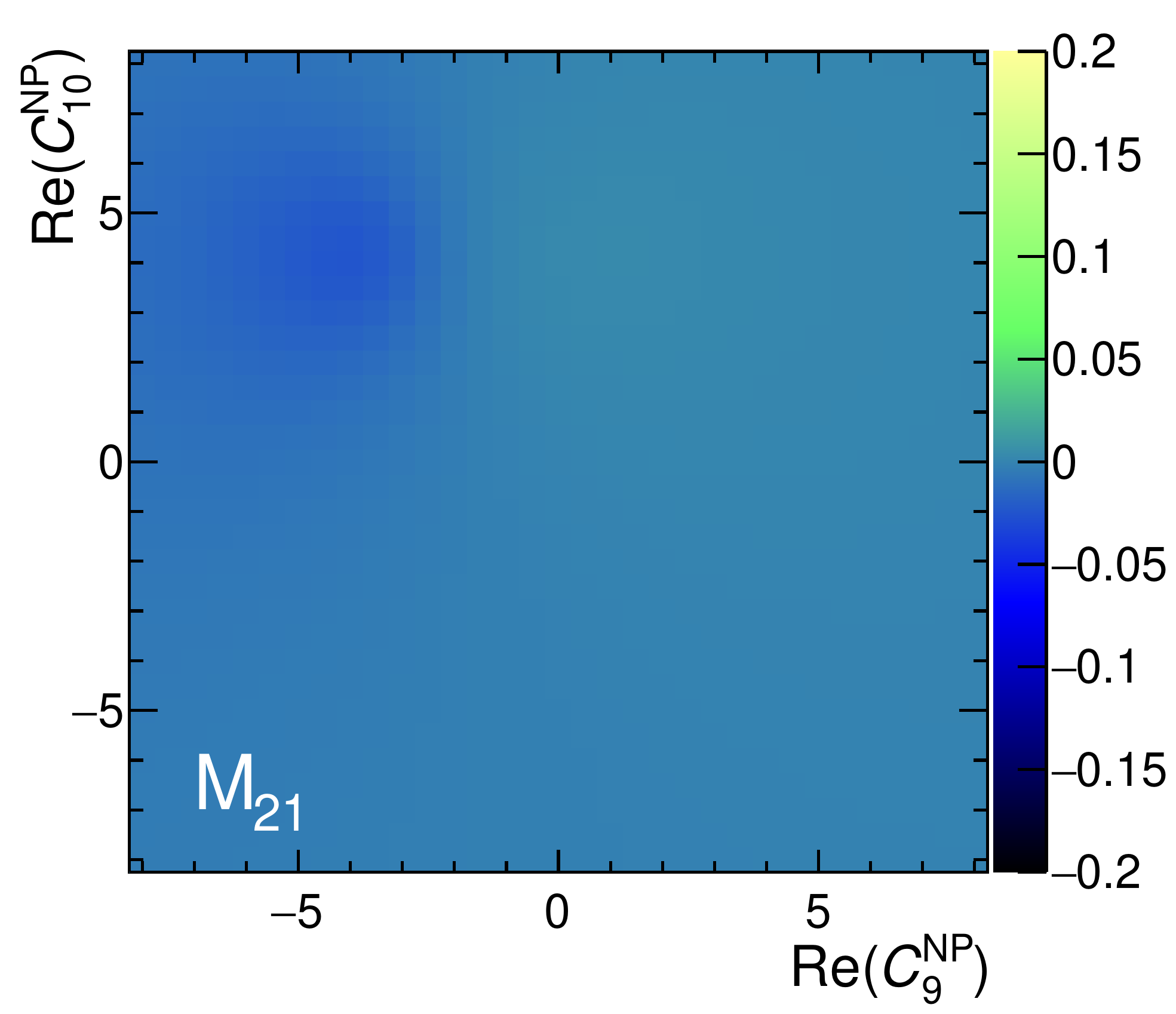}  
\includegraphics[width=0.24\linewidth]{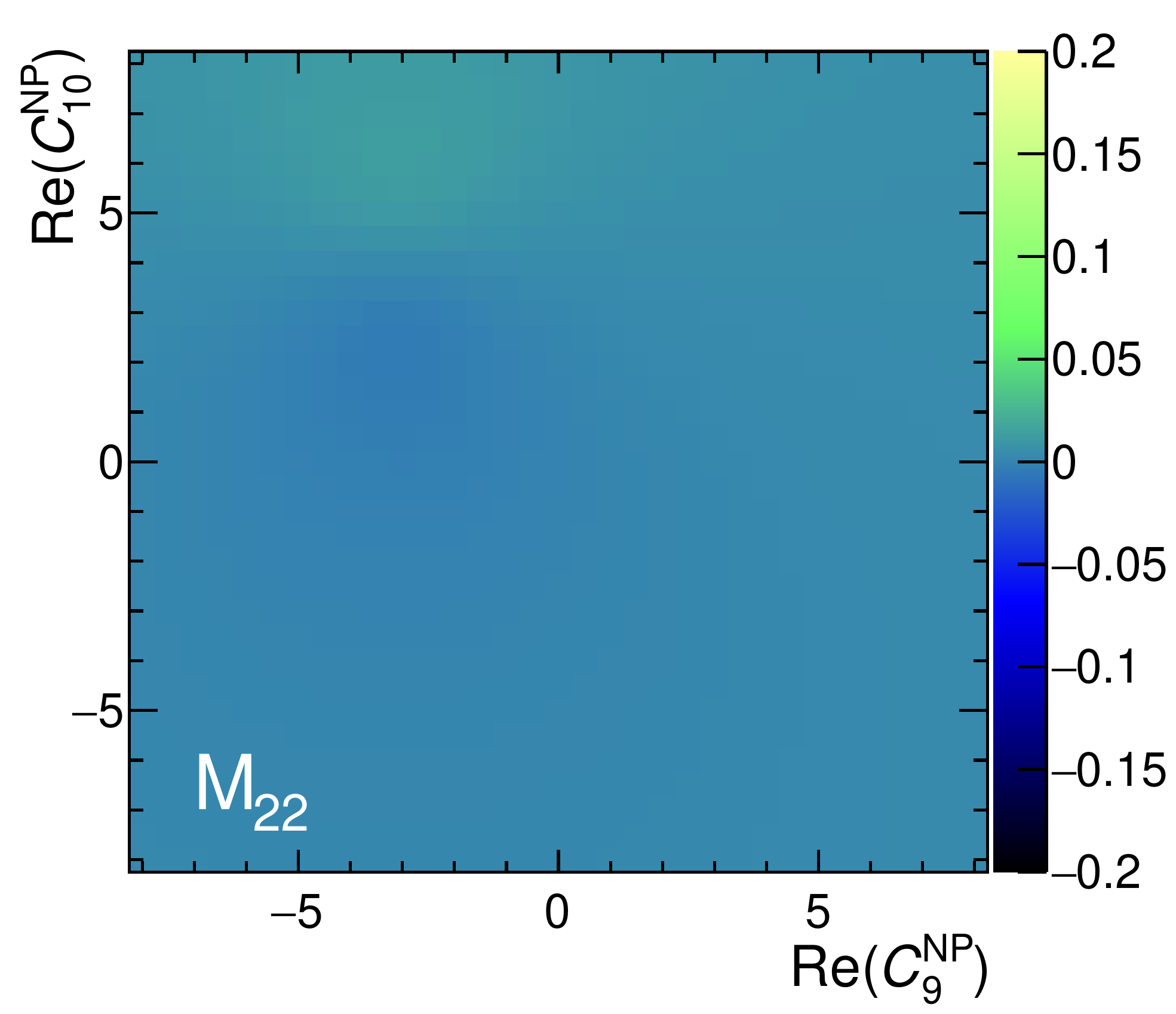} \\
\includegraphics[width=0.24\linewidth]{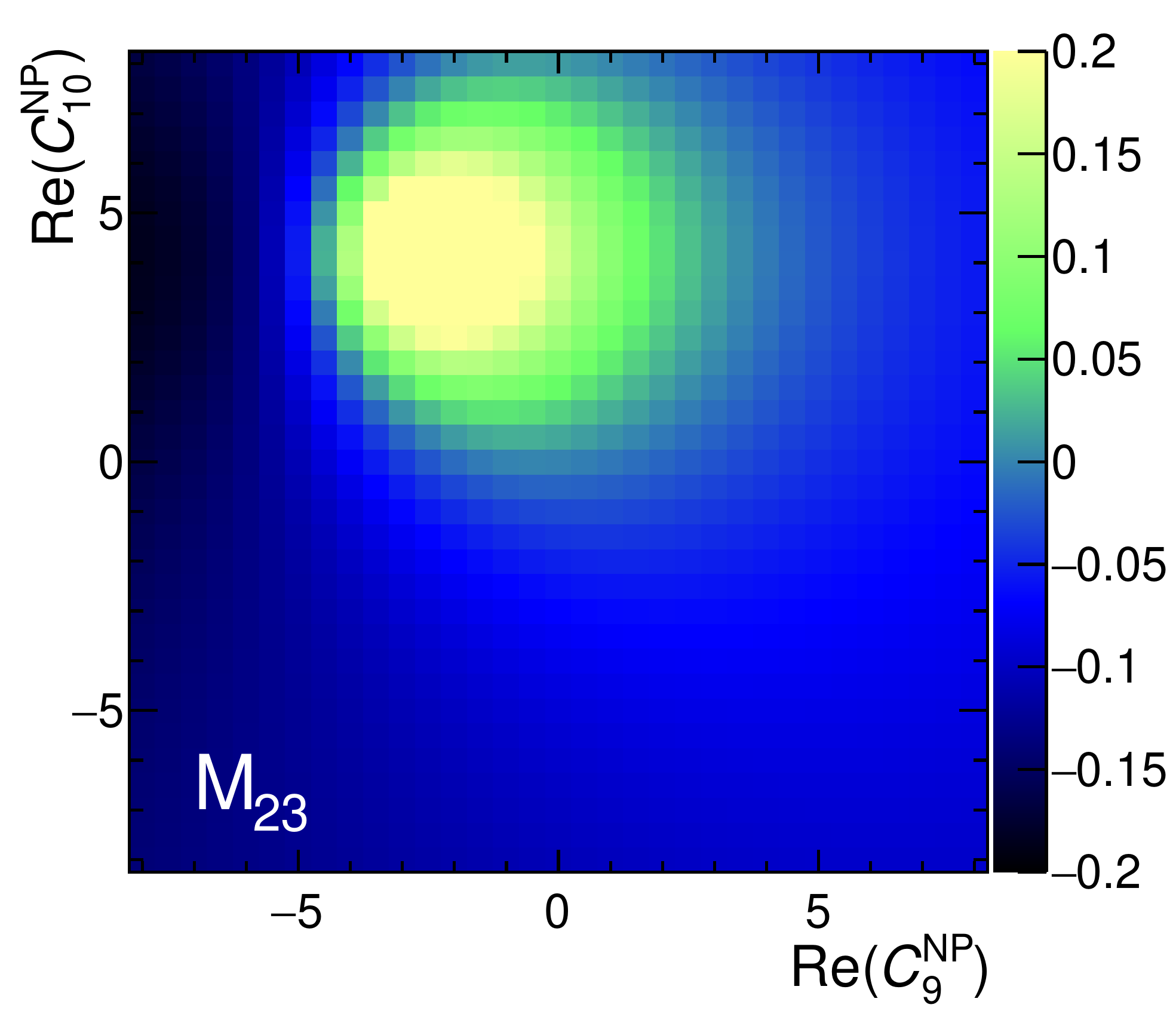} 
\includegraphics[width=0.24\linewidth]{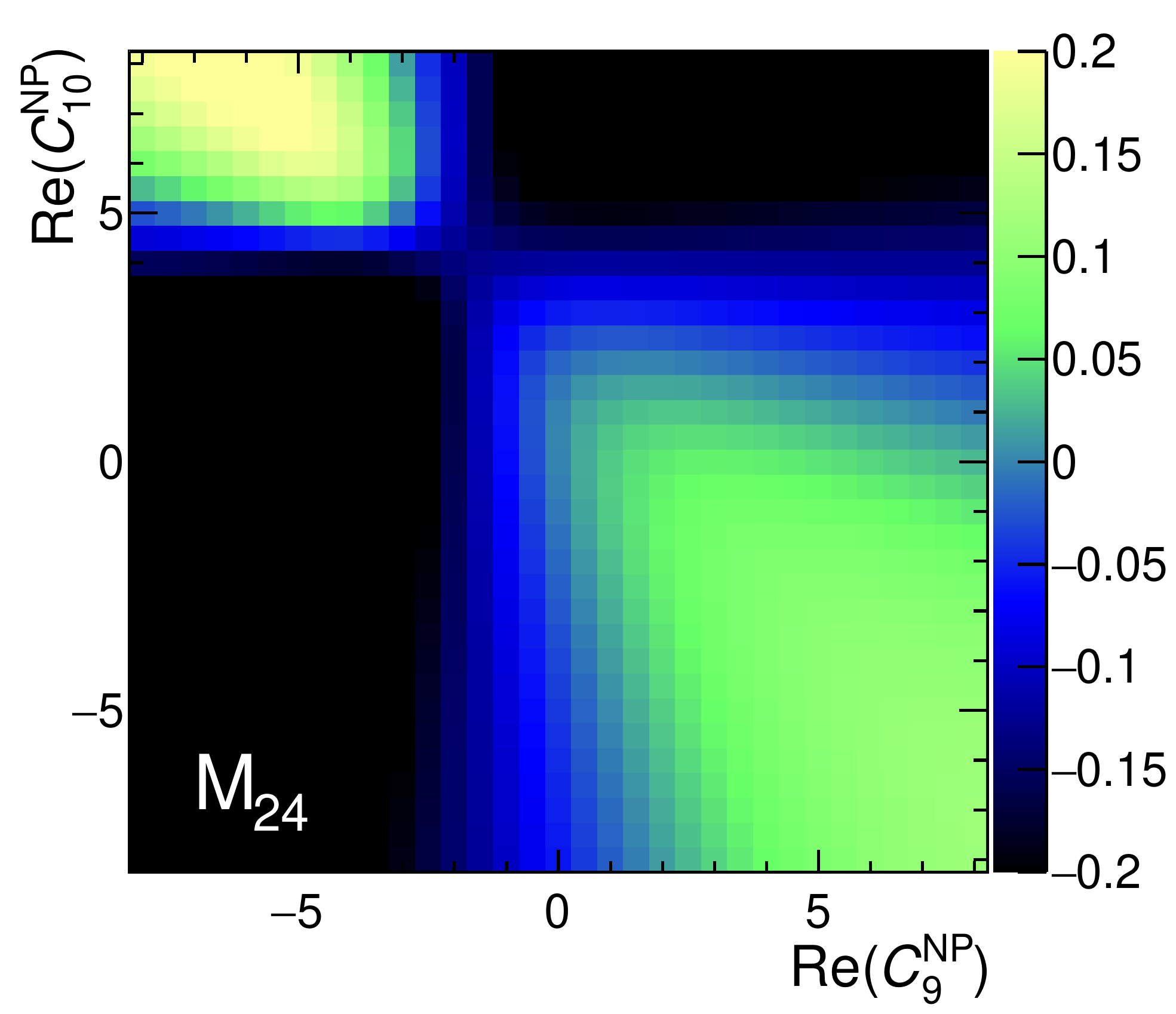}  
\includegraphics[width=0.24\linewidth]{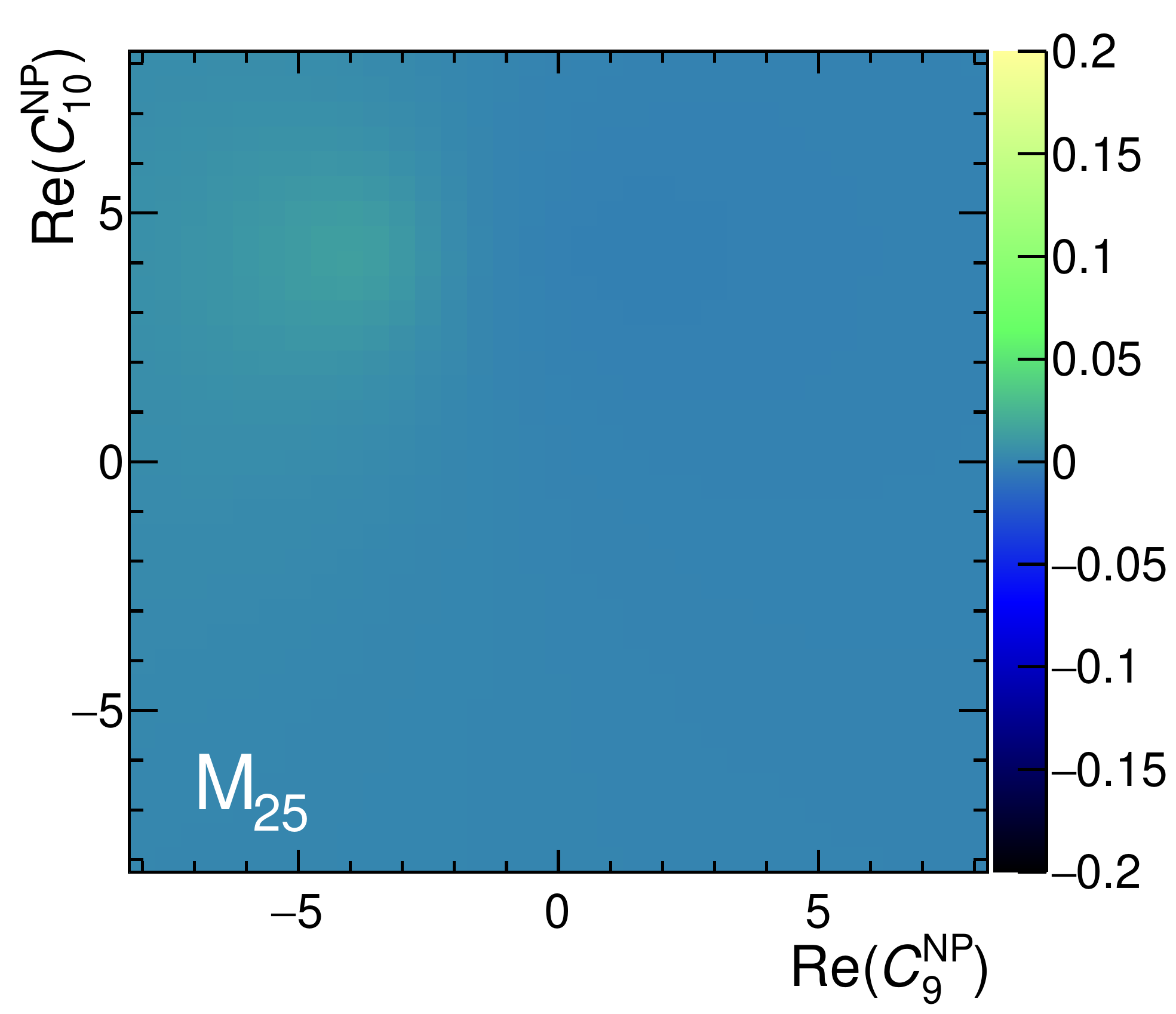} 
\includegraphics[width=0.24\linewidth]{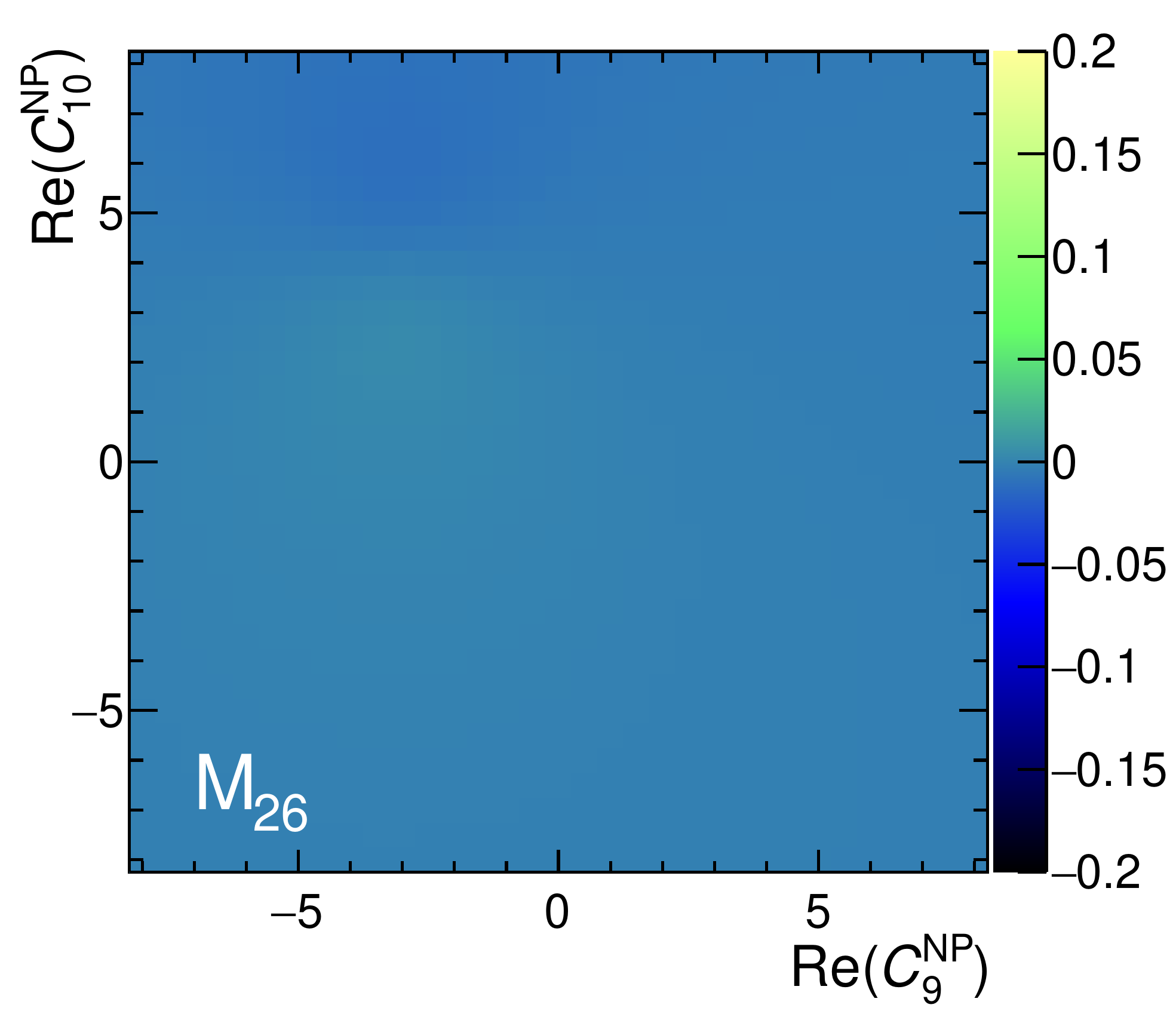} \\ 
\includegraphics[width=0.24\linewidth]{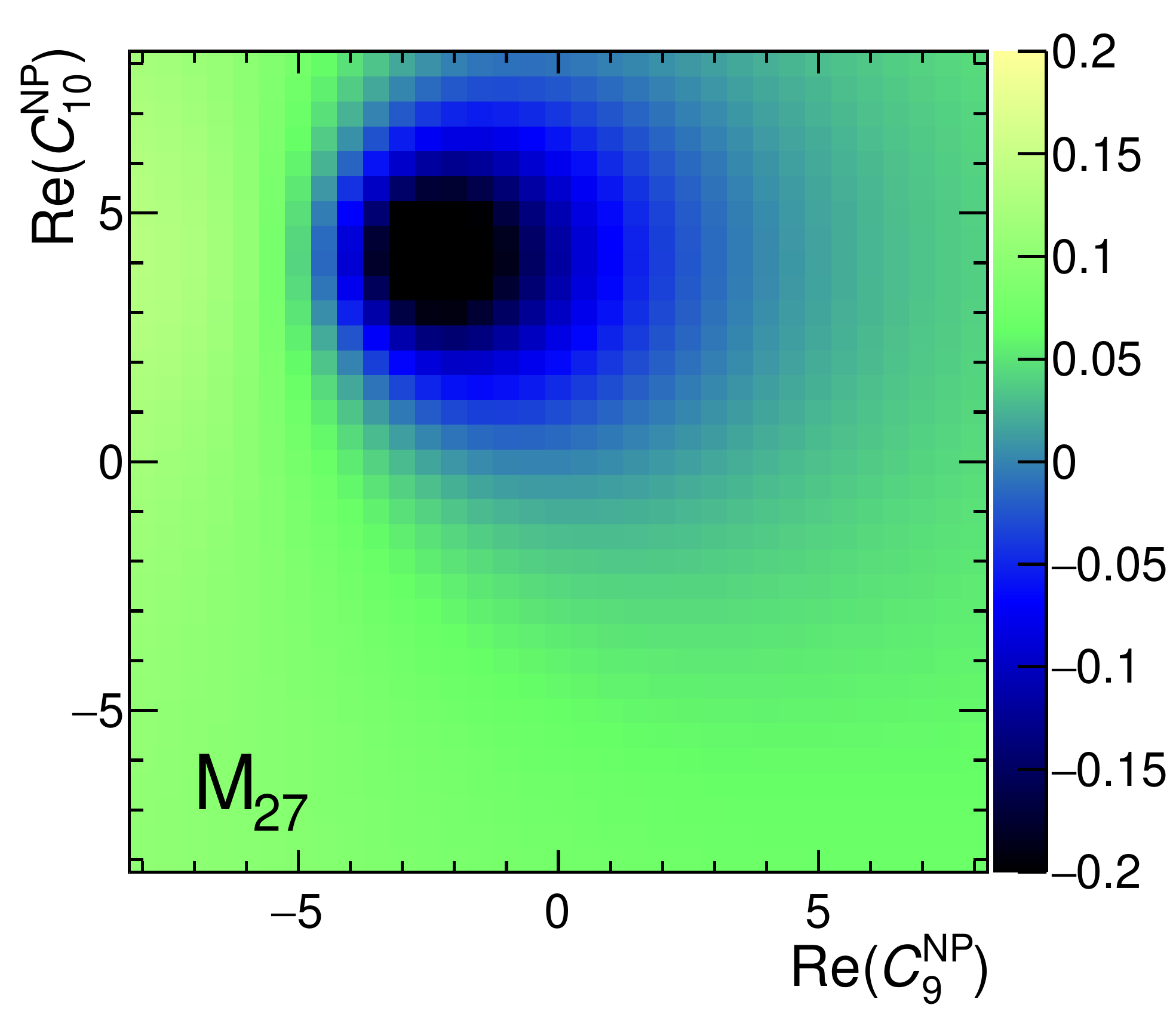}  
\includegraphics[width=0.24\linewidth]{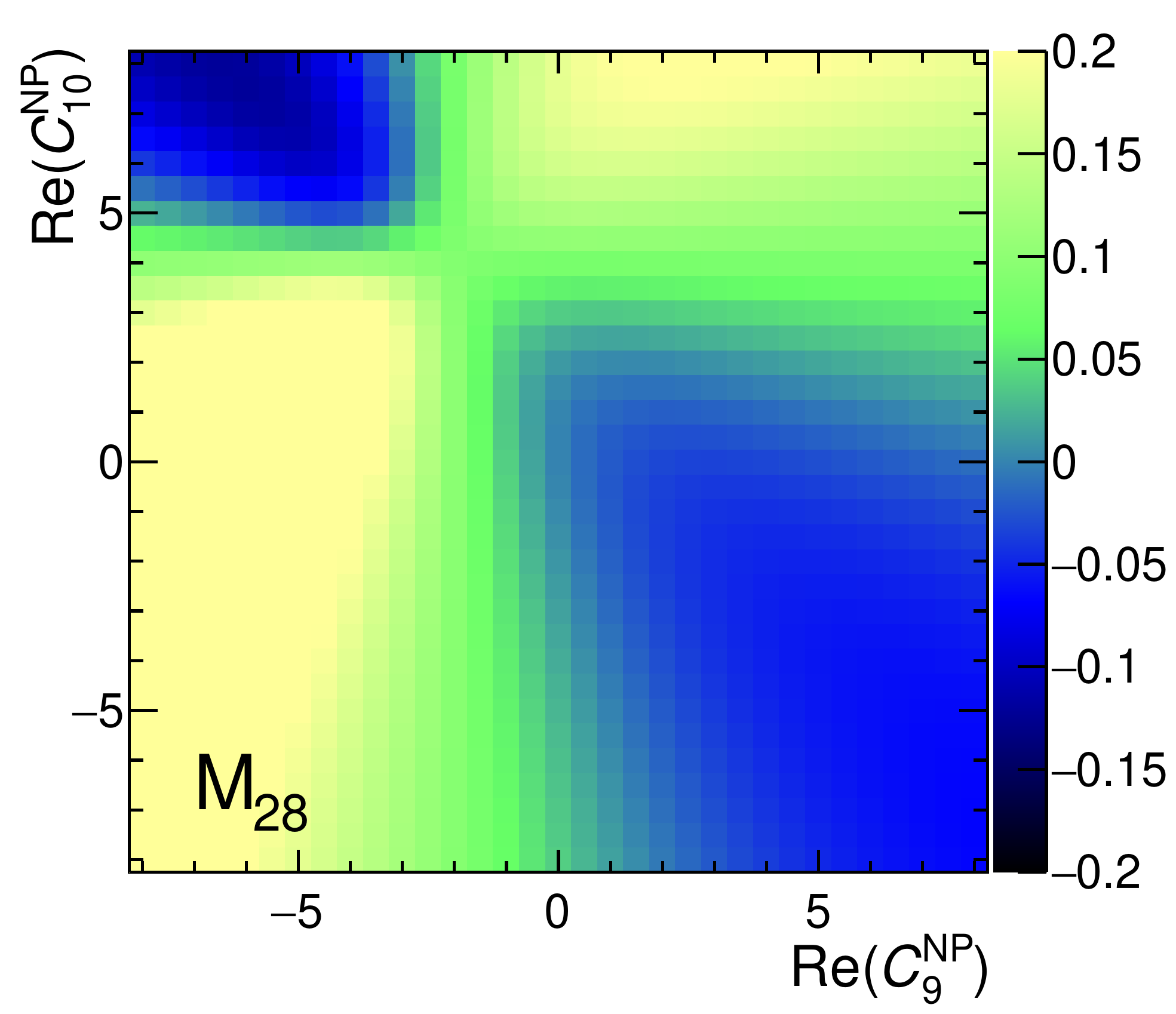} 
\includegraphics[width=0.24\linewidth]{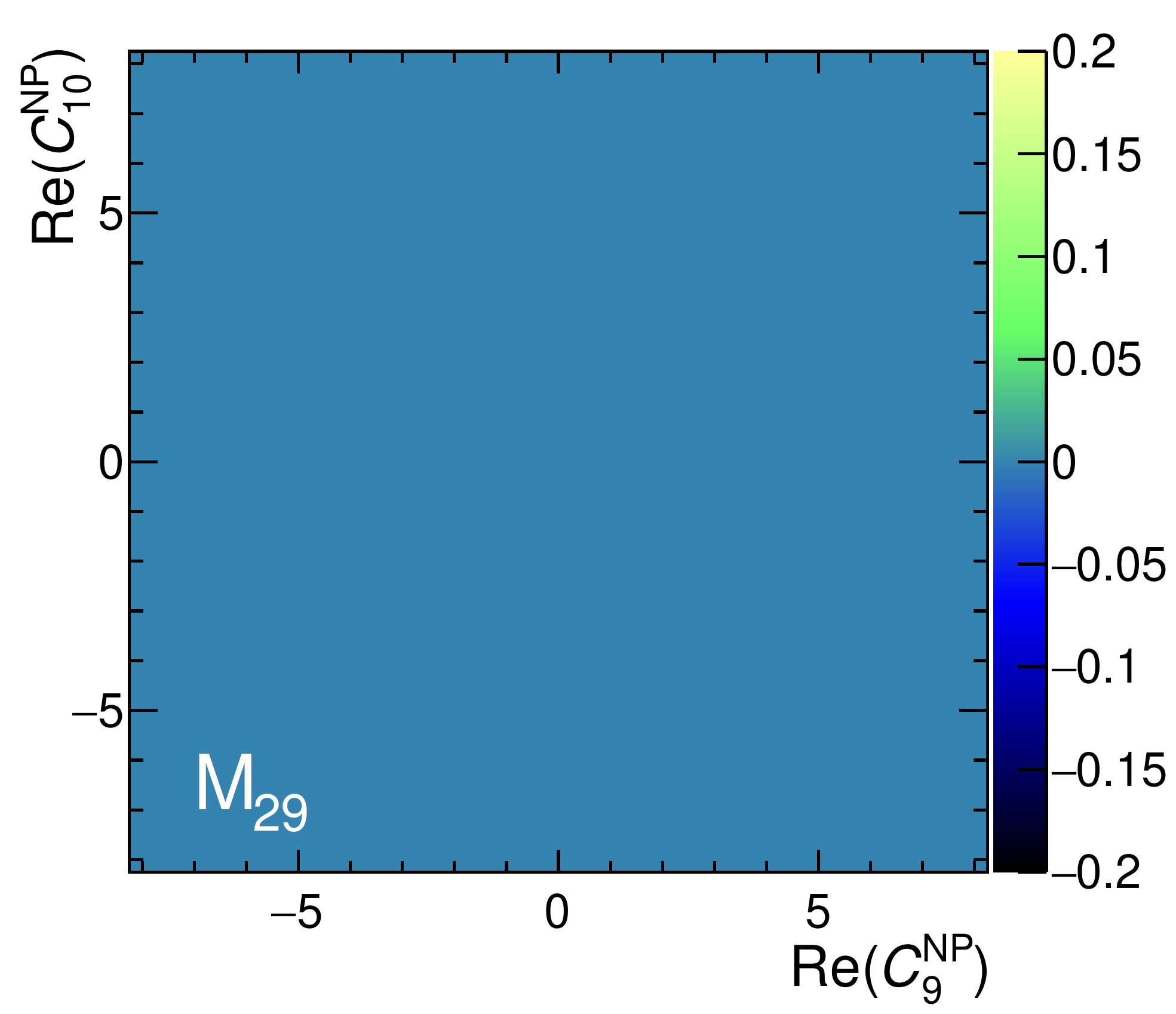} 
\includegraphics[width=0.24\linewidth]{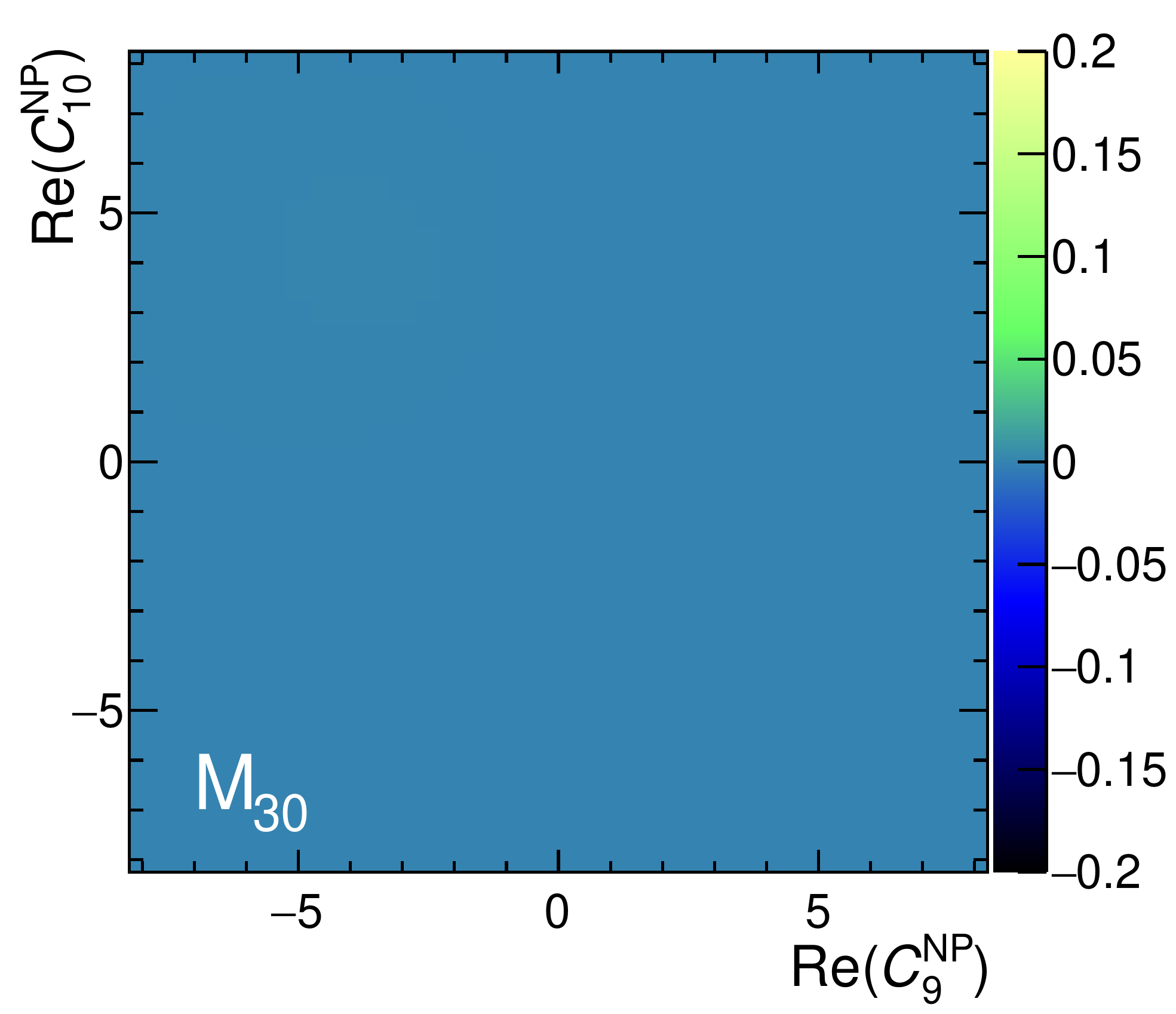} \\ 
\includegraphics[width=0.24\linewidth]{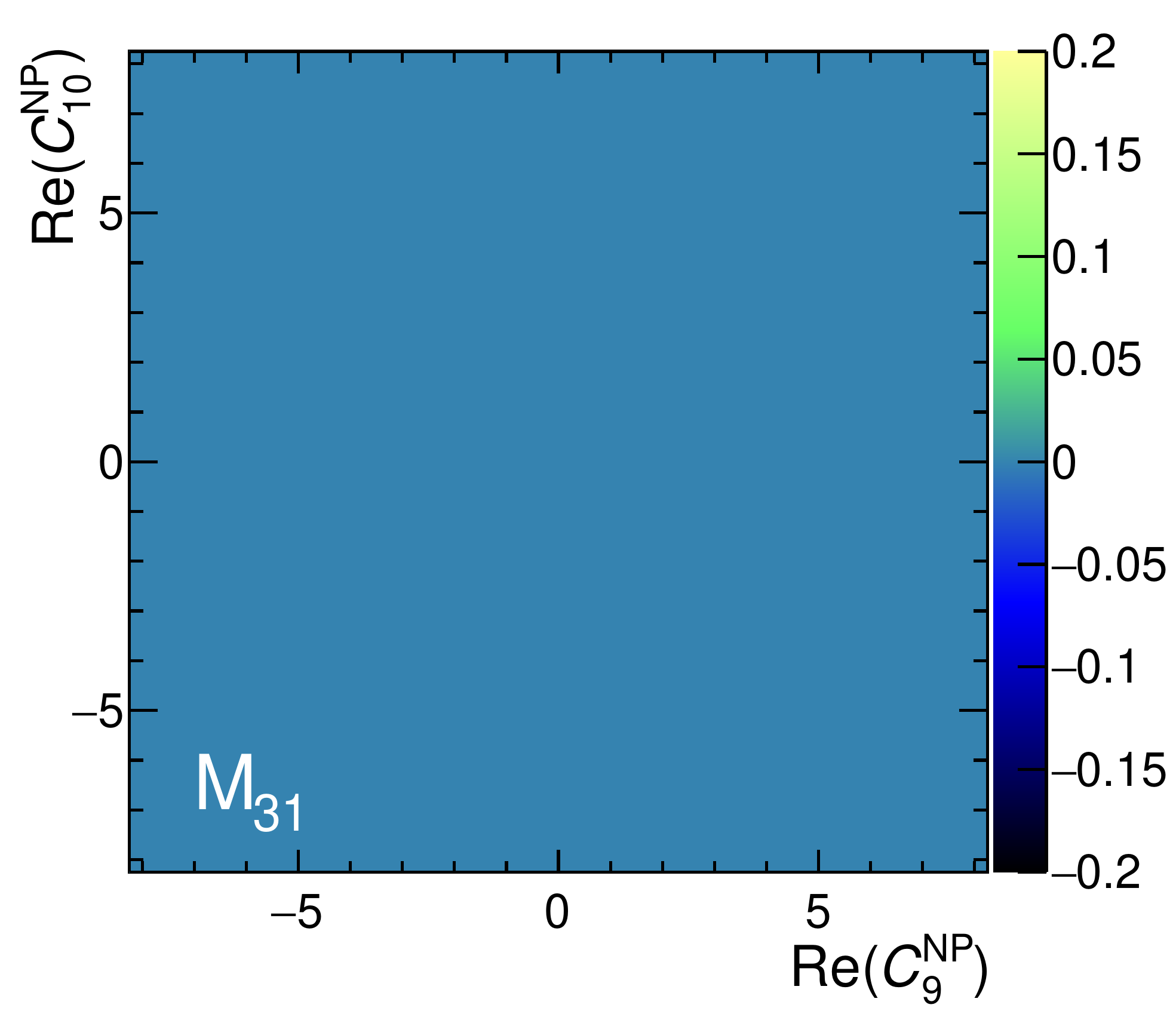}  
\includegraphics[width=0.24\linewidth]{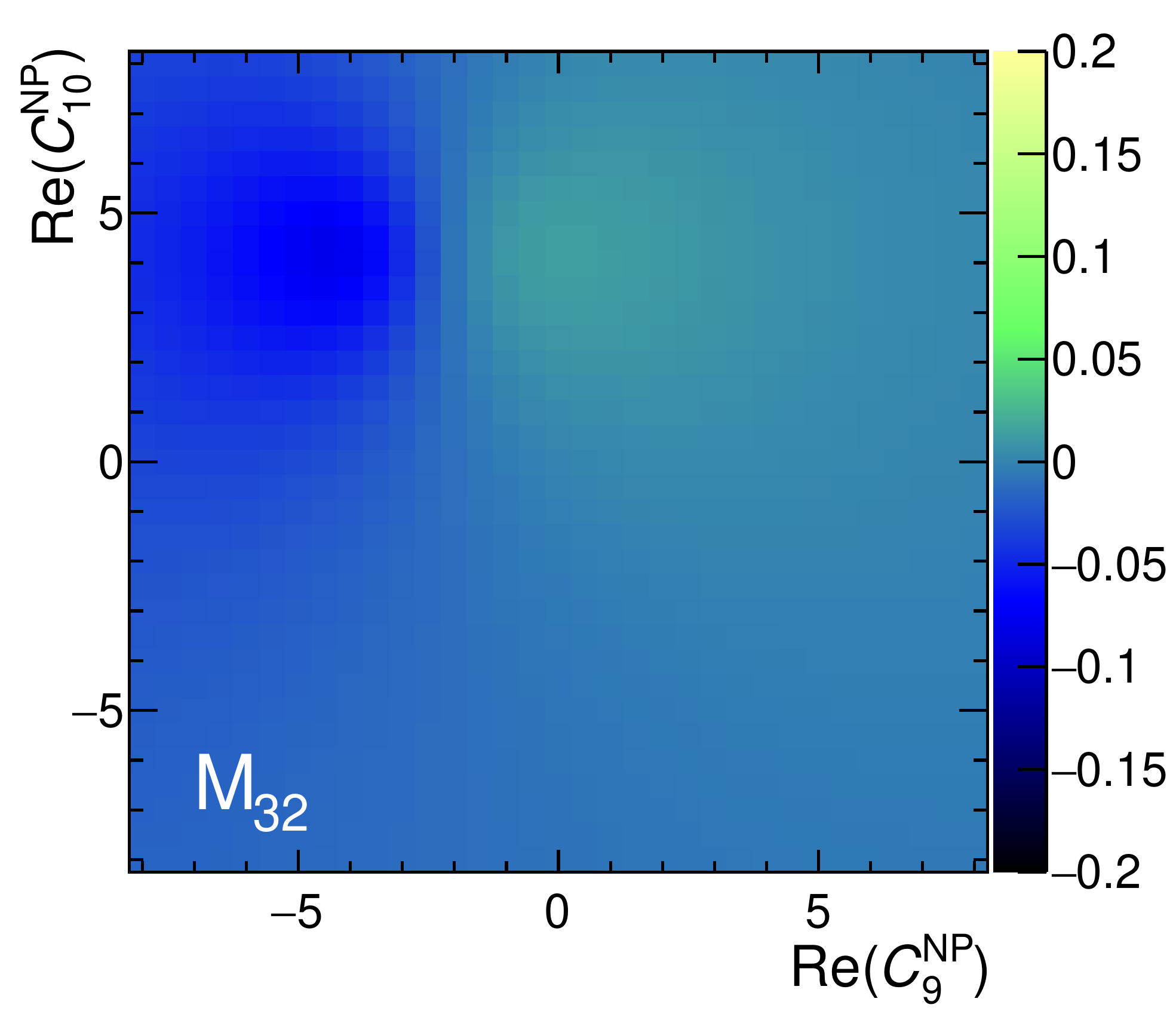} 
\includegraphics[width=0.24\linewidth]{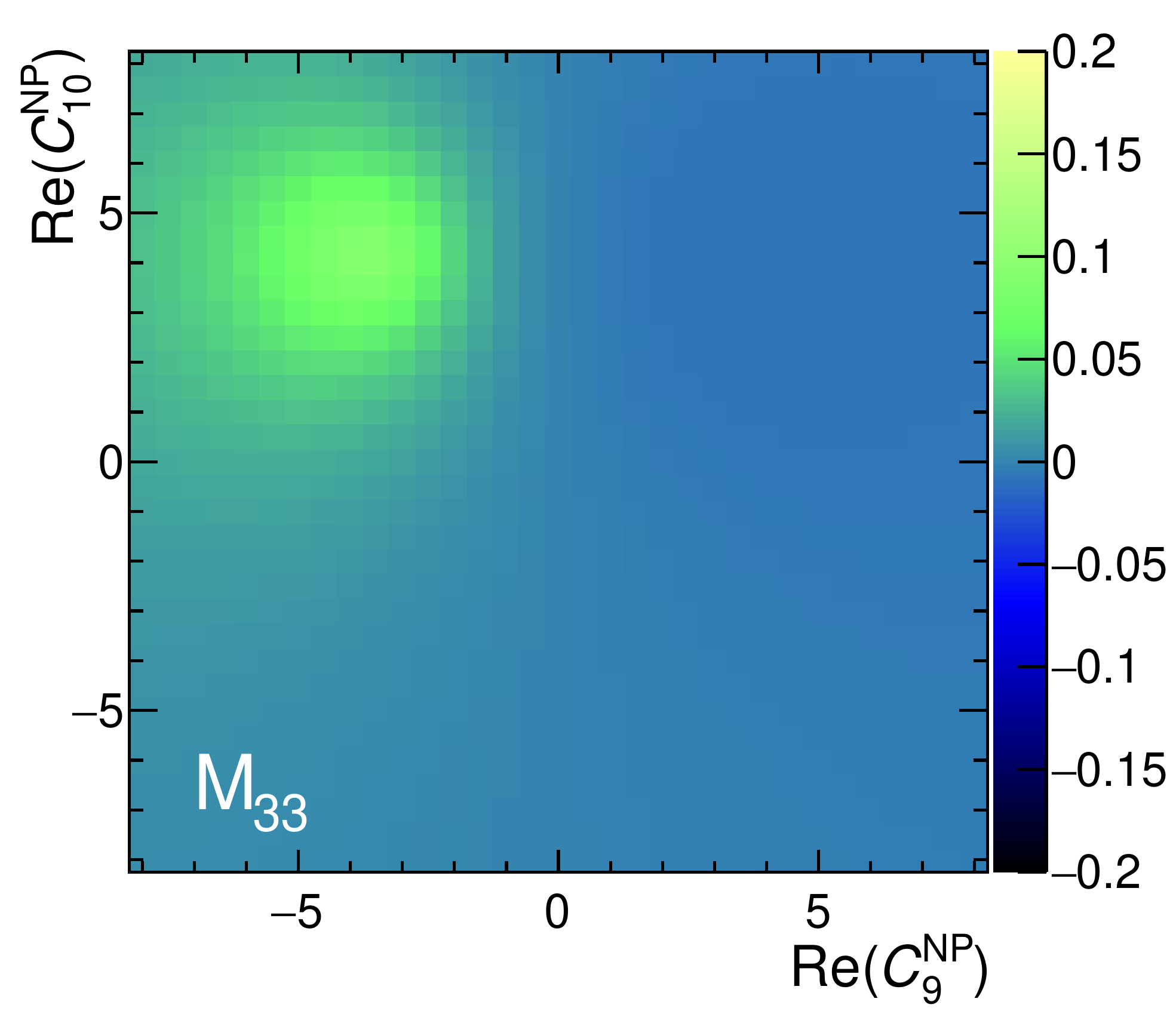} 
\includegraphics[width=0.24\linewidth]{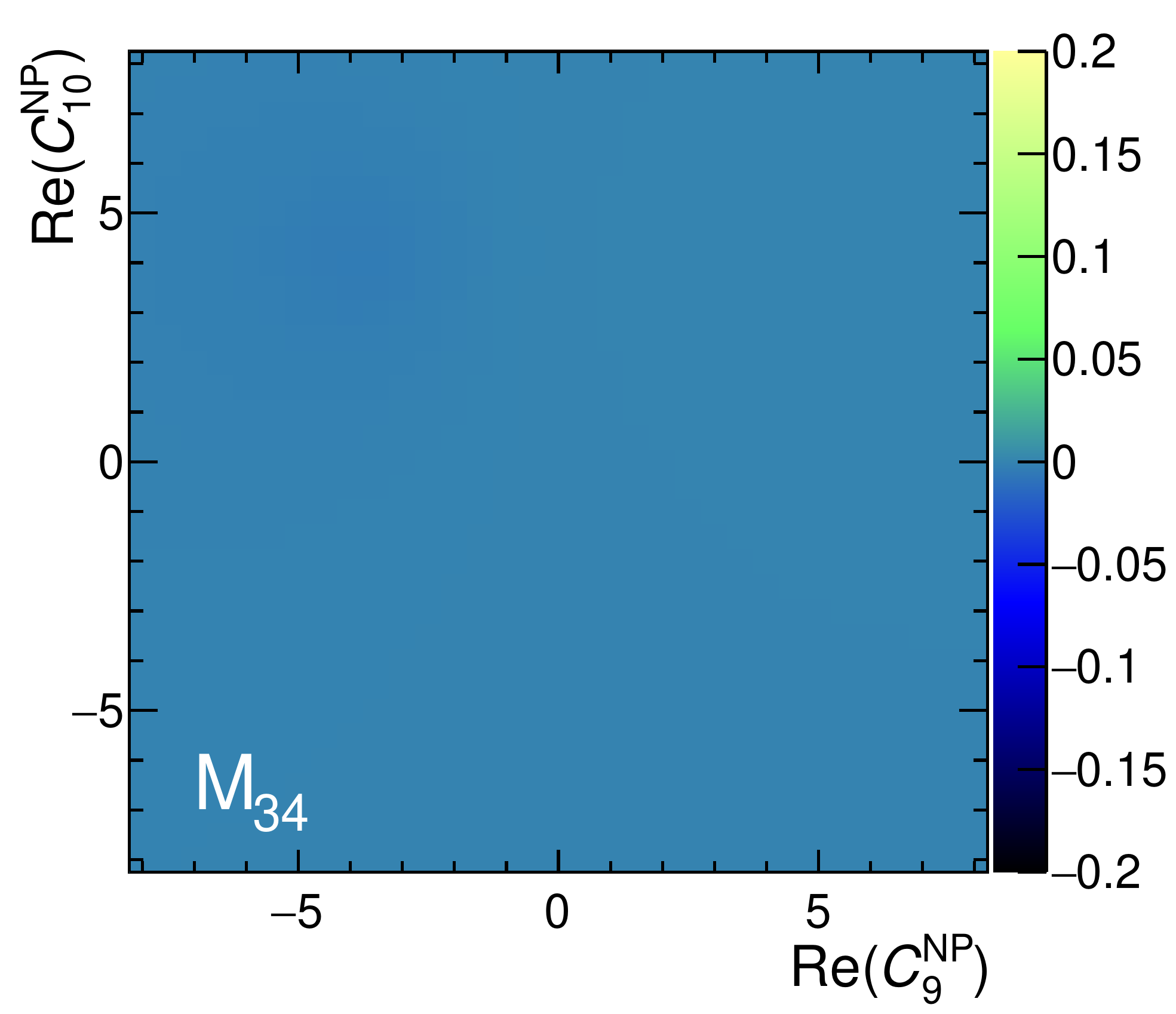}  
\caption{
Variation of the polarisation dependent angular observables of the \decay{\Lb}{\Lz\mumu} decay  from their SM central values in the large-recoil region ($1 < \qsq < 6\gev^{2}/c^{4}$) with a NP contribution to ${\rm Re}(C_9)$ or ${\rm Re}(C_{10})$. 
The SM point is at $(0,0)$.
To illustrate the size of the effects, $P_{\Lb}  = 1$ is used.
\label{fig:scan:c9:c10:largerecoil:pol} 
}
\end{figure}

\begin{figure}[!htb]
\centering
\includegraphics[width=0.24\linewidth]{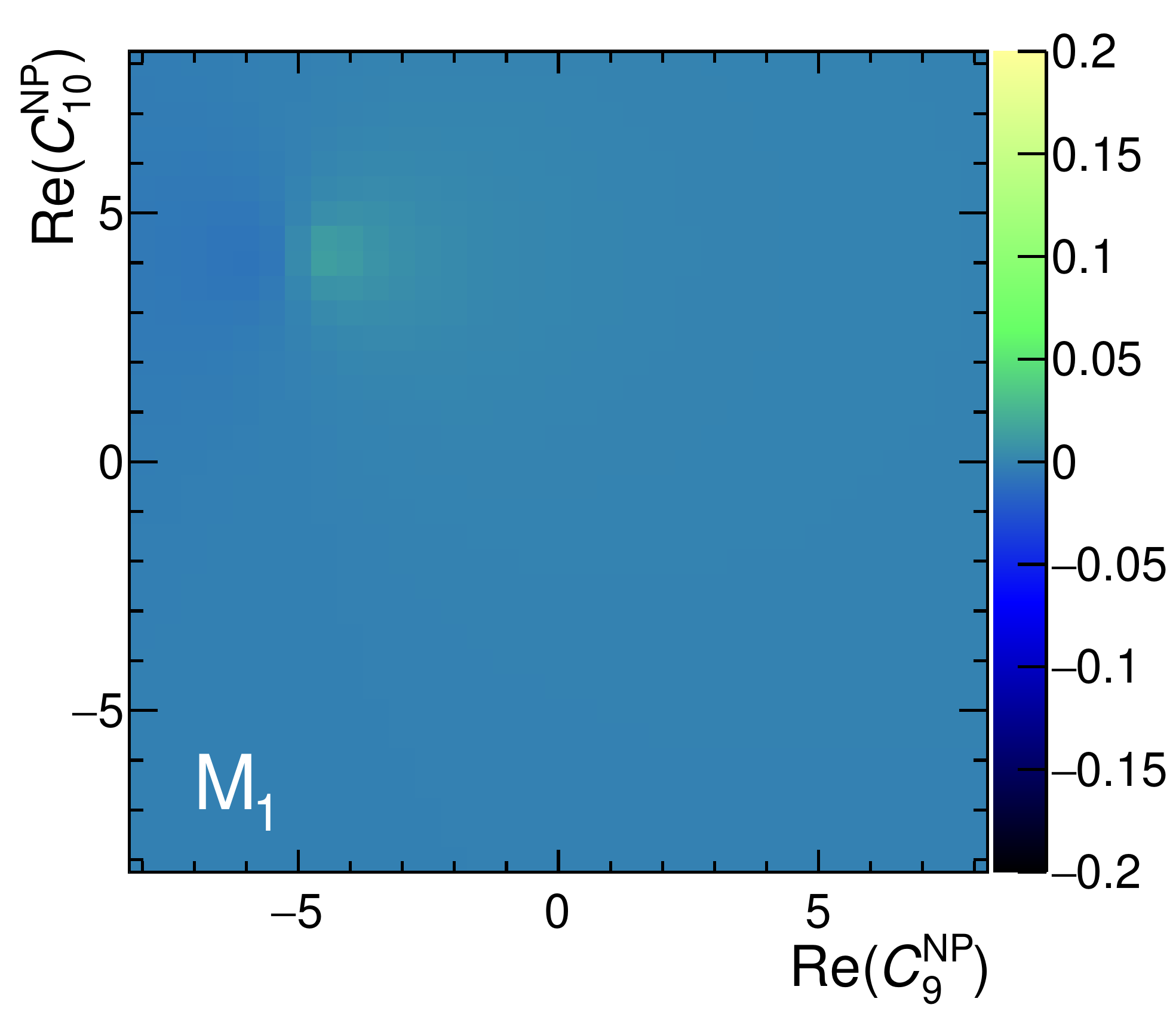}  
\includegraphics[width=0.24\linewidth]{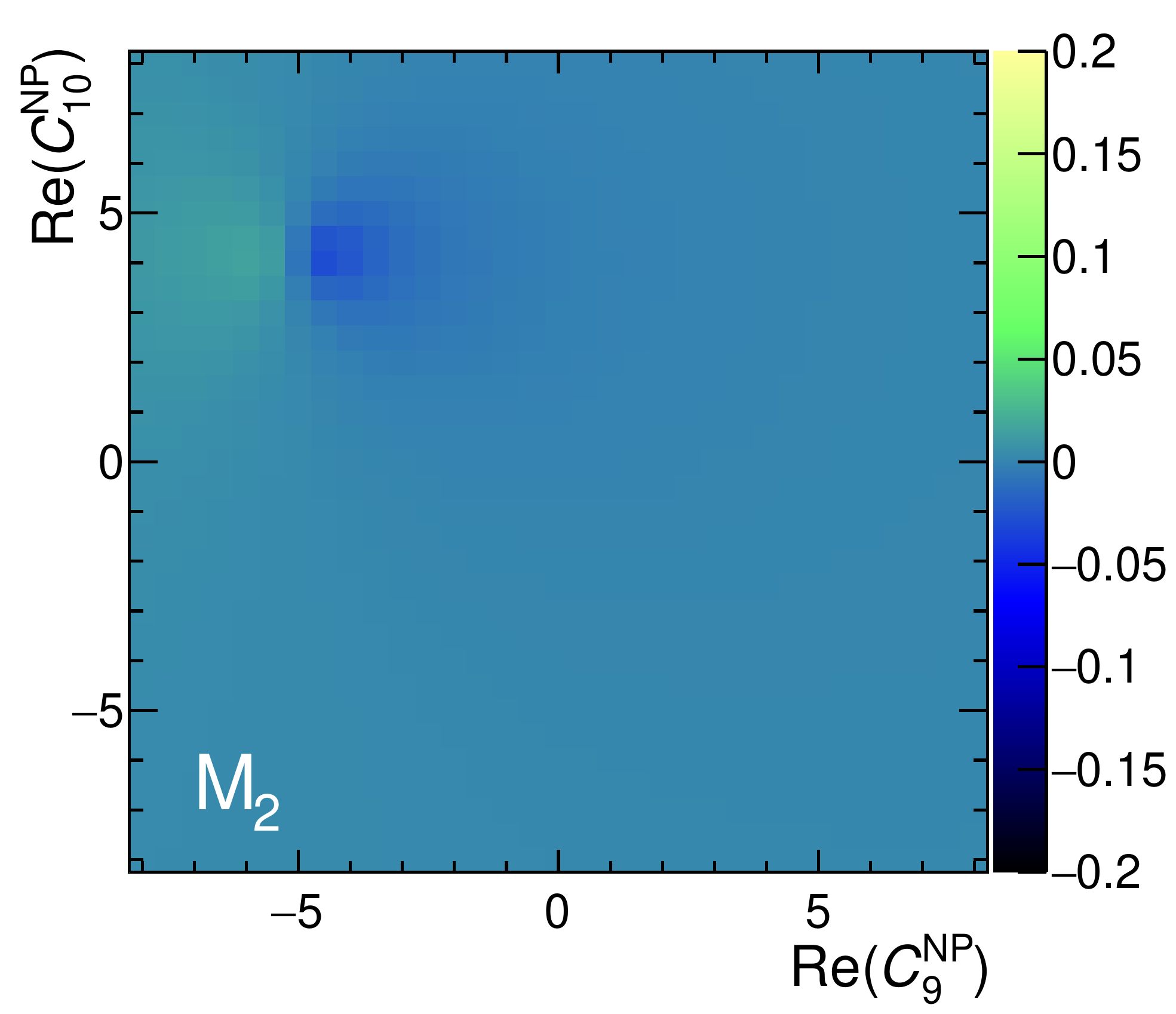} 
\includegraphics[width=0.24\linewidth]{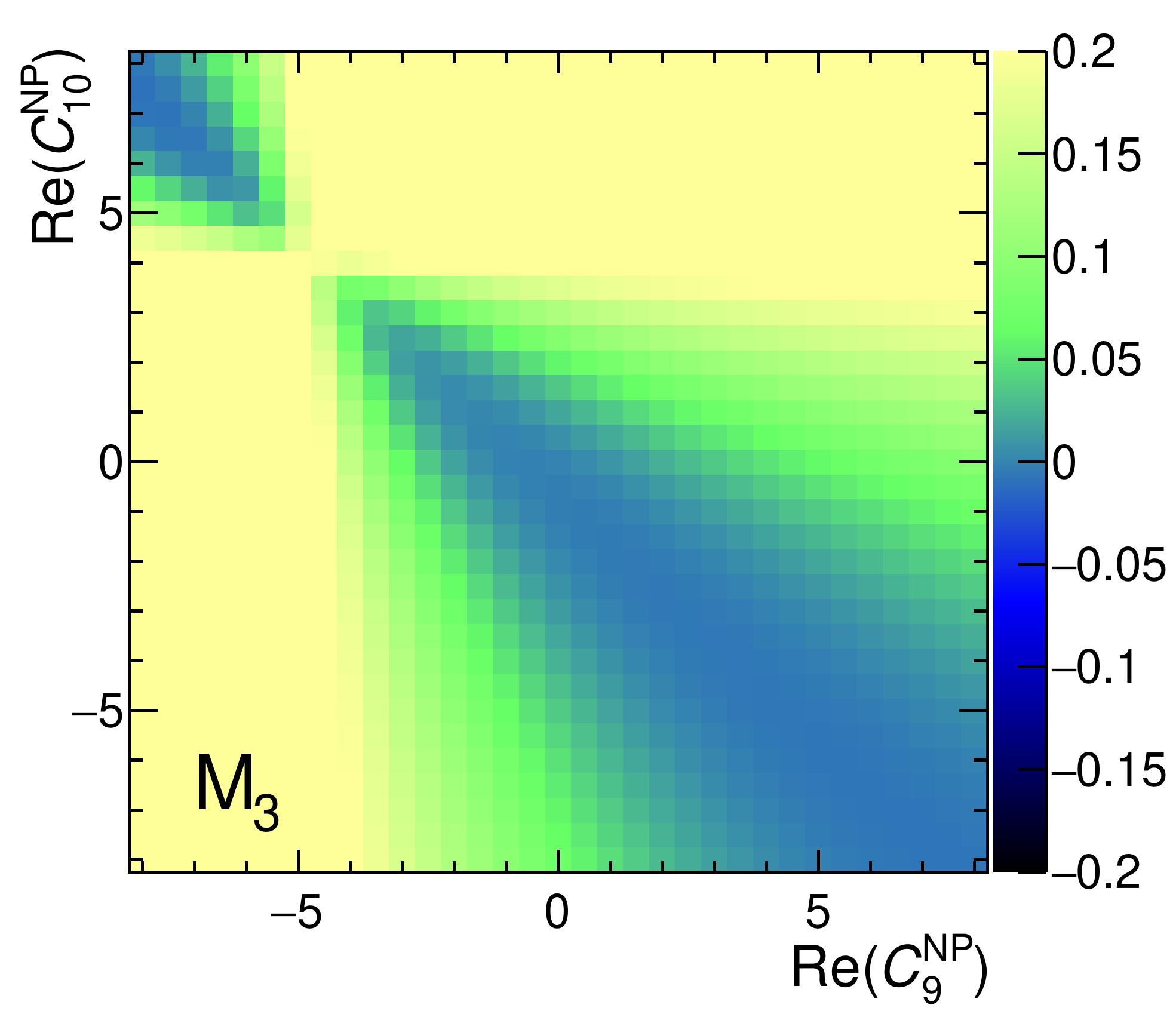} 
\includegraphics[width=0.24\linewidth]{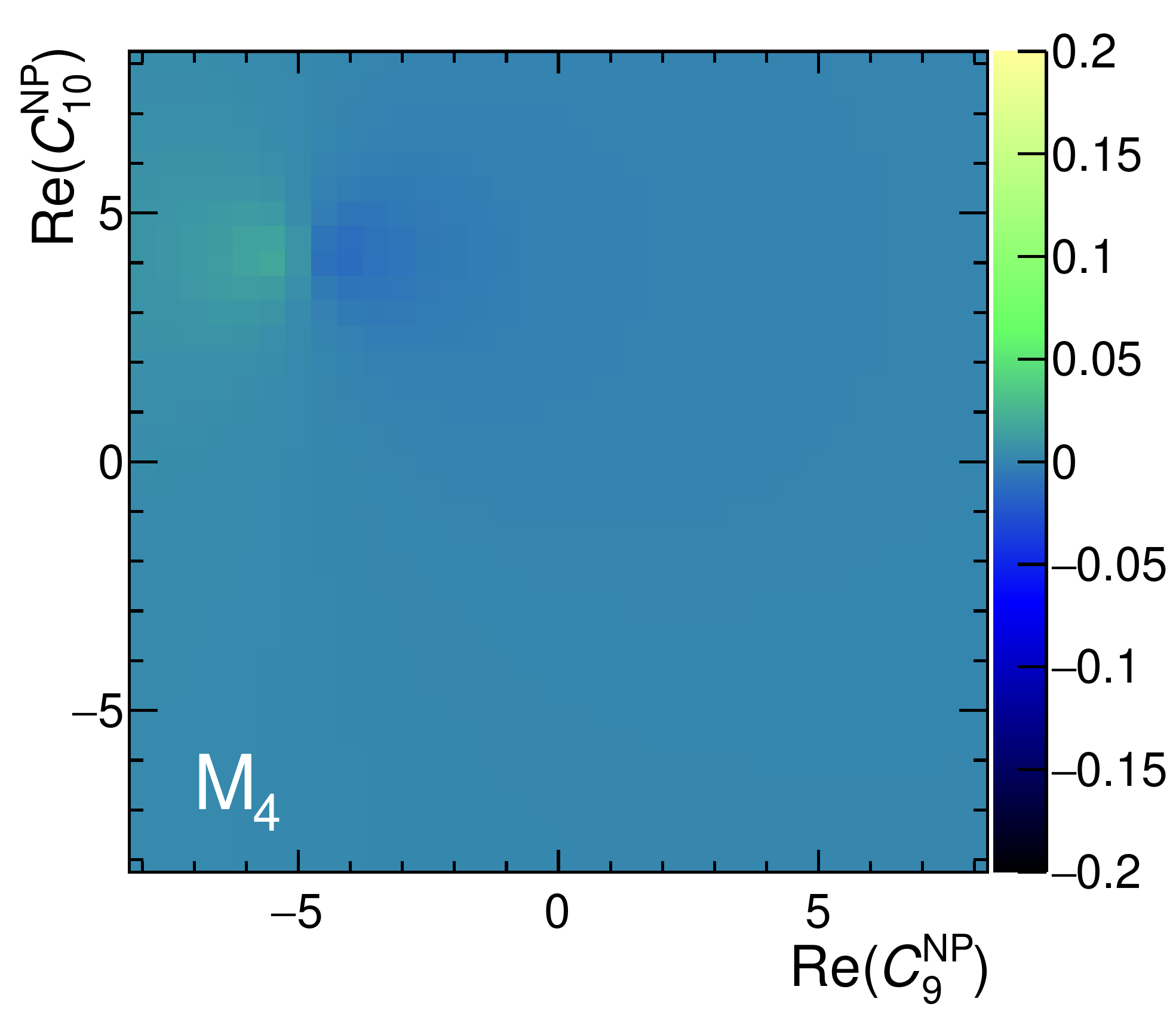} \\ 
\includegraphics[width=0.24\linewidth]{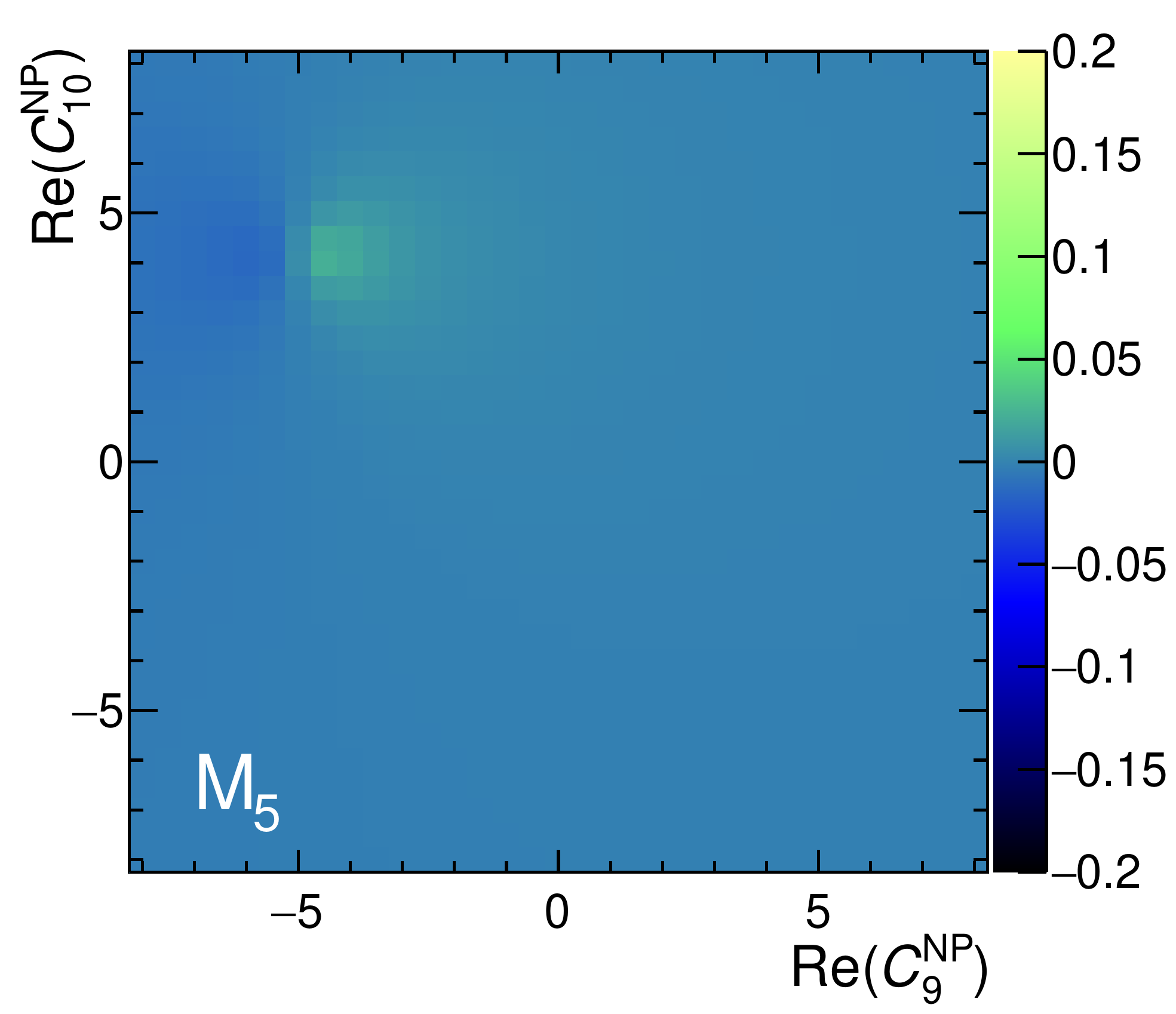}  
\includegraphics[width=0.24\linewidth]{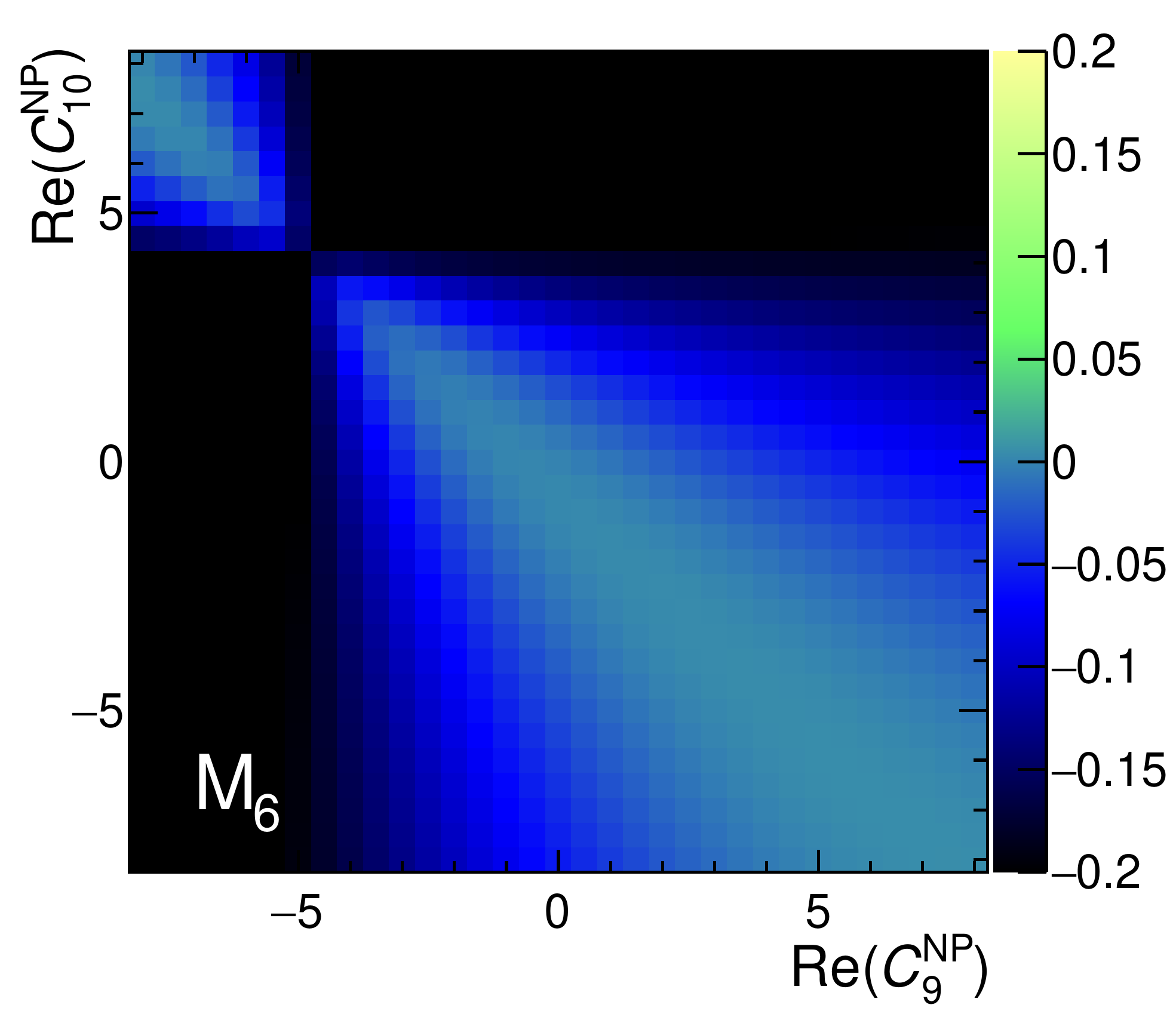} 
\includegraphics[width=0.24\linewidth]{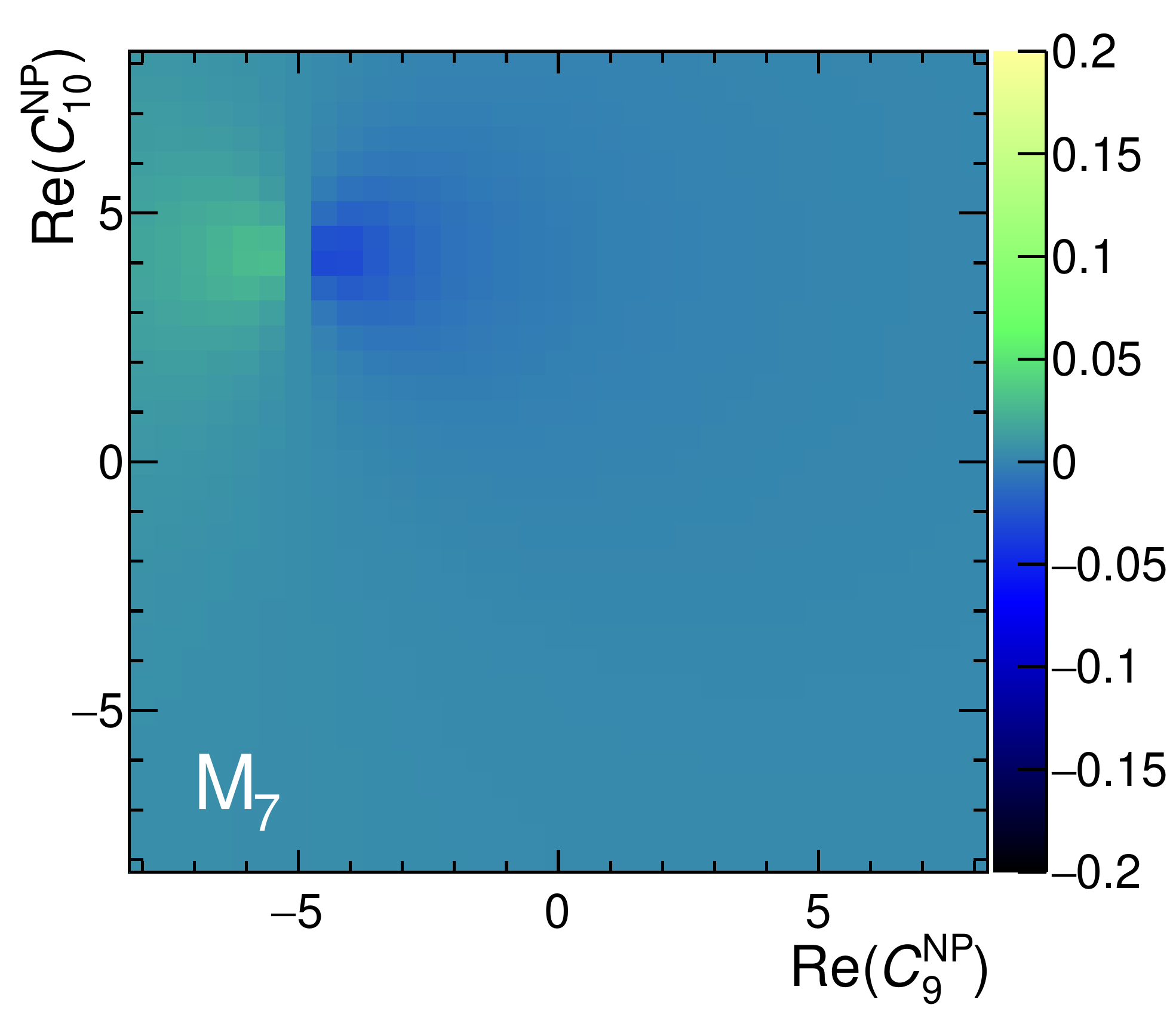} 
\includegraphics[width=0.24\linewidth]{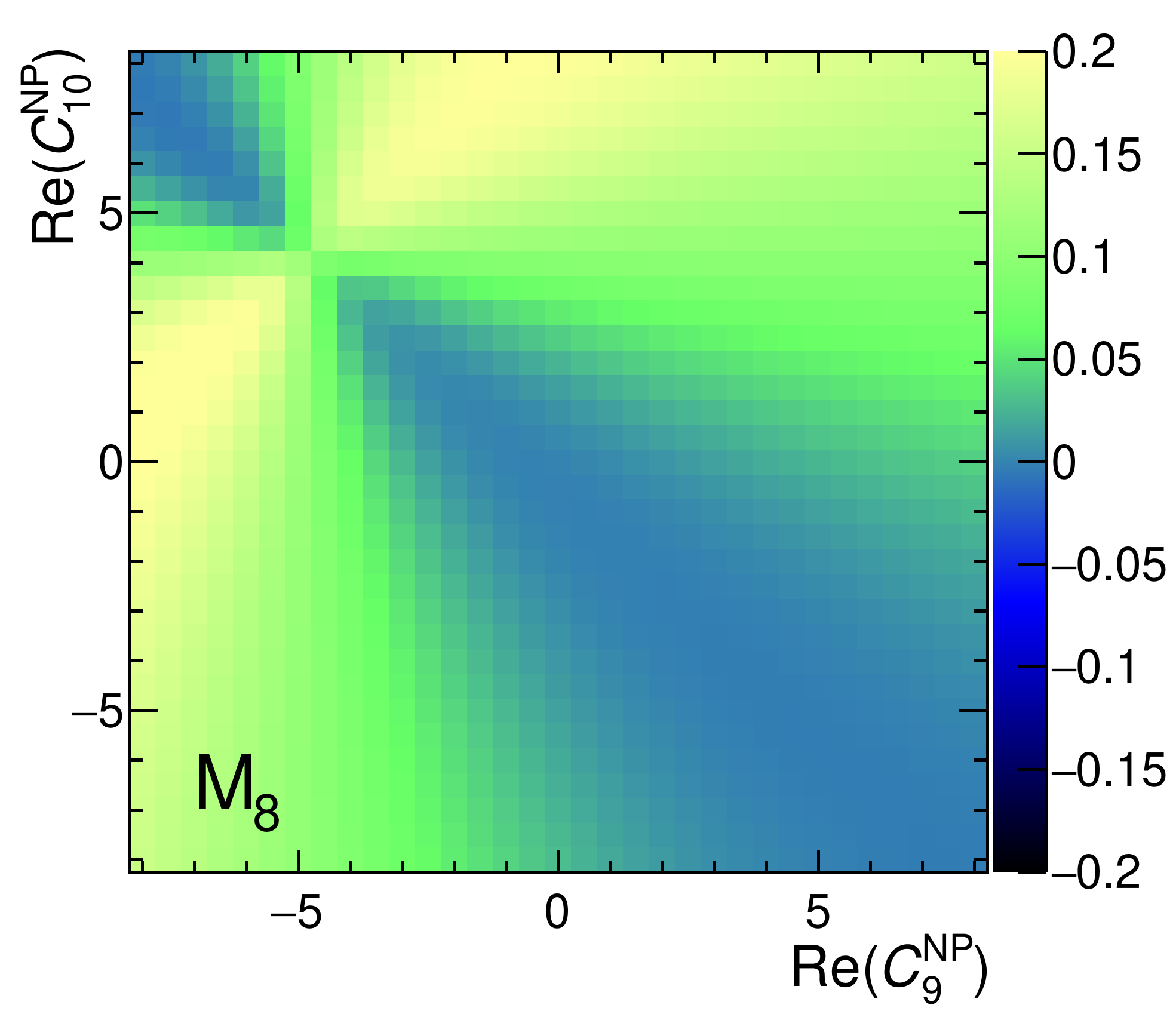} \\ 
\includegraphics[width=0.24\linewidth]{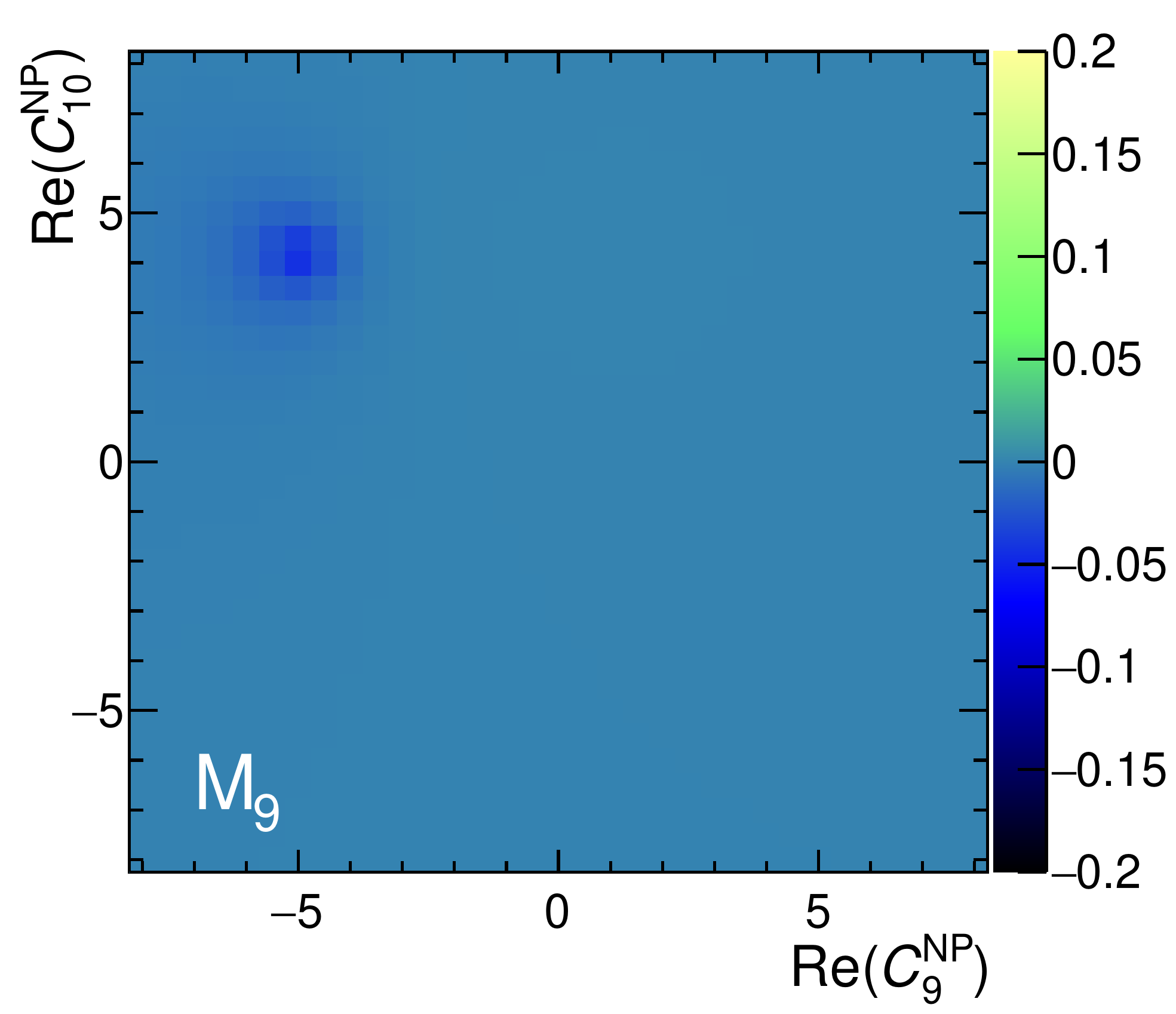}  
\includegraphics[width=0.24\linewidth]{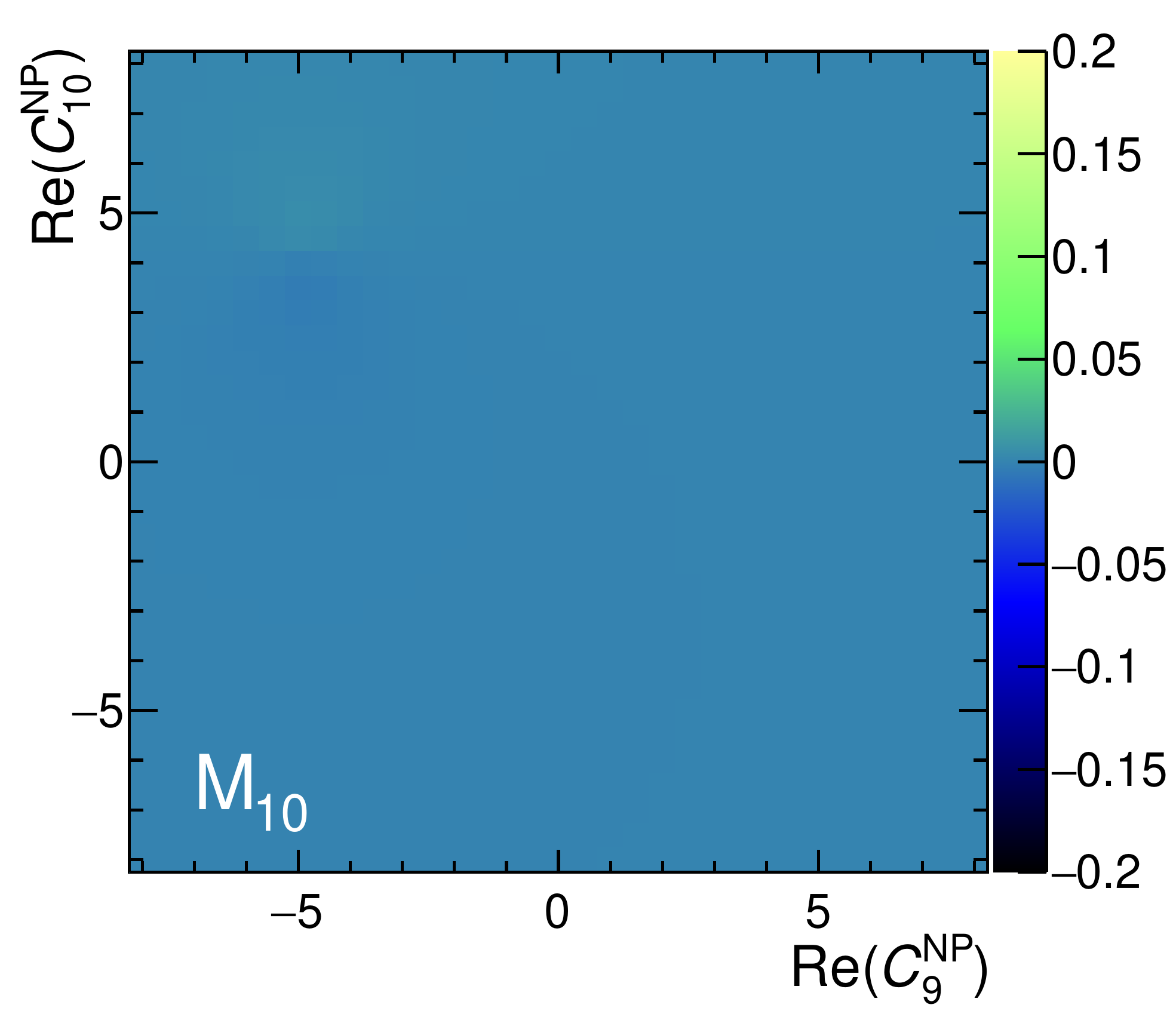} 
\caption{
Variation of the observables $M_{1}$--$M_{10}$  of the \decay{\Lb}{\Lz\mumu} decay  from their SM central values in the low-recoil region ($15 < \qsq < 20\gev^{2}/c^{4}$) with a NP contribution to ${\rm Re}(C_9)$ or ${\rm Re}(C_{10})$. 
The SM point is at $(0,0)$.
\label{fig:scan:c9:c10:lowrecoil} 
}
\end{figure}

\begin{figure}[!htb]
\centering
\includegraphics[width=0.24\linewidth]{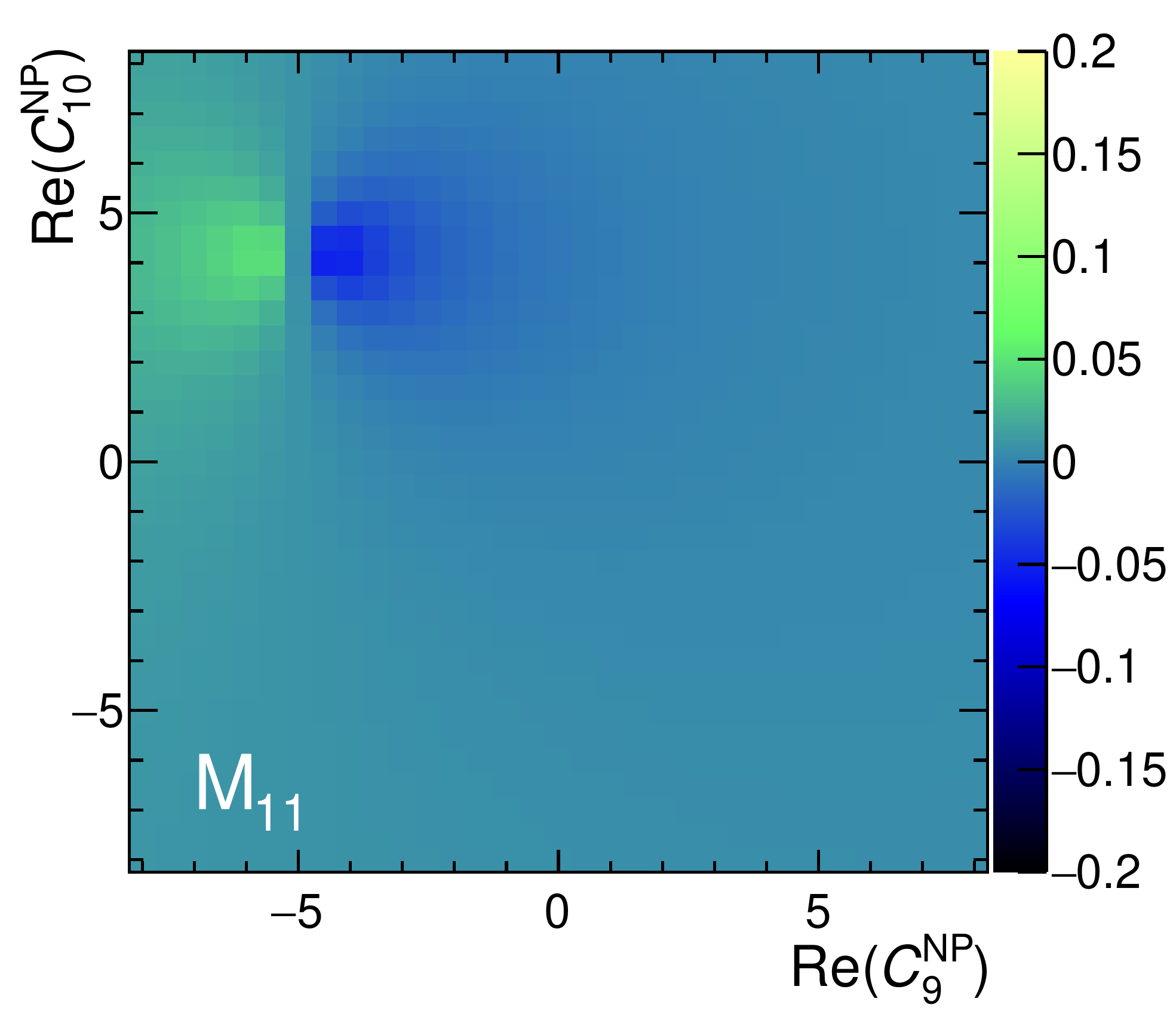} 
\includegraphics[width=0.24\linewidth]{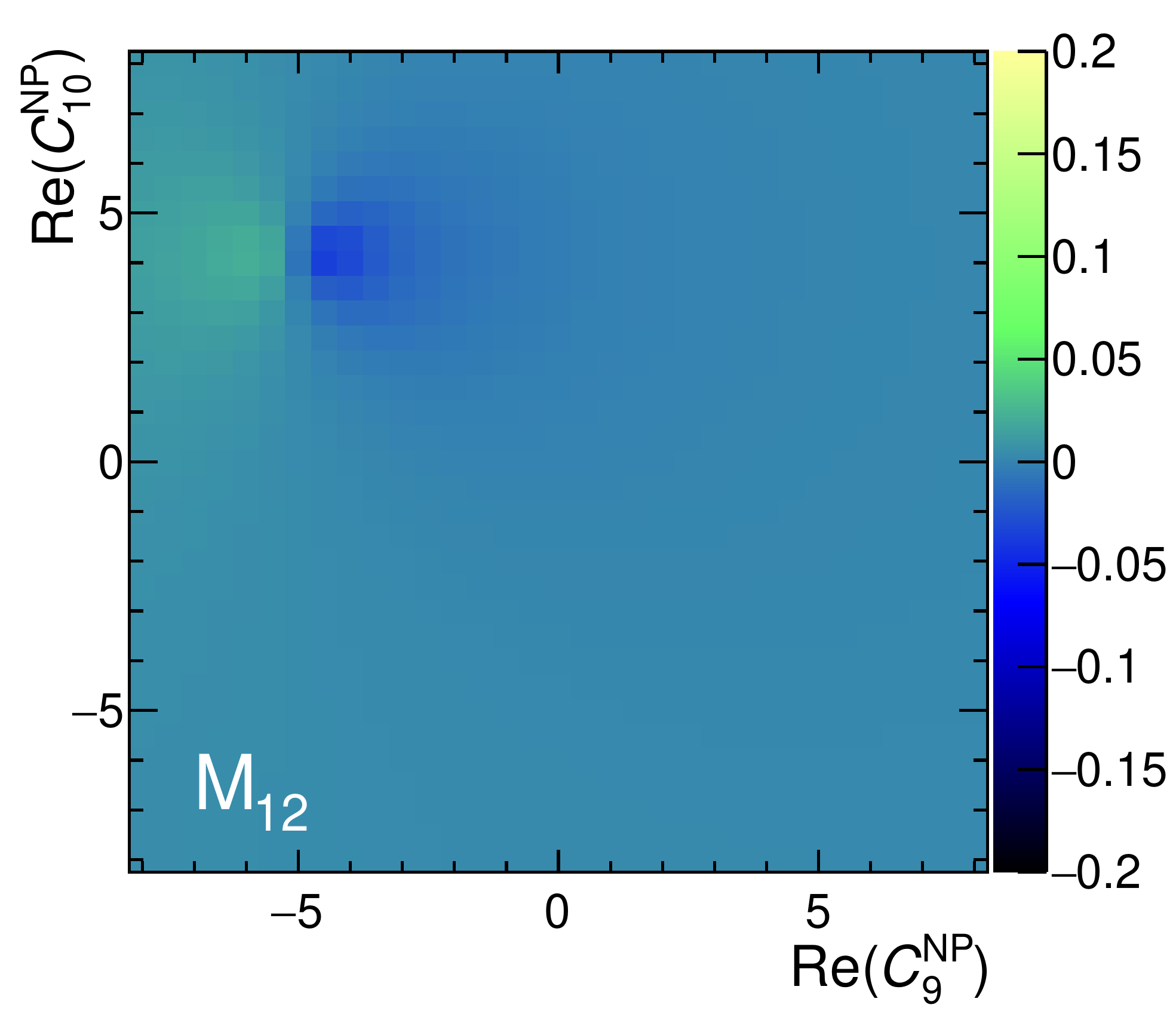} 
\includegraphics[width=0.24\linewidth]{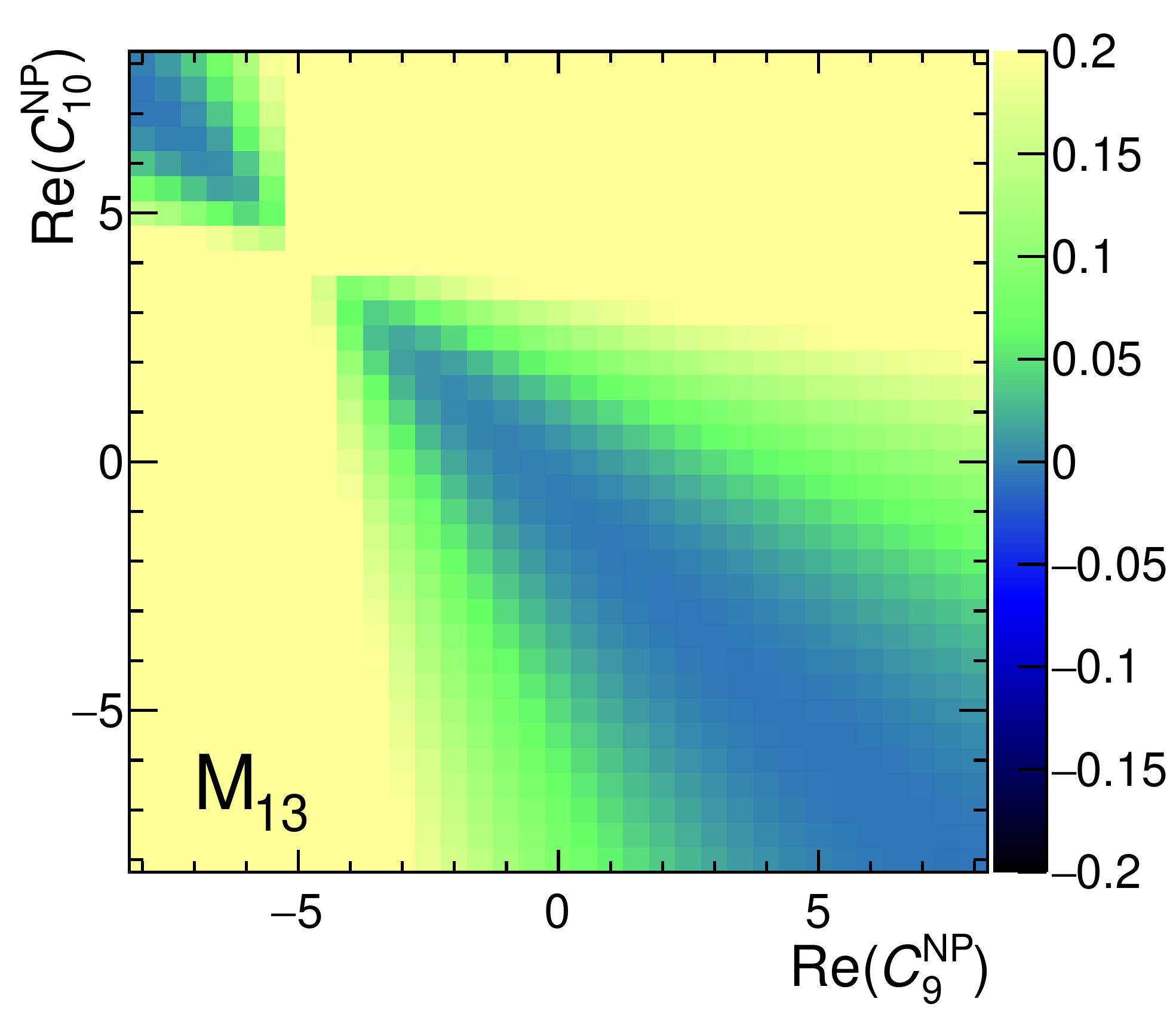}  
\includegraphics[width=0.24\linewidth]{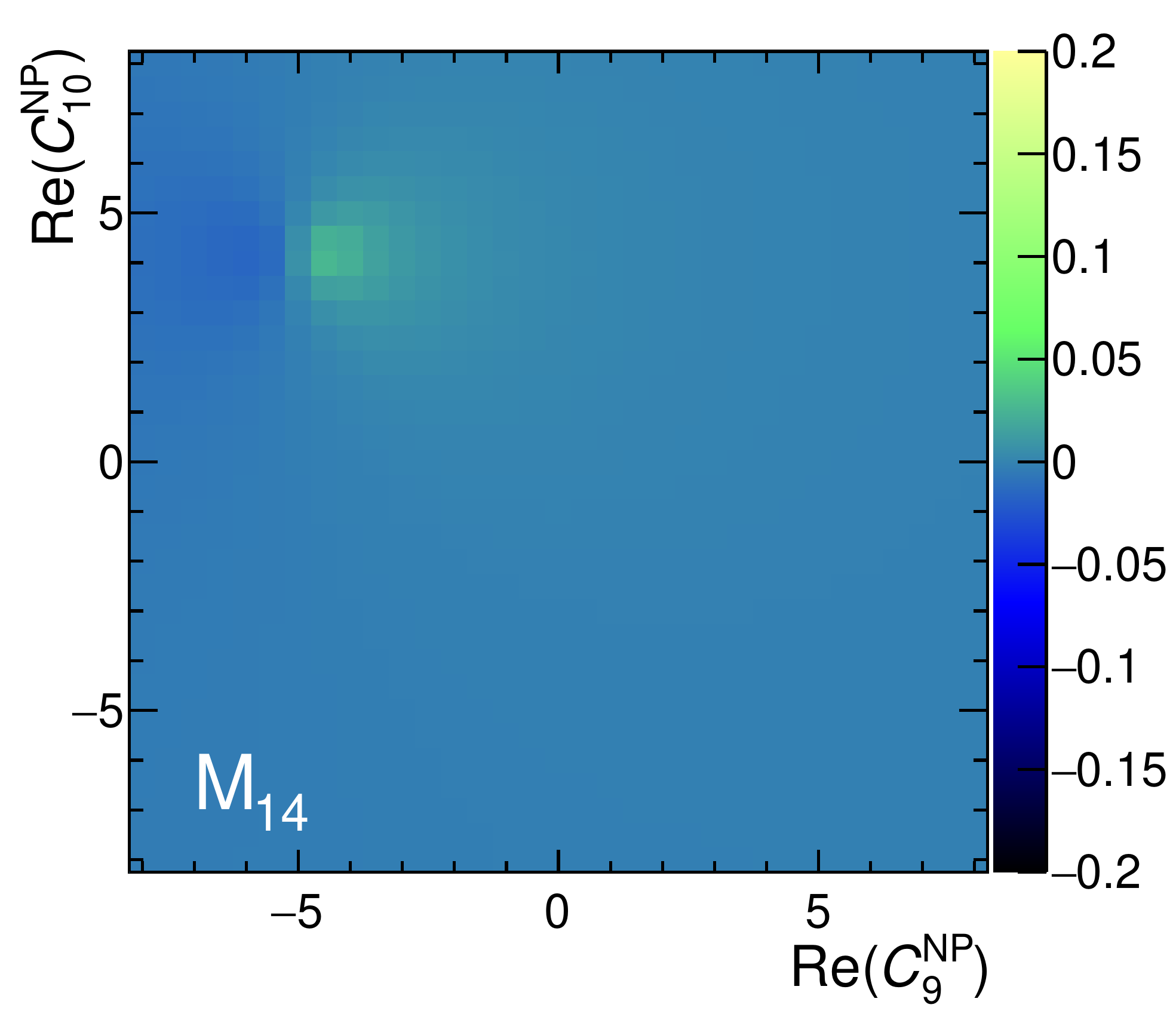}  \\ 
\includegraphics[width=0.24\linewidth]{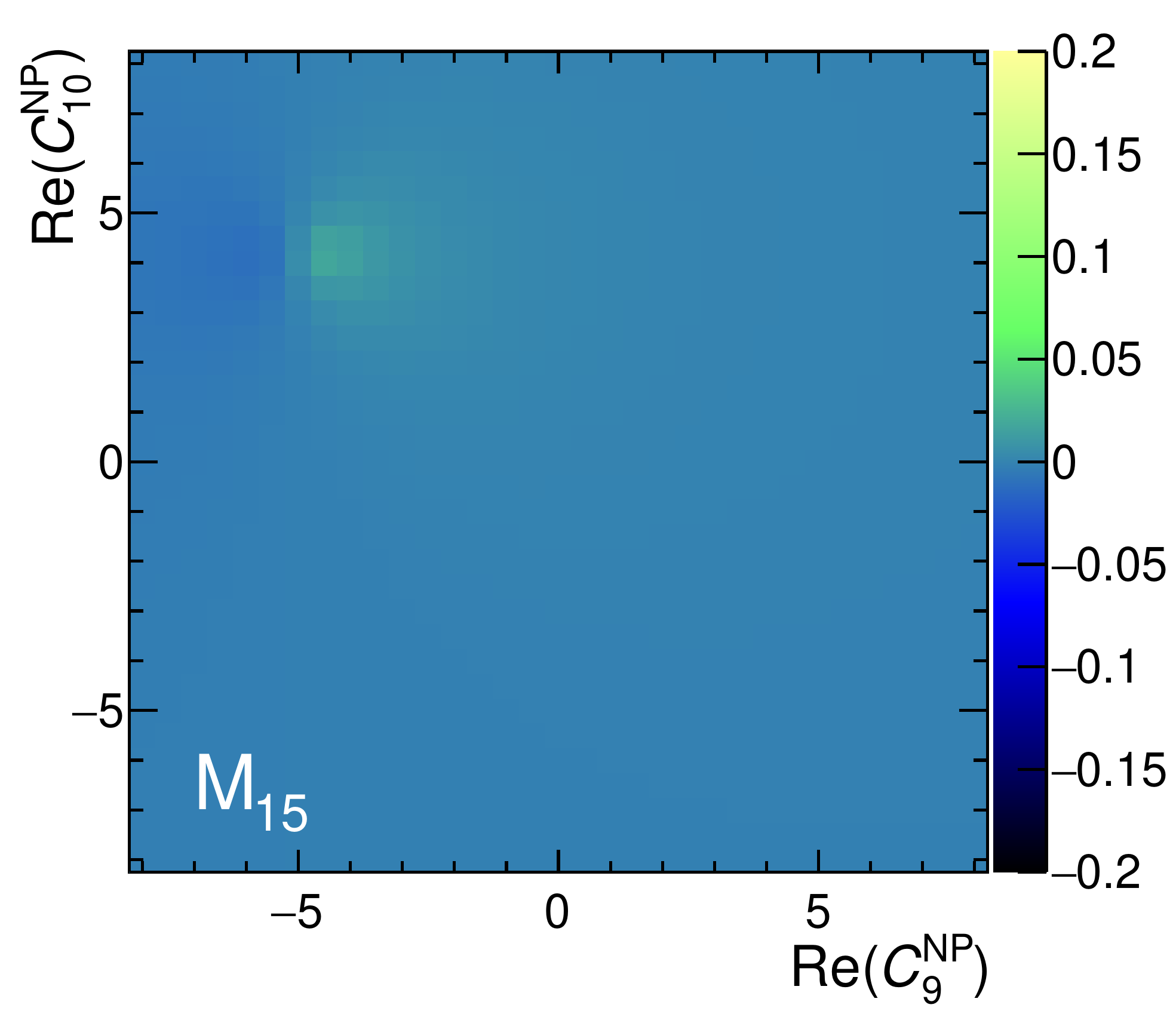} 
\includegraphics[width=0.24\linewidth]{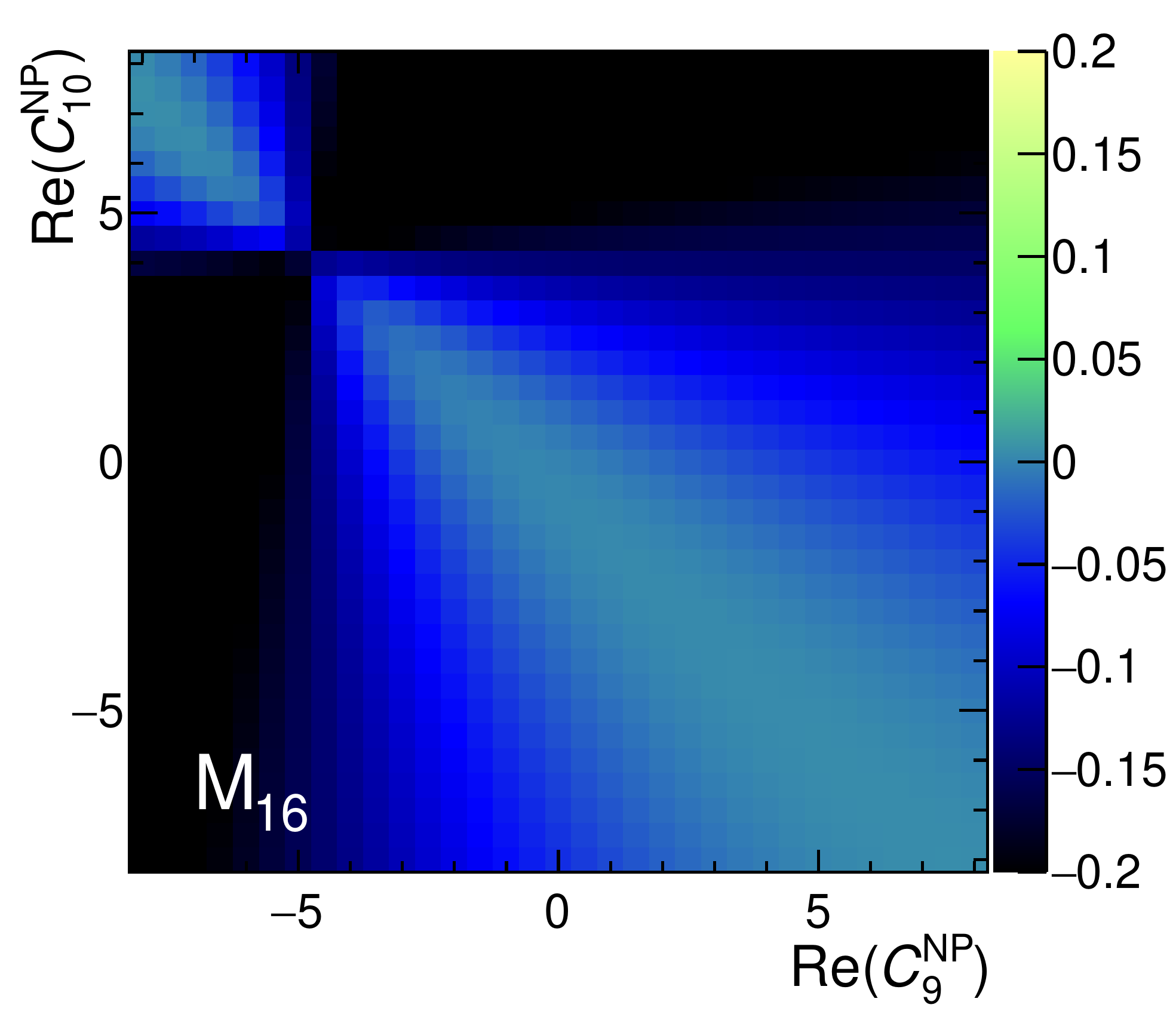} 
\includegraphics[width=0.24\linewidth]{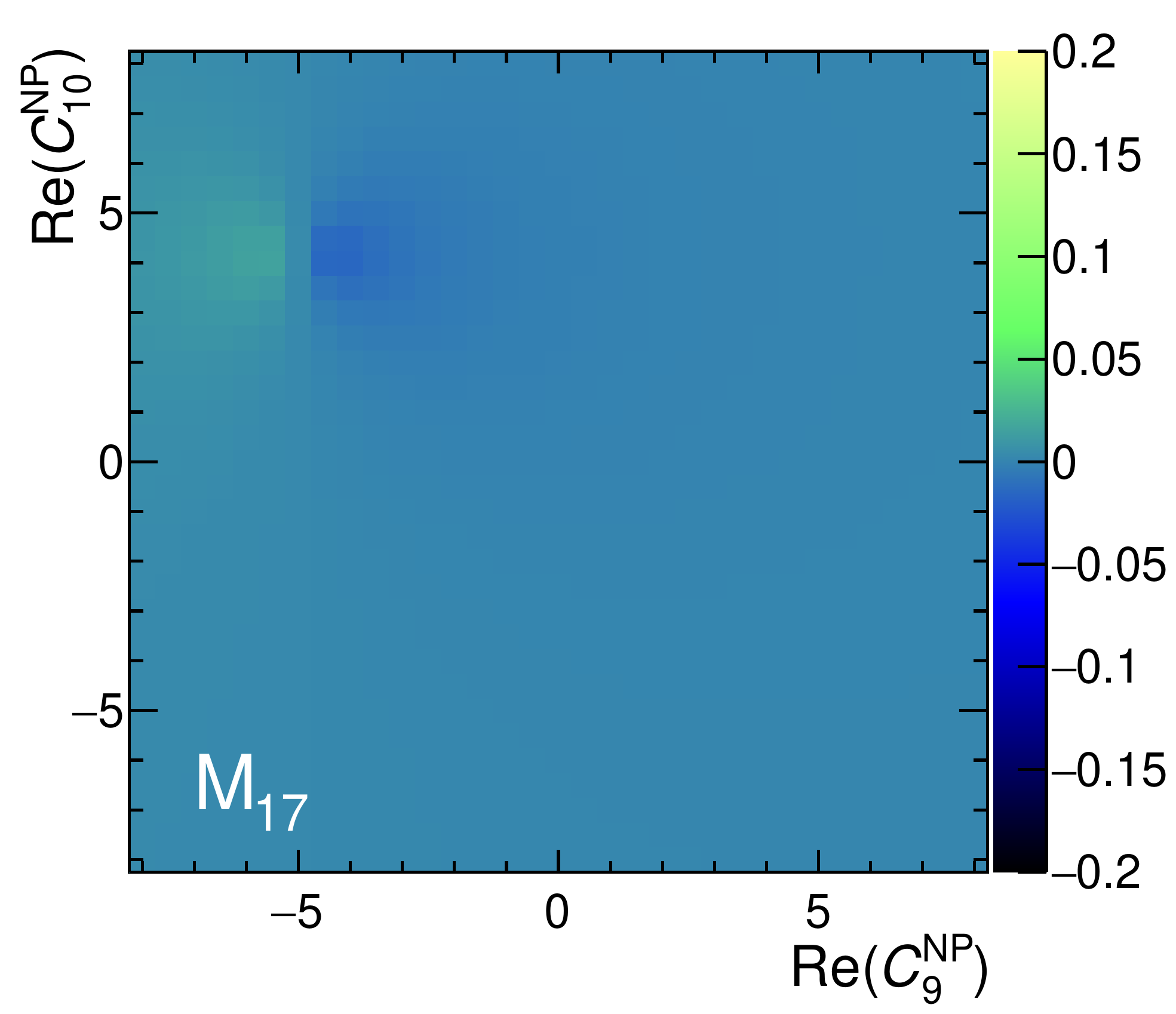}  
\includegraphics[width=0.24\linewidth]{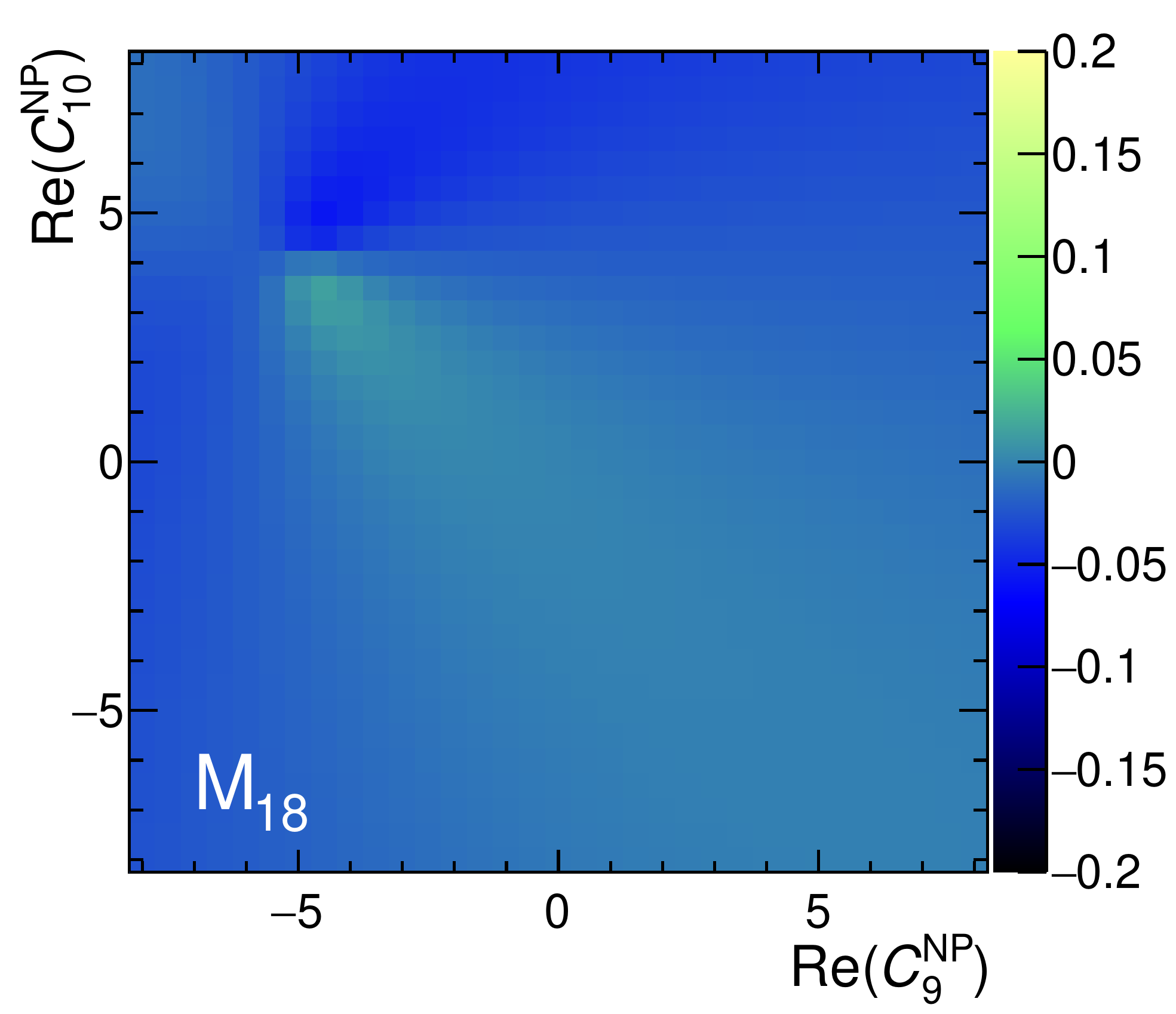} \\
\includegraphics[width=0.24\linewidth]{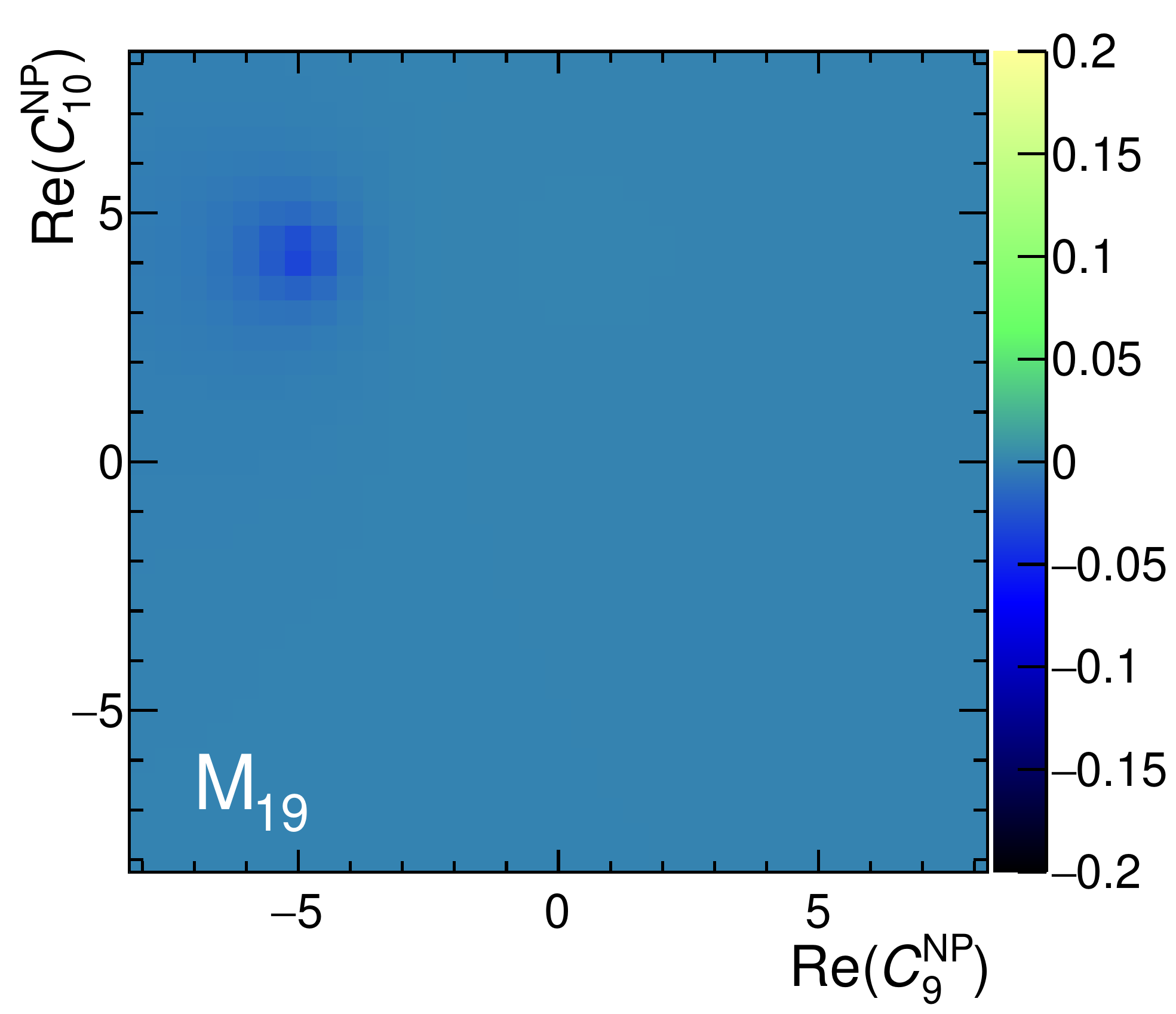} 
\includegraphics[width=0.24\linewidth]{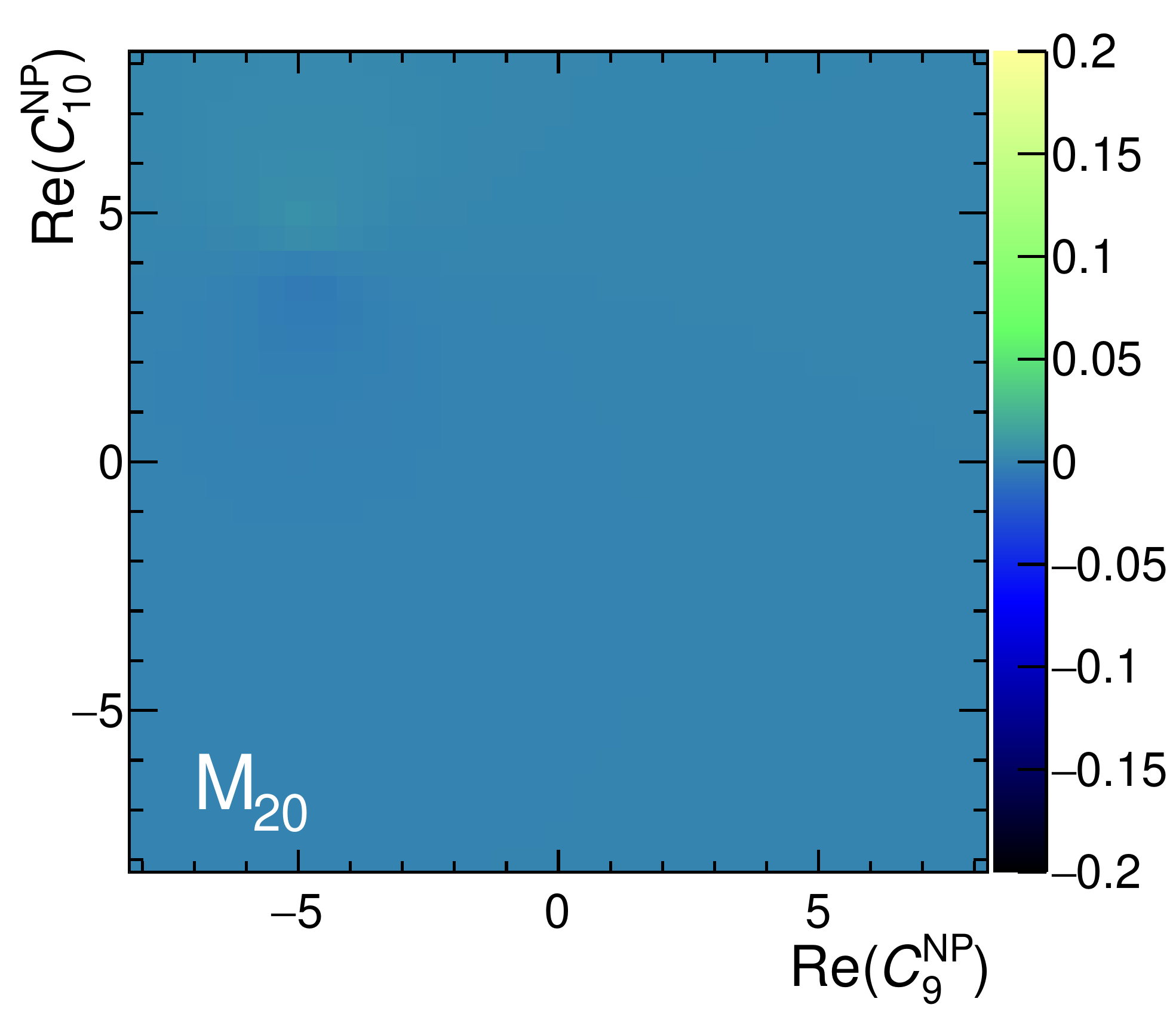}  
\includegraphics[width=0.24\linewidth]{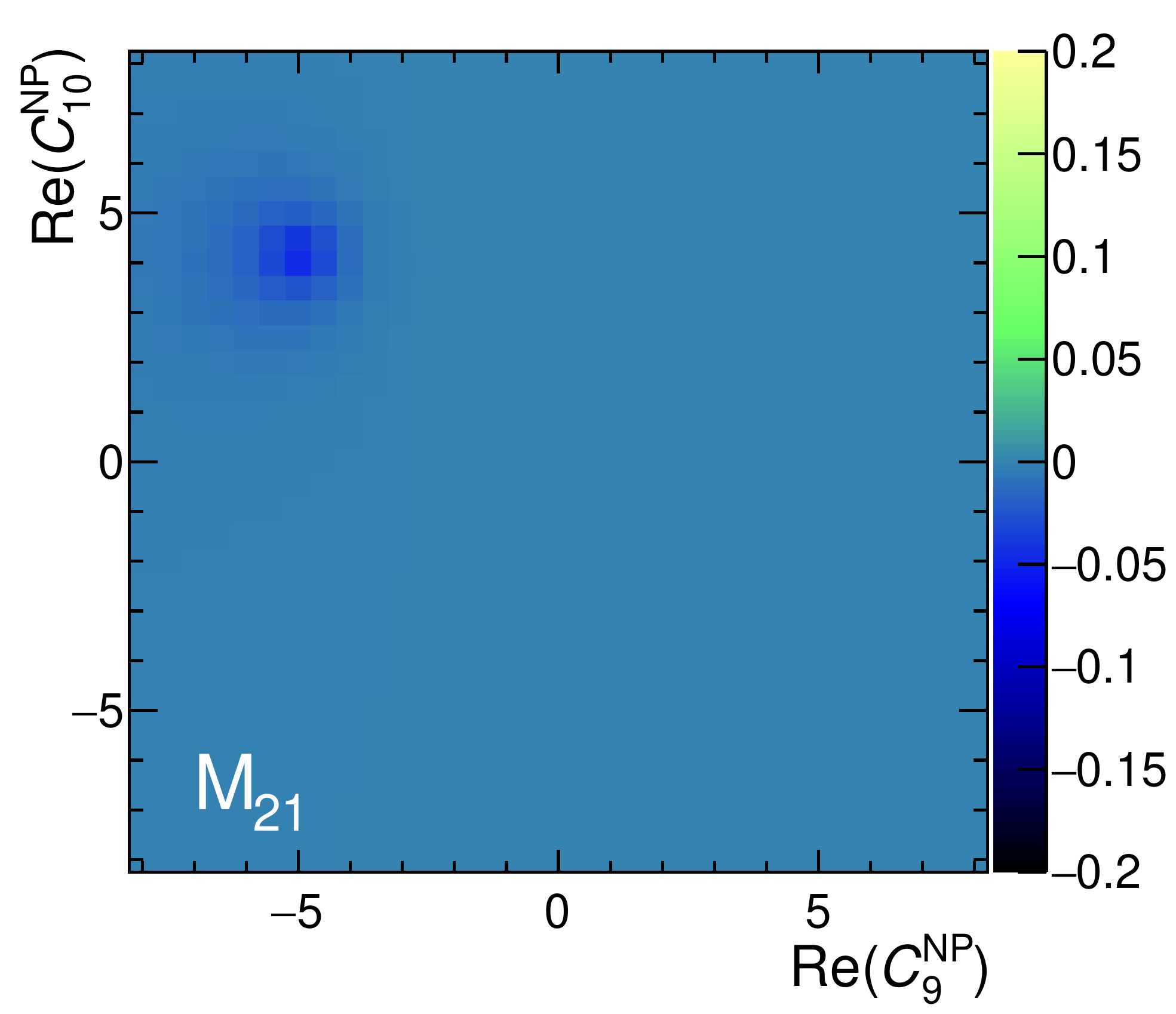}  
\includegraphics[width=0.24\linewidth]{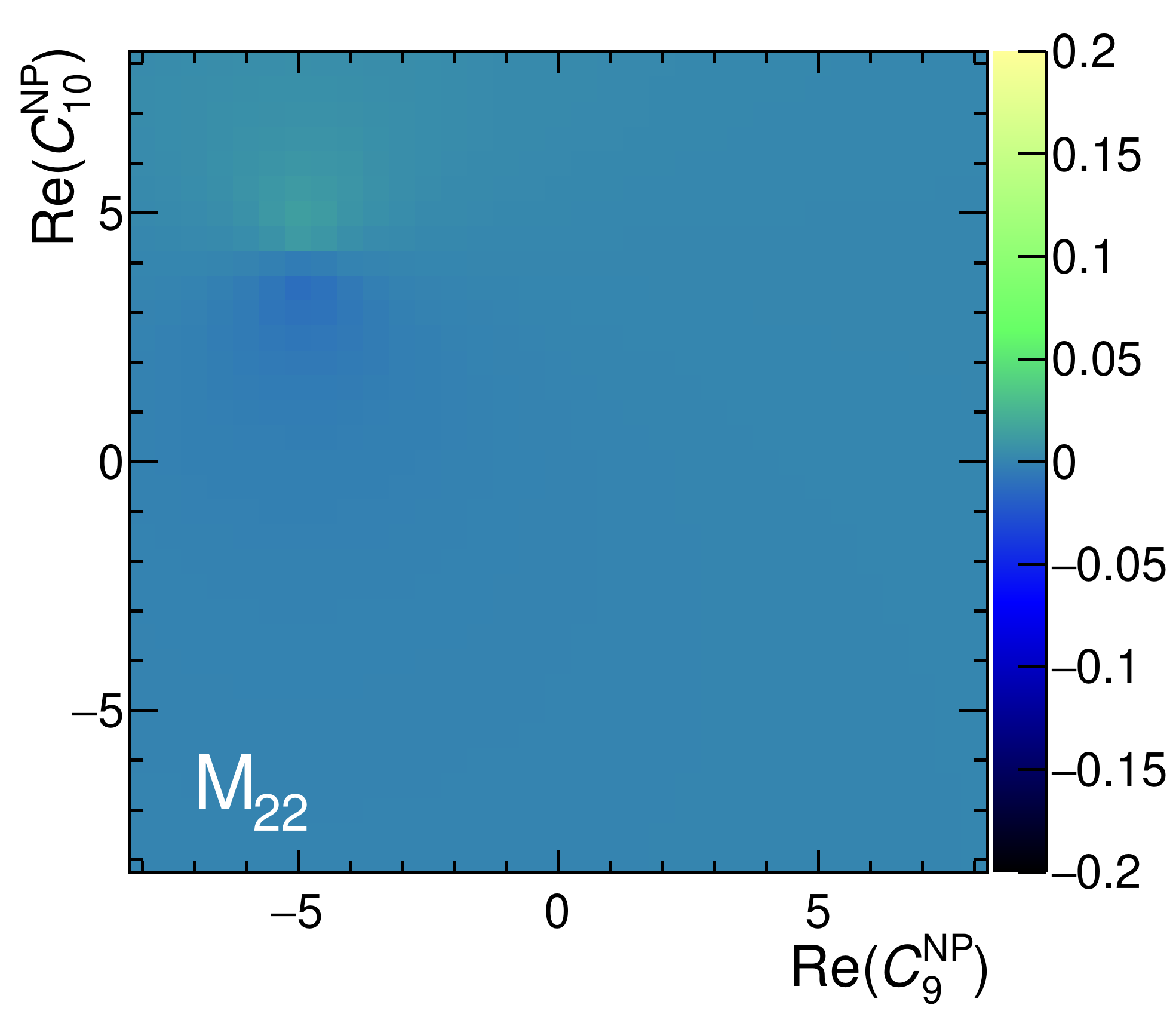} \\
\includegraphics[width=0.24\linewidth]{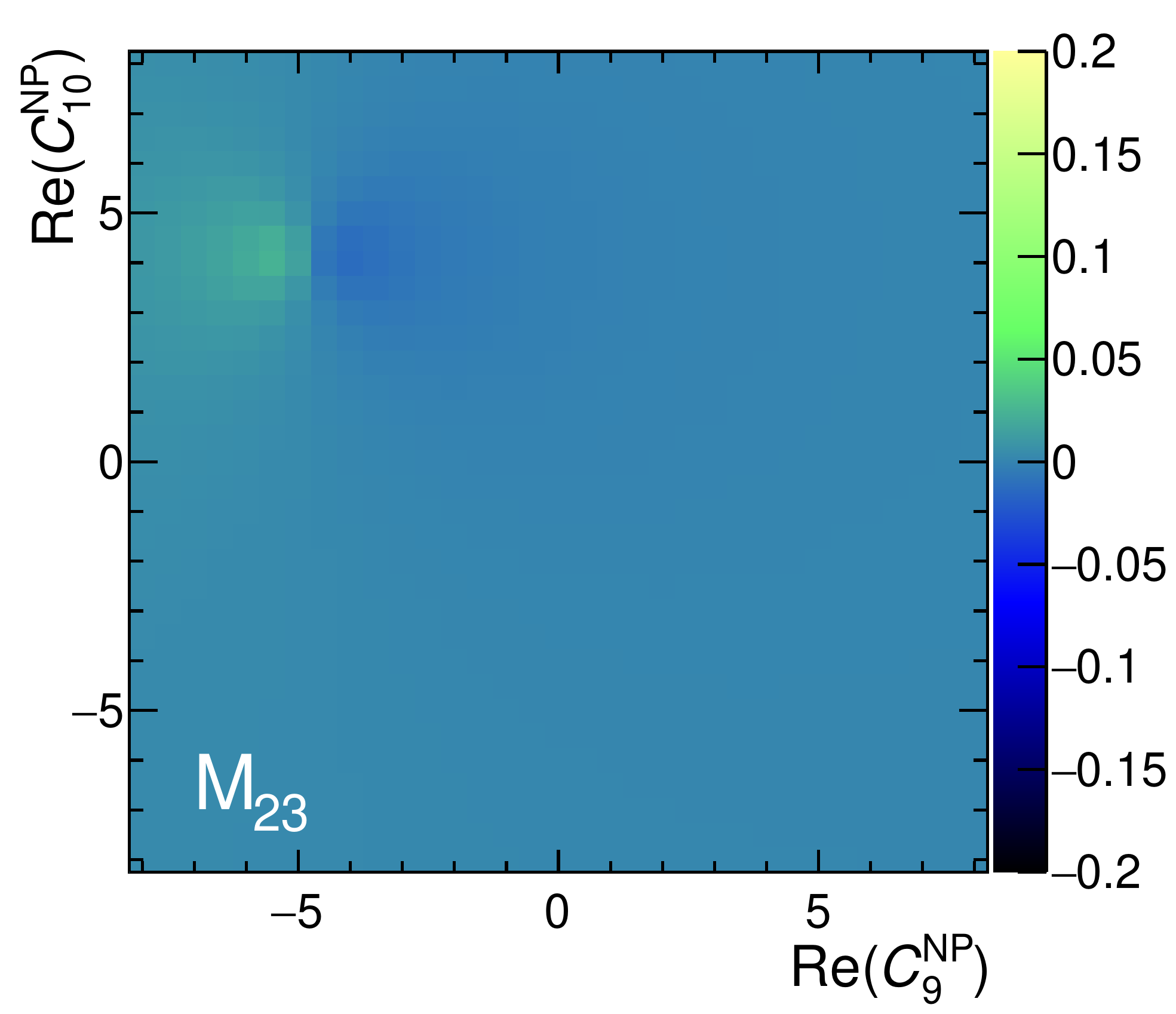} 
\includegraphics[width=0.24\linewidth]{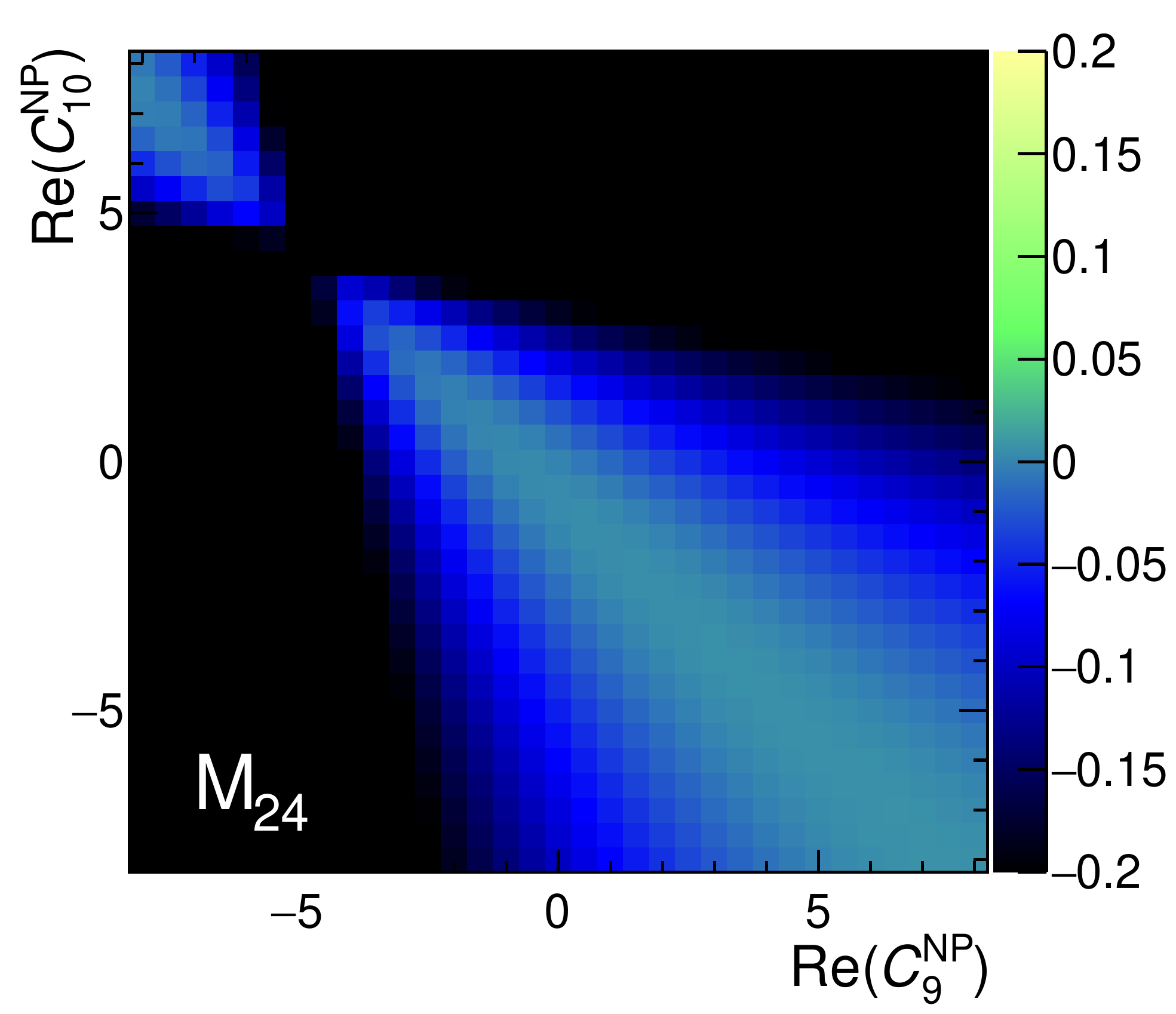}  
\includegraphics[width=0.24\linewidth]{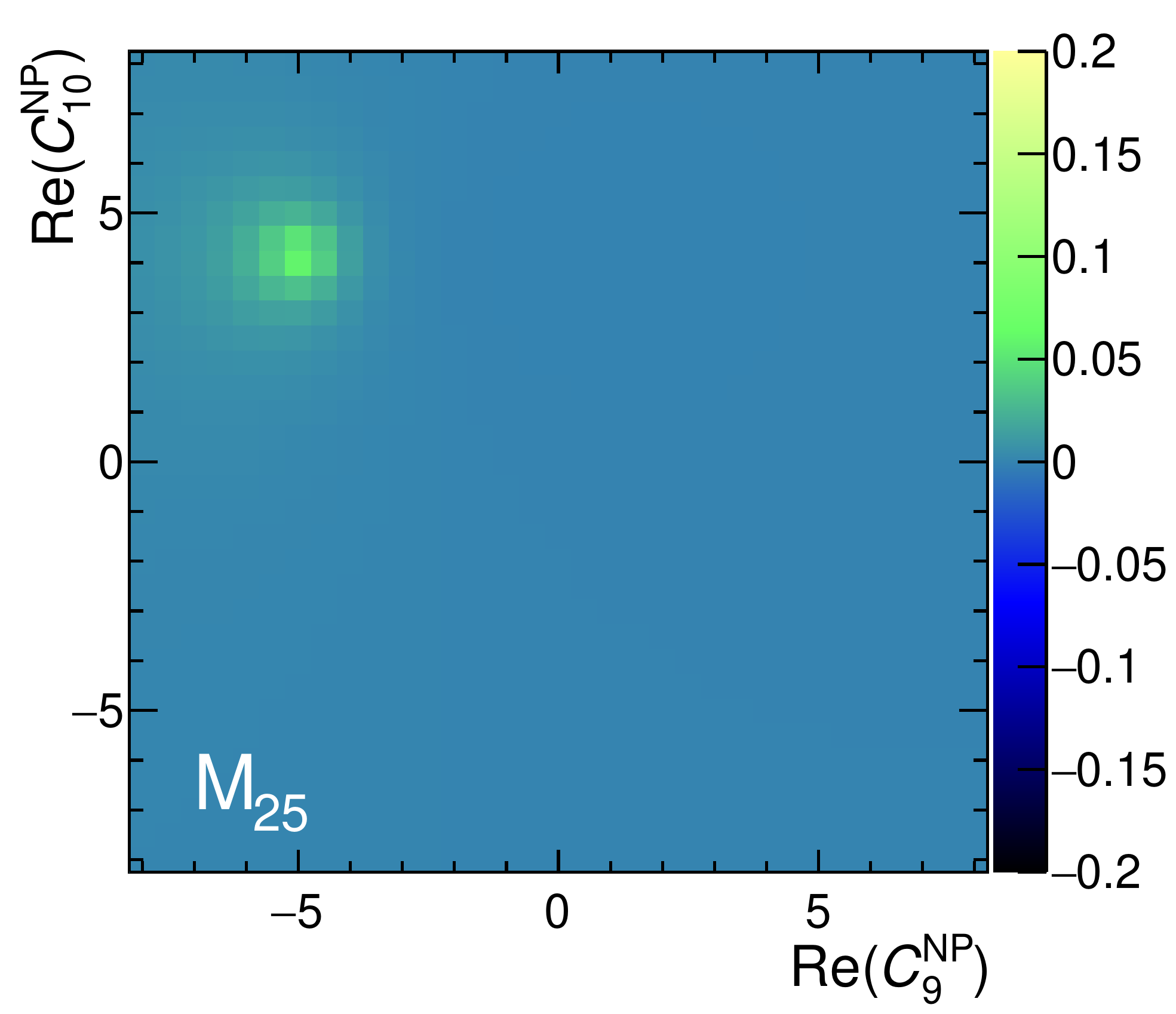} 
\includegraphics[width=0.24\linewidth]{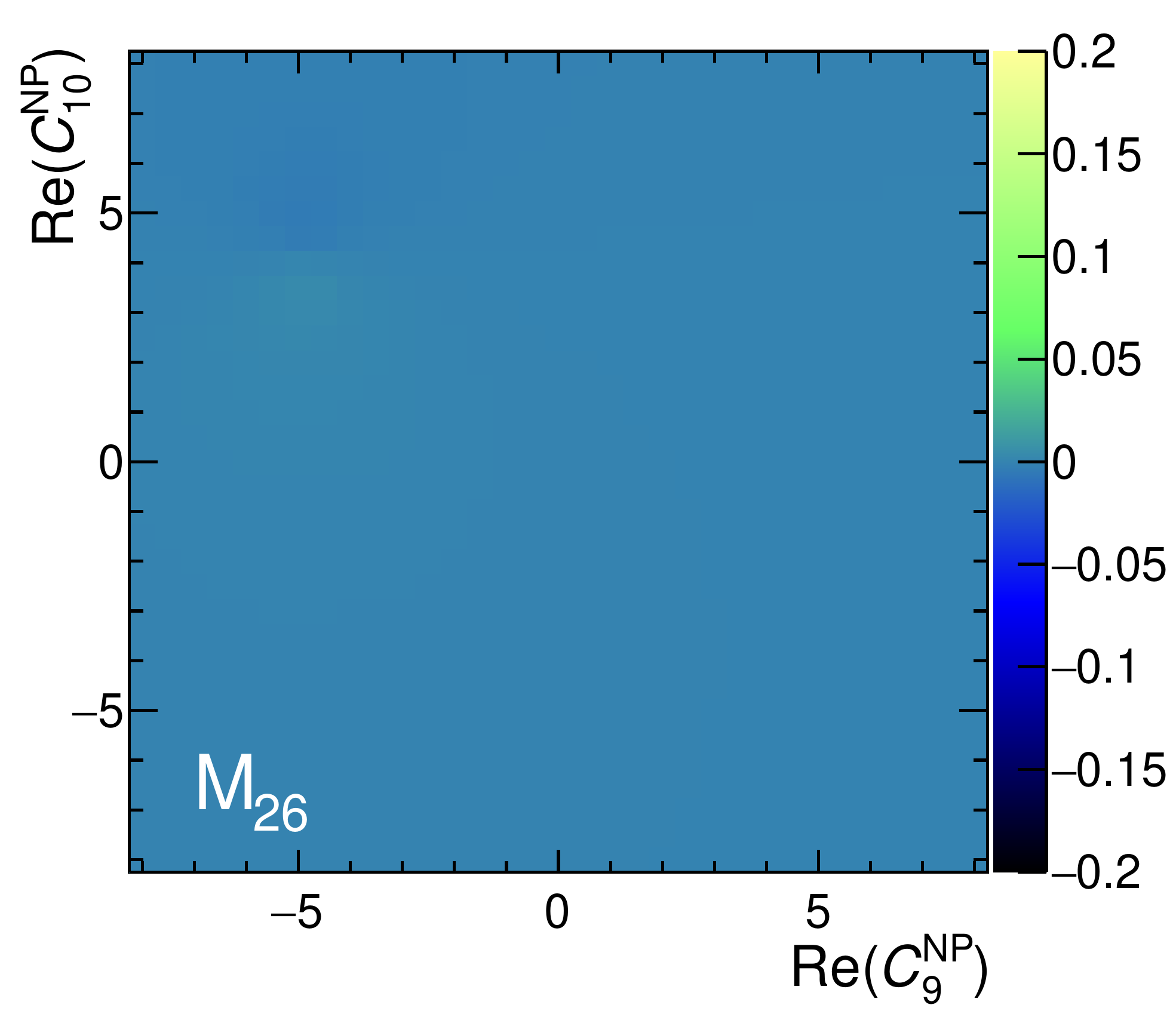} \\ 
\includegraphics[width=0.24\linewidth]{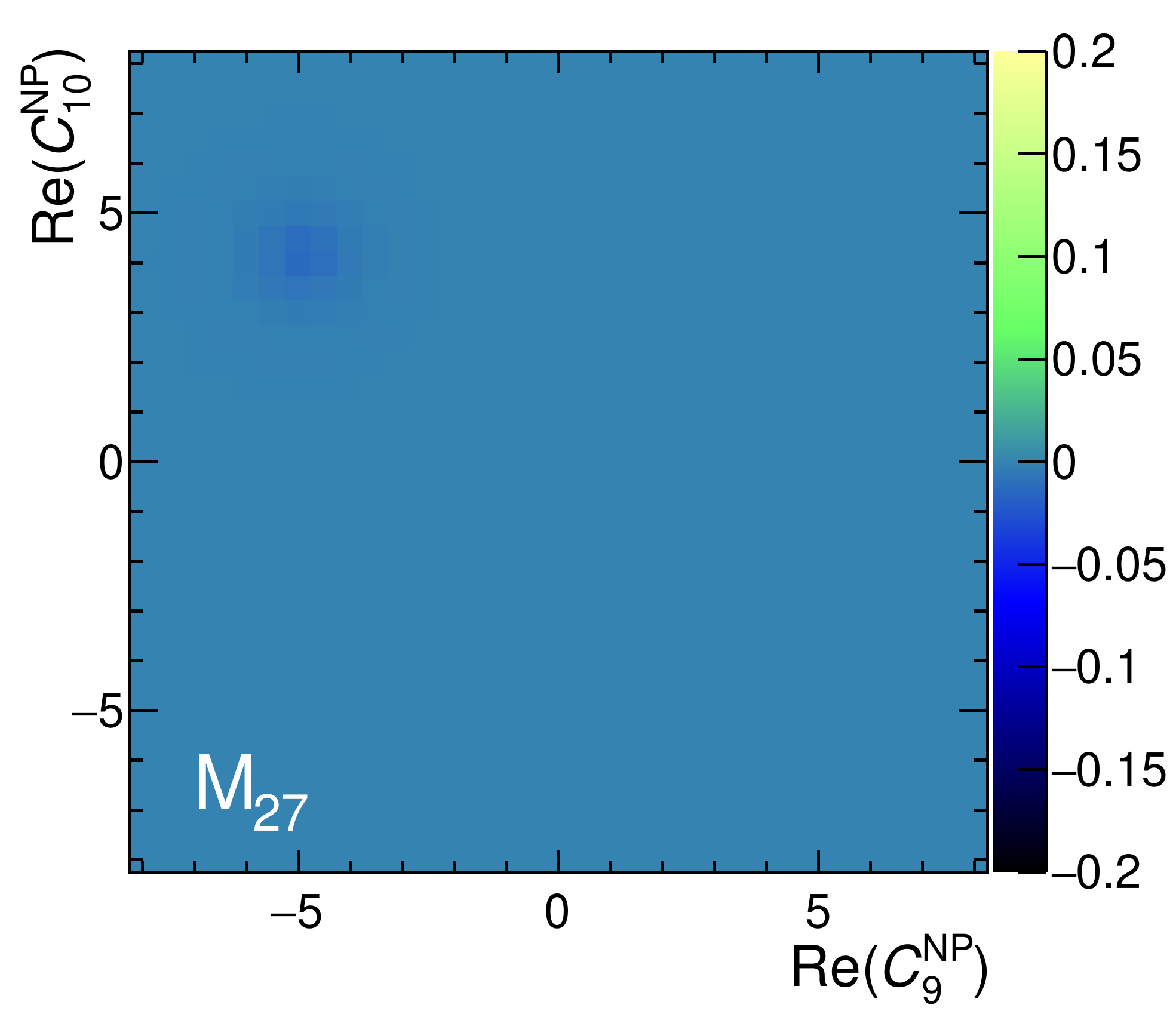}  
\includegraphics[width=0.24\linewidth]{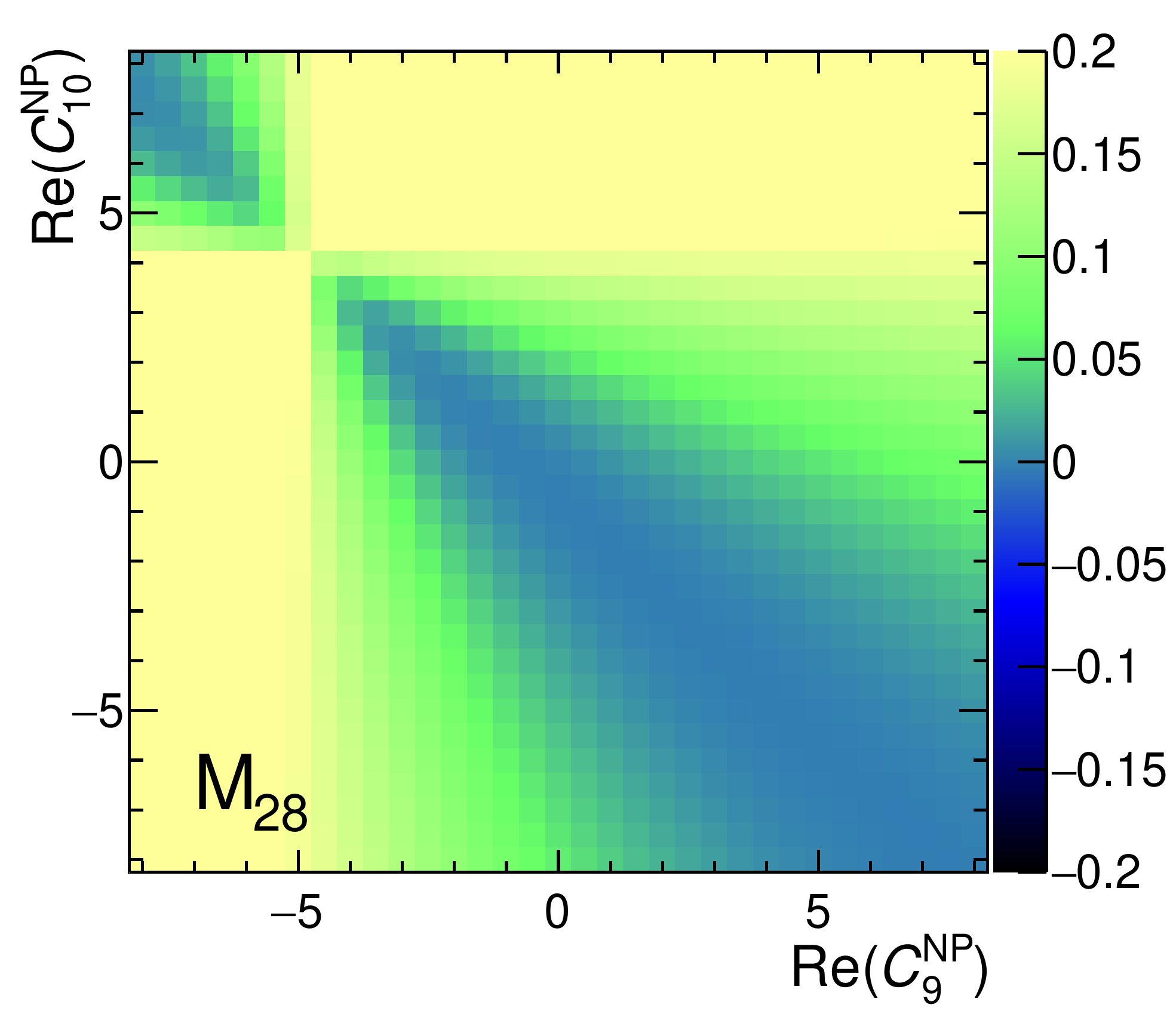} 
\includegraphics[width=0.24\linewidth]{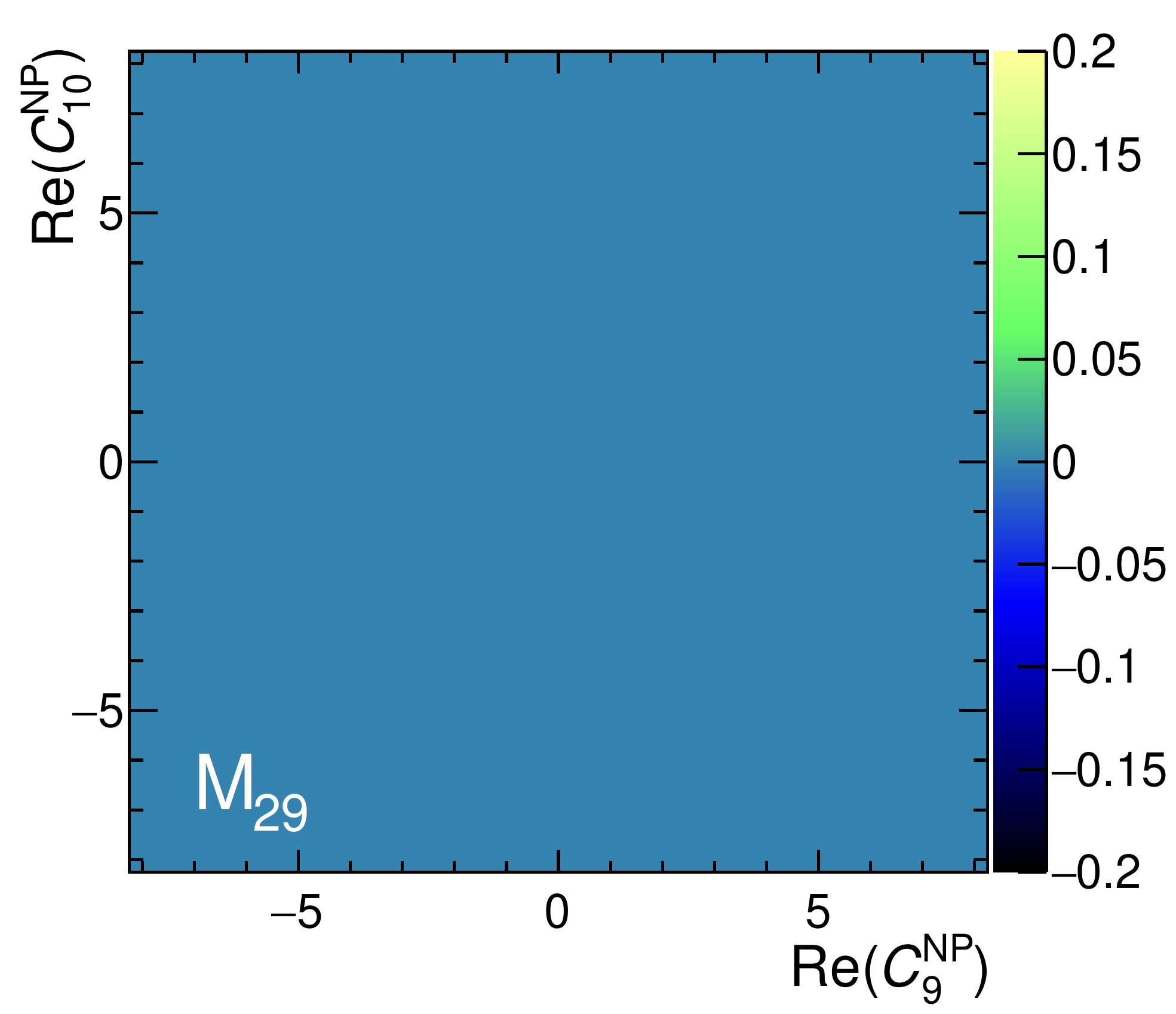} 
\includegraphics[width=0.24\linewidth]{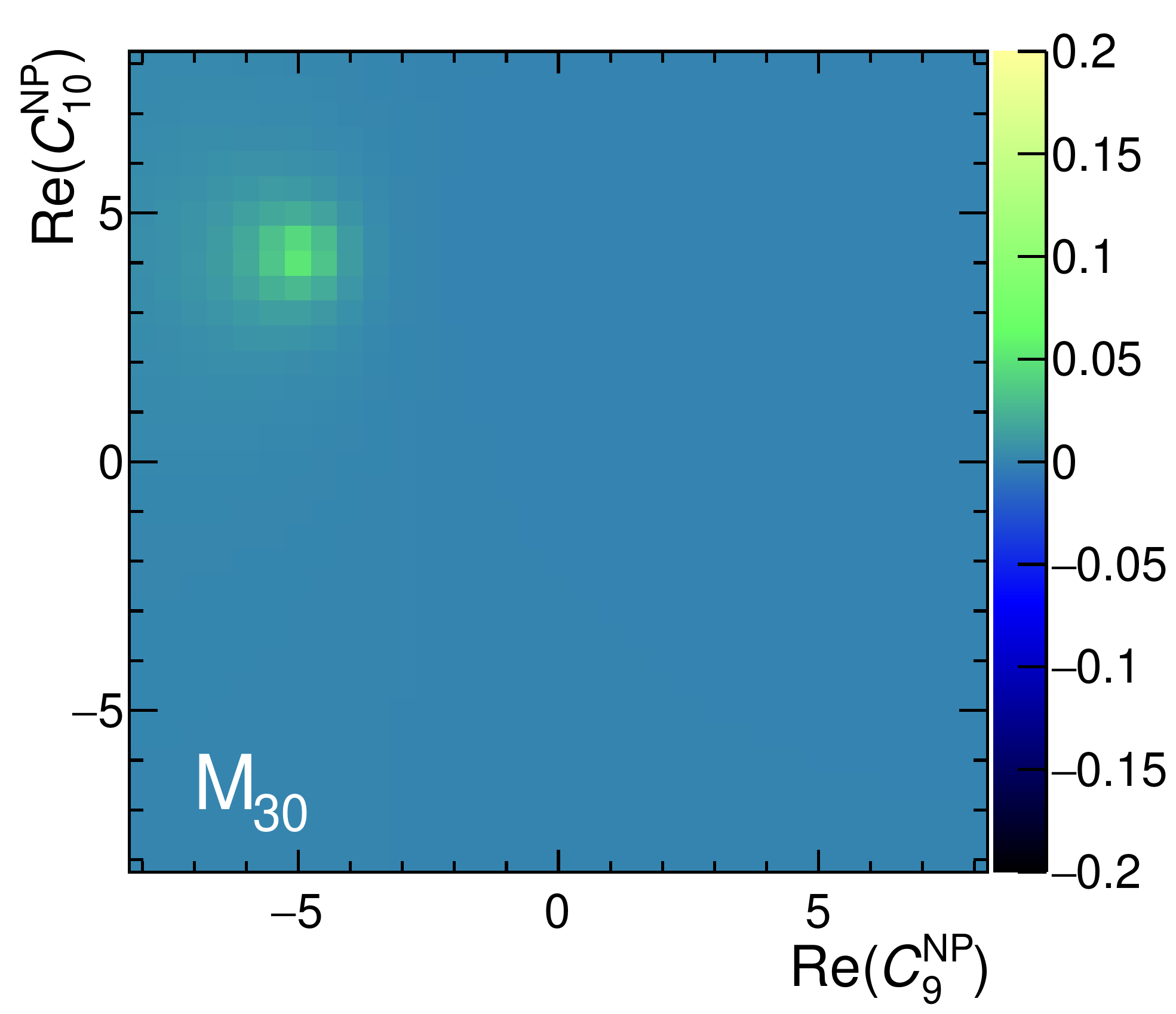} \\ 
\includegraphics[width=0.24\linewidth]{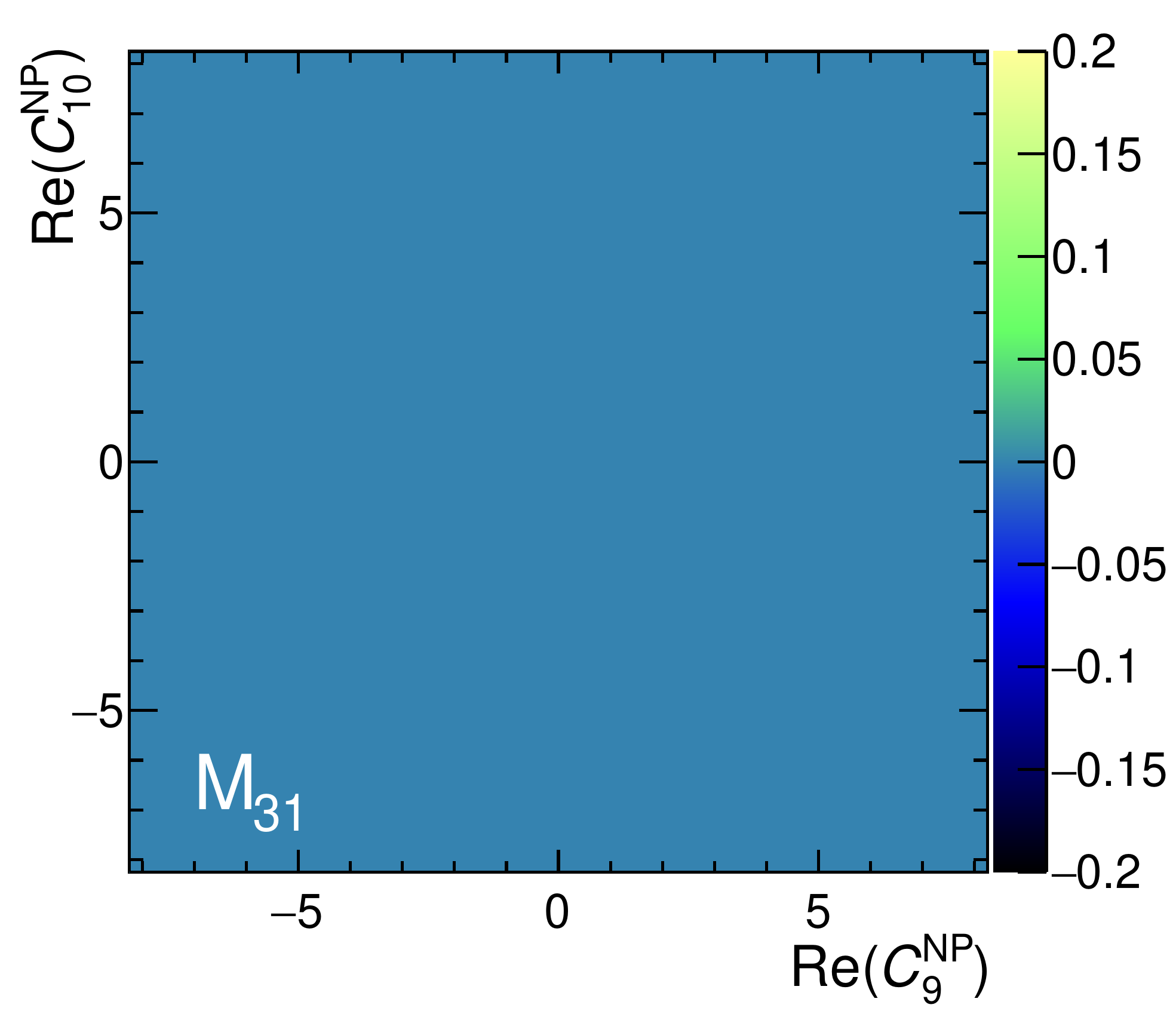}  
\includegraphics[width=0.24\linewidth]{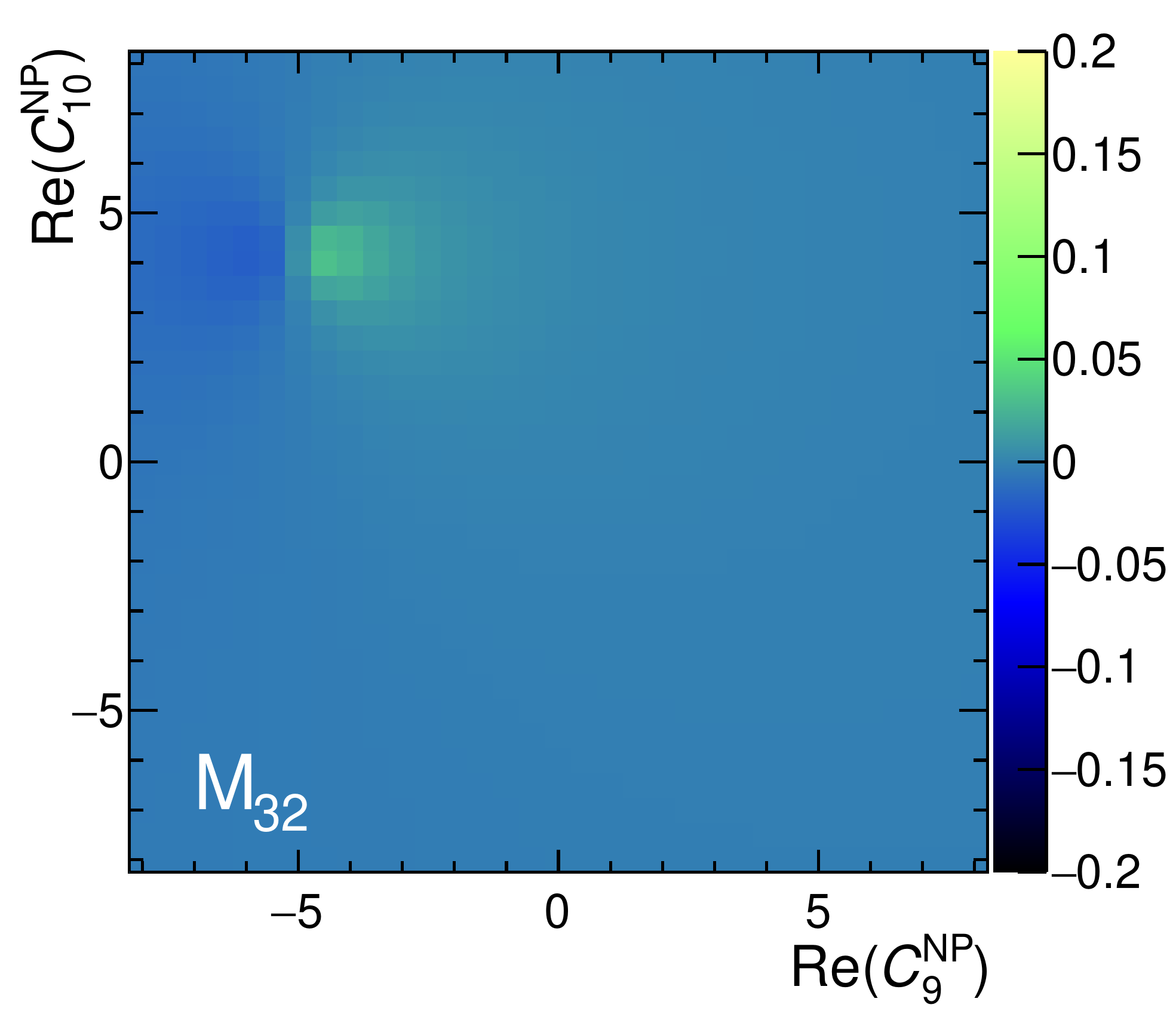} 
\includegraphics[width=0.24\linewidth]{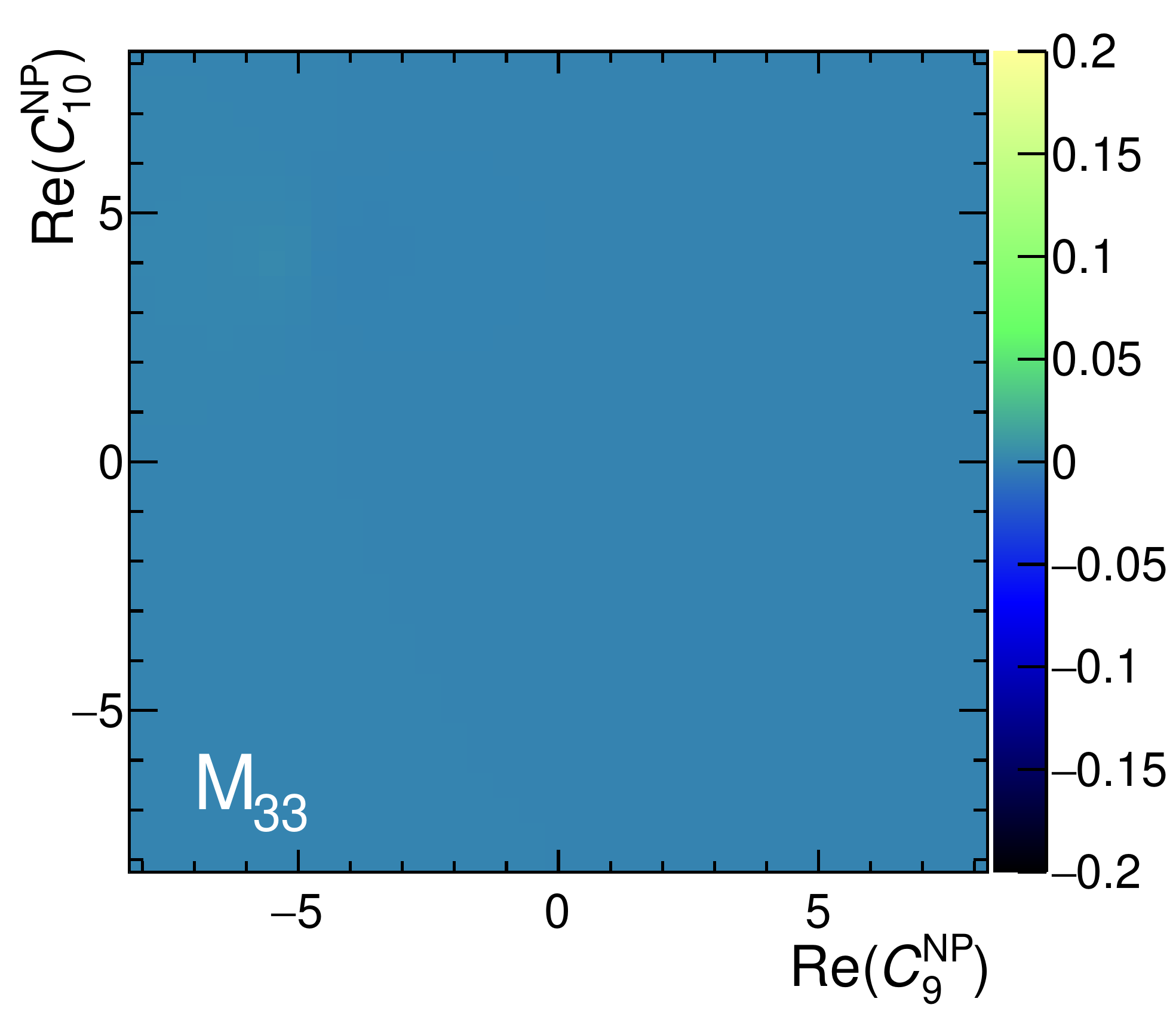} 
\includegraphics[width=0.24\linewidth]{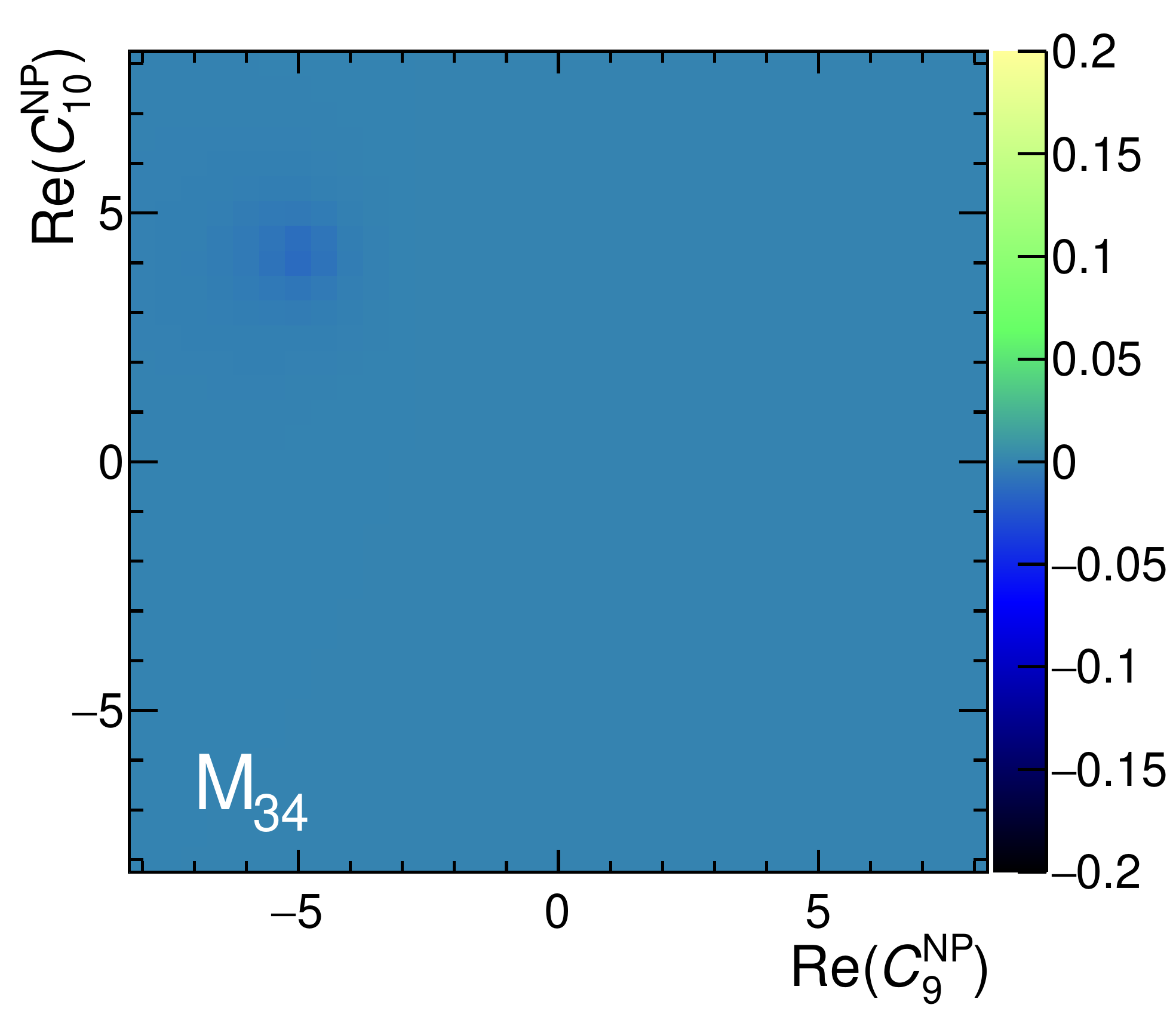}  
\caption{
Variation of the polarisation dependent angular observables of the \decay{\Lb}{\Lz\mumu} decay  from their SM central values in the low-recoil region ($15 < \qsq < 20\gev^{2}/c^{4}$) with a NP contribution to ${\rm Re}(C_9)$ or ${\rm Re}(C_{10})$. 
The SM point is at $(0,0)$.
To illustrate the size of the effects, $P_{\Lb}  = 1$ is used.
\label{fig:scan:c9:c10:lowrecoil:pol} 
}
\end{figure}

\begin{figure}[!htb]
\centering
\includegraphics[width=0.24\linewidth]{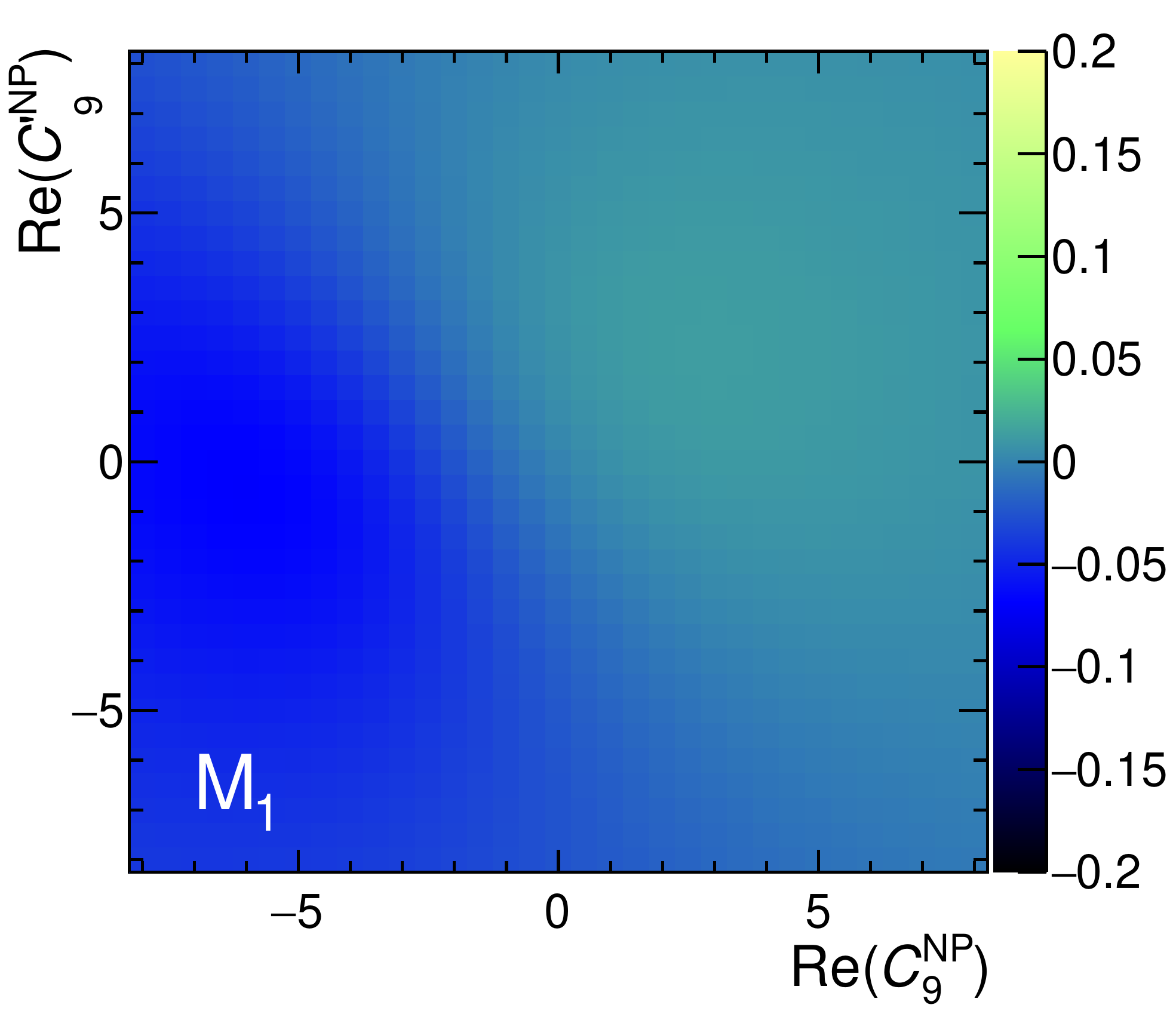}  
\includegraphics[width=0.24\linewidth]{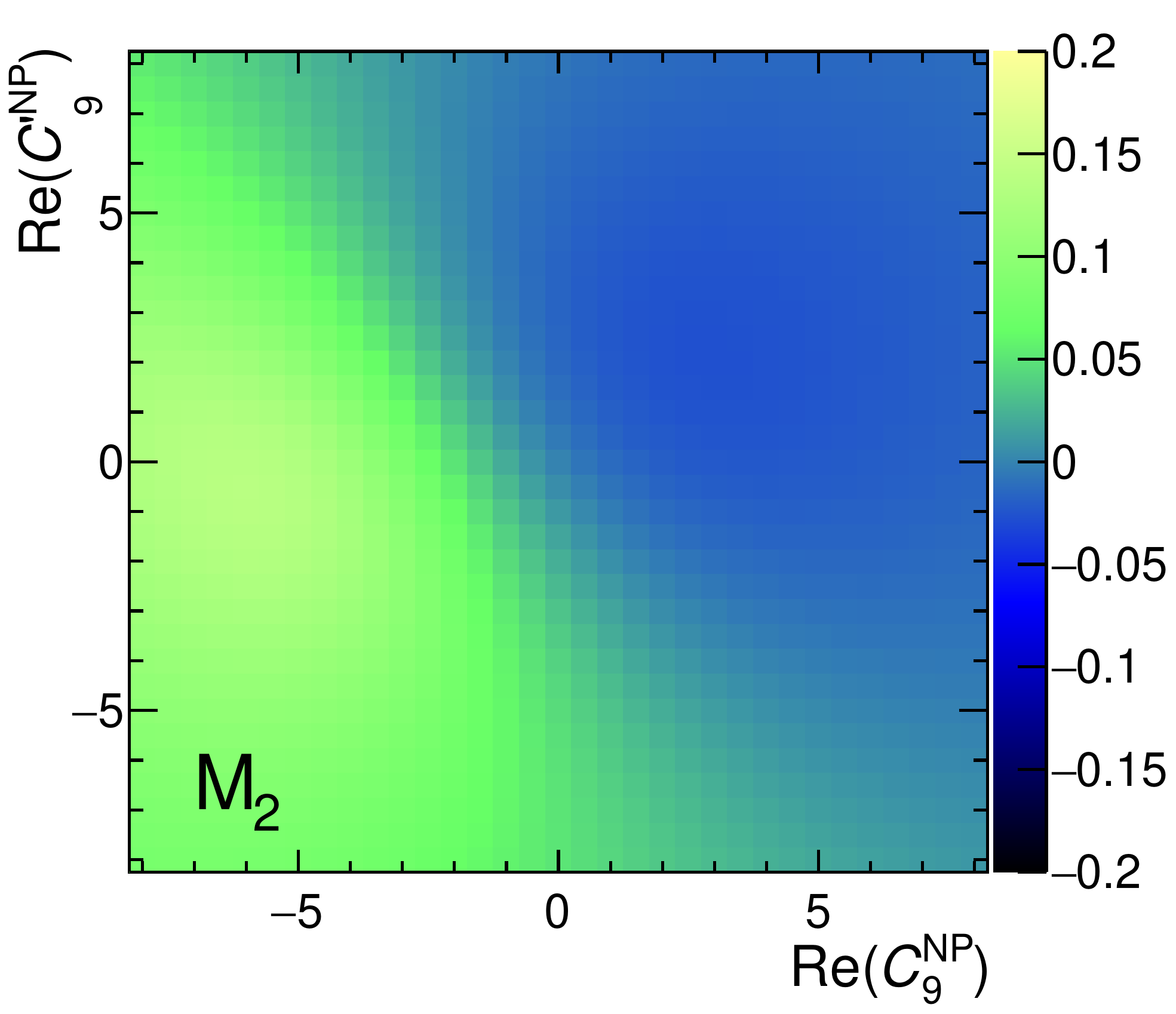} 
\includegraphics[width=0.24\linewidth]{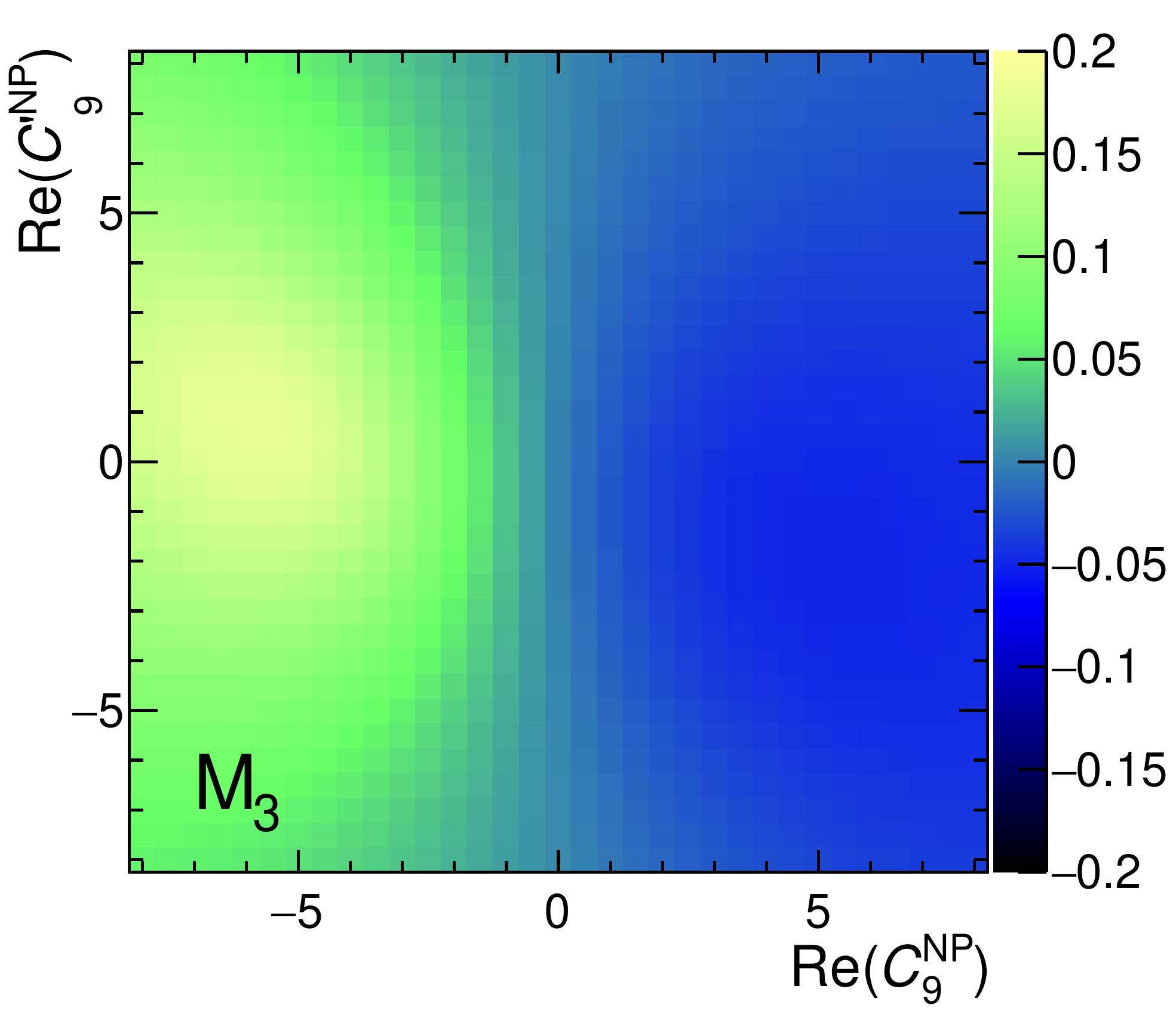} 
\includegraphics[width=0.24\linewidth]{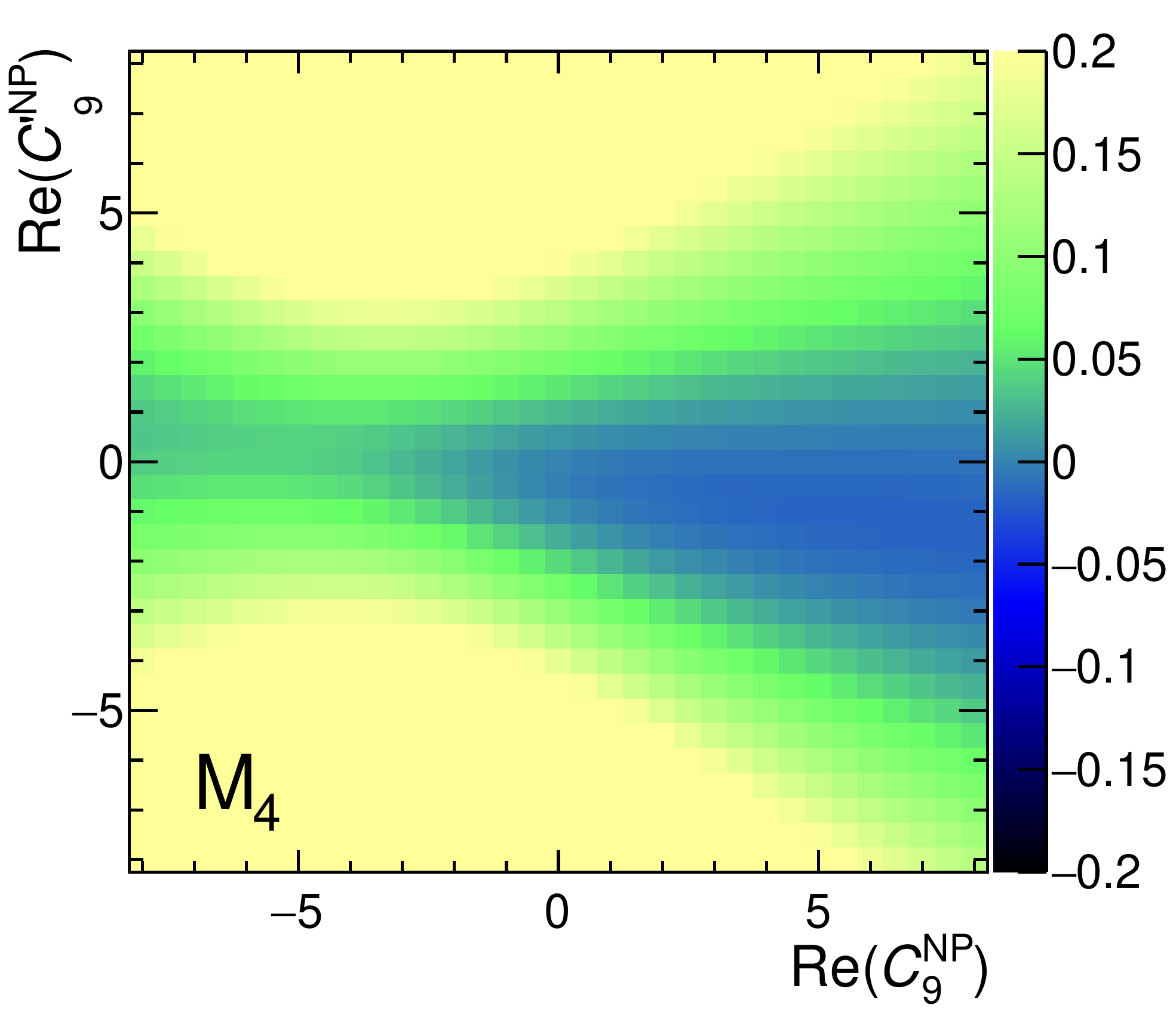} \\ 
\includegraphics[width=0.24\linewidth]{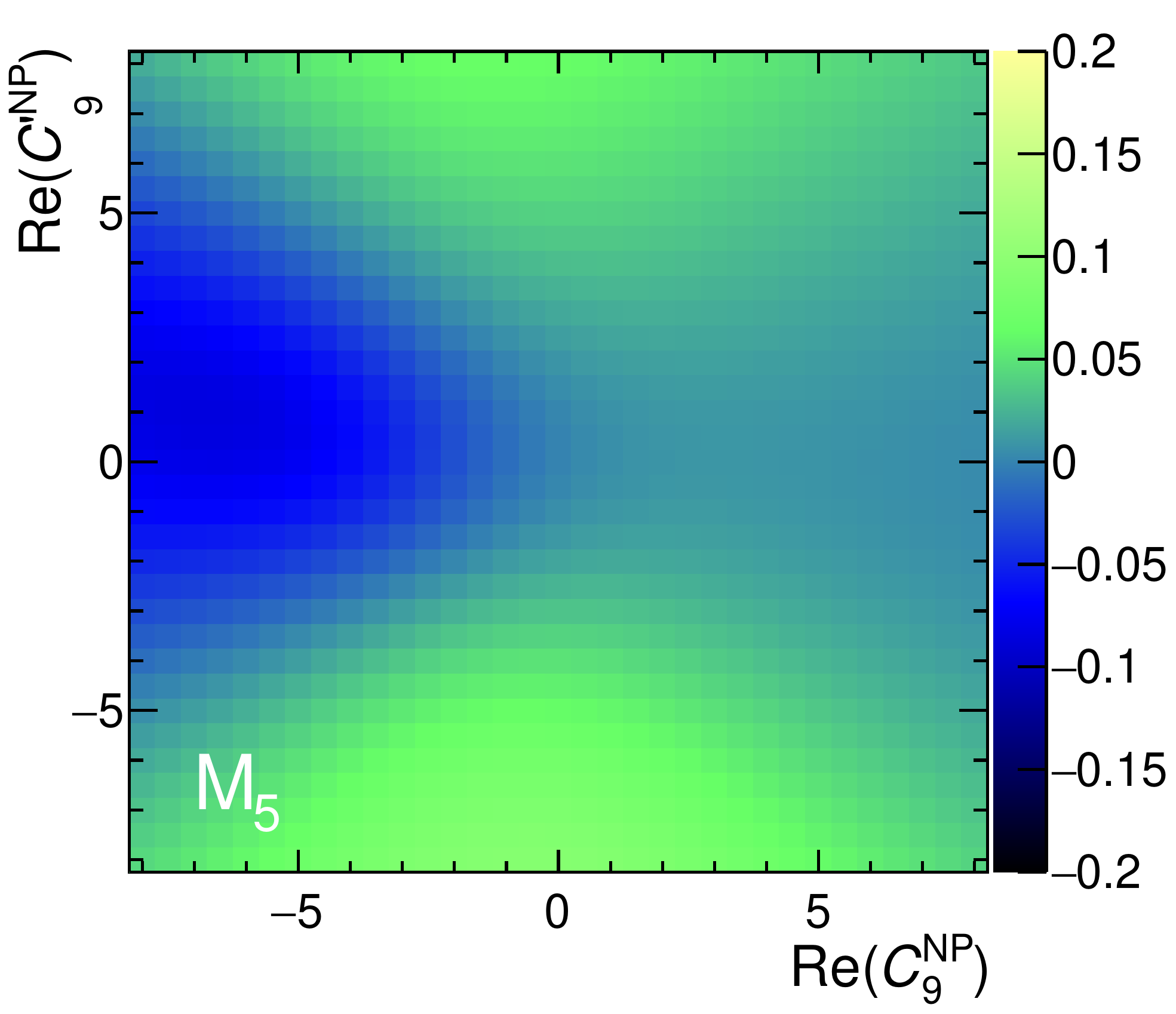}  
\includegraphics[width=0.24\linewidth]{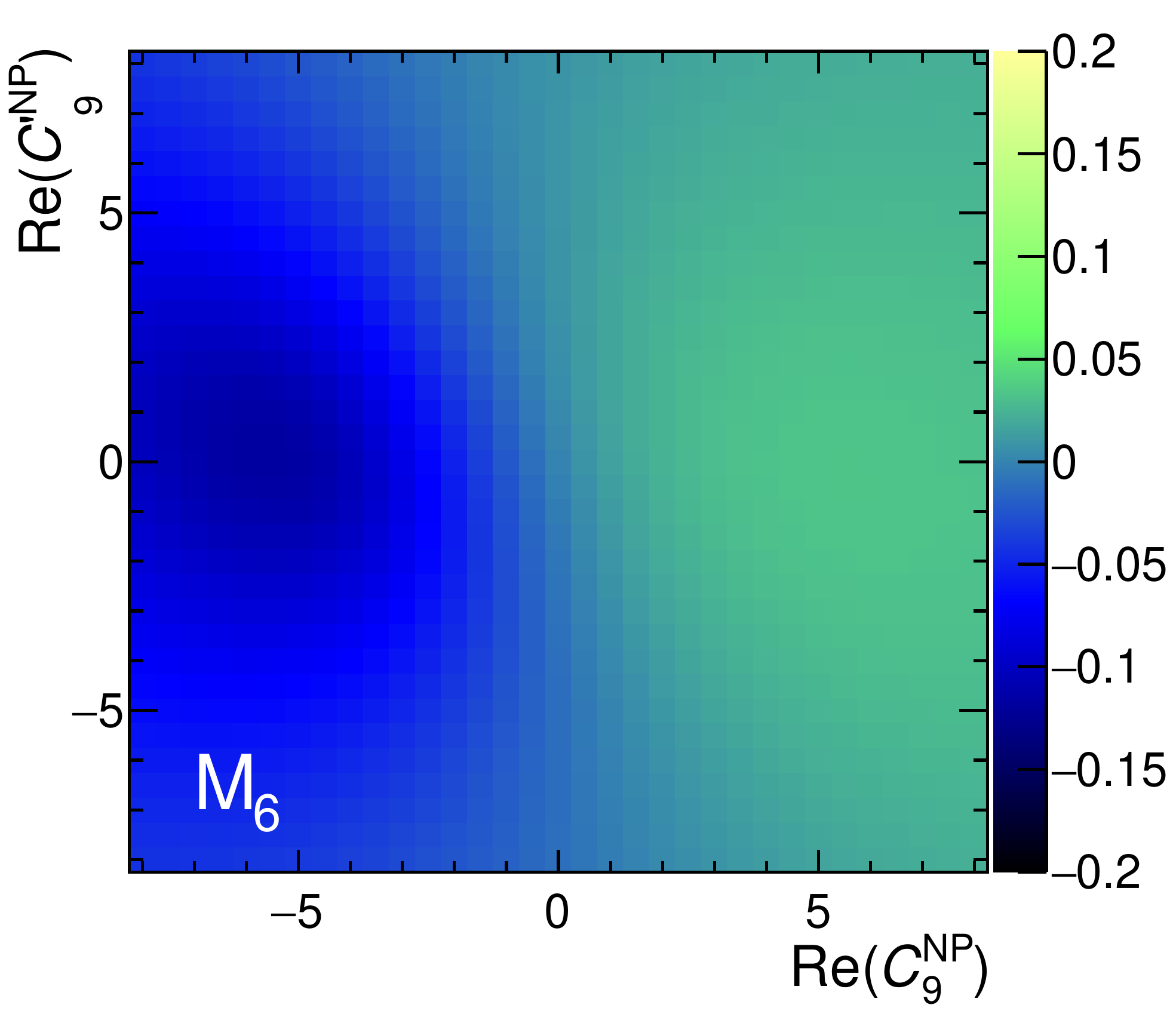} 
\includegraphics[width=0.24\linewidth]{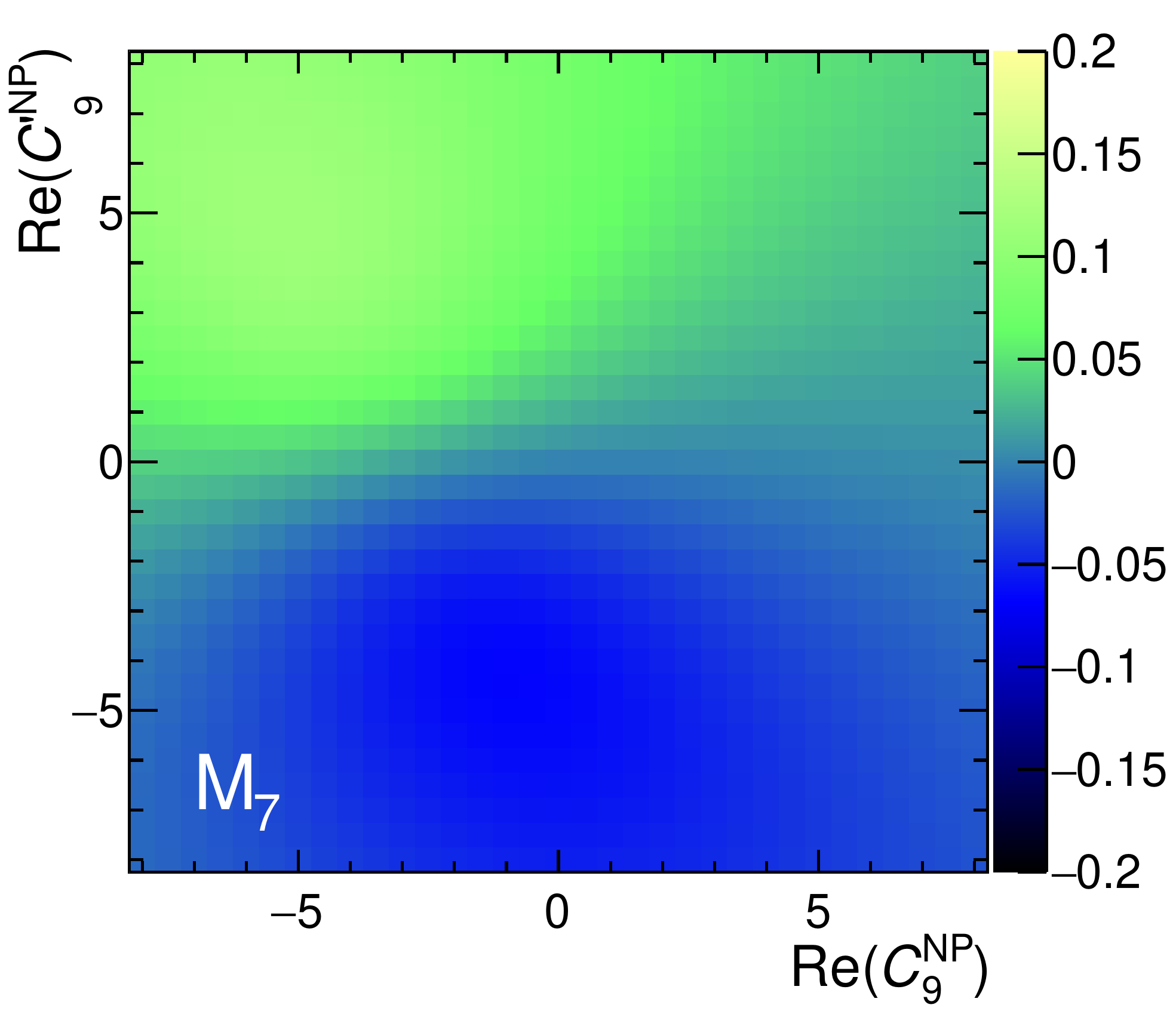} 
\includegraphics[width=0.24\linewidth]{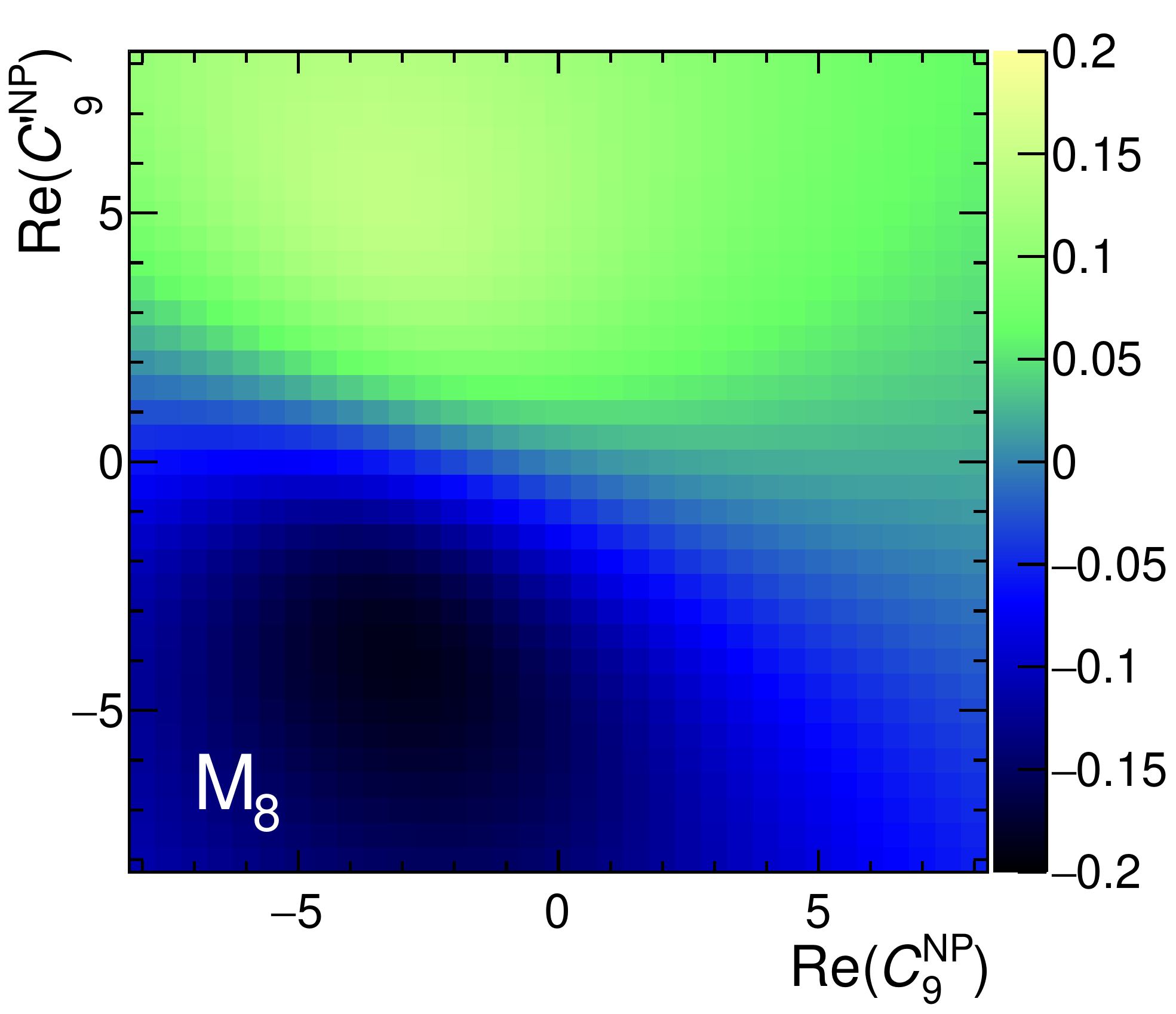} \\ 
\includegraphics[width=0.24\linewidth]{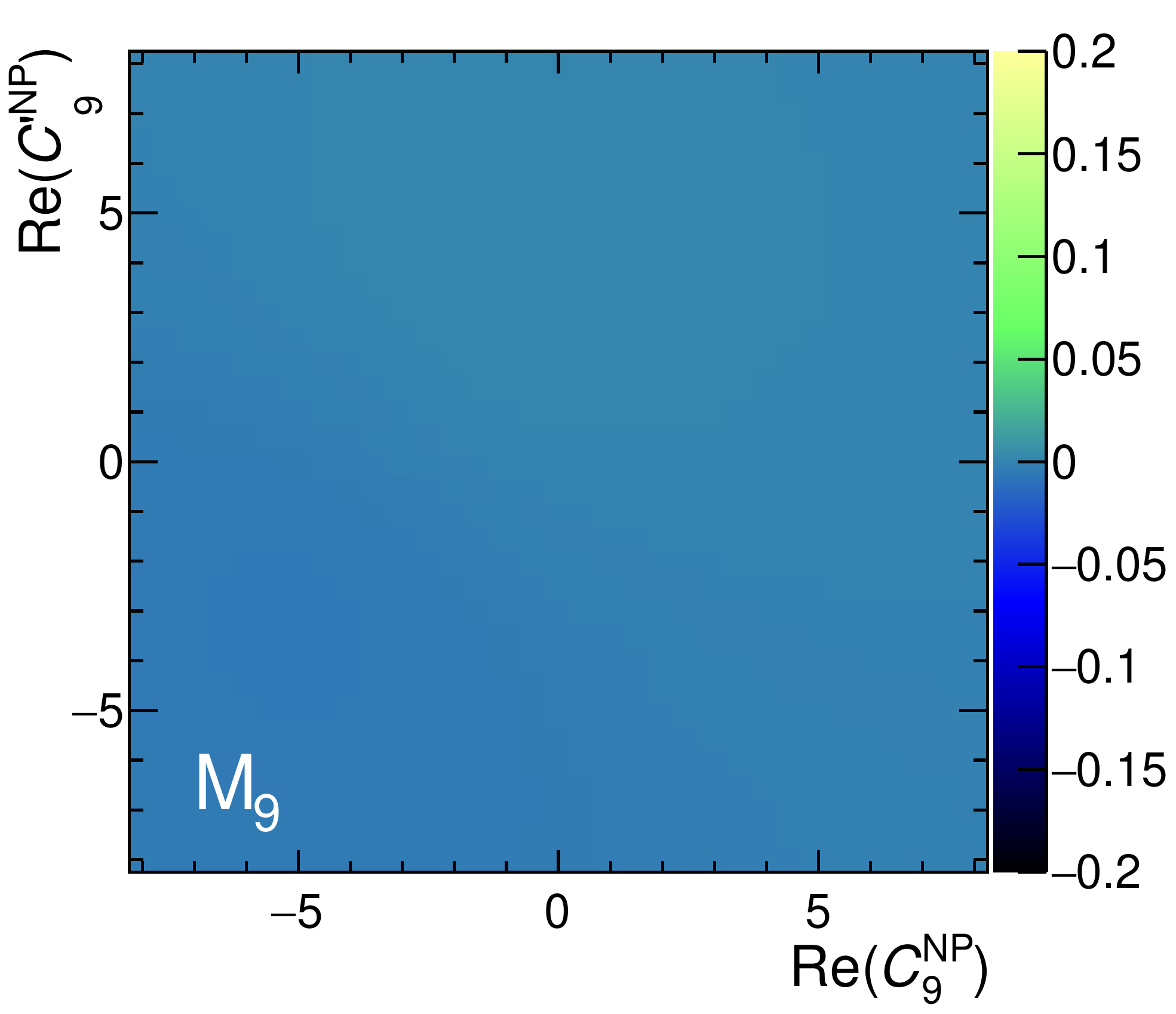}  
\includegraphics[width=0.24\linewidth]{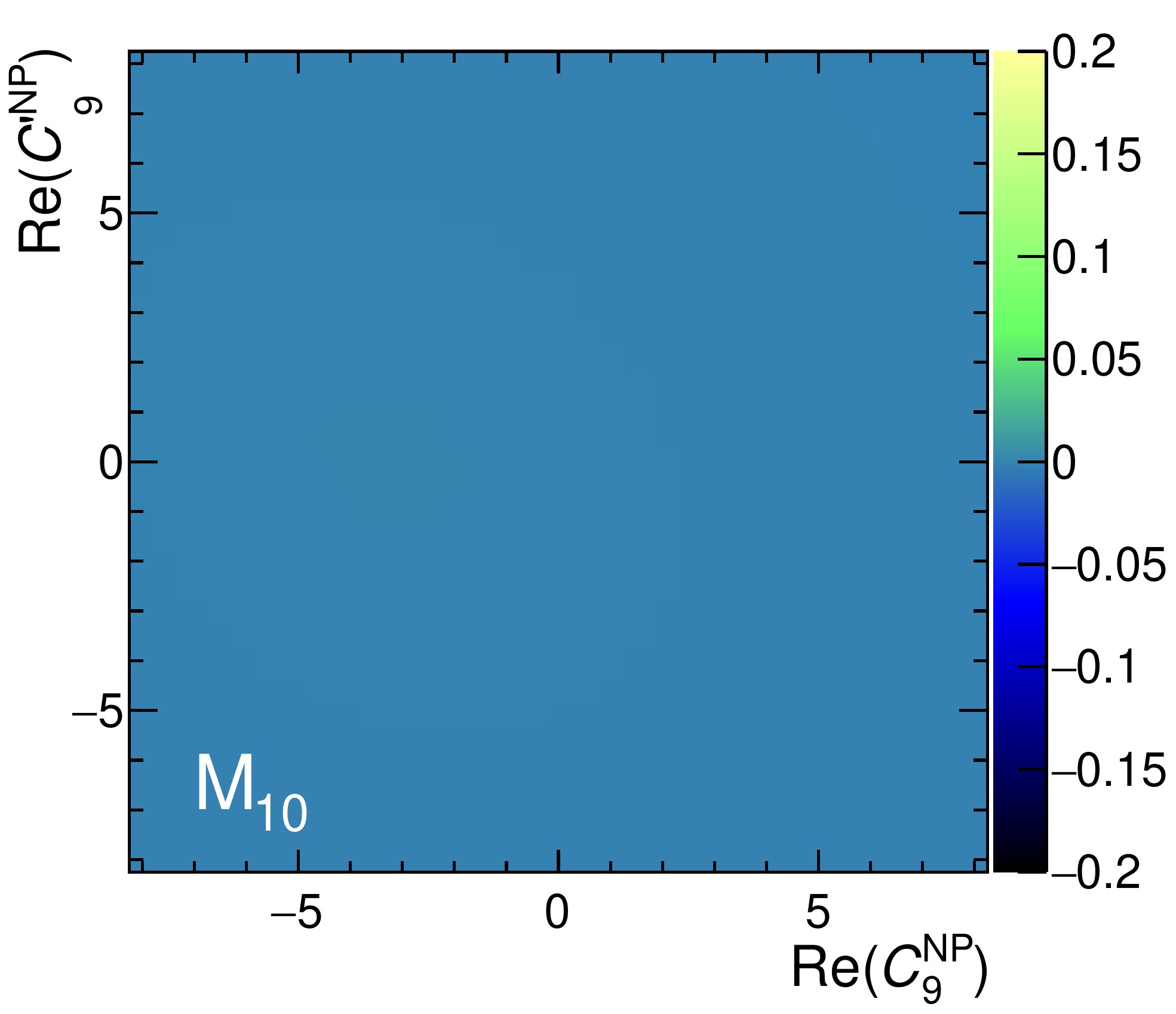} 
\caption{
Variation of the observables $M_{1}$--$M_{10}$  of the \decay{\Lb}{\Lz\mumu} decay from their SM central values in the large-recoil region ($1 < \qsq < 6\gev^{2}/c^{4}$) with a NP contribution to ${\rm Re}(C_9)$ or ${\rm Re}(C'_{9})$. 
The SM point is at $(0,0)$.
\label{fig:scan:c9:c9p:largerecoil} 
}
\end{figure}

\begin{figure}[!htb]
\centering
\includegraphics[width=0.24\linewidth]{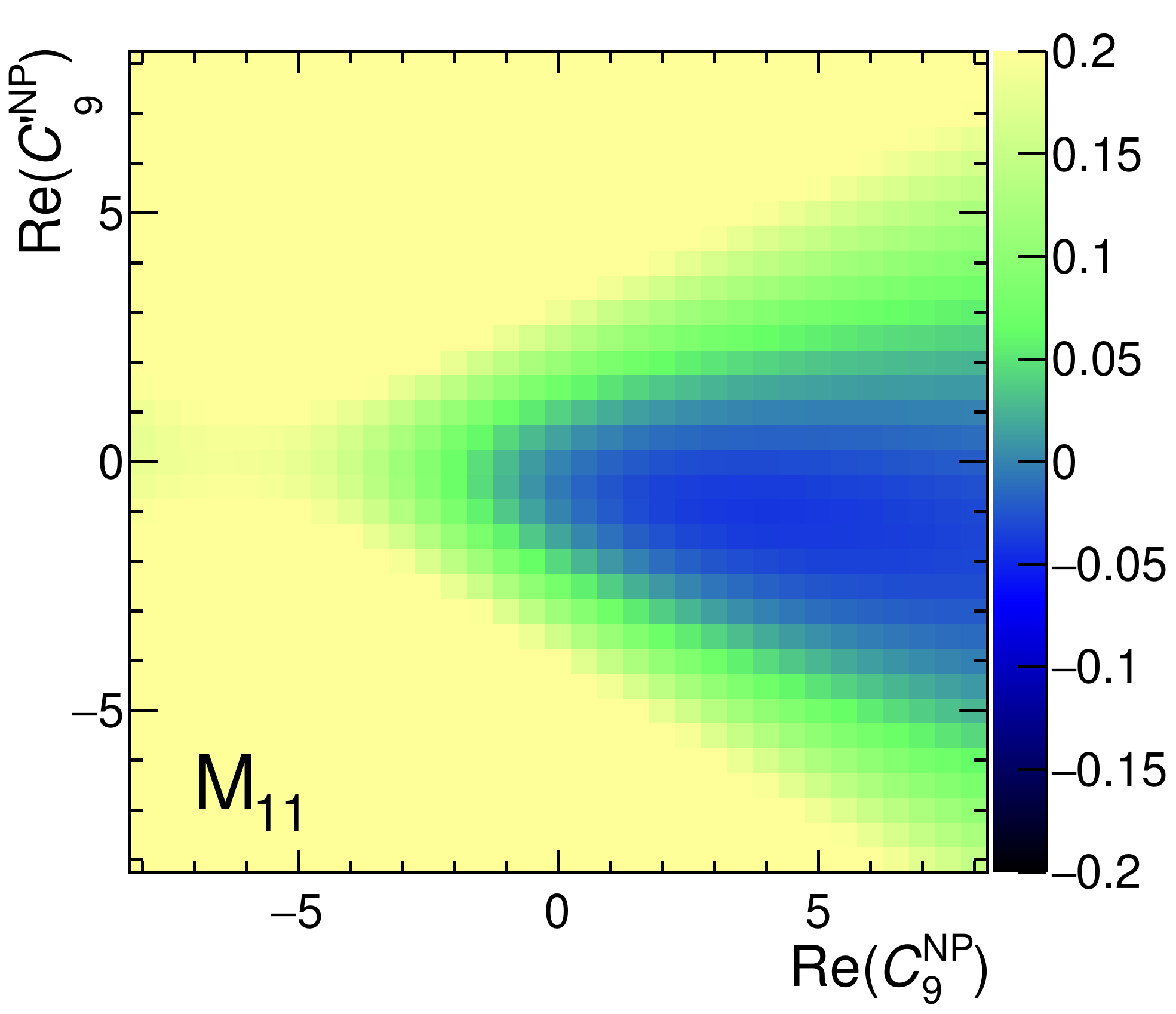} 
\includegraphics[width=0.24\linewidth]{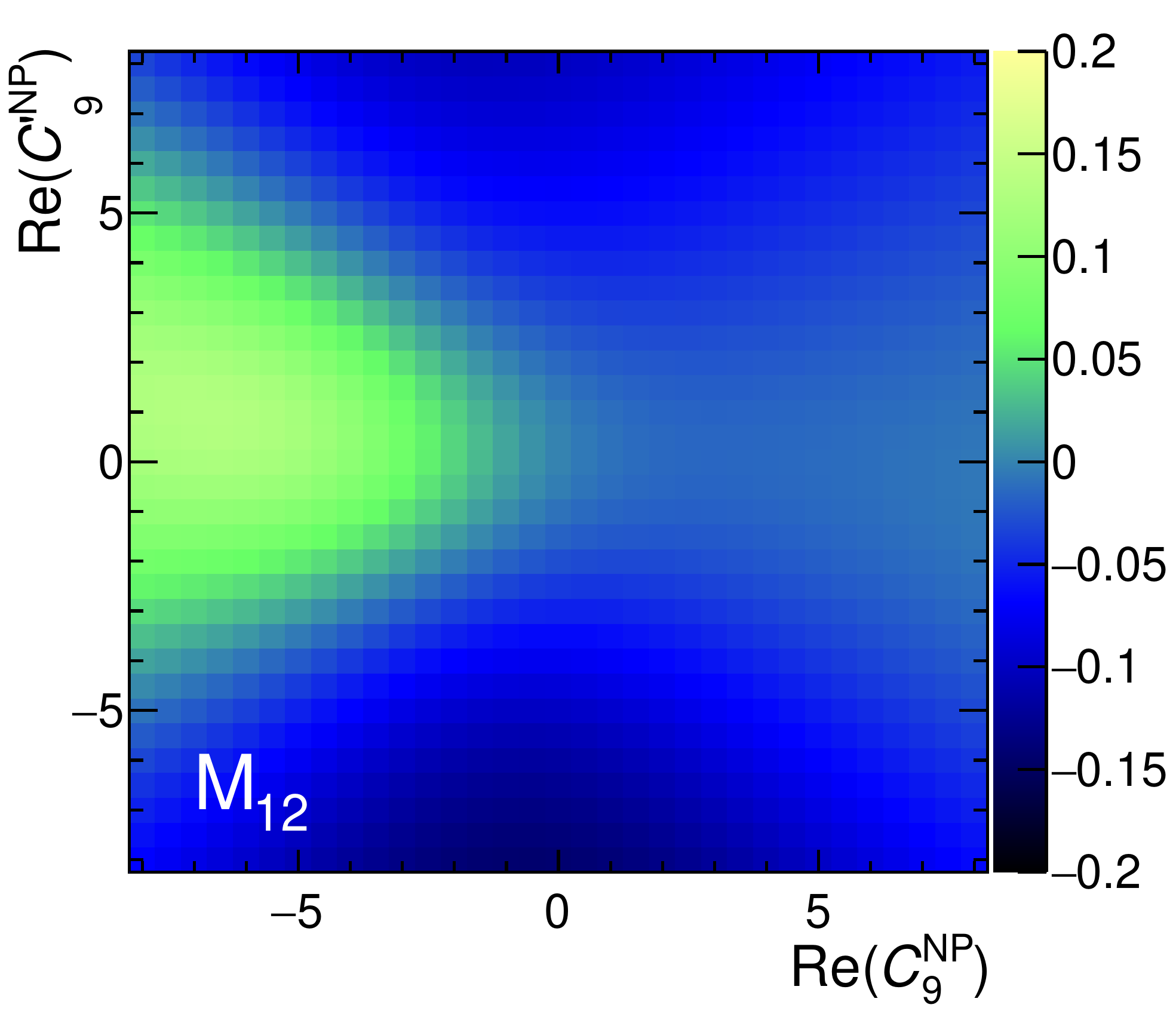} 
\includegraphics[width=0.24\linewidth]{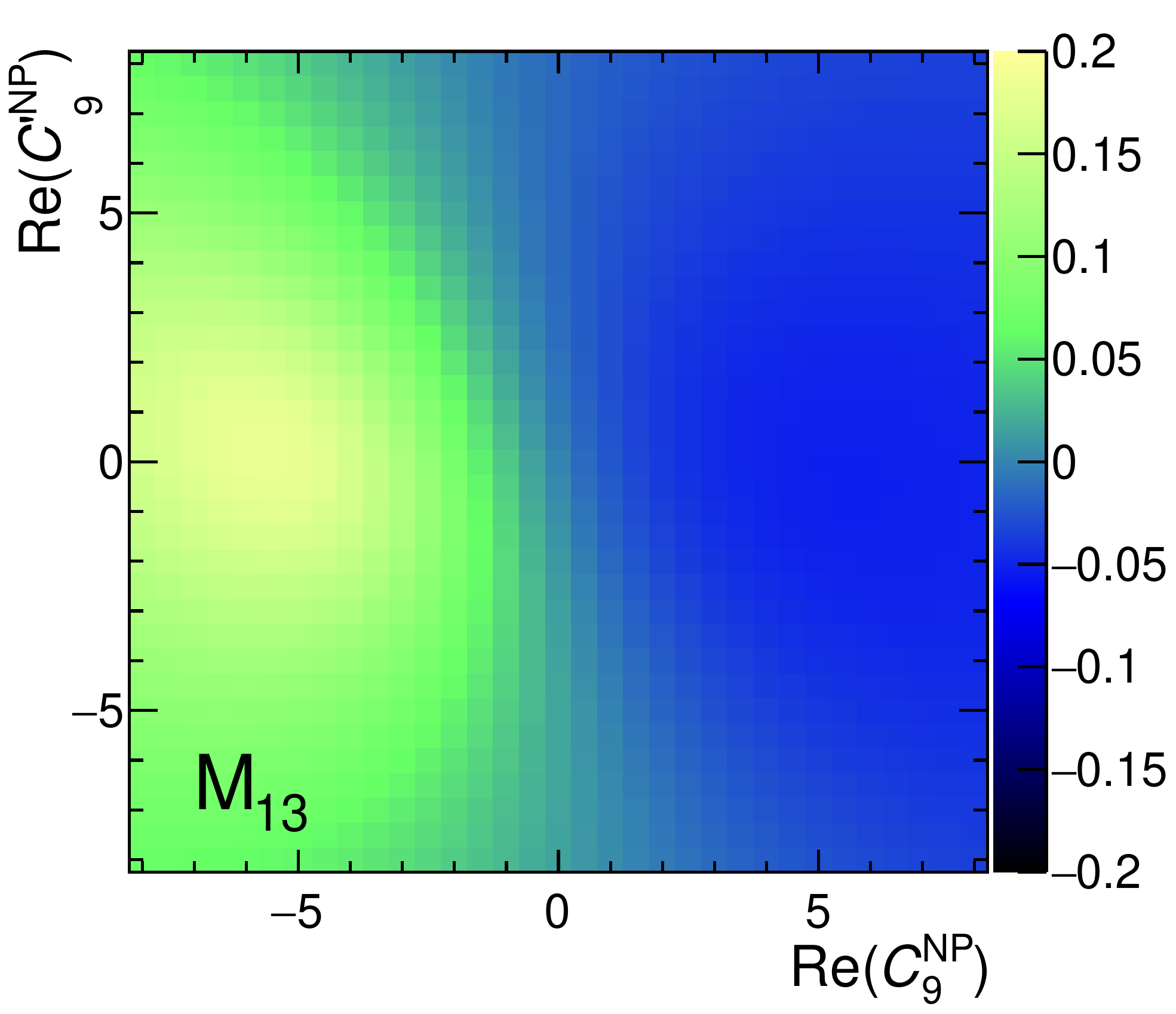}  
\includegraphics[width=0.24\linewidth]{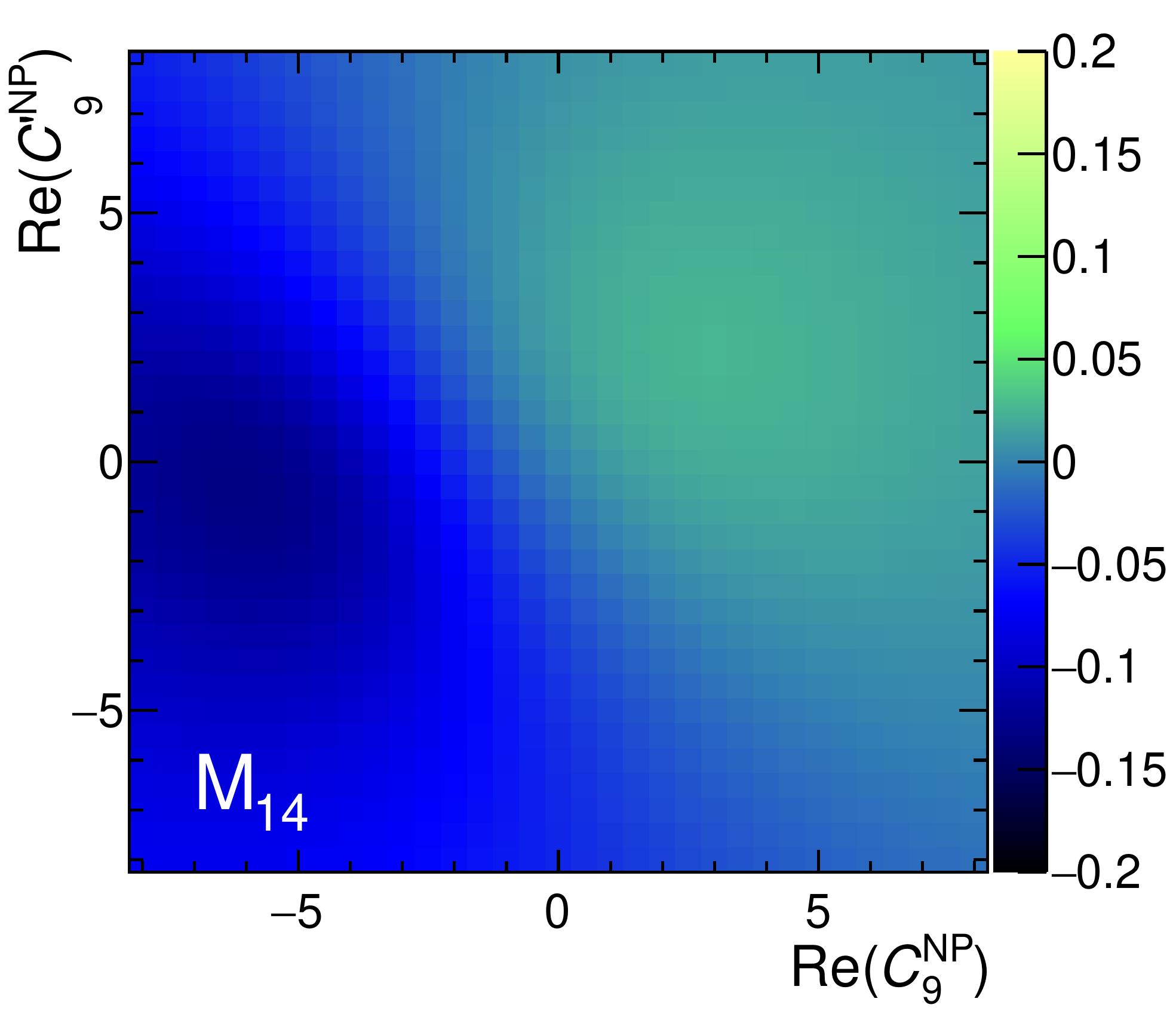}  \\ 
\includegraphics[width=0.24\linewidth]{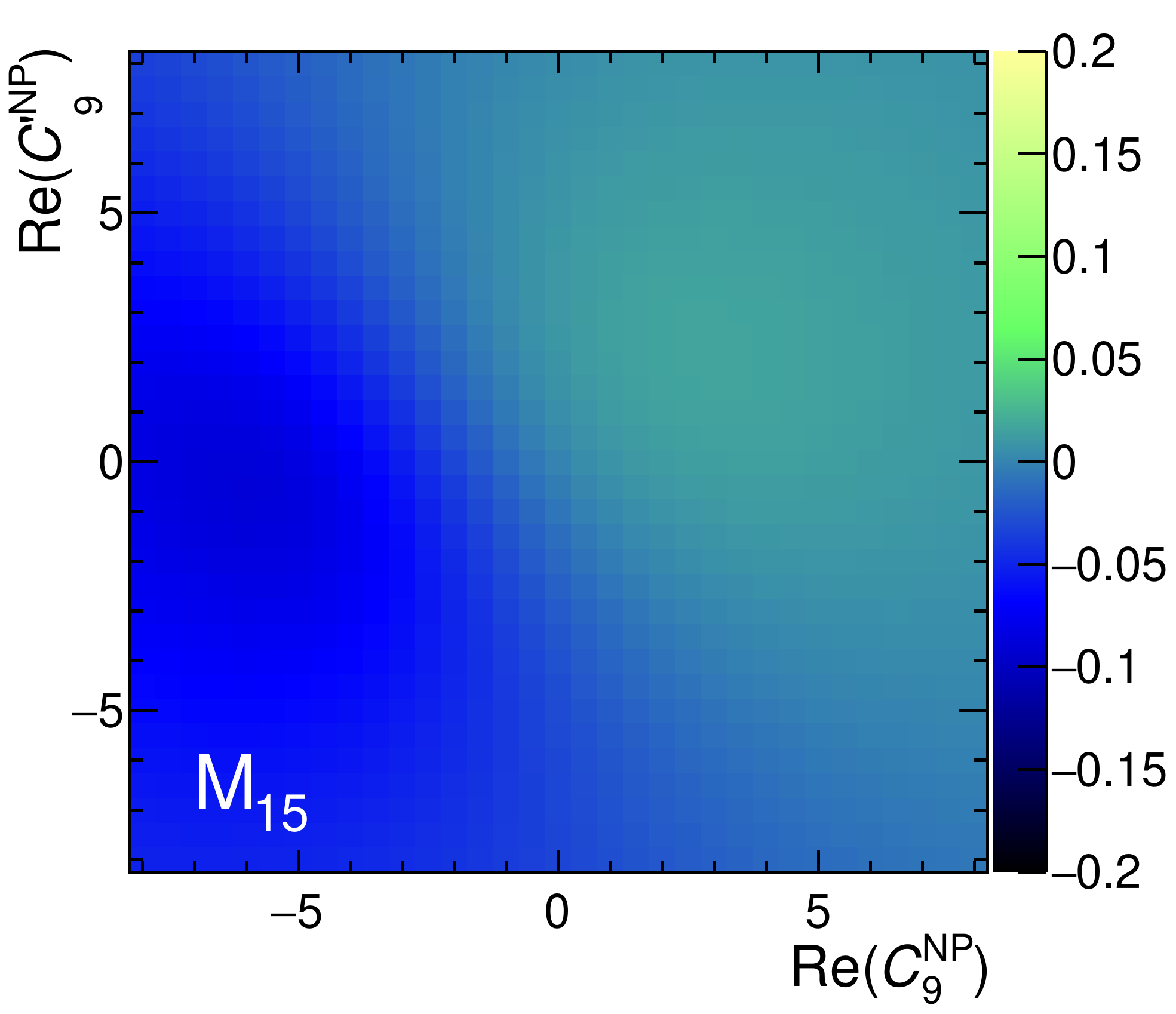} 
\includegraphics[width=0.24\linewidth]{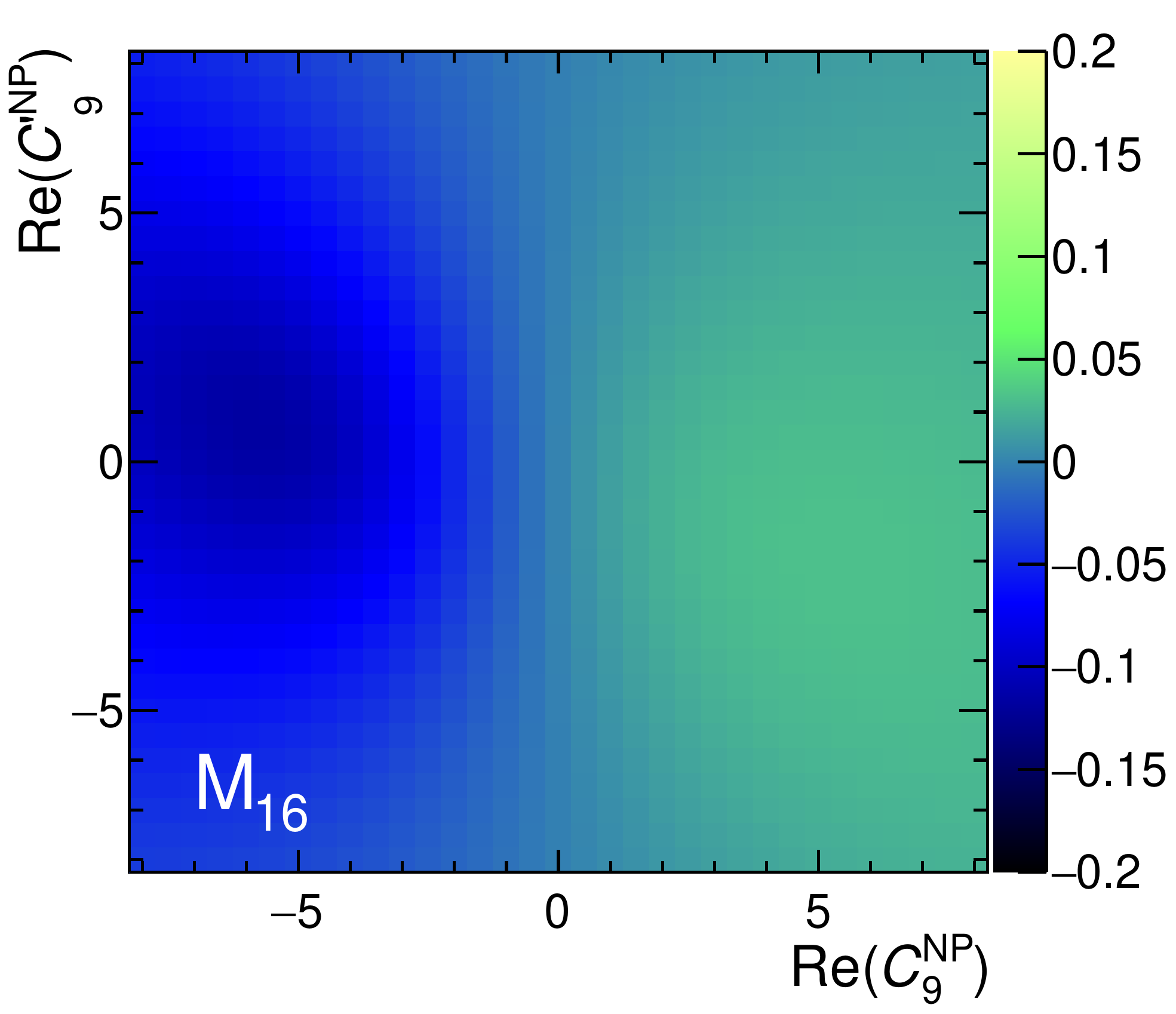} 
\includegraphics[width=0.24\linewidth]{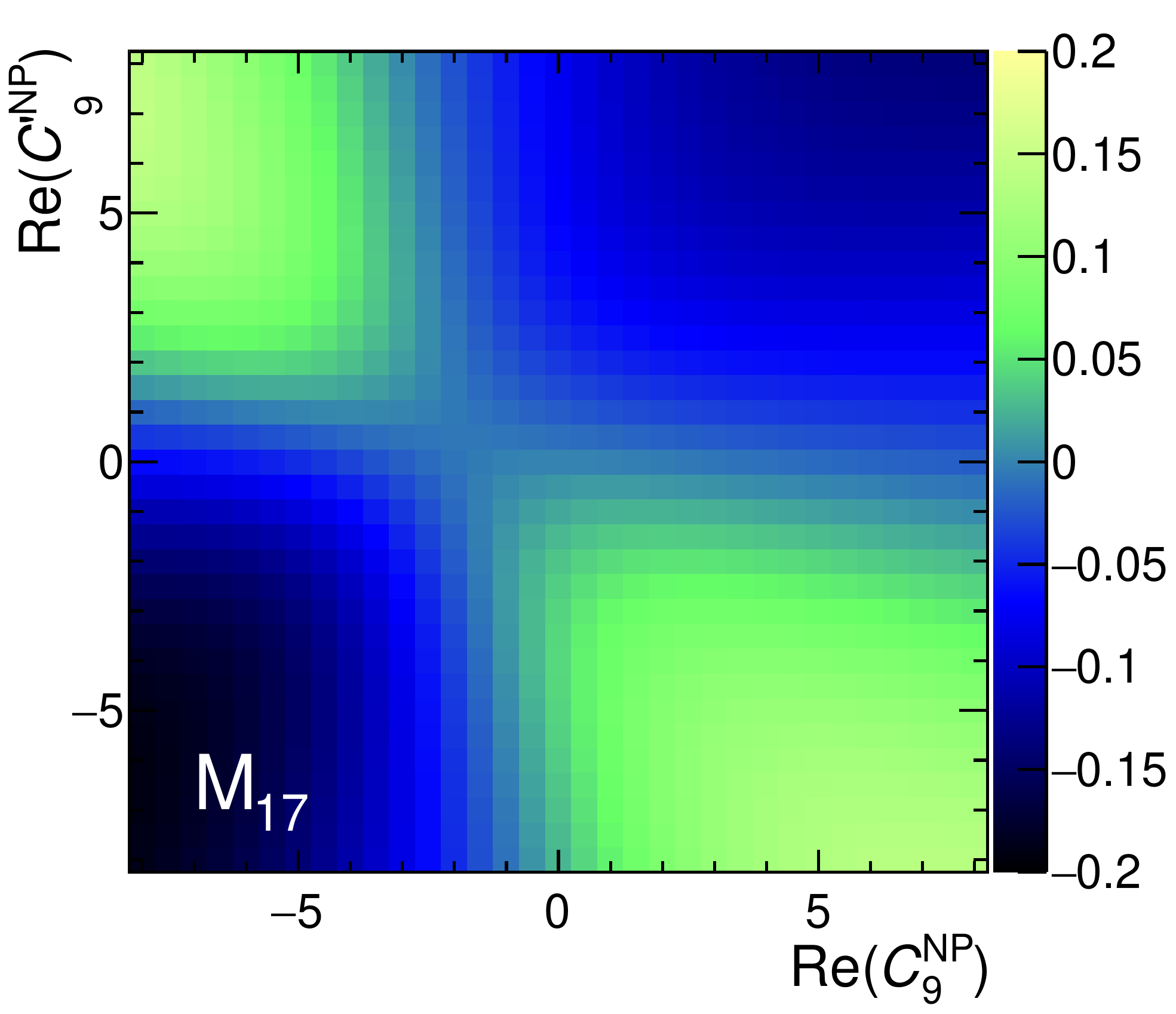}  
\includegraphics[width=0.24\linewidth]{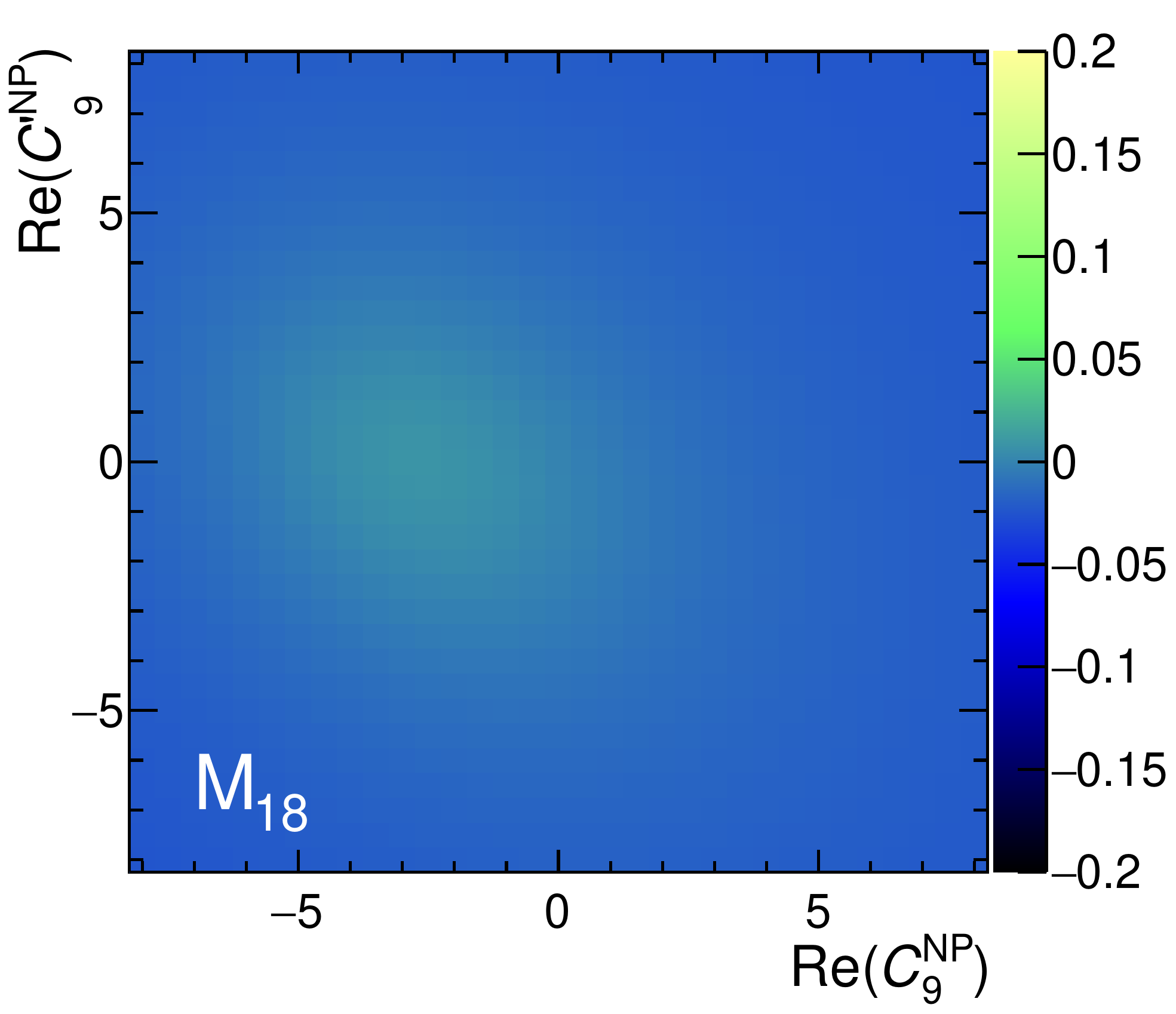} \\
\includegraphics[width=0.24\linewidth]{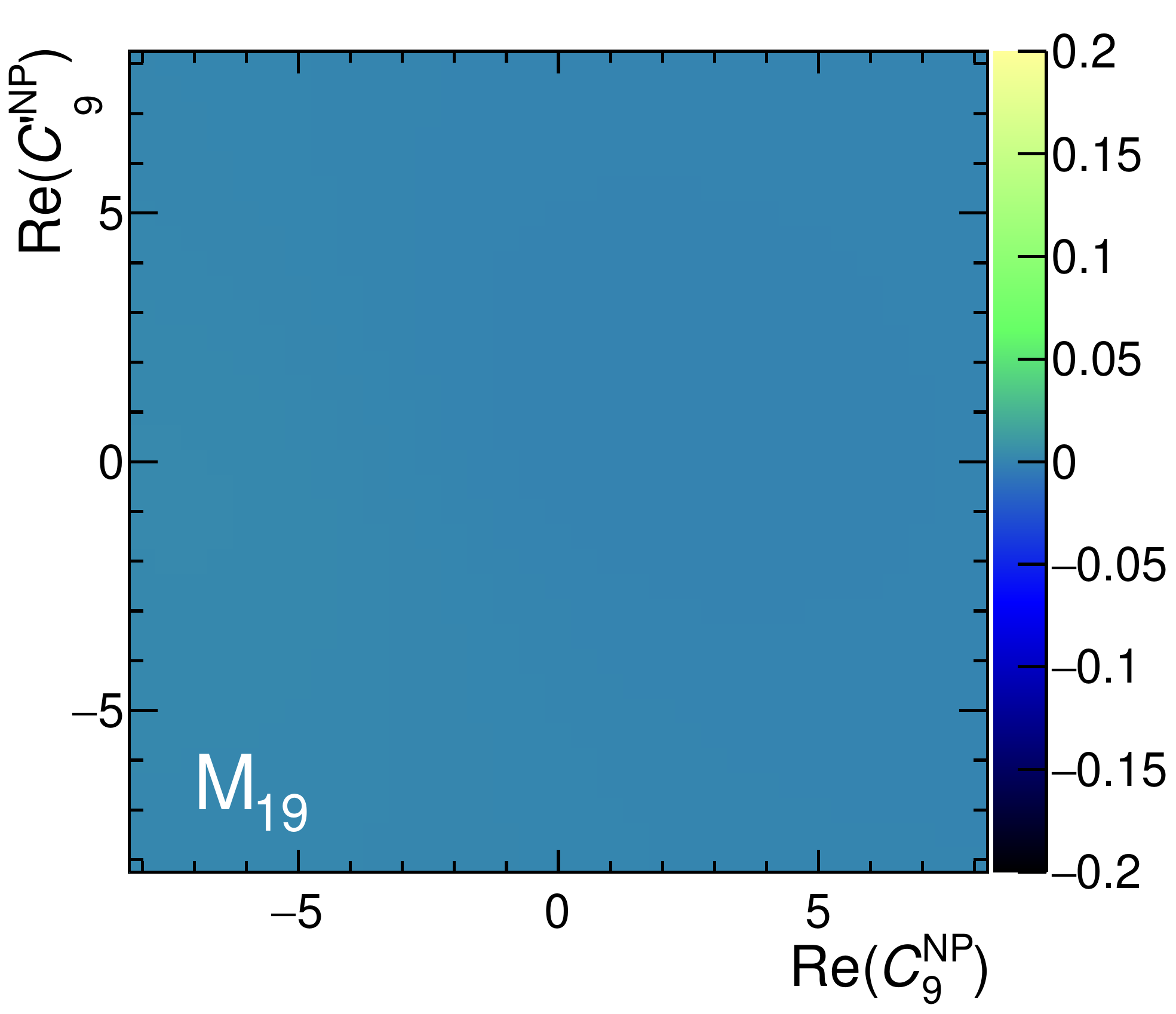} 
\includegraphics[width=0.24\linewidth]{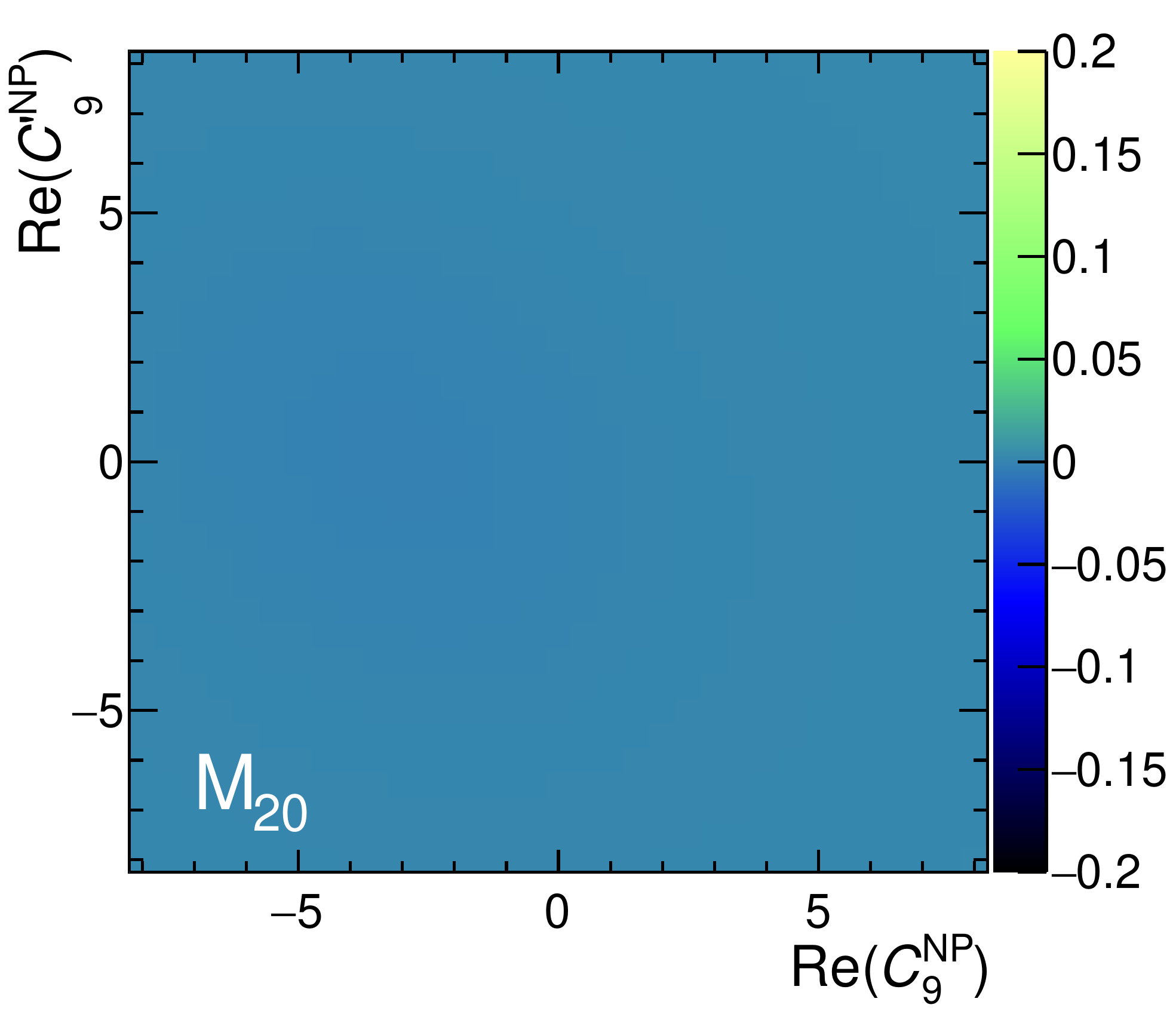}  
\includegraphics[width=0.24\linewidth]{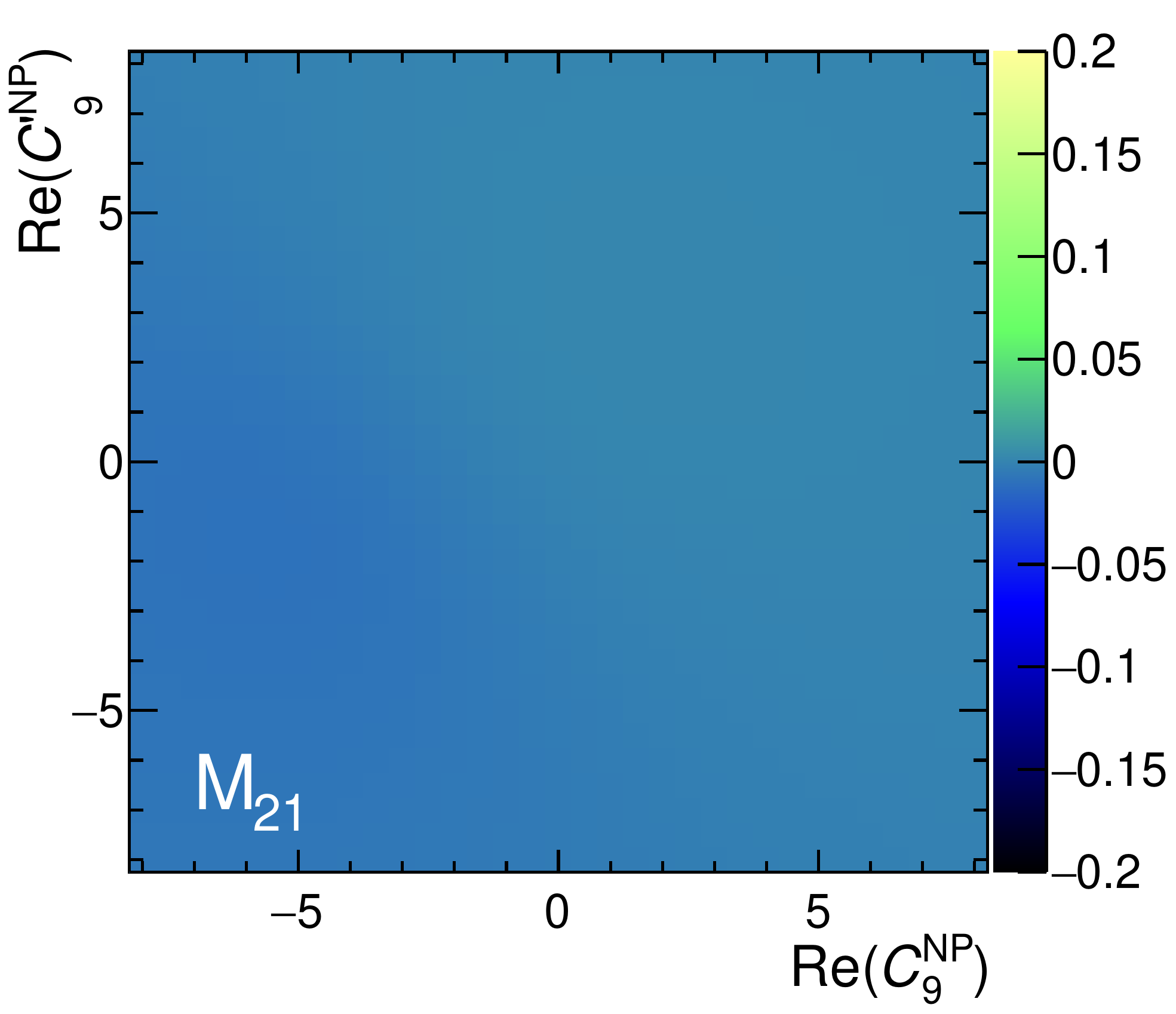}  
\includegraphics[width=0.24\linewidth]{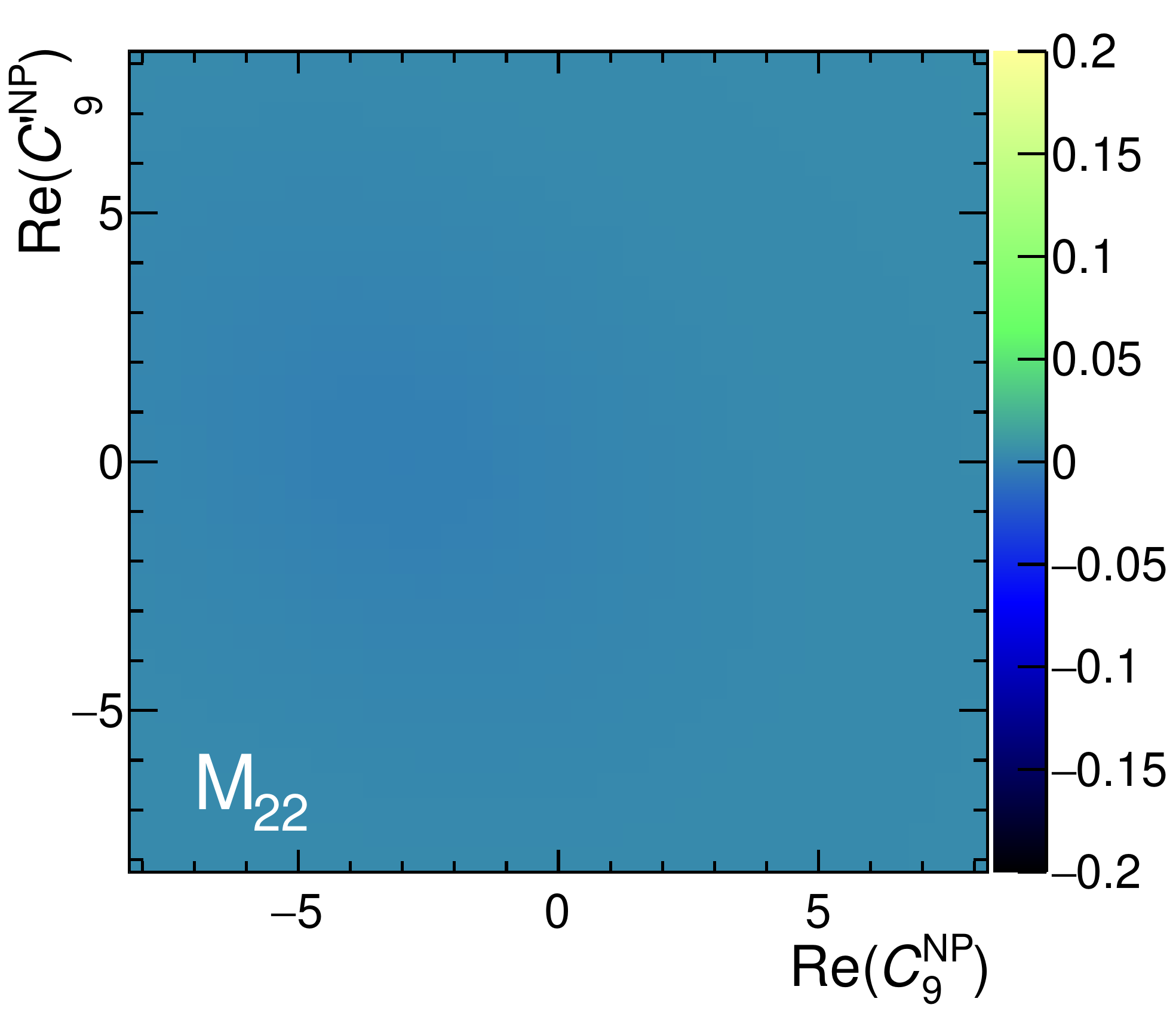} \\
\includegraphics[width=0.24\linewidth]{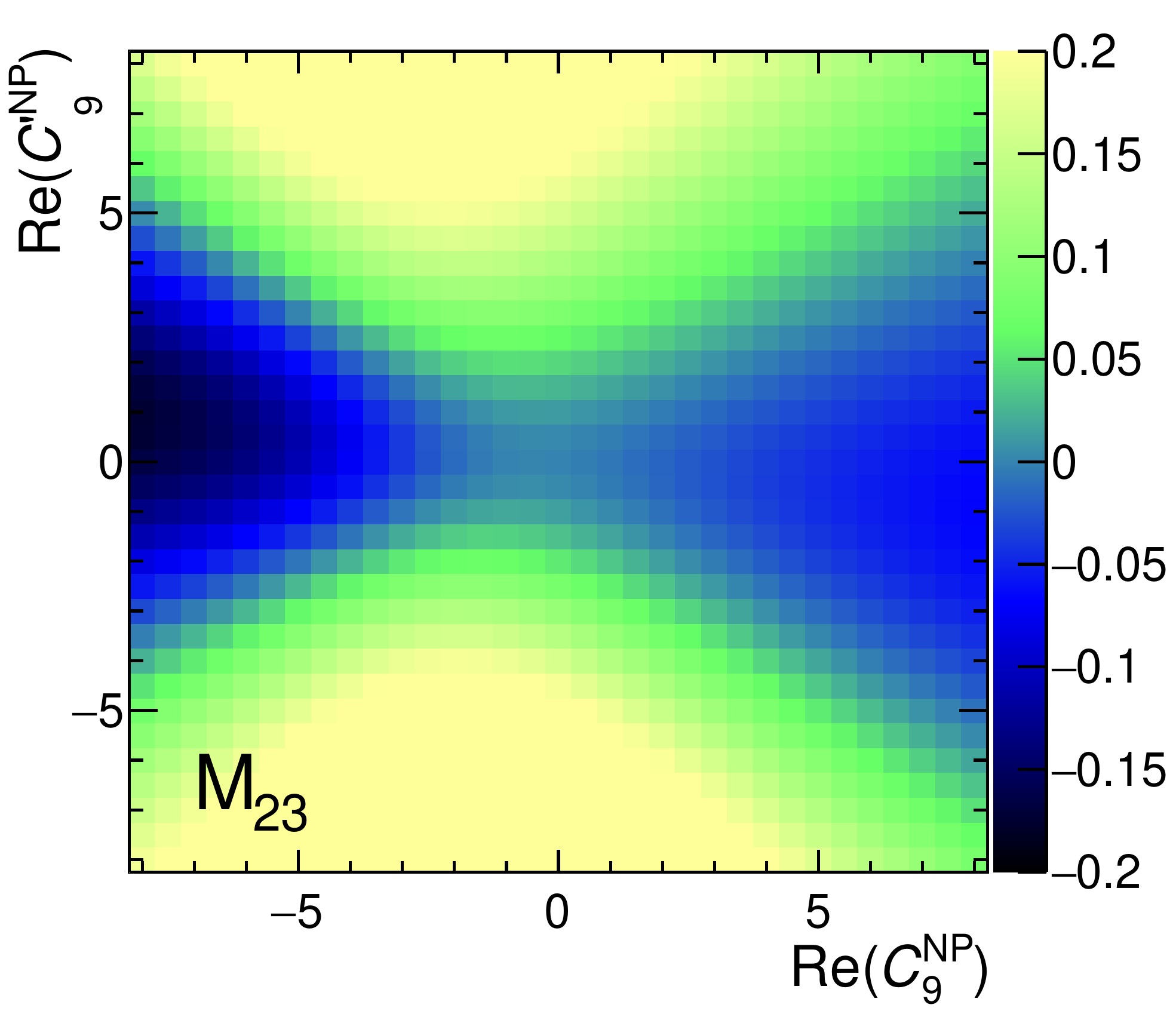} 
\includegraphics[width=0.24\linewidth]{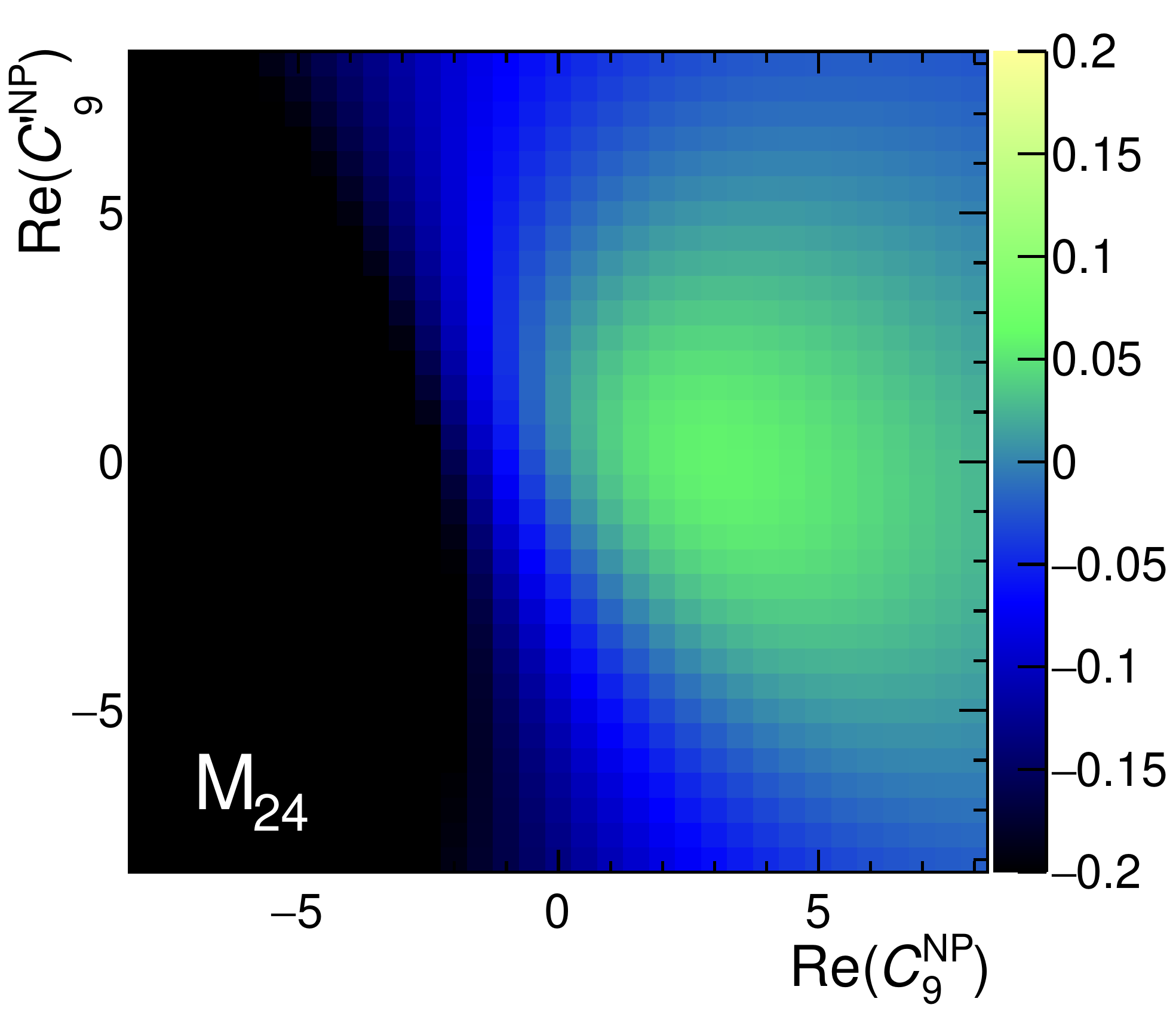}  
\includegraphics[width=0.24\linewidth]{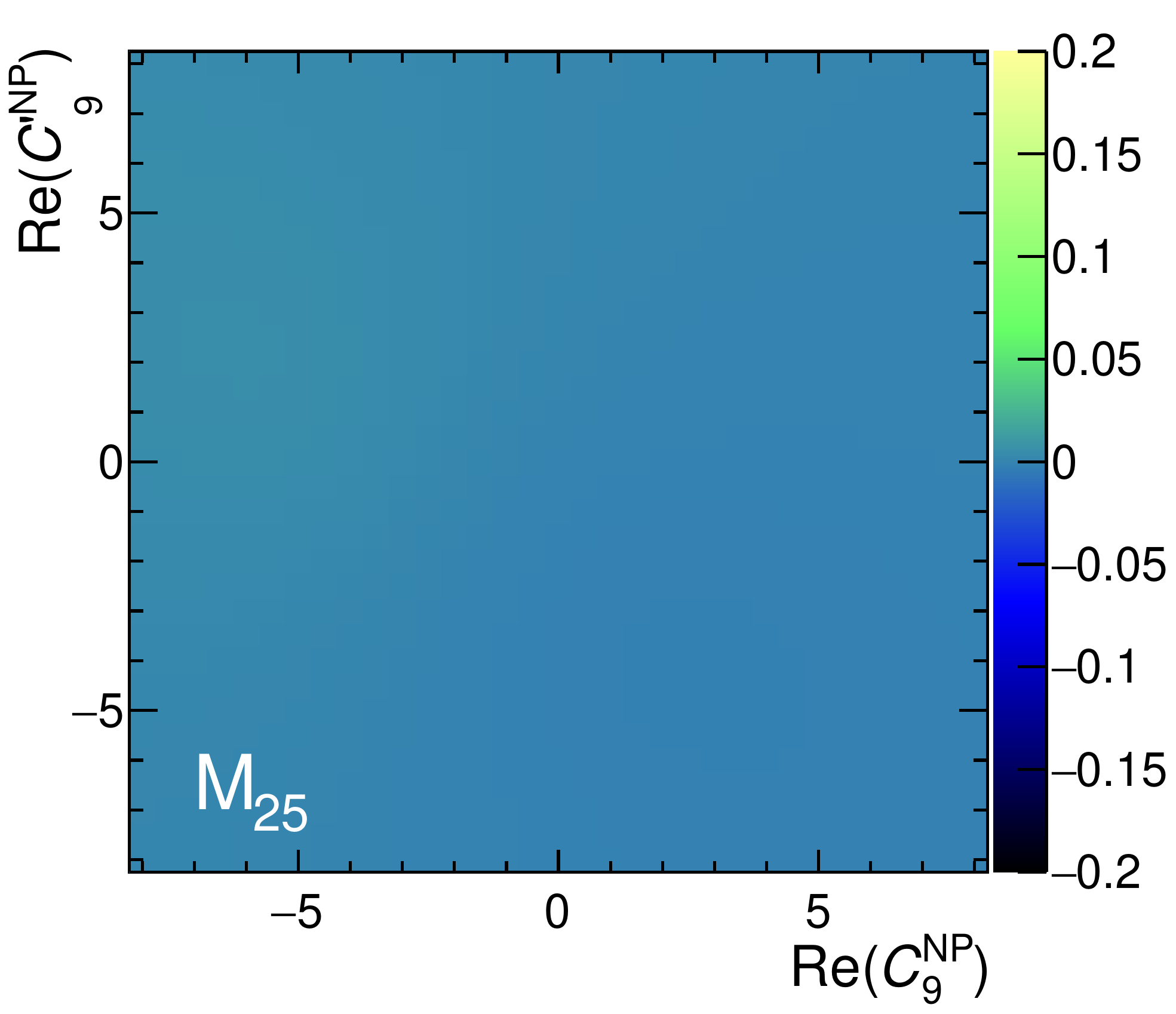} 
\includegraphics[width=0.24\linewidth]{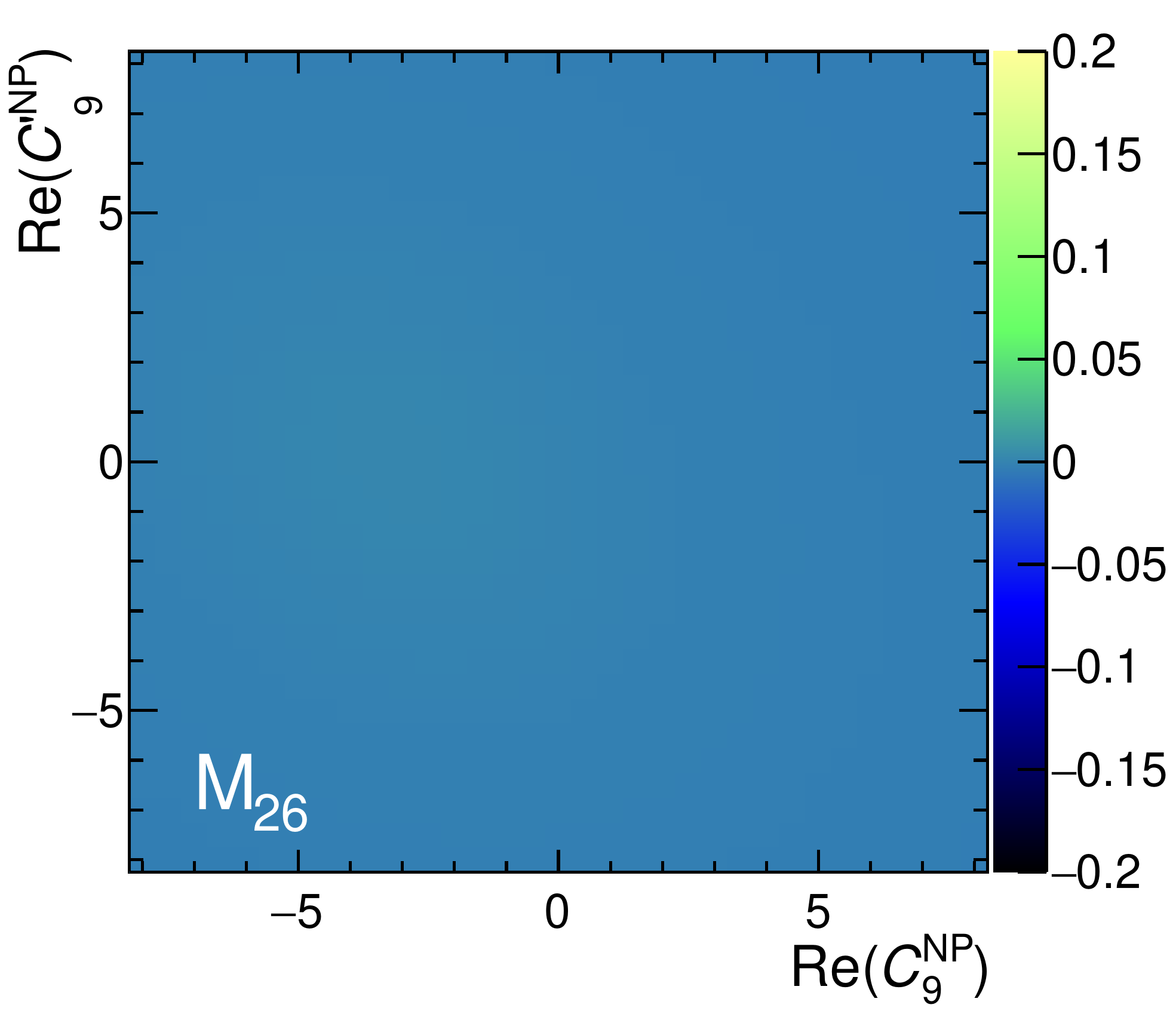} \\ 
\includegraphics[width=0.24\linewidth]{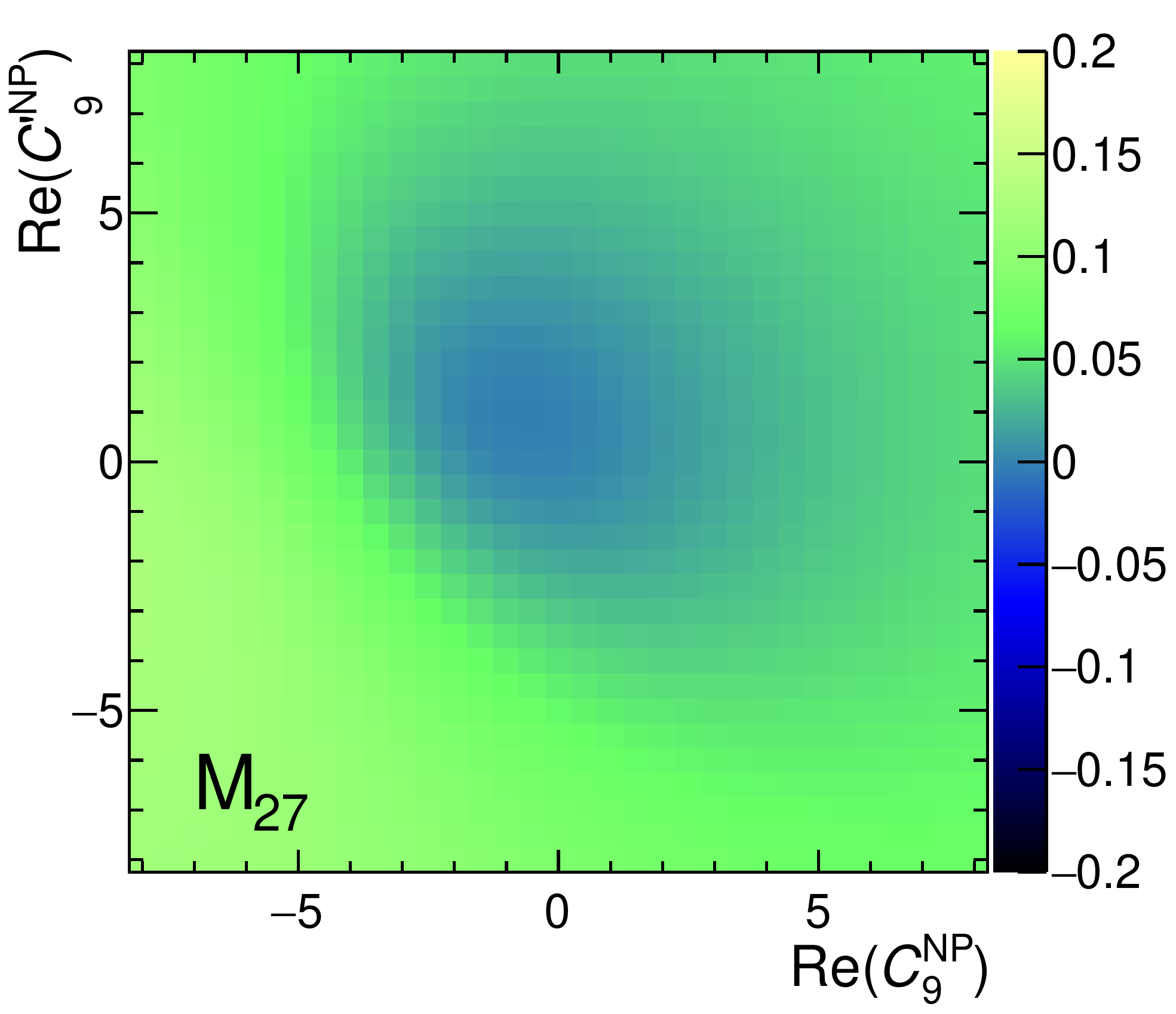}  
\includegraphics[width=0.24\linewidth]{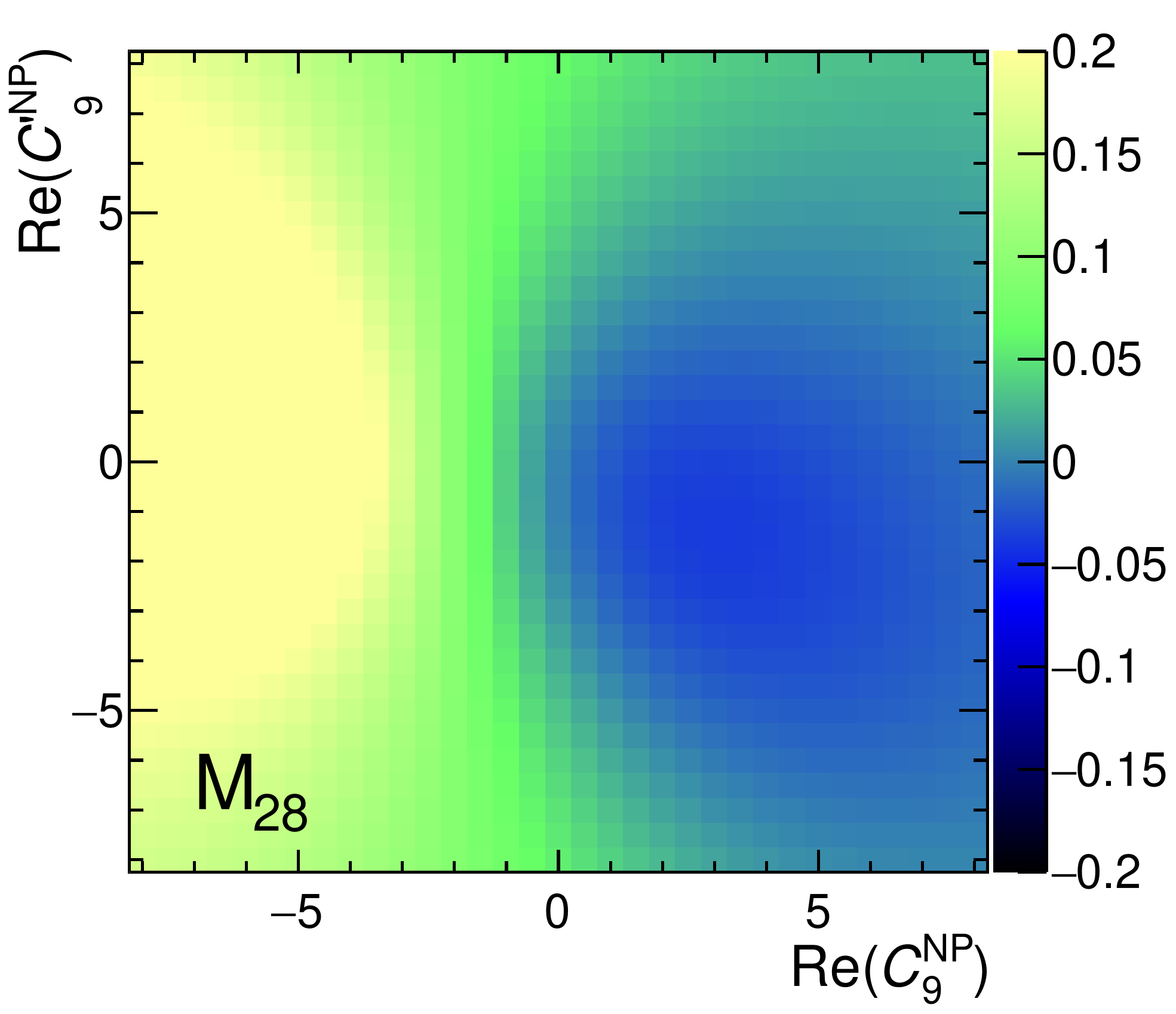} 
\includegraphics[width=0.24\linewidth]{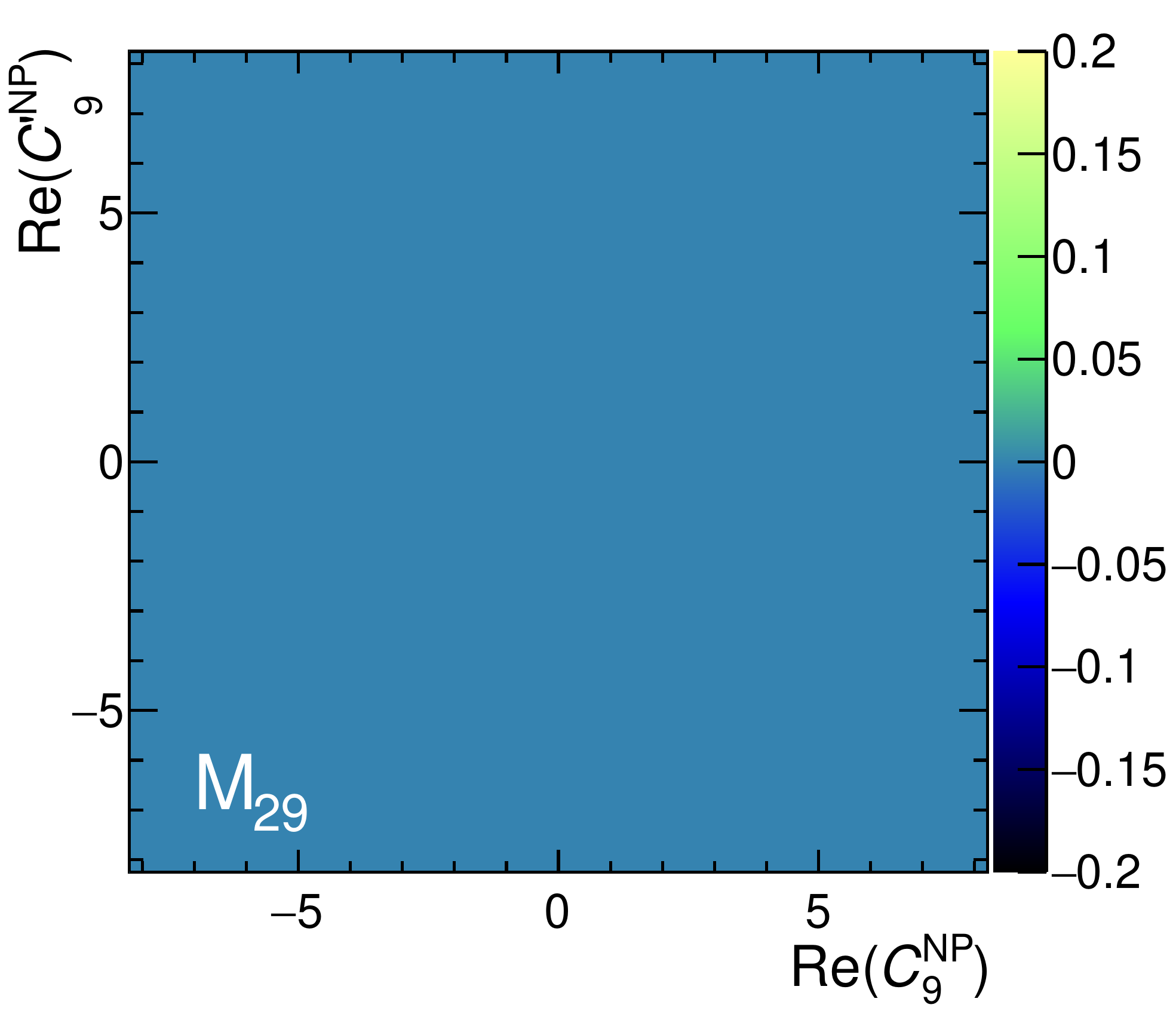} 
\includegraphics[width=0.24\linewidth]{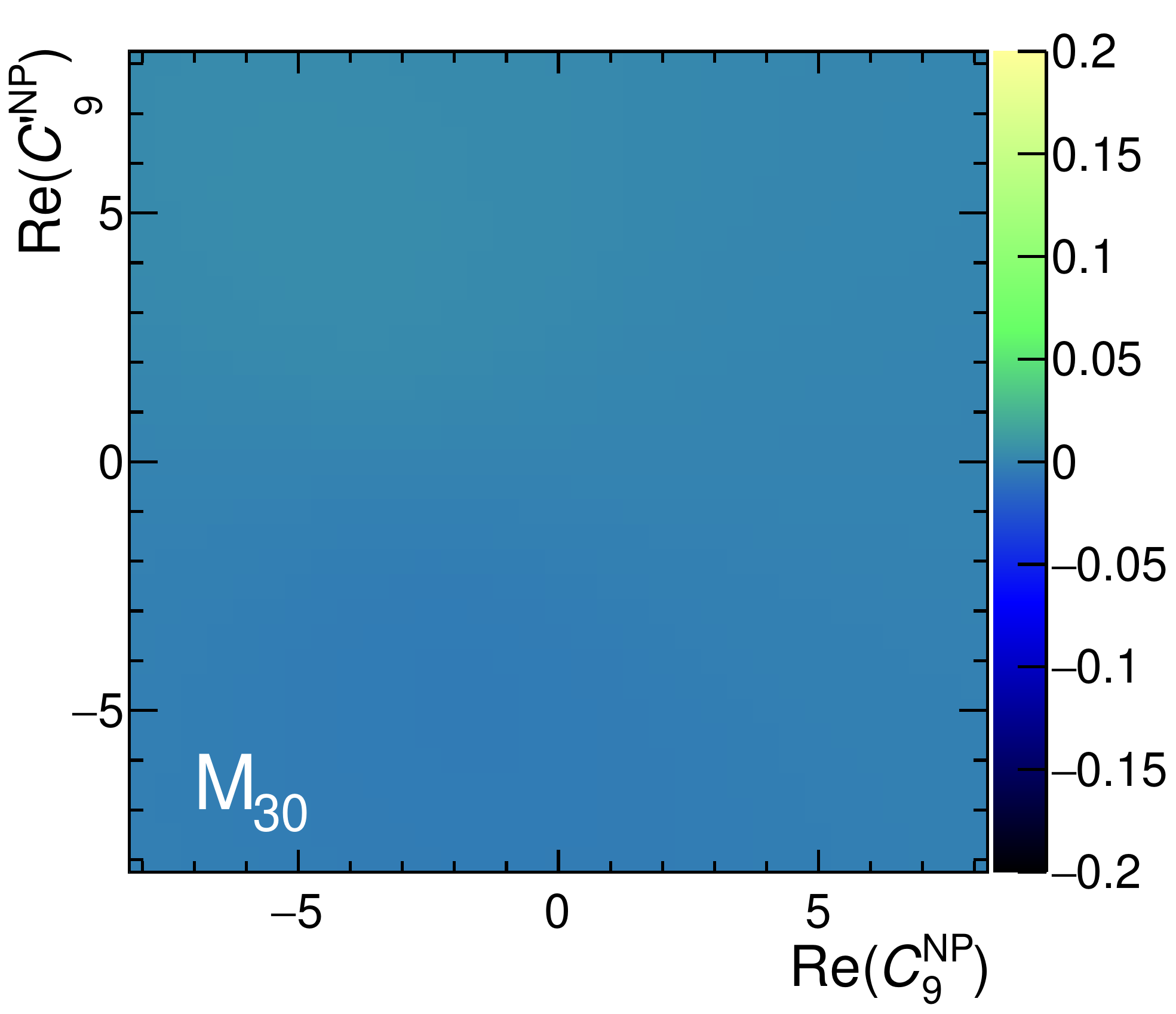} \\ 
\includegraphics[width=0.24\linewidth]{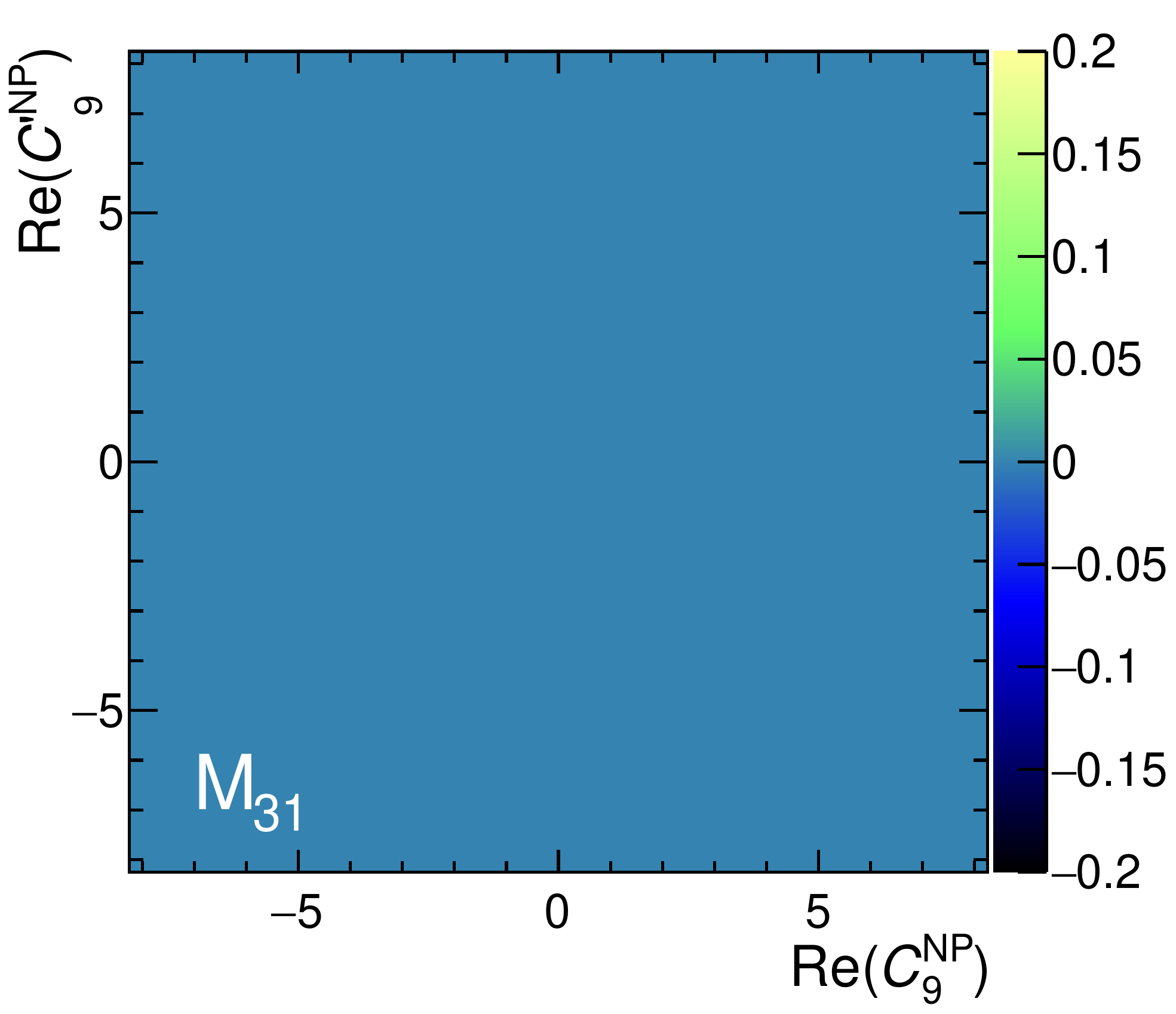}  
\includegraphics[width=0.24\linewidth]{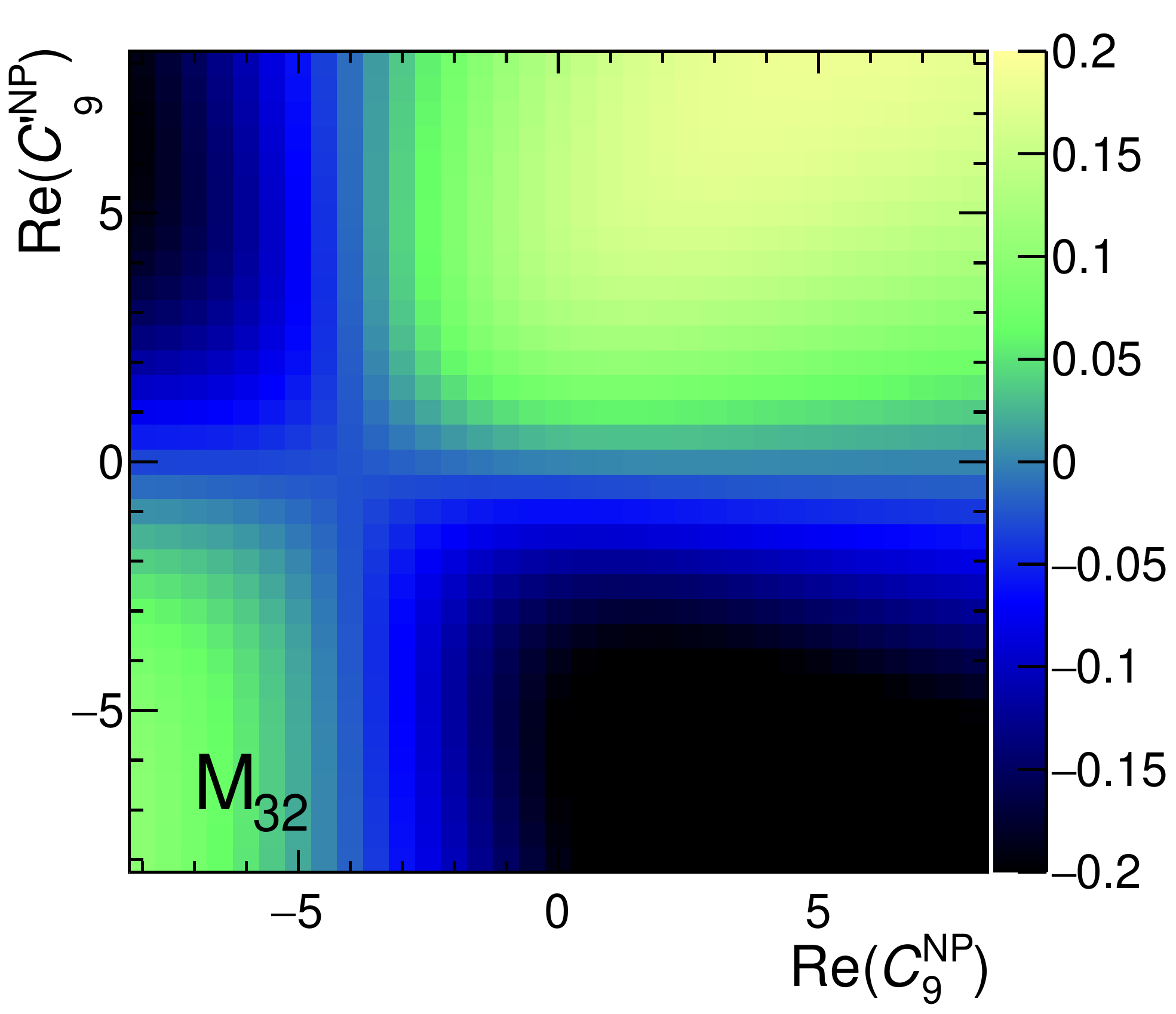} 
\includegraphics[width=0.24\linewidth]{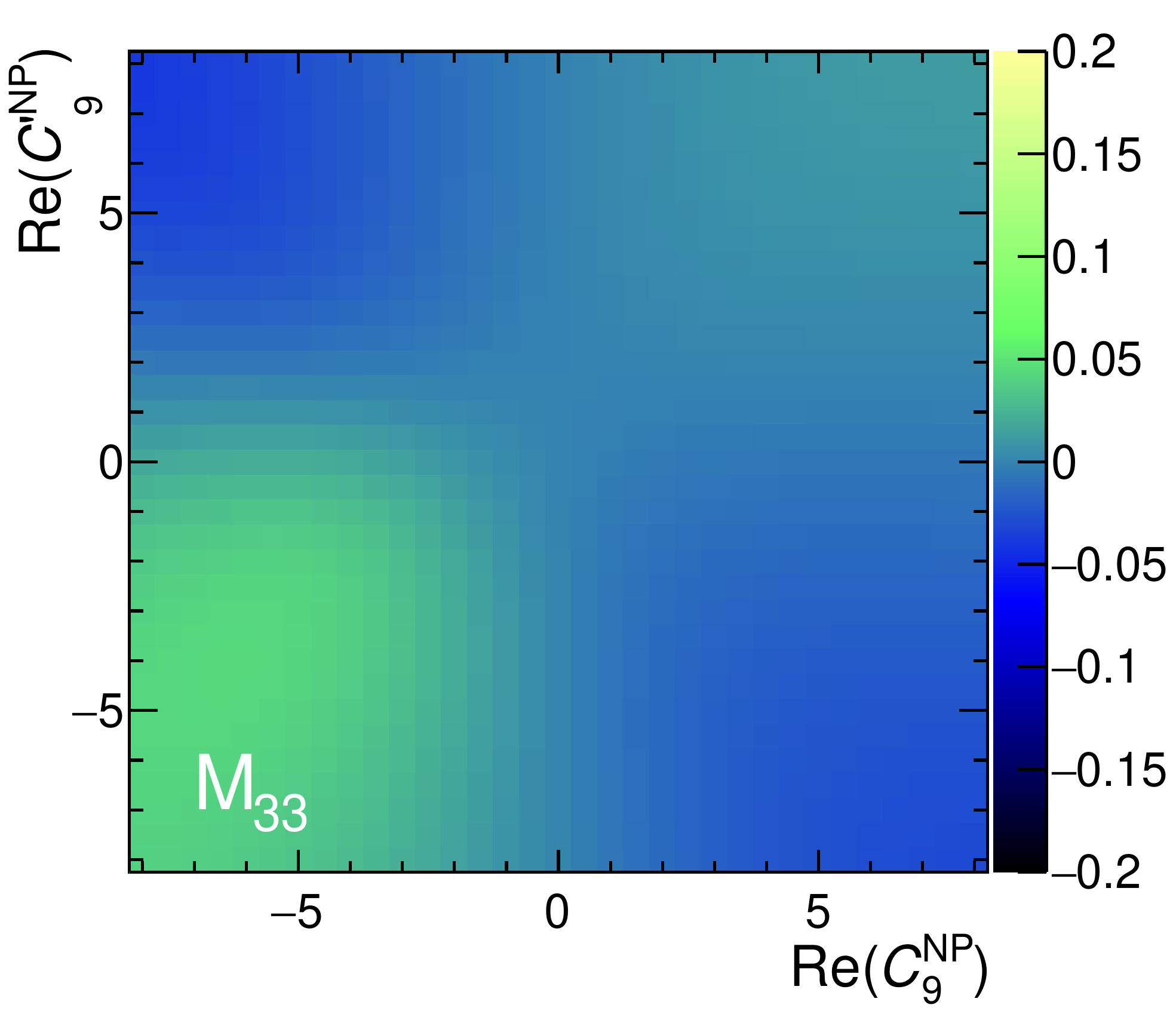} 
\includegraphics[width=0.24\linewidth]{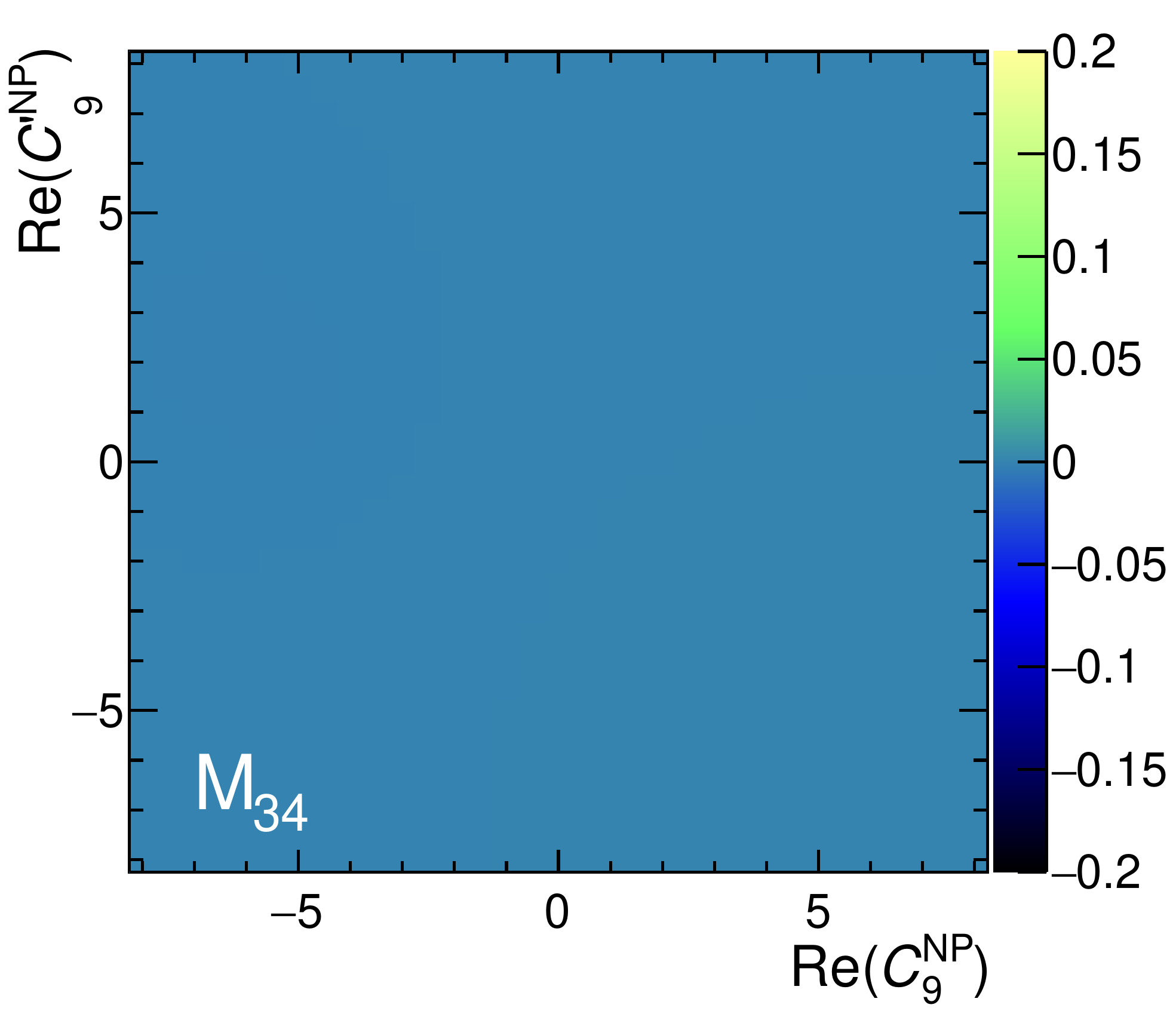}  
\caption{
Variation of the polarisation dependent angular observables  of the \decay{\Lb}{\Lz\mumu} decay from their SM central values in the large-recoil region ($1 < \qsq < 6\gev^{2}/c^{4}$) with a NP contribution to ${\rm Re}(C_9)$ or ${\rm Re}(C'_{9})$. 
The SM point is at $(0,0)$.
To illustrate the size of the effects, $P_{\Lb}  = 1$ is used.
\label{fig:scan:c9:c9p:largerecoil:pol} 
}
\end{figure}

\begin{figure}[!htb]
\centering
\includegraphics[width=0.24\linewidth]{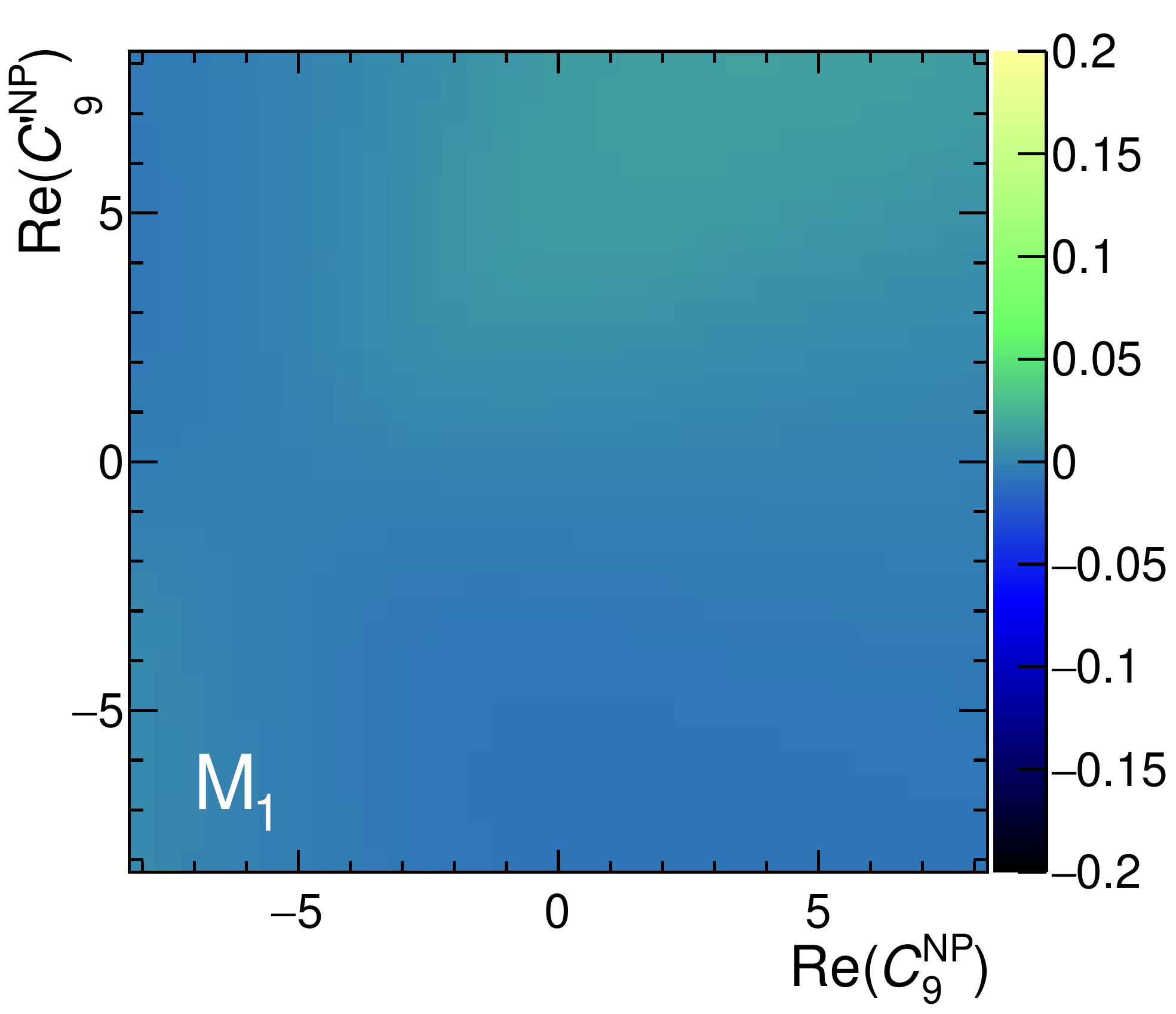}  
\includegraphics[width=0.24\linewidth]{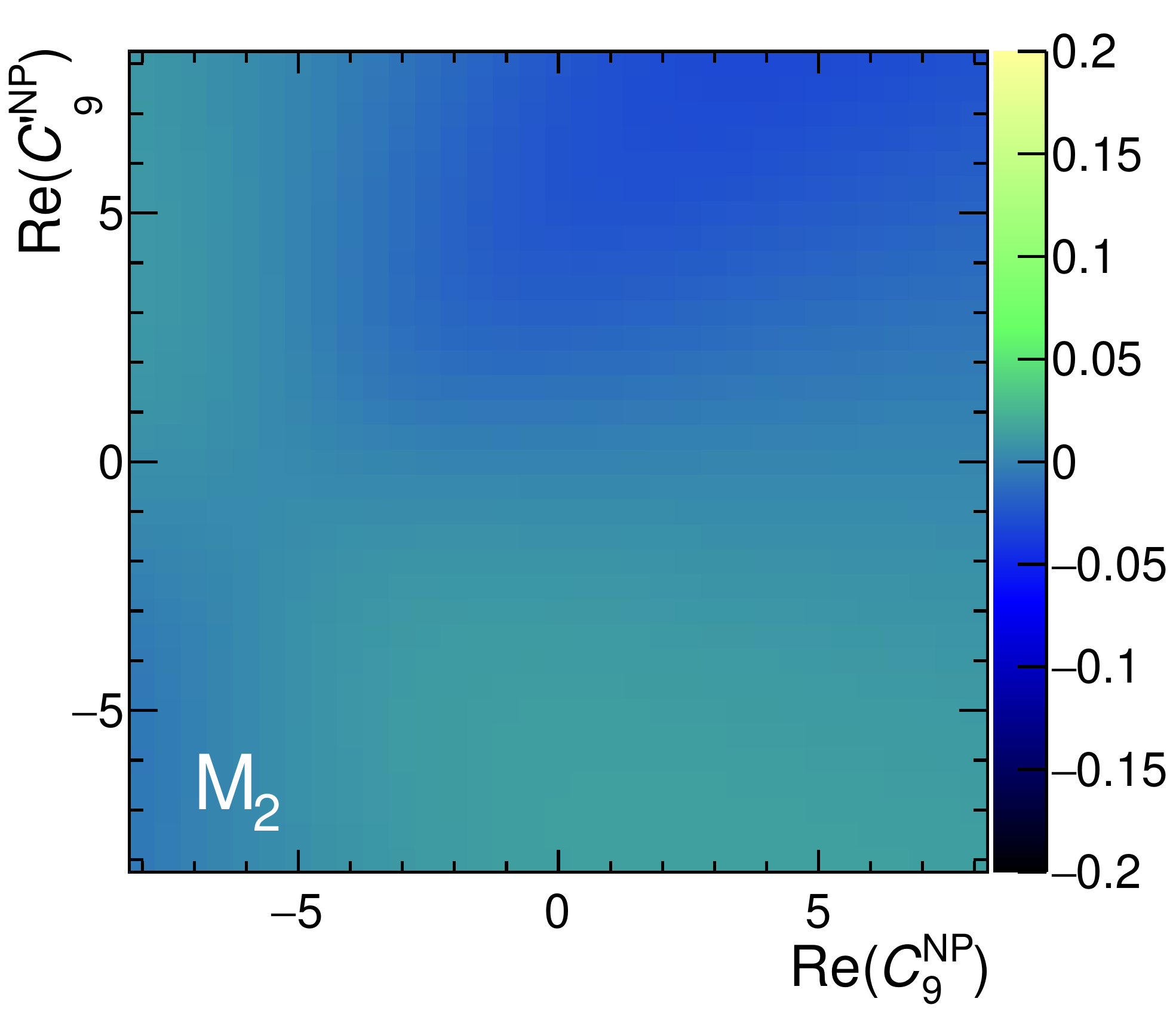} 
\includegraphics[width=0.24\linewidth]{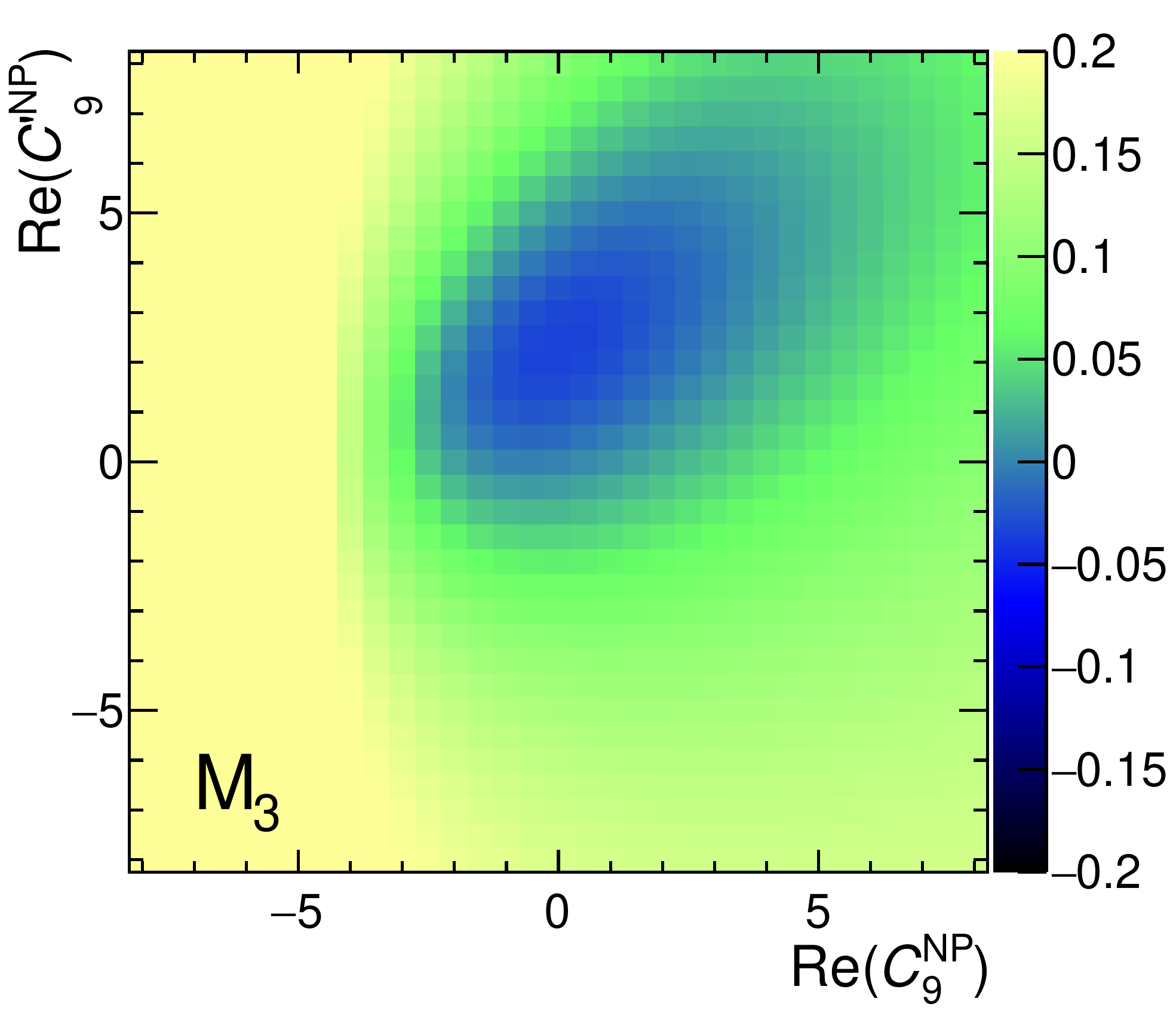} 
\includegraphics[width=0.24\linewidth]{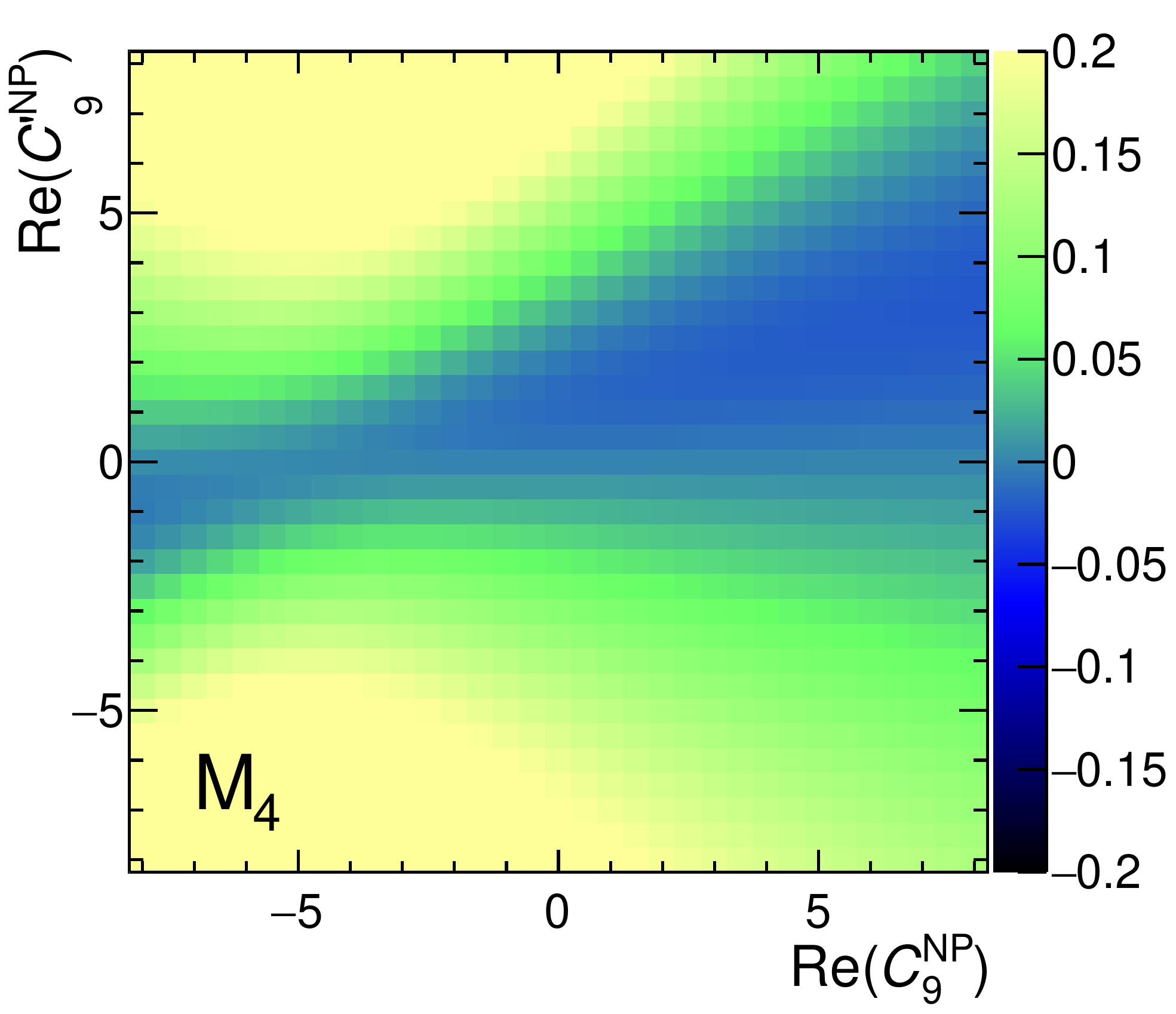} \\ 
\includegraphics[width=0.24\linewidth]{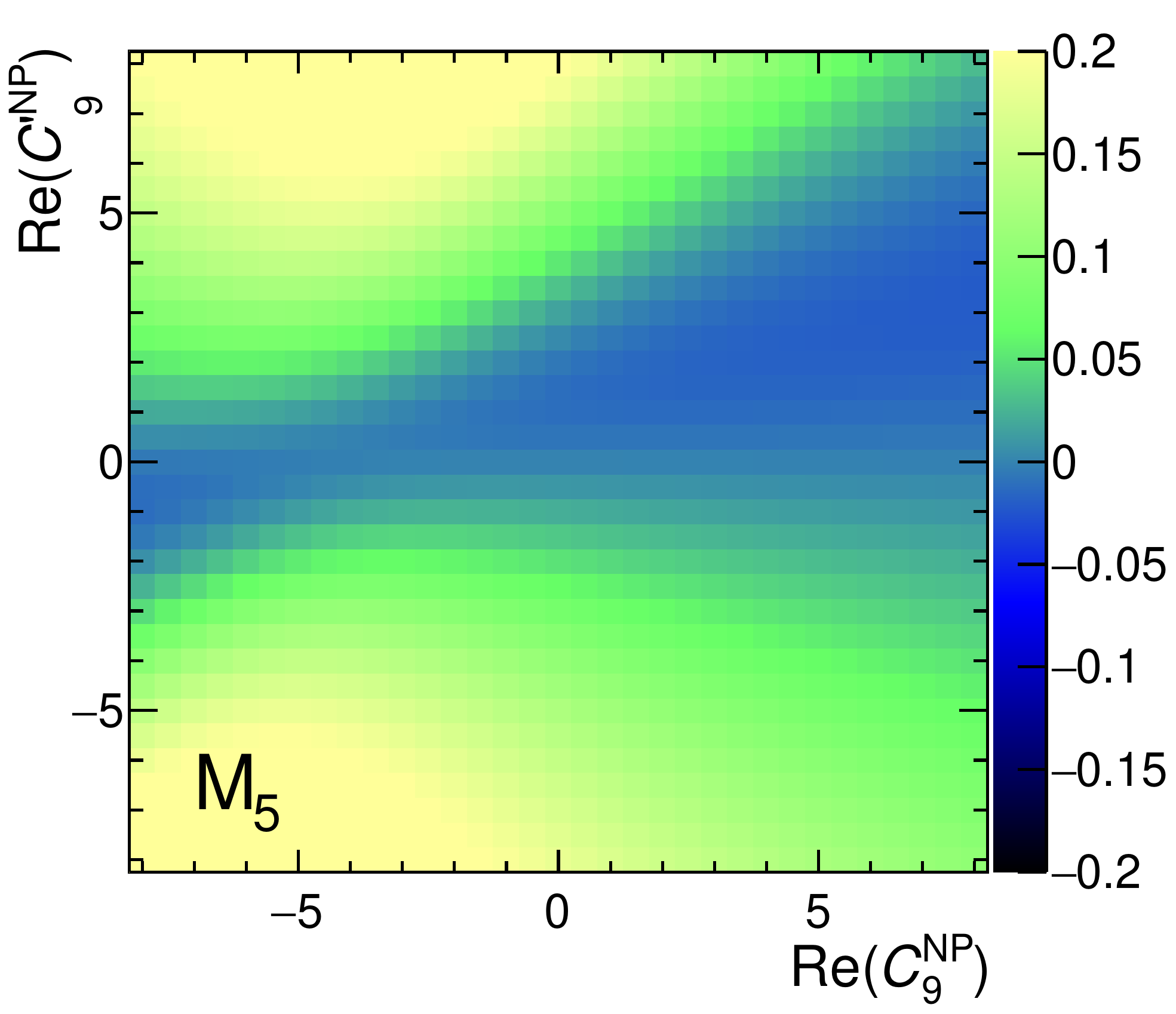}  
\includegraphics[width=0.24\linewidth]{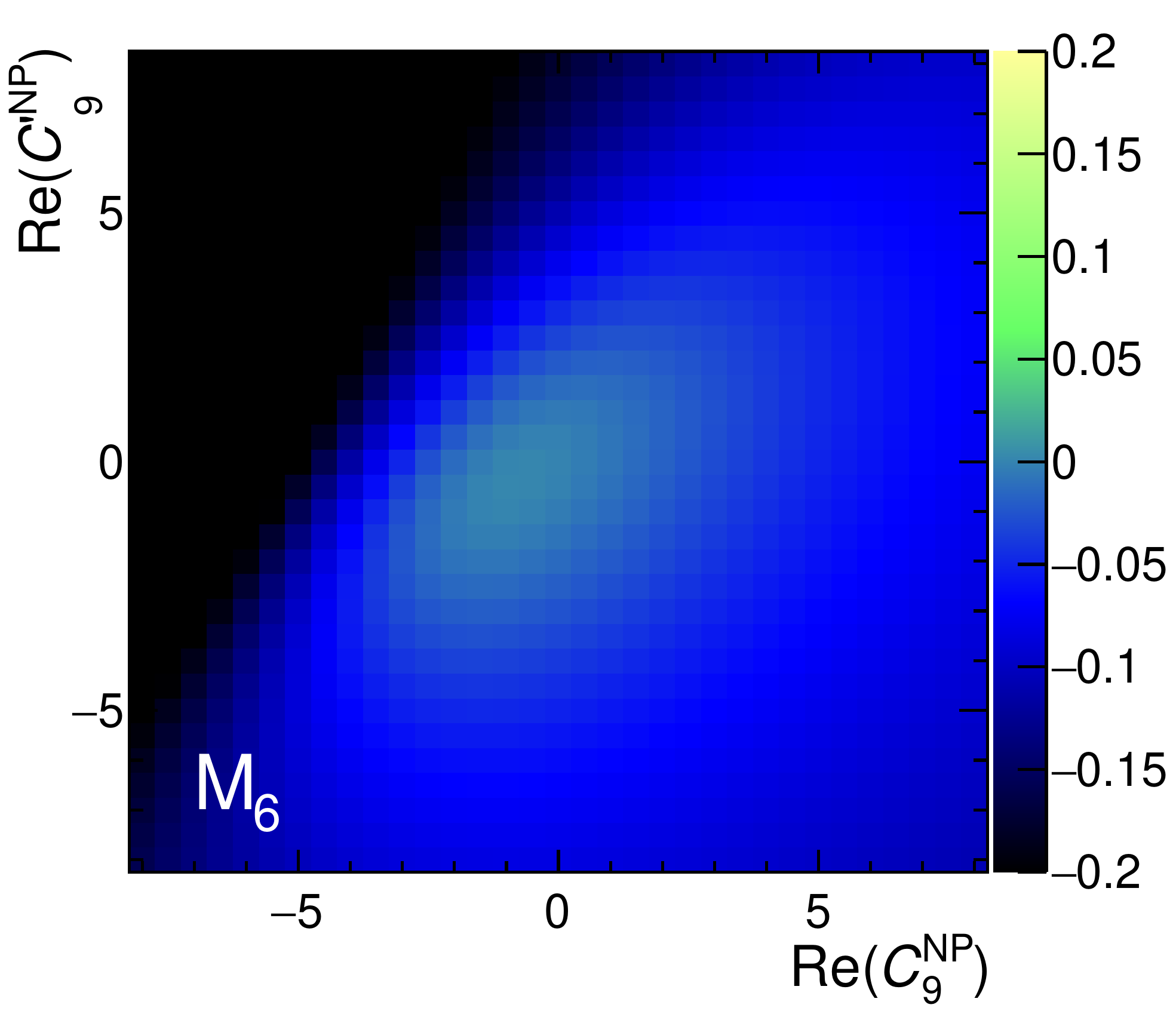} 
\includegraphics[width=0.24\linewidth]{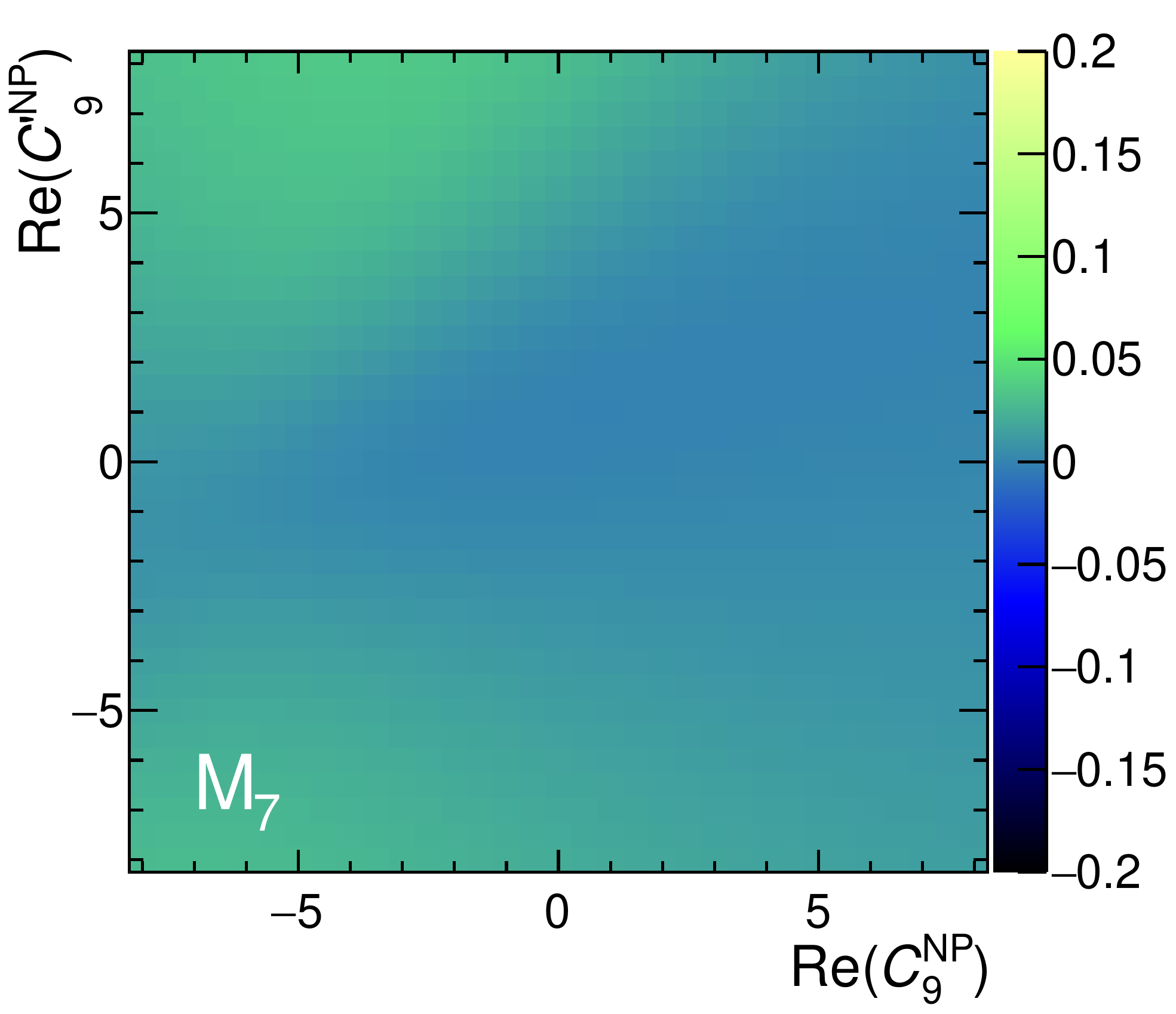} 
\includegraphics[width=0.24\linewidth]{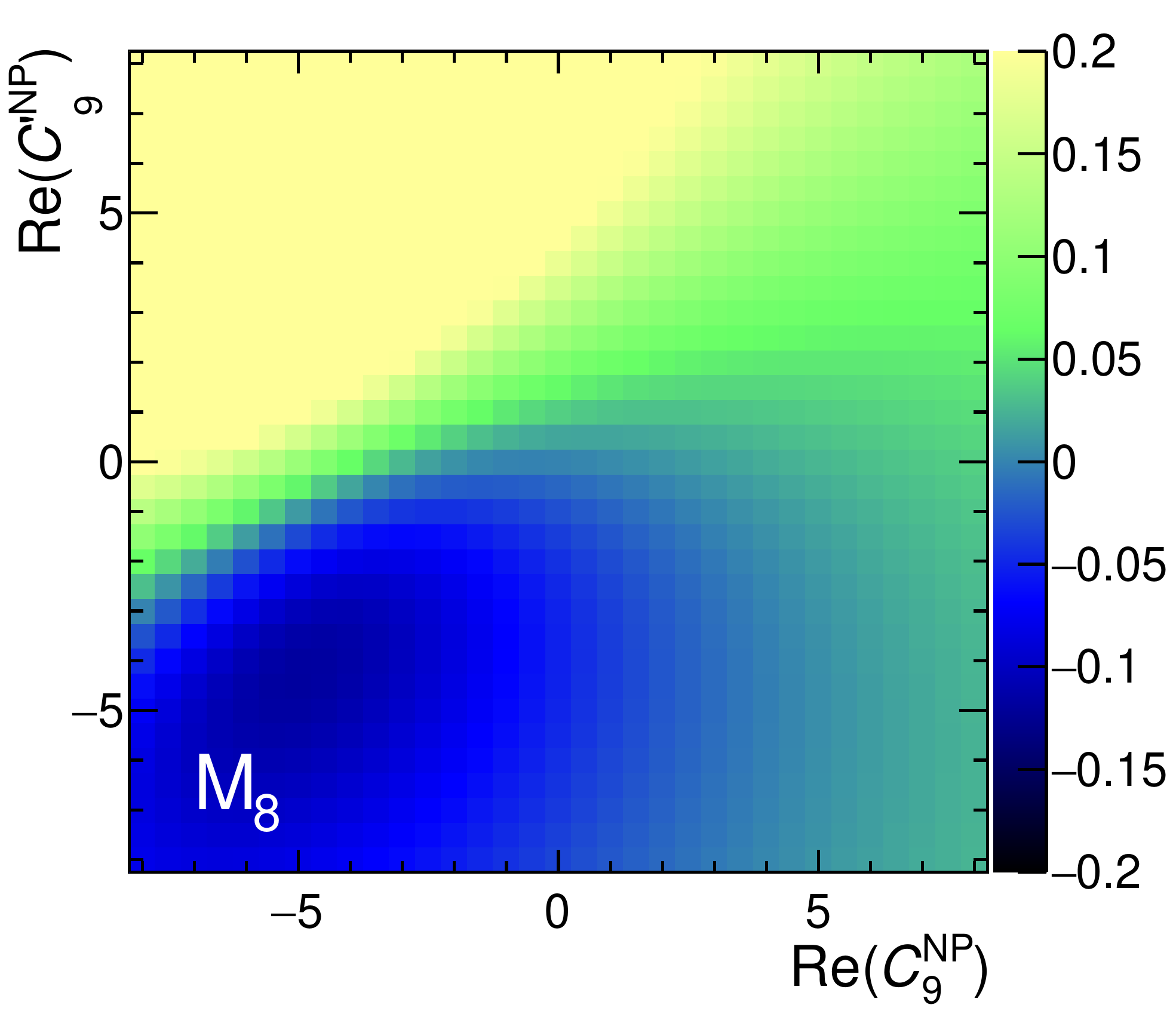} \\ 
\includegraphics[width=0.24\linewidth]{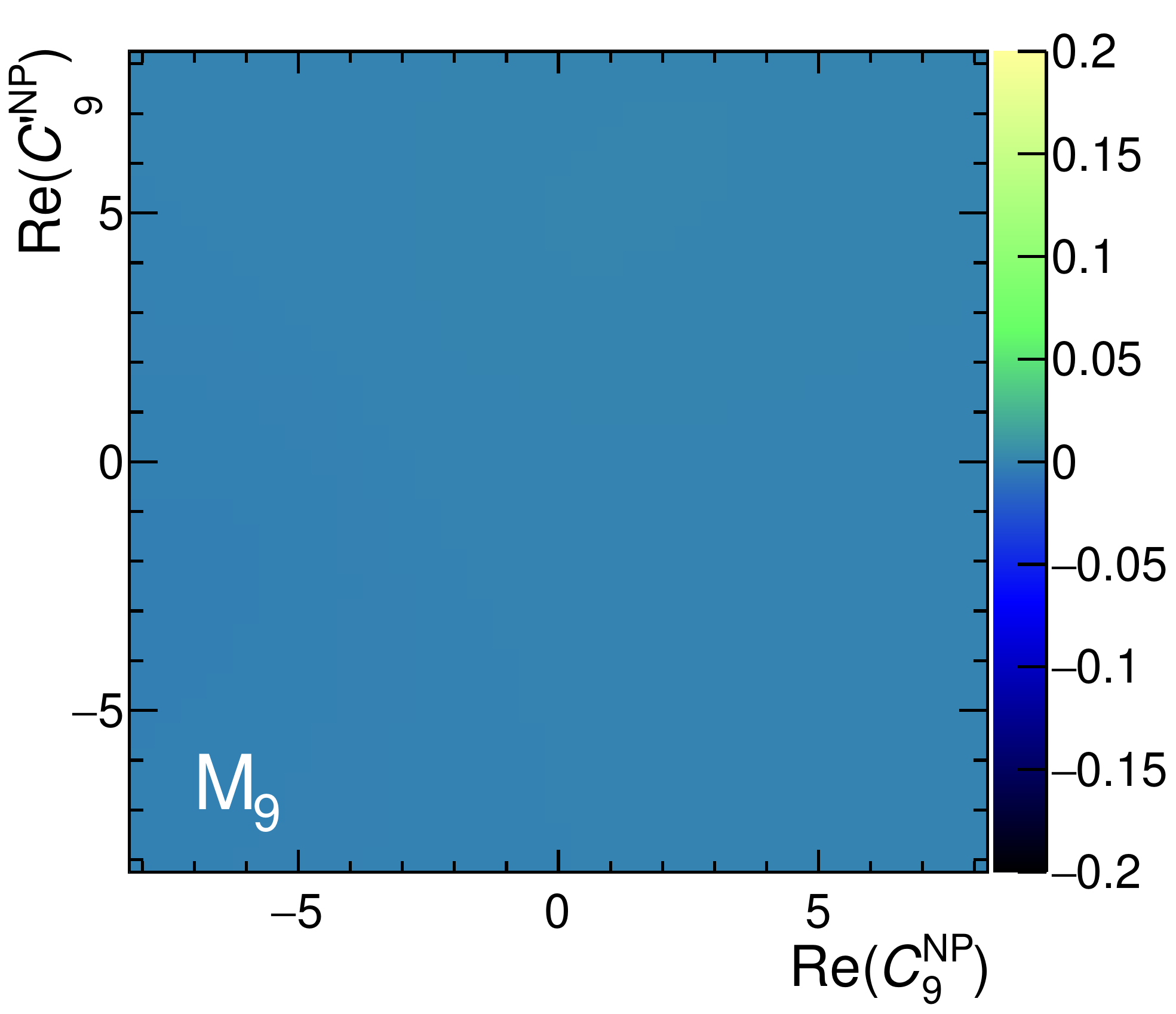}  
\includegraphics[width=0.24\linewidth]{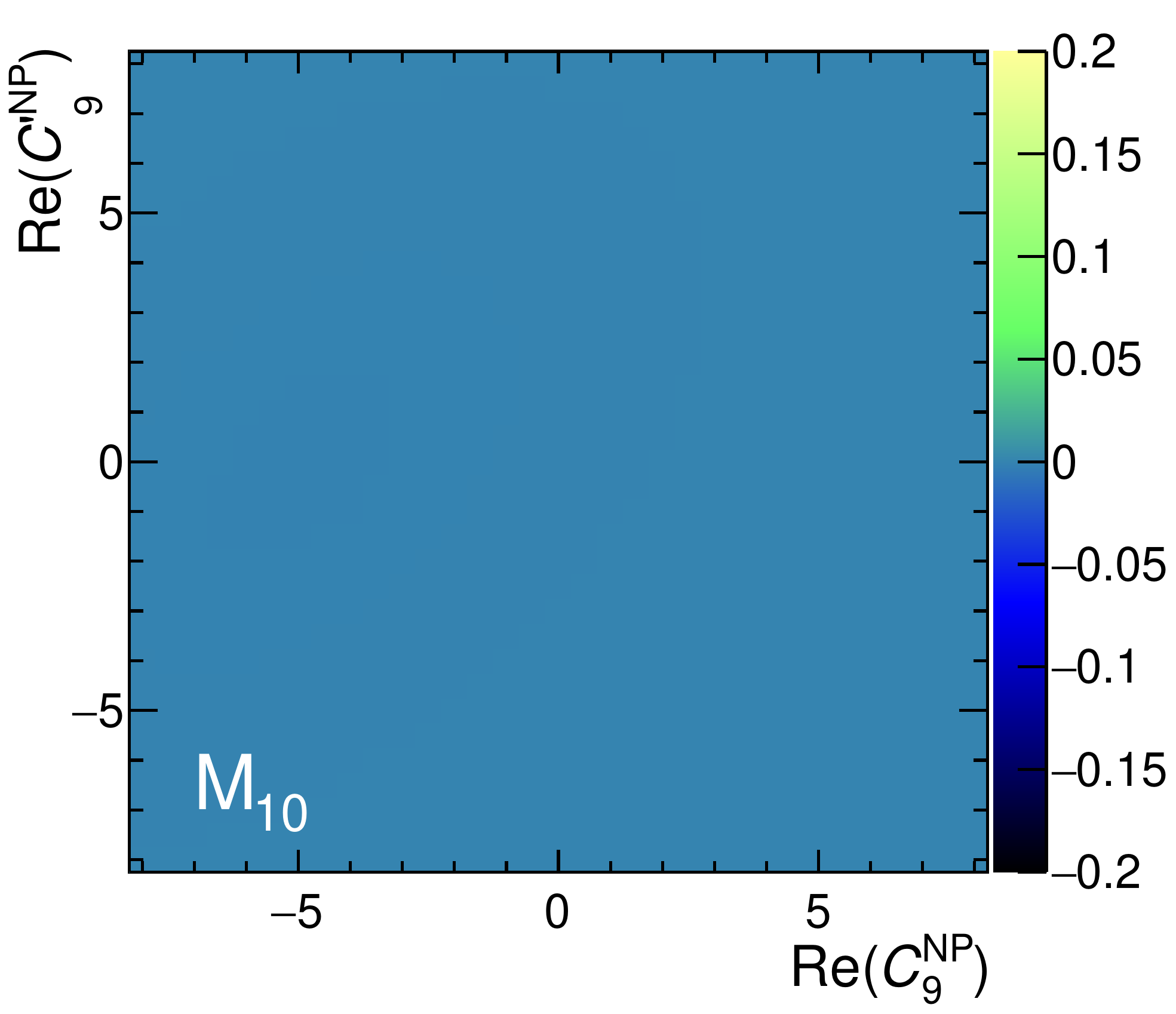} 
\caption{
Variation of the observables $M_{1}$--$M_{10}$  of the \decay{\Lb}{\Lz\mumu} decay  from their SM central values in the low-recoil region ($15 < \qsq < 20\gev^{2}/c^{4}$) with a NP contribution to ${\rm Re}(C_9)$ or ${\rm Re}(C'_{9})$. 
The SM point is at $(0,0)$.
\label{fig:scan:c9:c9p:lowrecoil} 
}
\end{figure}

\begin{figure}[!htb]
\centering
\includegraphics[width=0.24\linewidth]{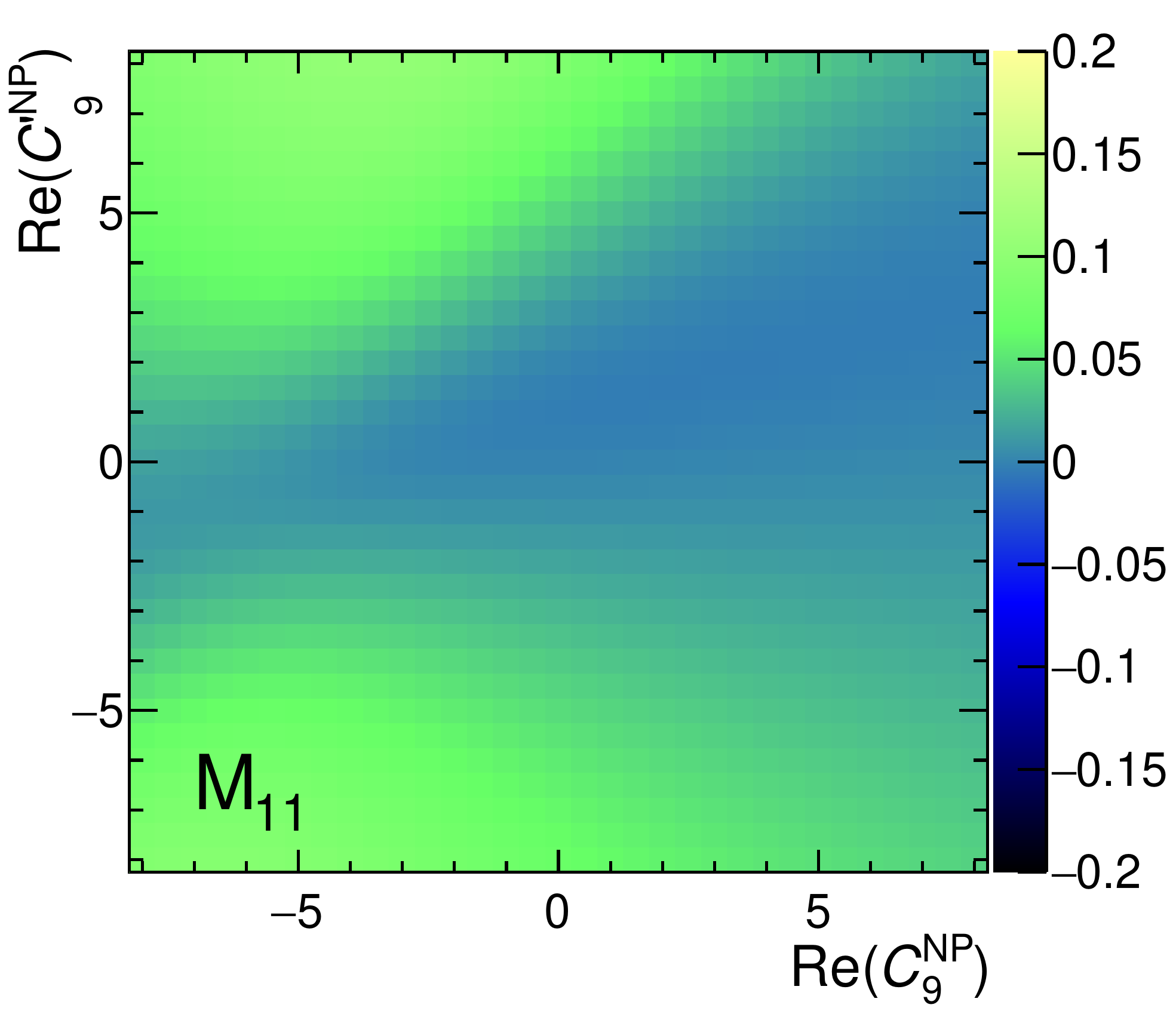} 
\includegraphics[width=0.24\linewidth]{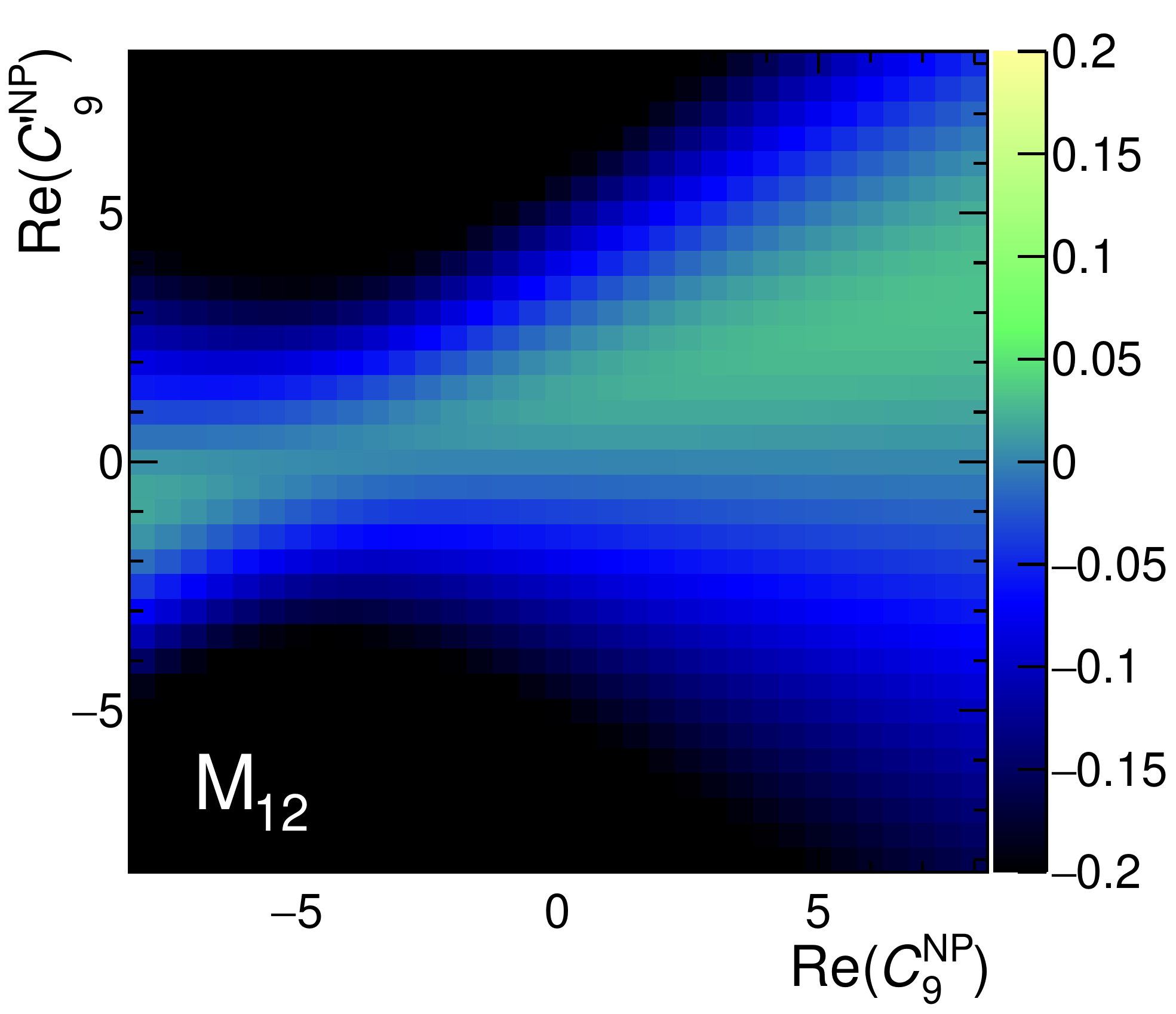} 
\includegraphics[width=0.24\linewidth]{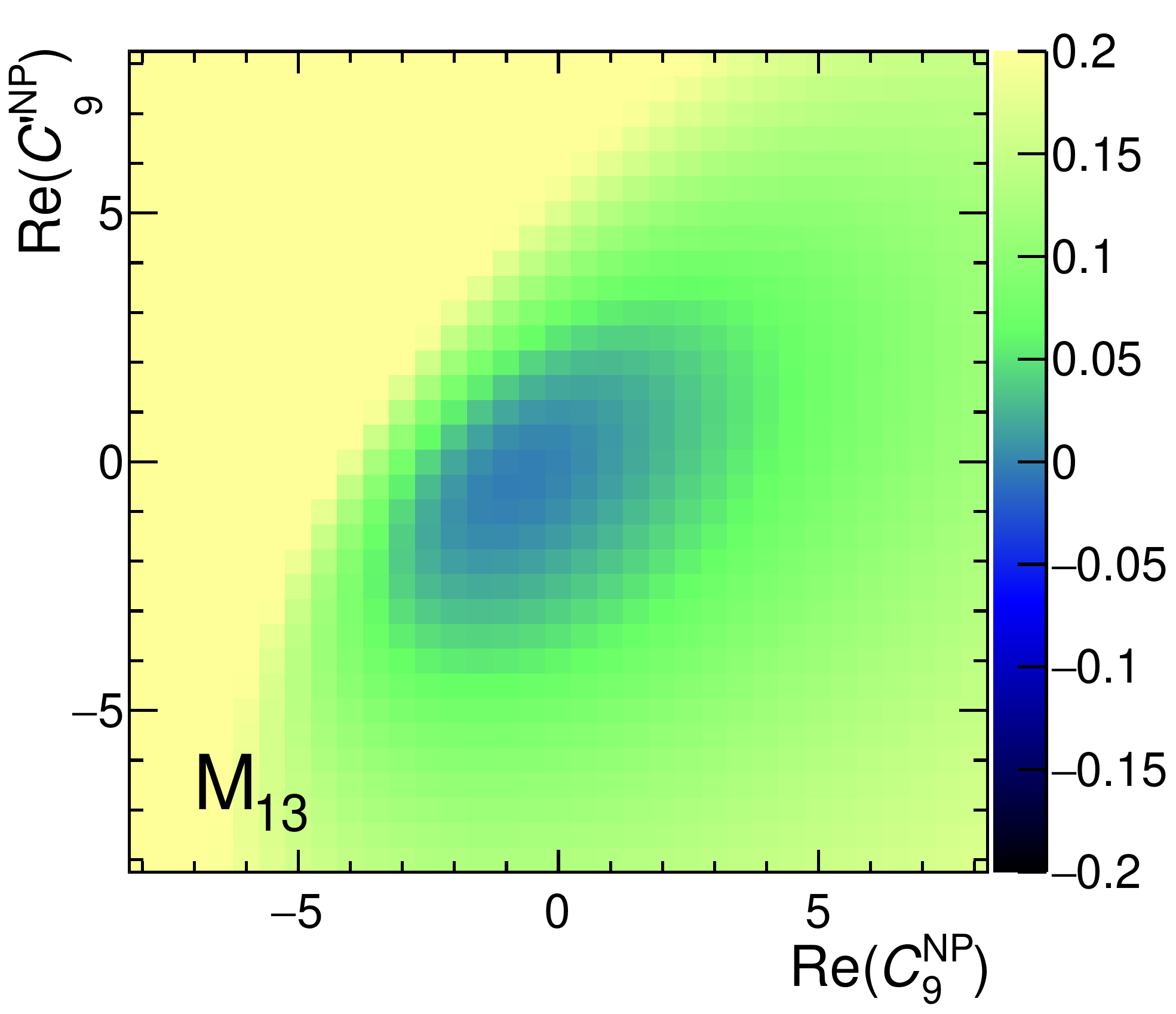}  
\includegraphics[width=0.24\linewidth]{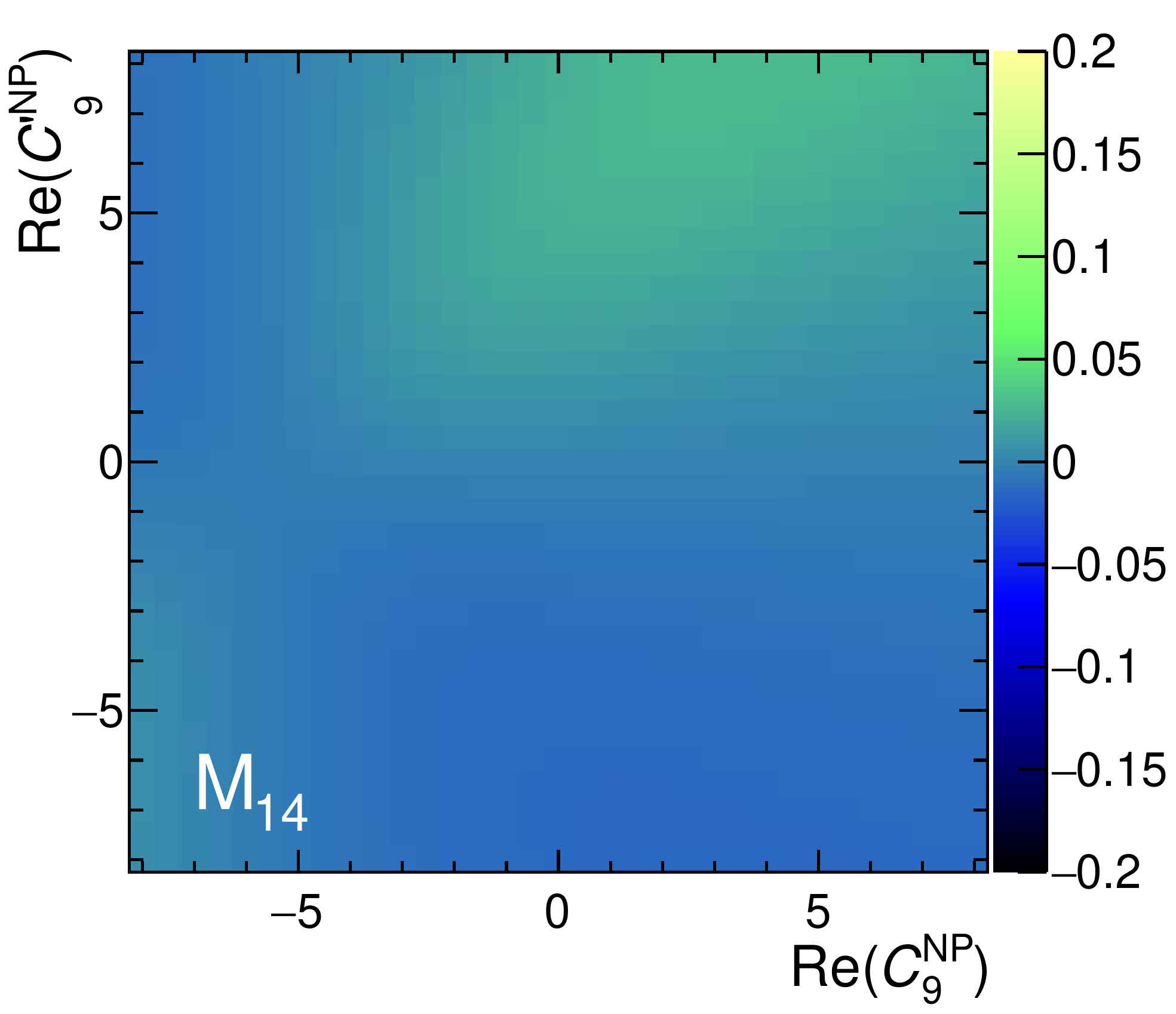}  \\ 
\includegraphics[width=0.24\linewidth]{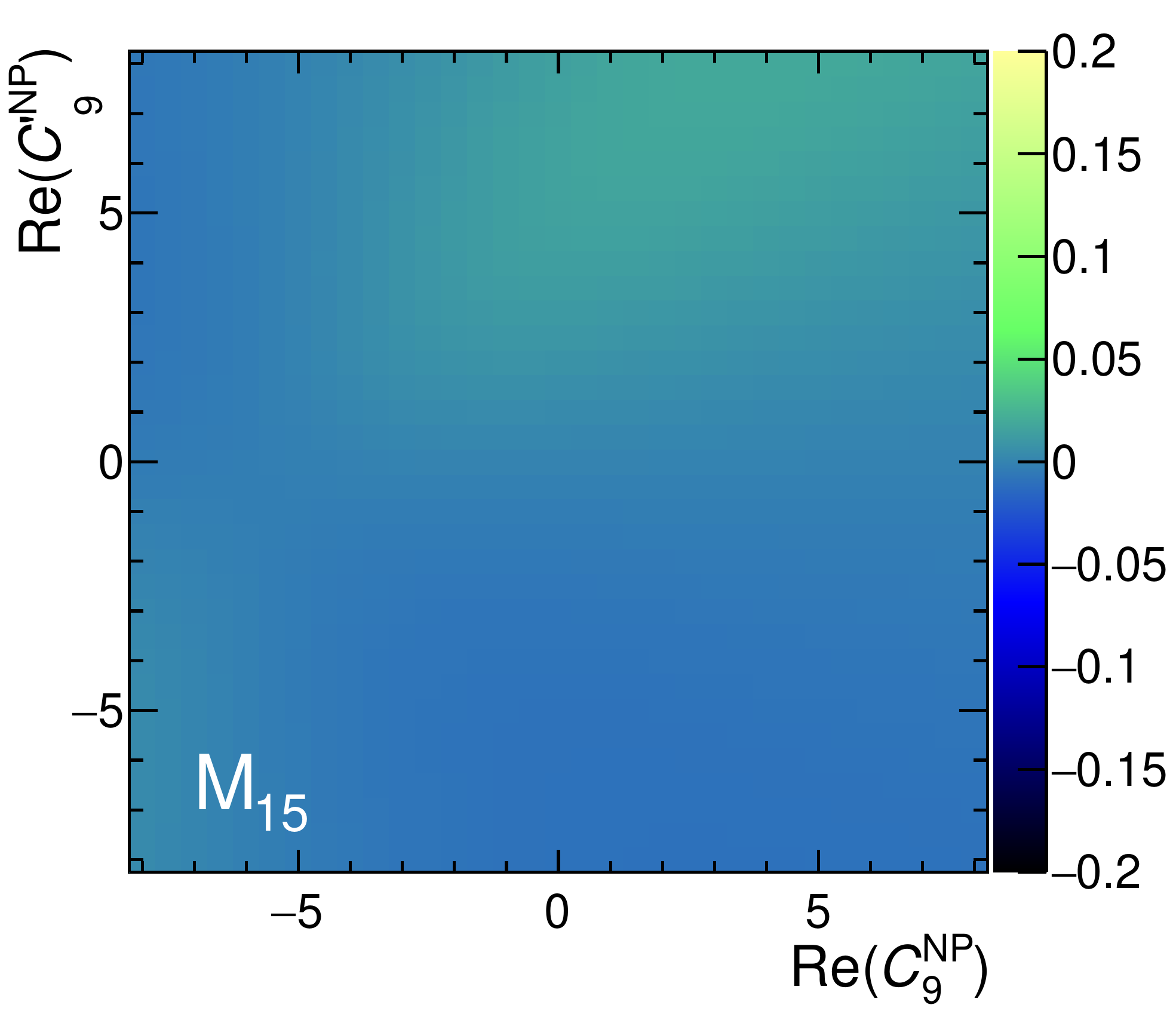} 
\includegraphics[width=0.24\linewidth]{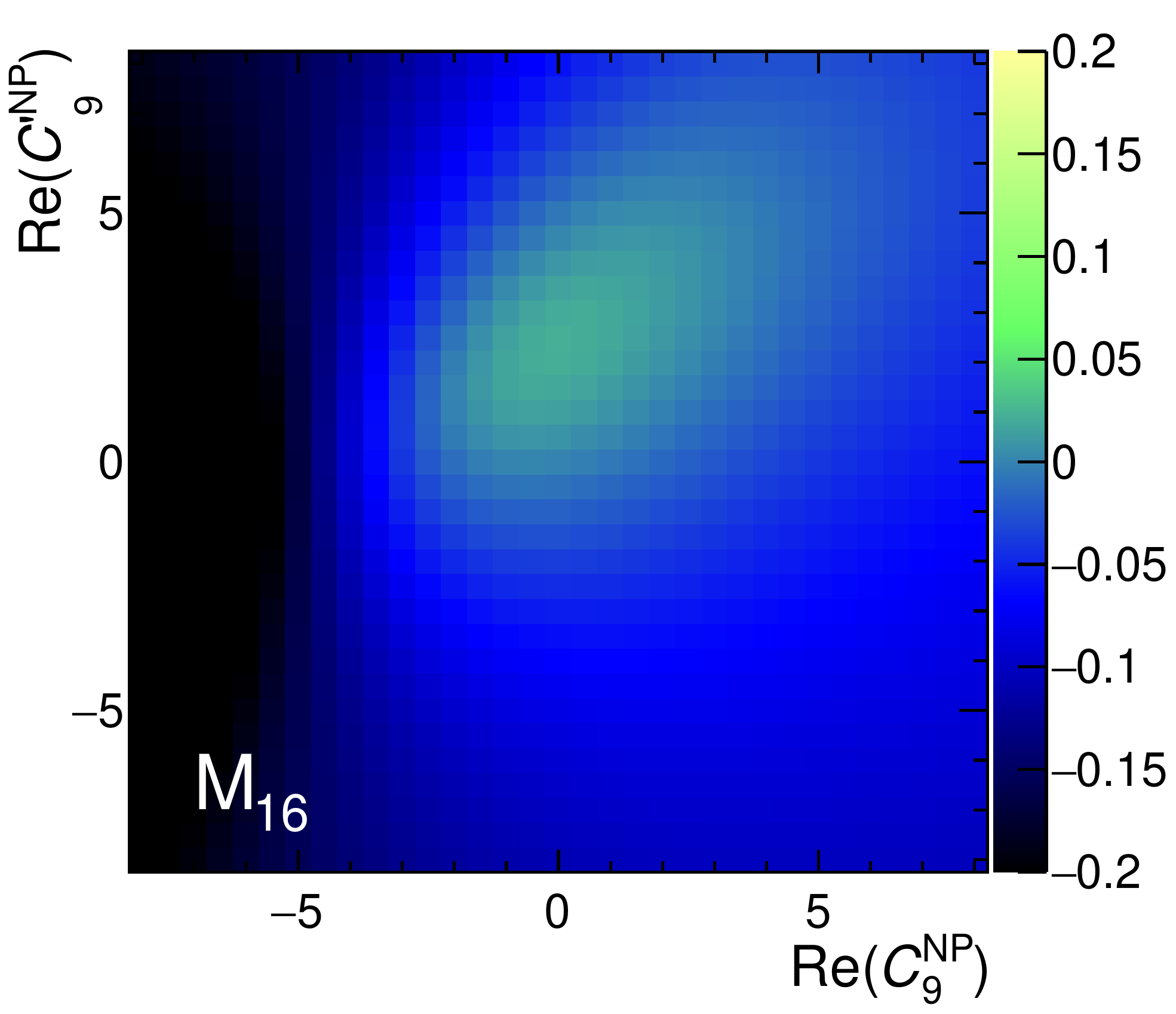} 
\includegraphics[width=0.24\linewidth]{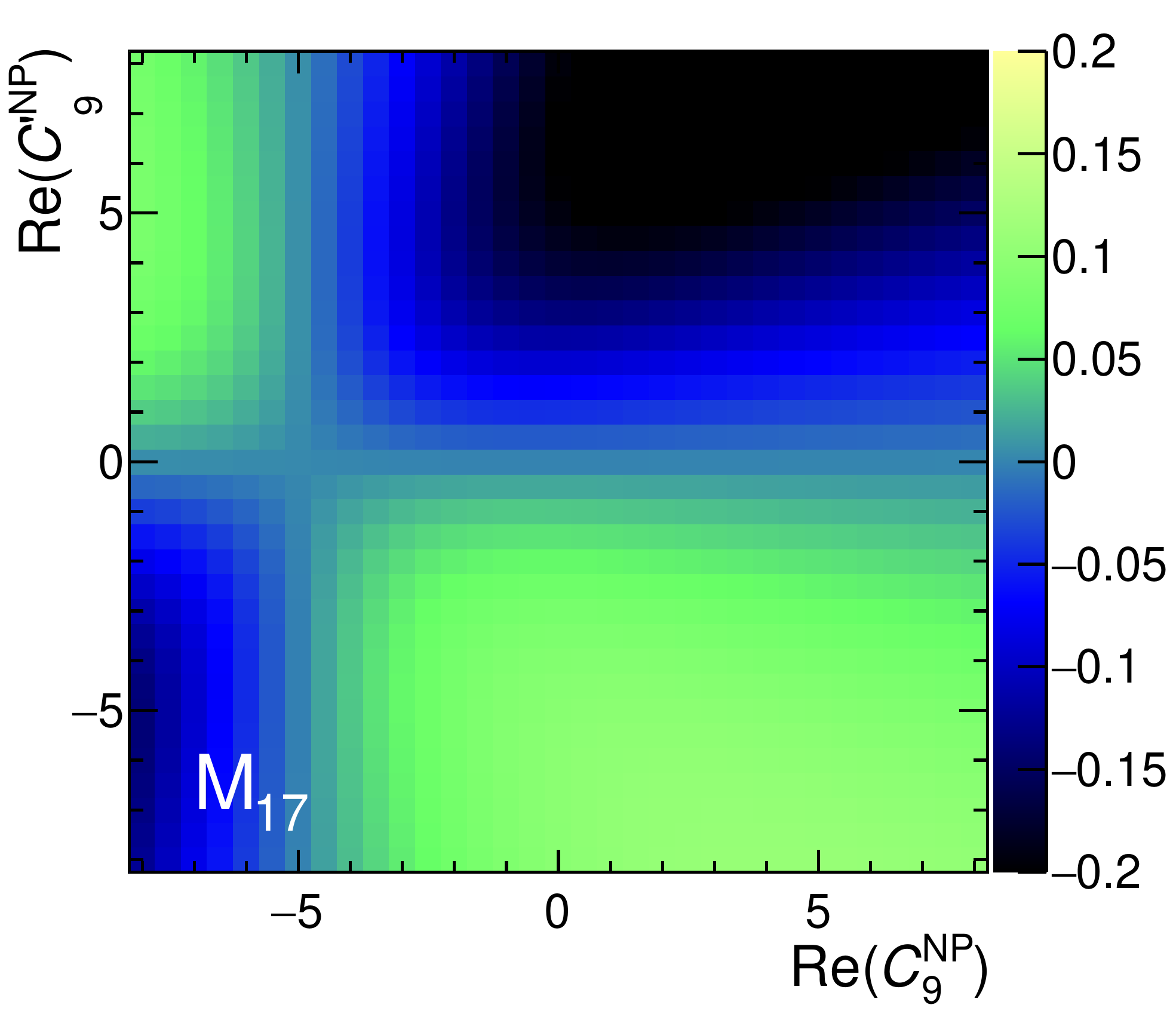}  
\includegraphics[width=0.24\linewidth]{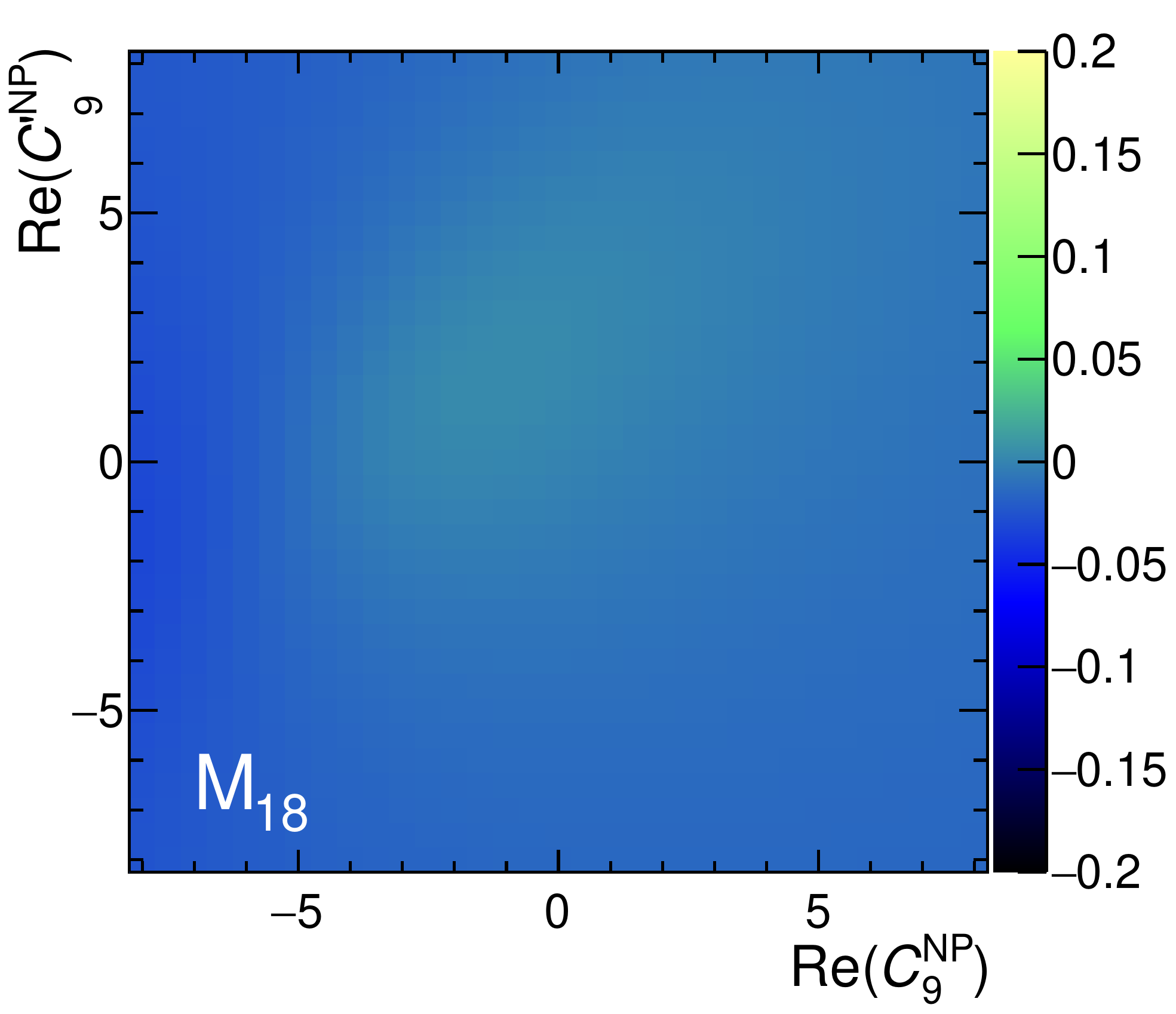} \\
\includegraphics[width=0.24\linewidth]{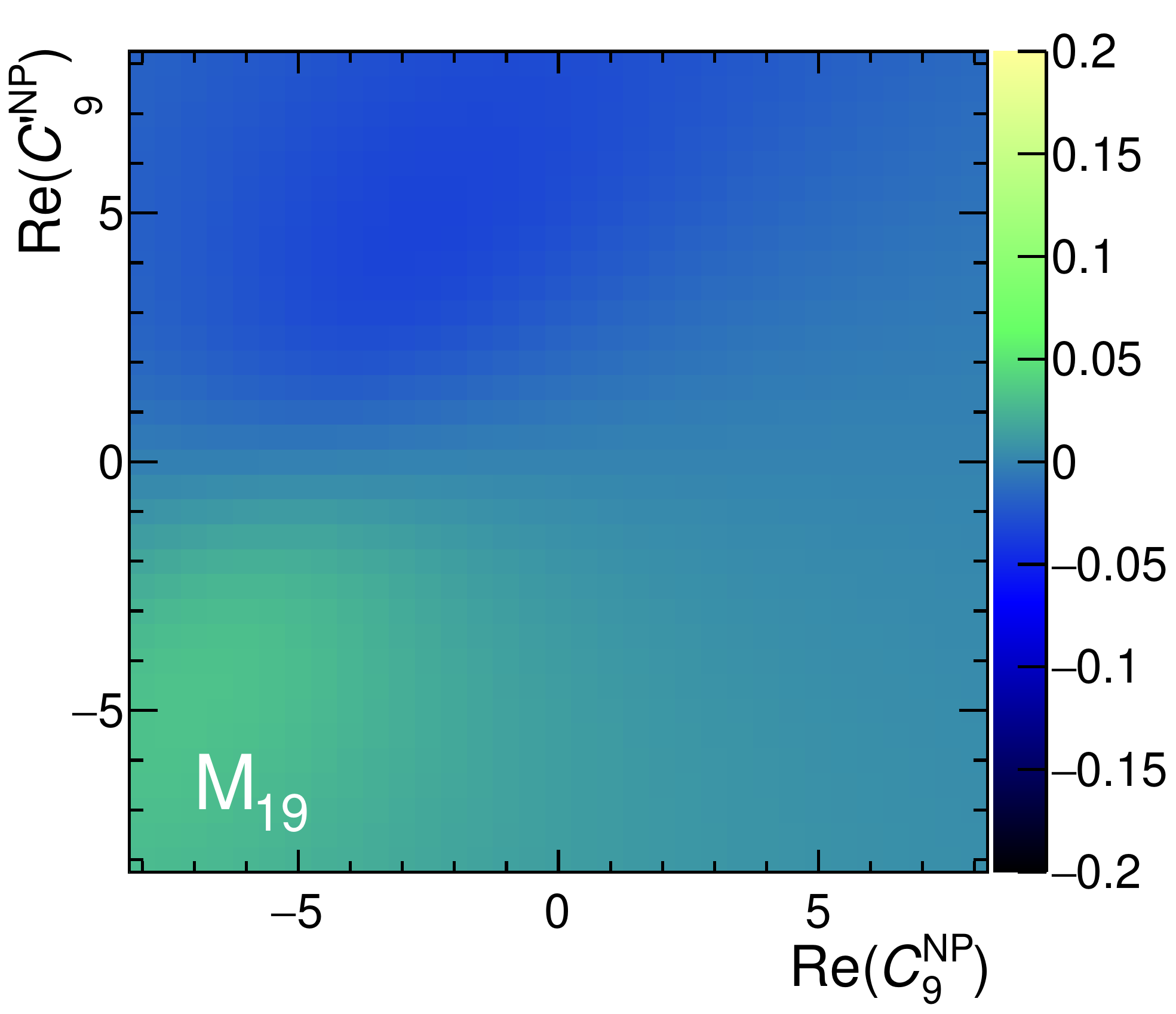} 
\includegraphics[width=0.24\linewidth]{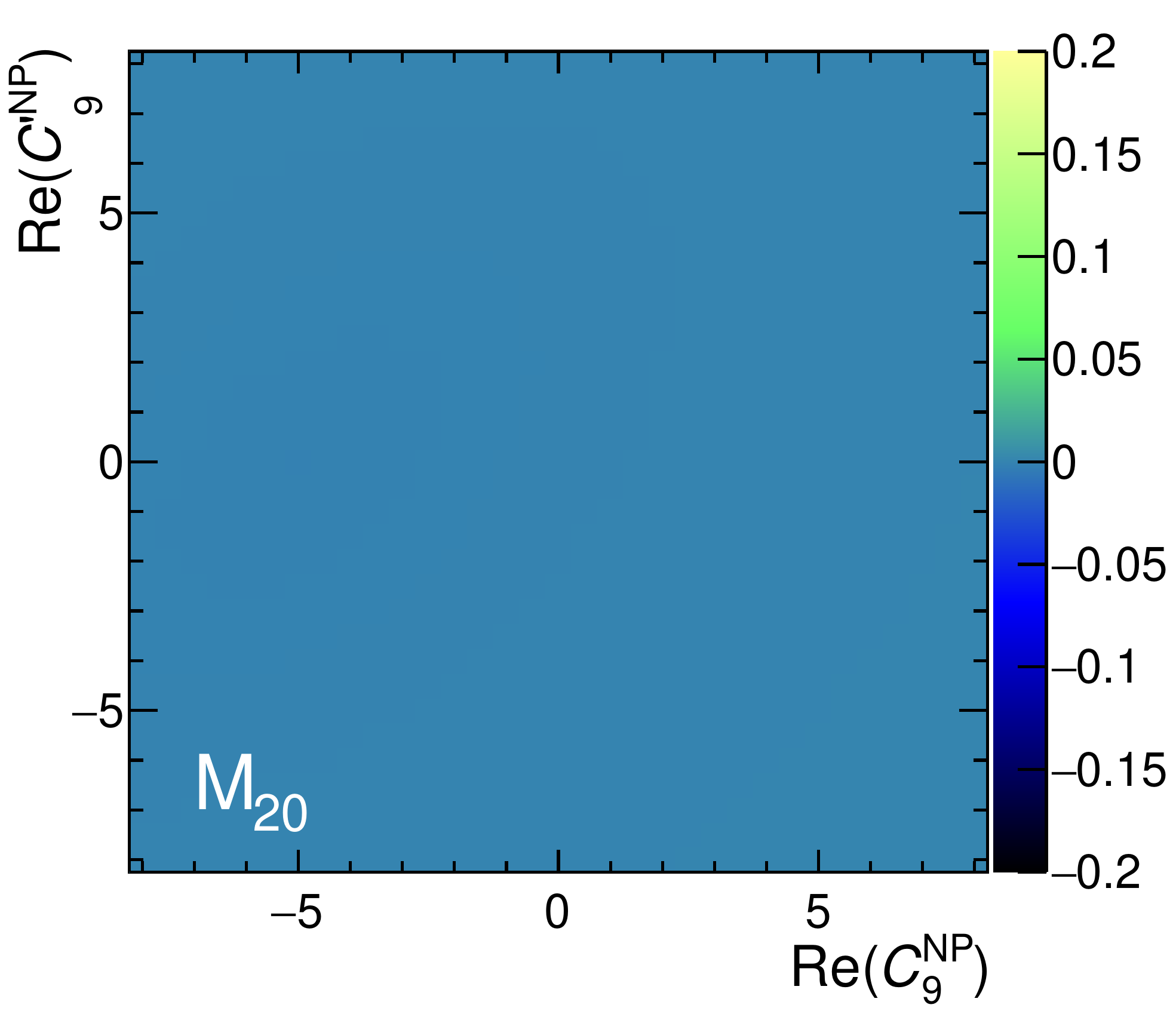}  
\includegraphics[width=0.24\linewidth]{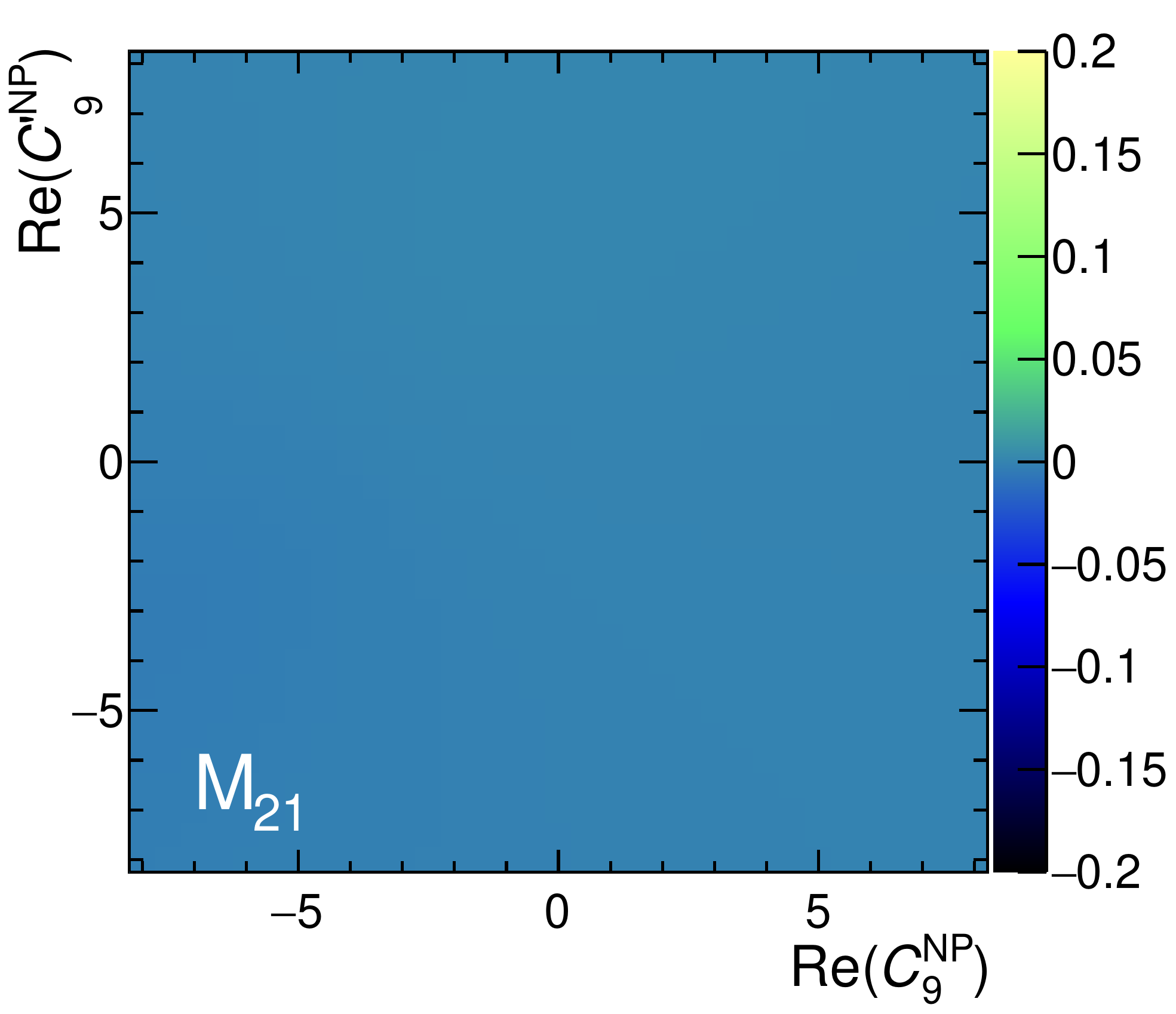}  
\includegraphics[width=0.24\linewidth]{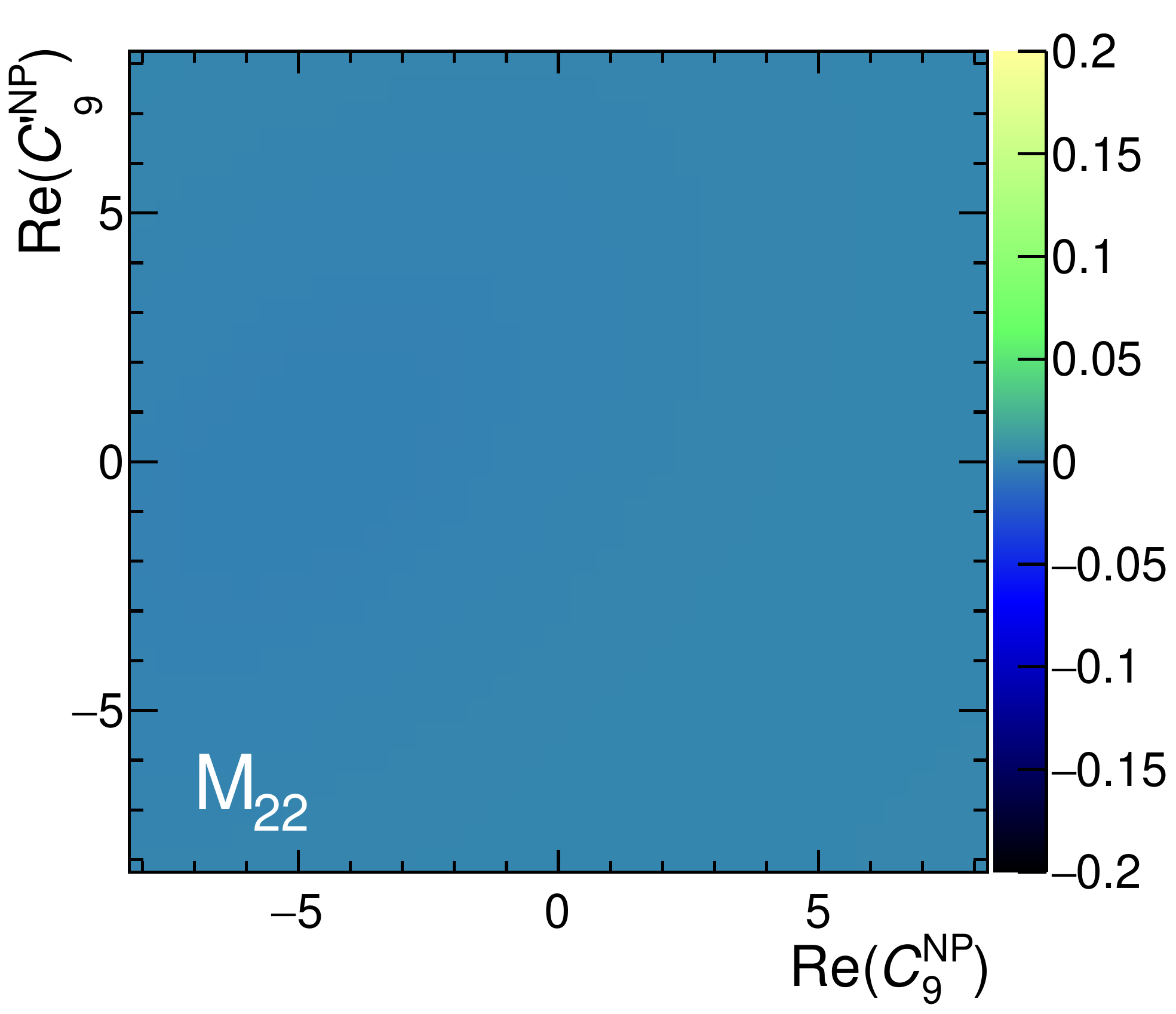} \\
\includegraphics[width=0.24\linewidth]{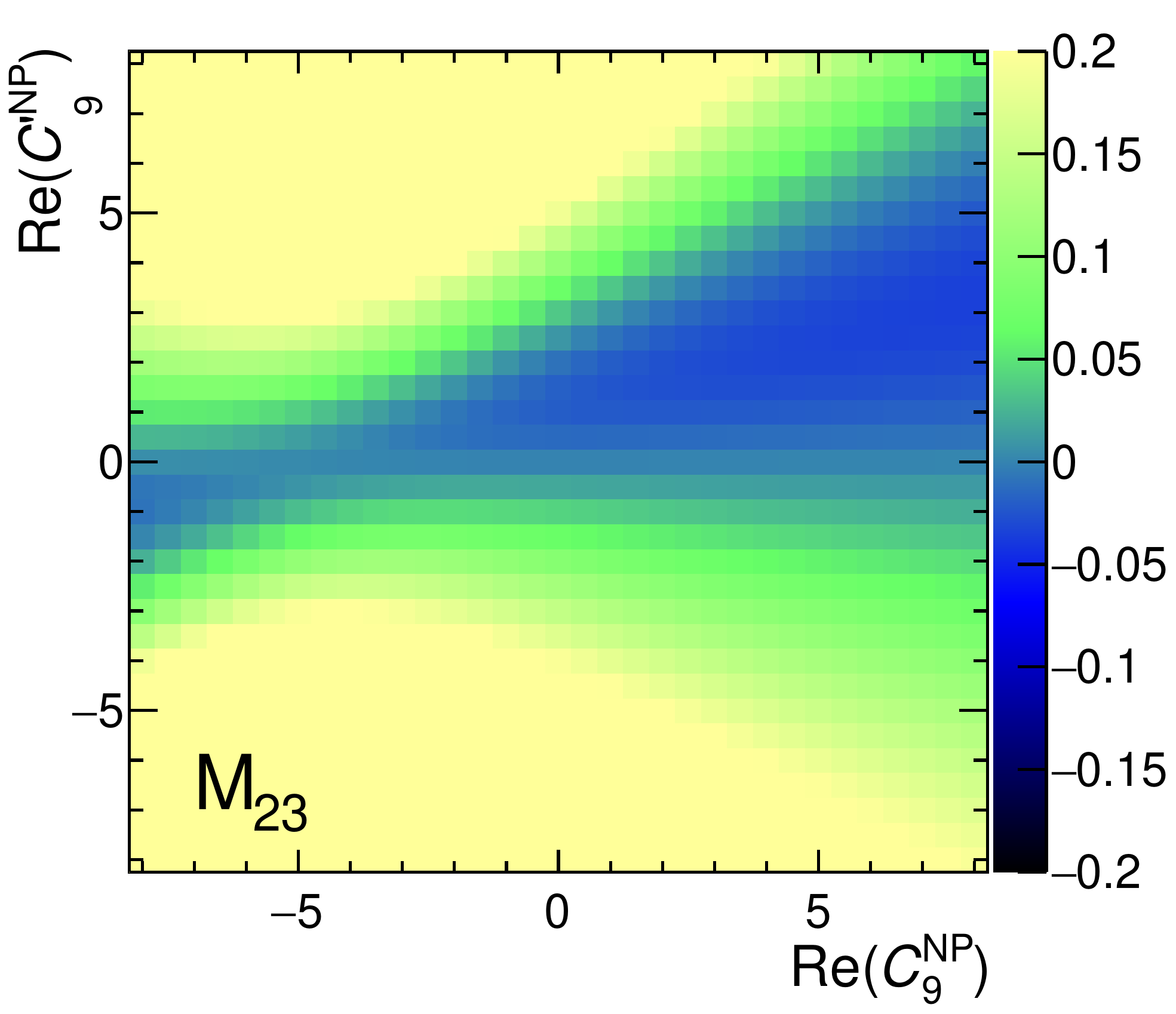} 
\includegraphics[width=0.24\linewidth]{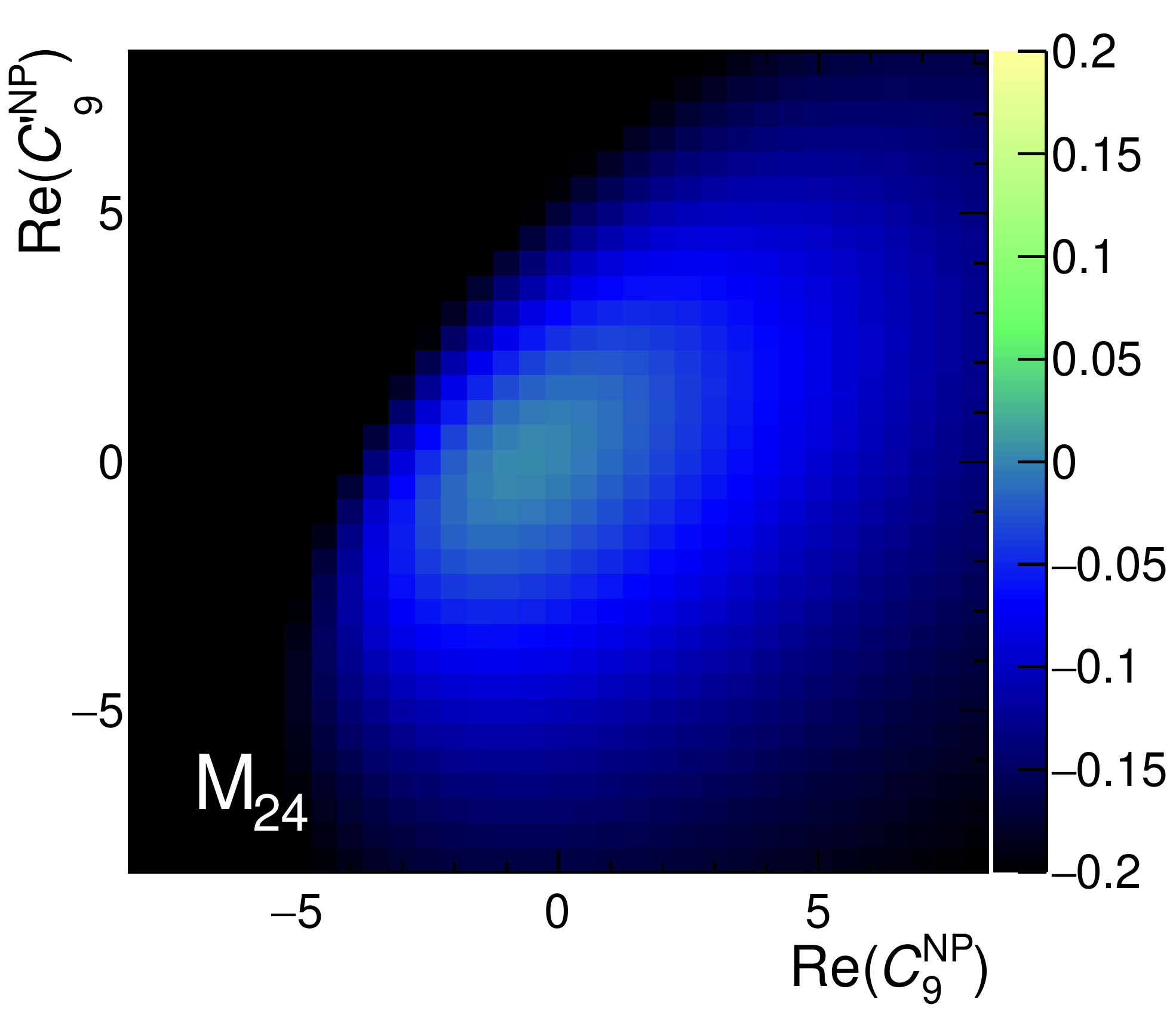}  
\includegraphics[width=0.24\linewidth]{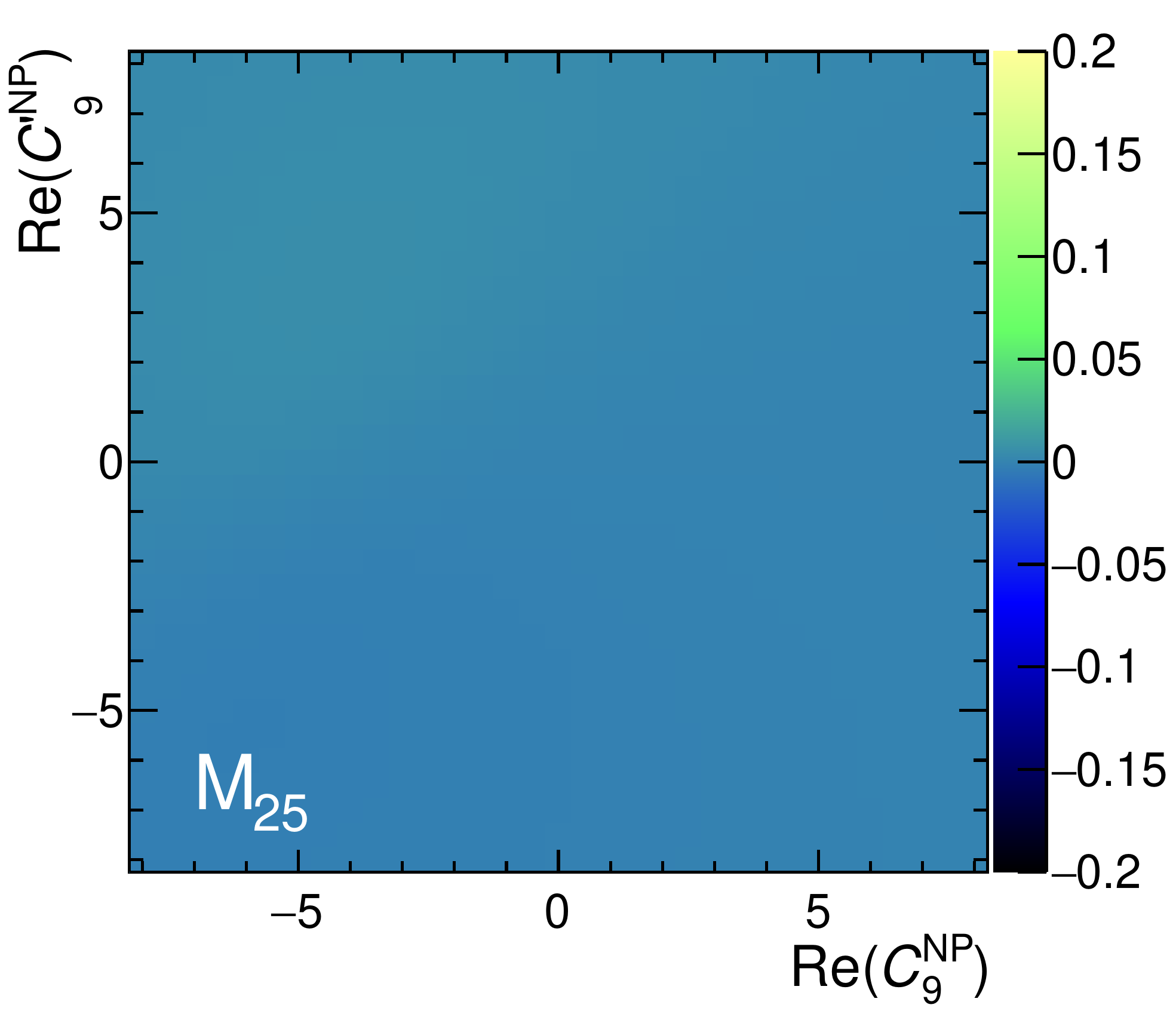} 
\includegraphics[width=0.24\linewidth]{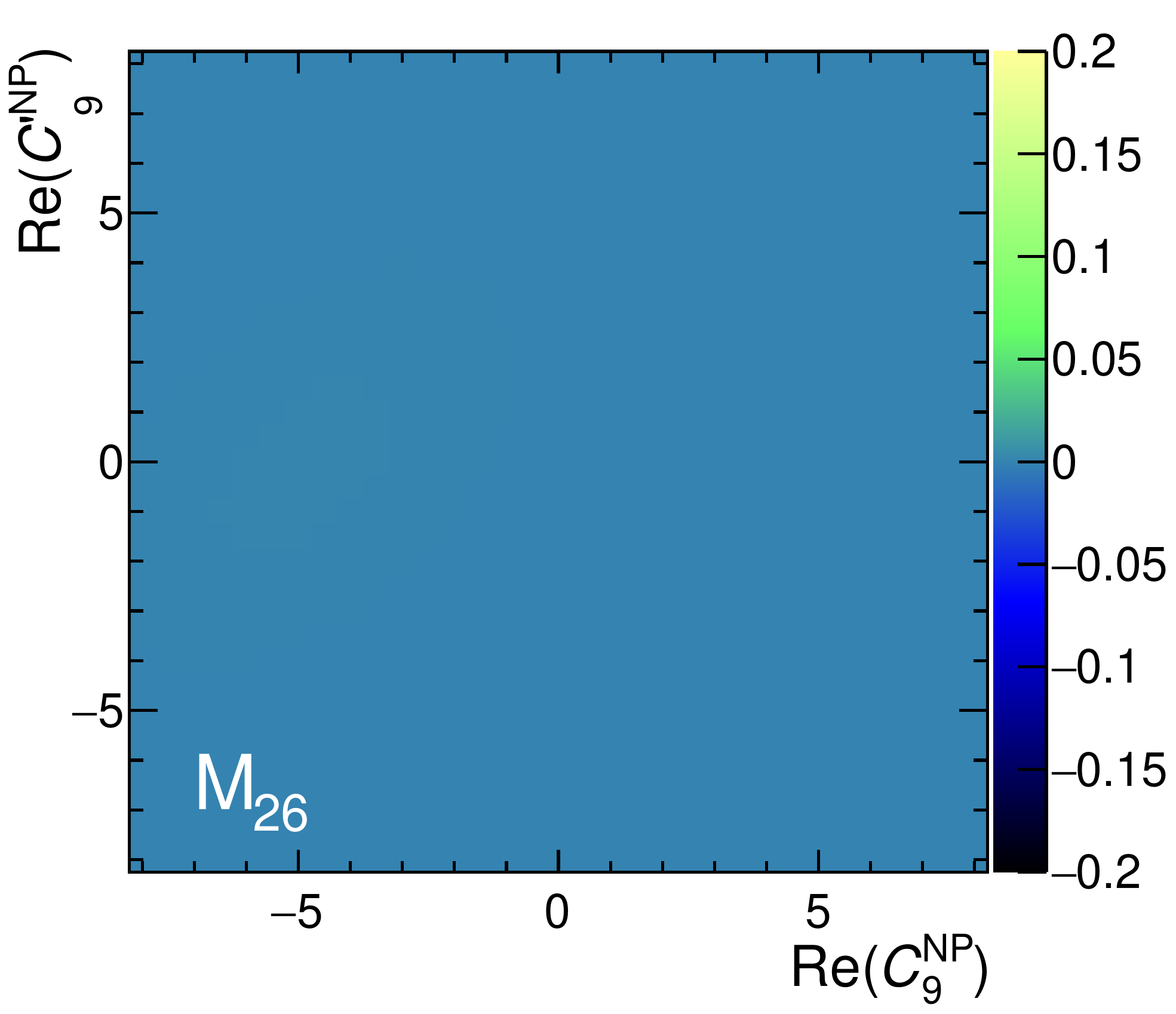} \\ 
\includegraphics[width=0.24\linewidth]{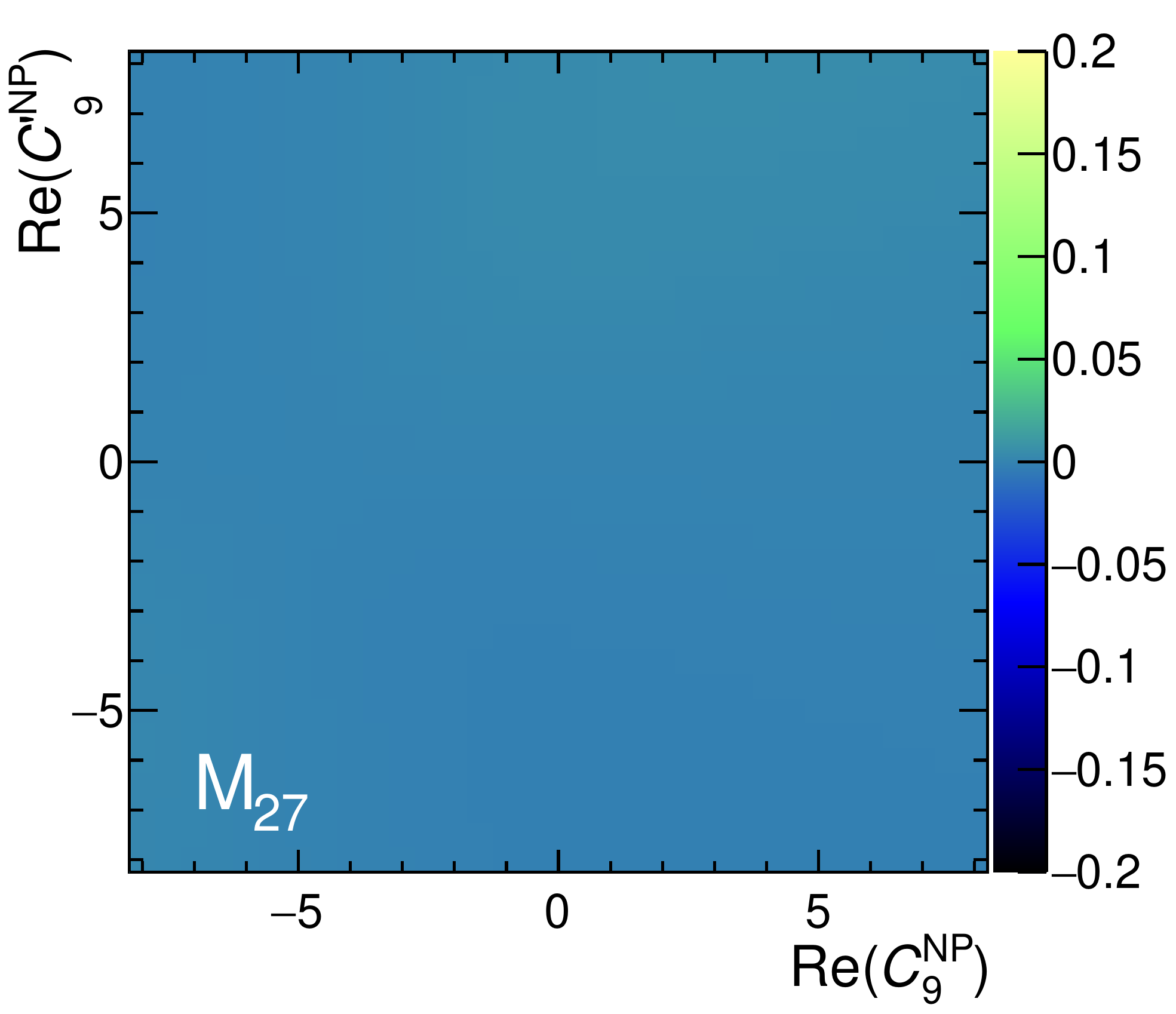}  
\includegraphics[width=0.24\linewidth]{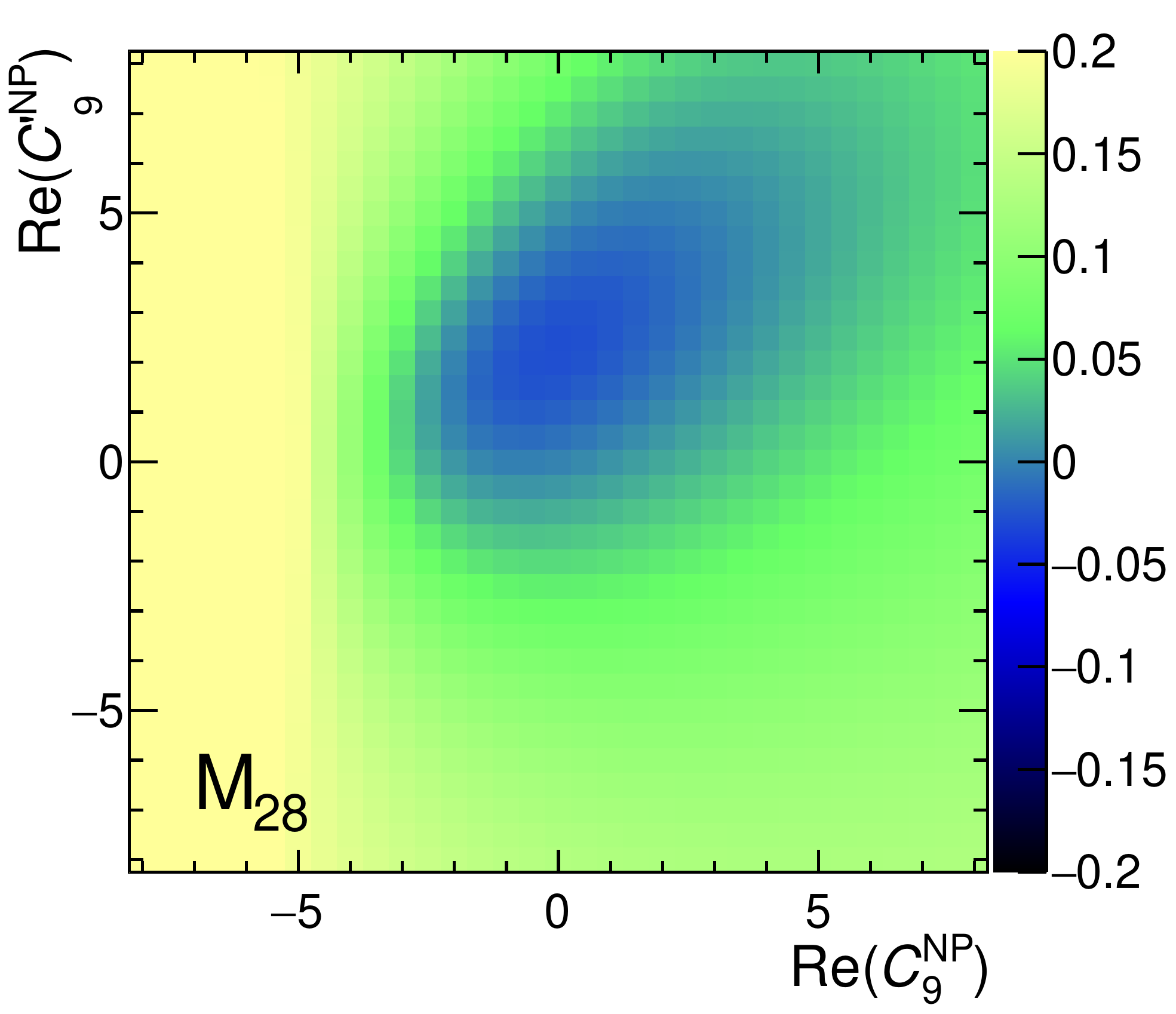} 
\includegraphics[width=0.24\linewidth]{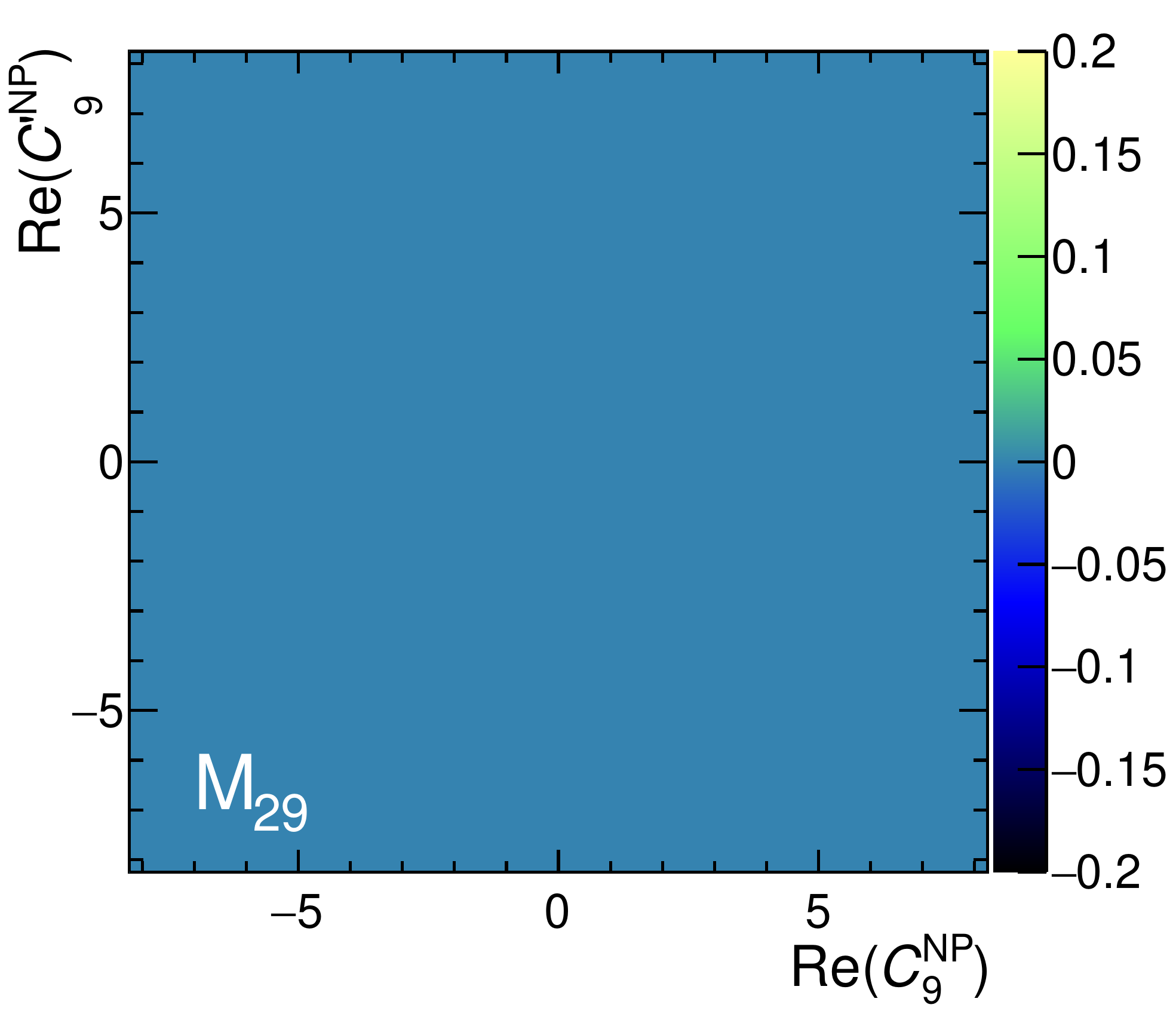} 
\includegraphics[width=0.24\linewidth]{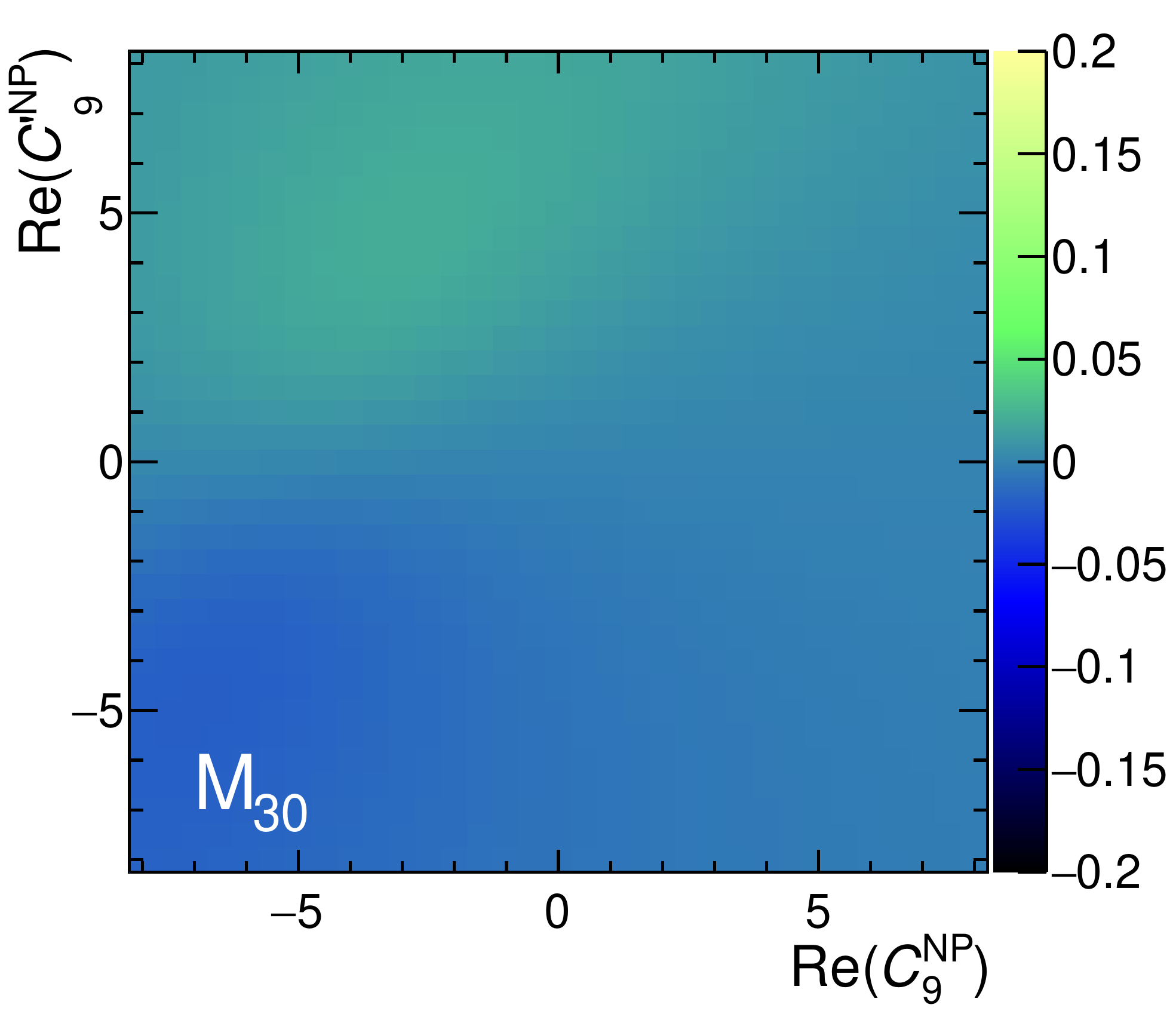} \\ 
\includegraphics[width=0.24\linewidth]{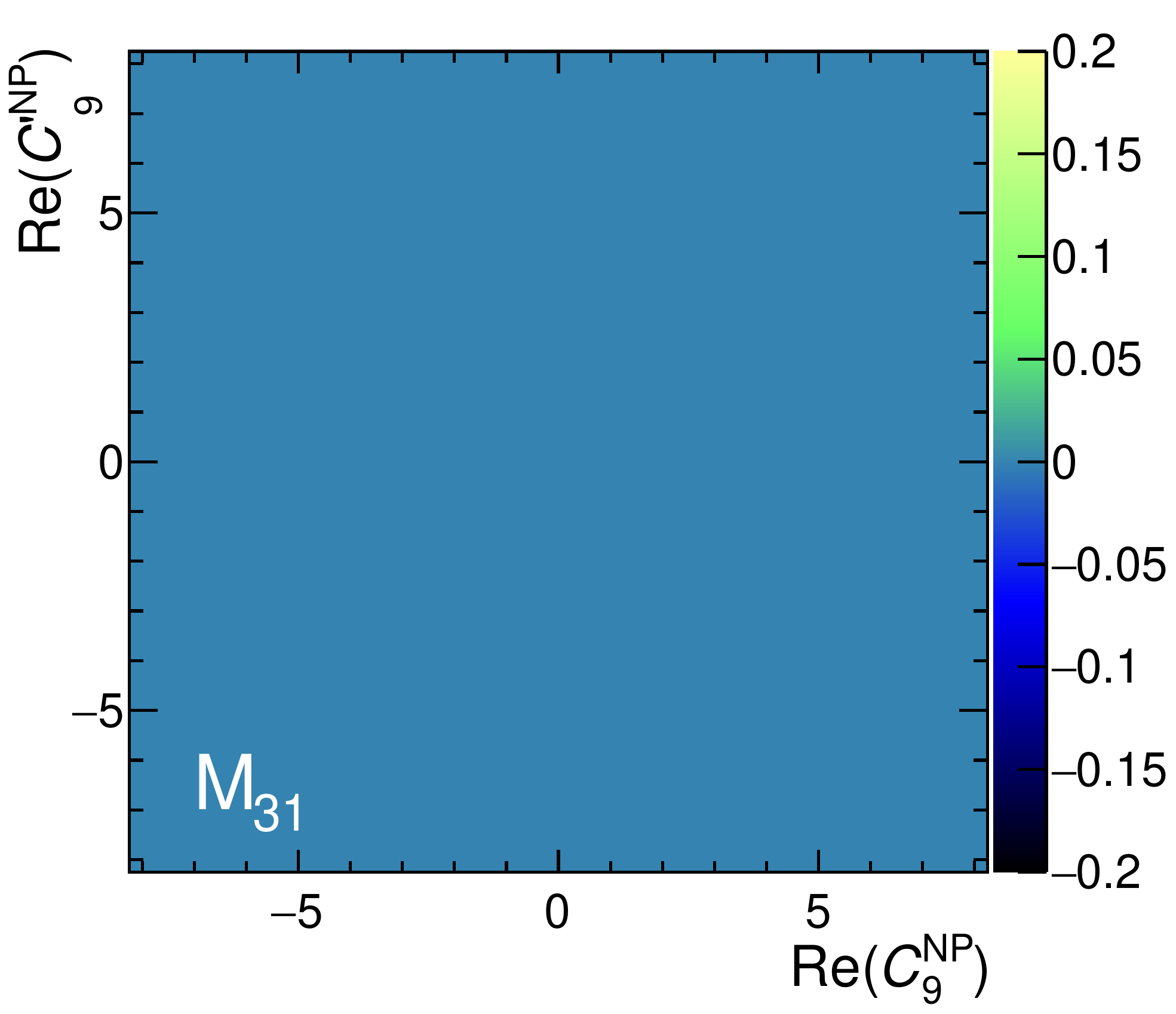}  
\includegraphics[width=0.24\linewidth]{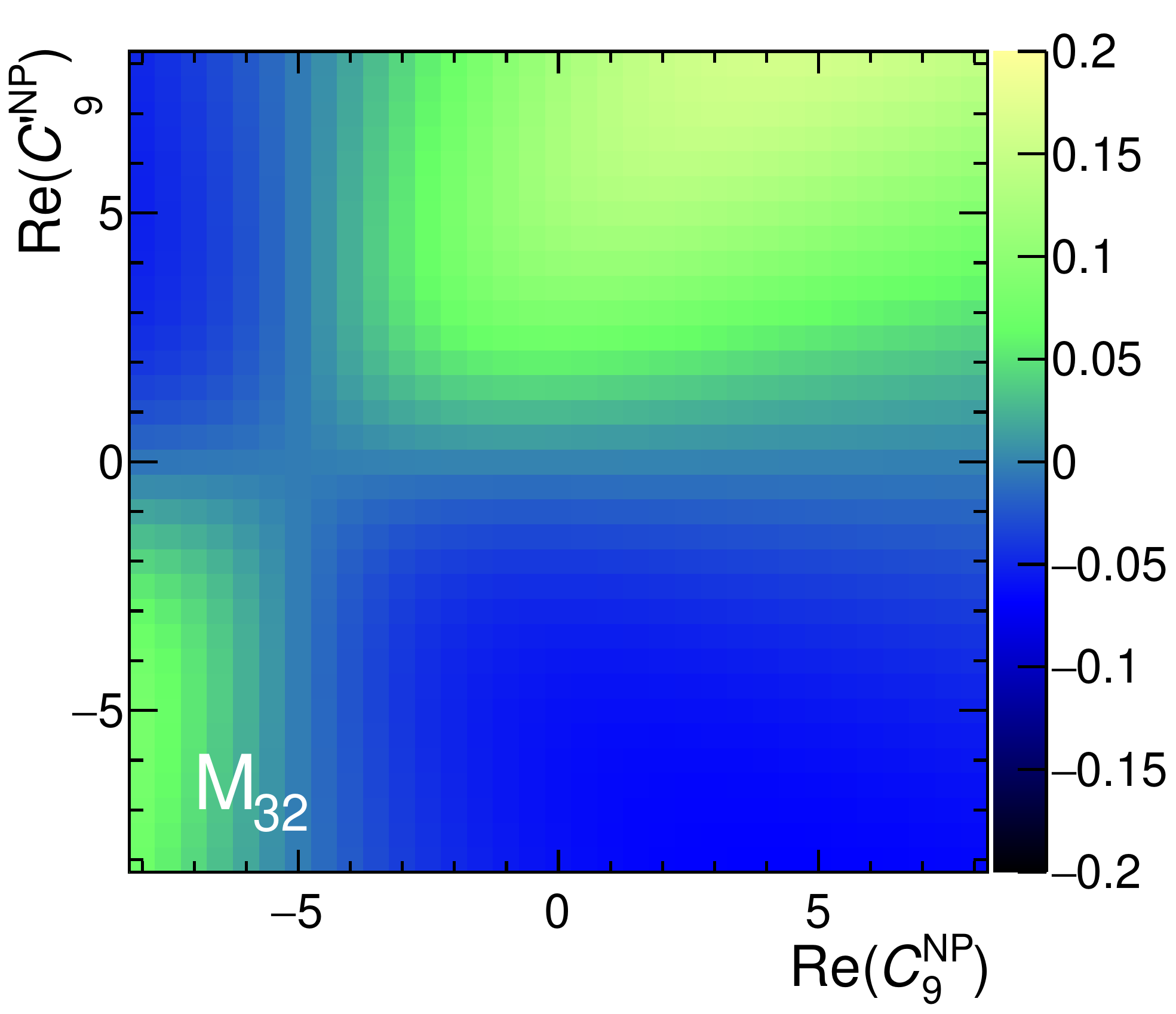} 
\includegraphics[width=0.24\linewidth]{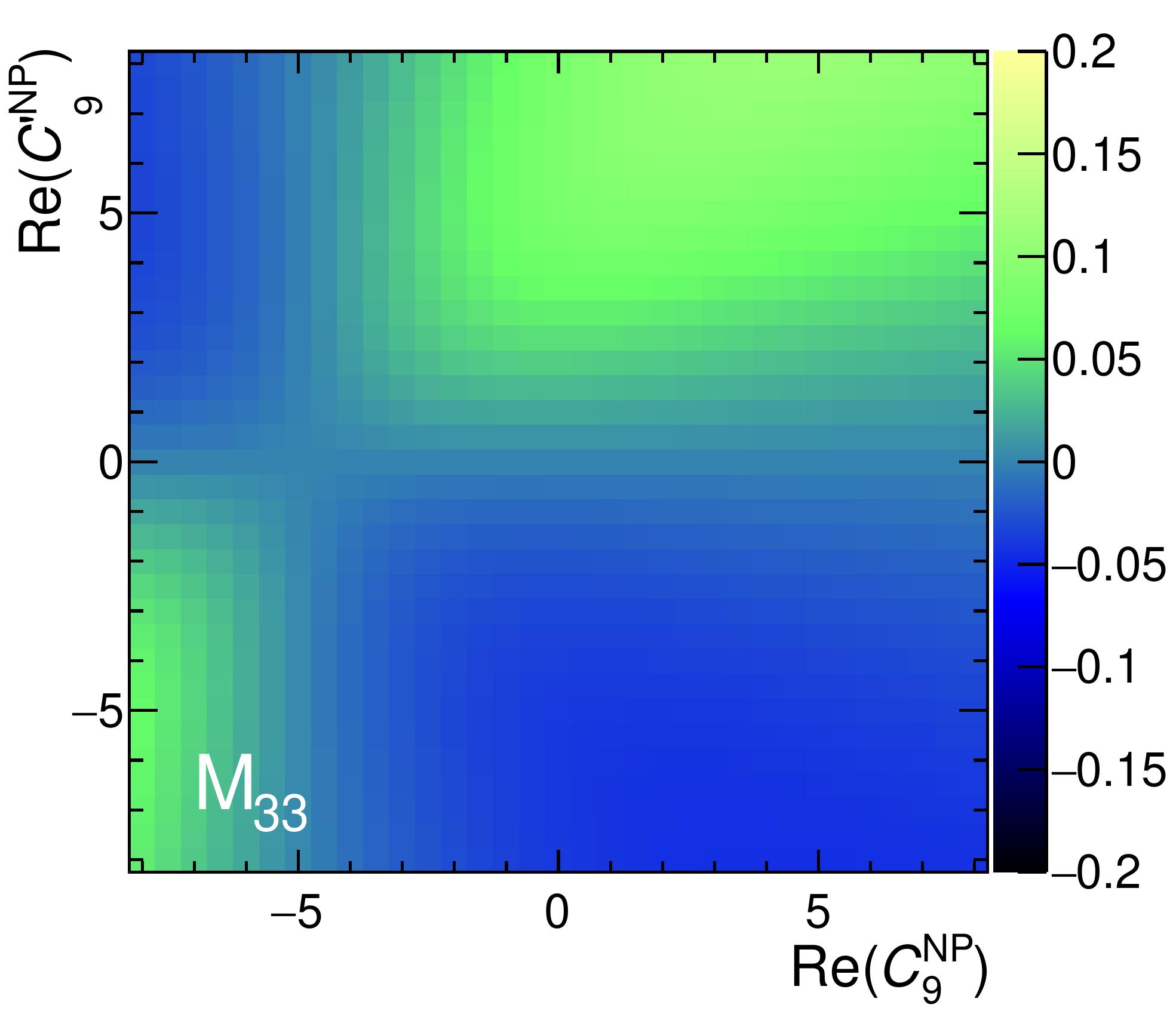} 
\includegraphics[width=0.24\linewidth]{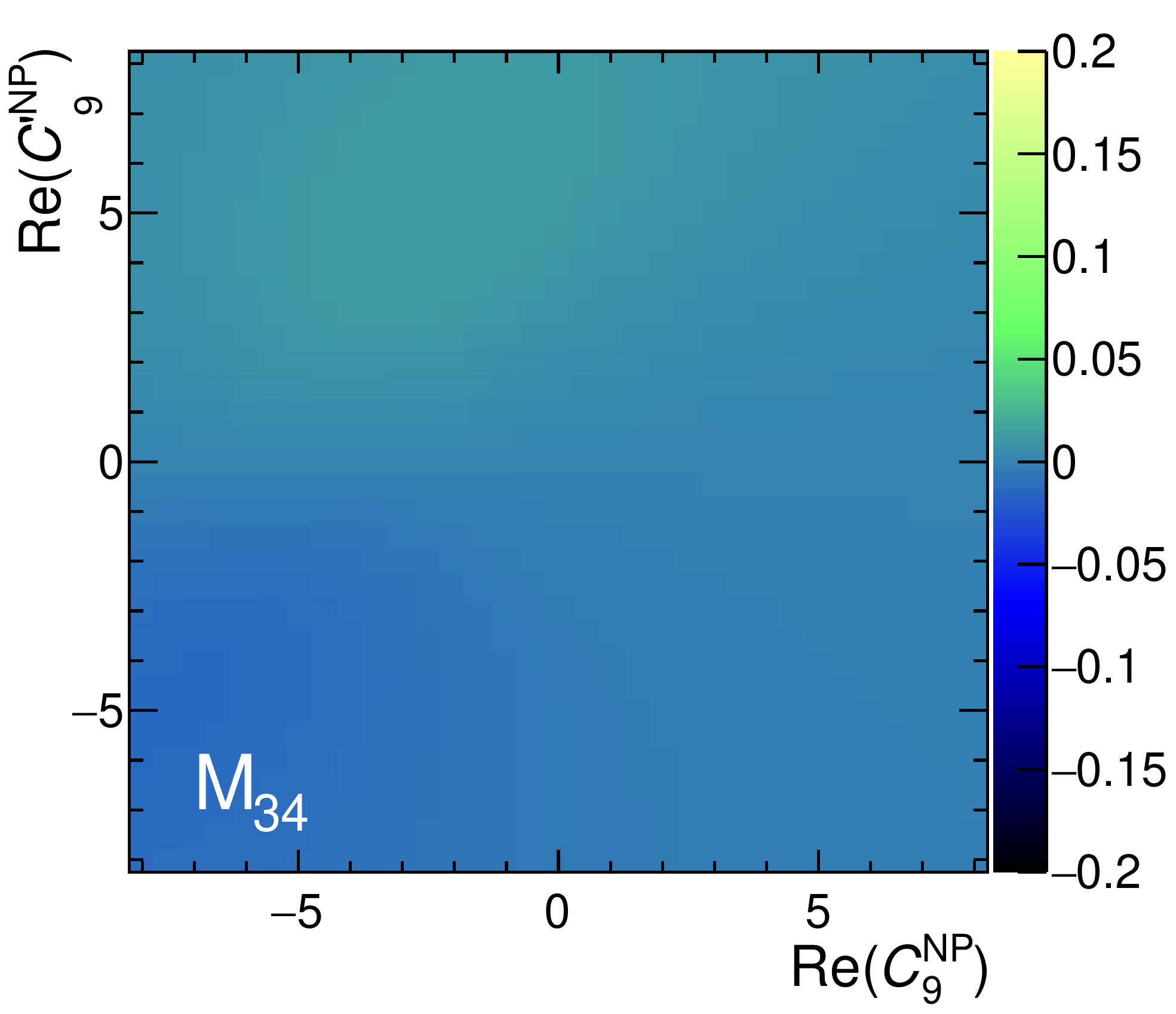}  
\caption{
\label{fig:scan:c9:c9p:lowrecoil:pol} 
Variation of the polarisation dependent angular observables  of the \decay{\Lb}{\Lz\mumu} decay from their SM central values in the low-recoil region ($15 < \qsq < 20\gev^{2}/c^{4}$) with a NP contribution to ${\rm Re}(C_9)$ or ${\rm Re}(C'_{9})$. 
The SM point is at $(0,0)$.
To illustrate the size of the effects, $P_{\Lb}  = 1$ is used.
}
\end{figure}

Figures~\ref{fig:scan:c9:c10:largerecoil}--\ref{fig:scan:c9:c9p:lowrecoil:pol} show the variation of $M_{1}$--$M_{34}$ under two possible modifications of the SM Wilson coefficients: a scenario where there is a NP contribution to ${\rm Re}(C_{9})$ or ${\rm Re}(C_{10})$; and a scenario where there is a NP contribution to ${\rm Re}(C_{9})$ or ${\rm Re}(C'_{9})$.

\clearpage

\section{Short-distance dependence at low hadronic recoil}
\label{appendix:rhodependence}

In the limit of  low hadronic recoil, and neglecting lepton mass dependent effects, the $K_{i}$ functions can be written in terms of the short-distance dependent $\rho$-functions of Ref.~\cite{Boer:2014kda} as
\begin{align} 
\begin{split}
K_{1} 
& = 4 s_{+} \left( |f_{\perp}^{A}|^{2} + \frac{(m_{\Lb} - m_{\Lz})^{2}}{q^{2}} |f_{0}^{A}|^{2} \right) \rho_{1}^{-}  \\ 
& + 4 s_{-} \left( |f_{\perp}^{V}|^{2} + \frac{(m_{\Lb} + m_{\Lz})^{2}}{q^{2}} |f_{0}^{V}|^{2}  \right) \rho_{1}^{+} ~,\\ 
K_{2} & = 8 s_{+} |f_{\perp}^{A}|^{2} \rho_{1}^{-} + 8 s_{-} |f_{\perp}^{V}|^{2} \rho_{1}^{+}~, \\
K_{3} & = 32 \sqrt{s_+ s_-} f_{\perp}^{A} f_{\perp}^{V} {\rm Re}(\rho_2) ~, \\
K_{4} & = -16 \alpha_{\Lz} \sqrt{s_{+} s_{-}} 
\left( f_{\perp}^{A} f_{\perp}^{V}  + \frac{(m_{\Lb}^{2} - m_{\Lz}^{2})}{q^{2}} f_{0}^{V} f_{0}^{A} \right) {\rm Re}(\rho_{4}) ~, \\
K_{5} & = -32 \alpha_{\Lz} \sqrt{s_{+} s_{-}} f_{\perp}^{A} f_{\perp}^{V} {\rm Re}(\rho_{4})~, \\
K_{6} & = 
-8 \alpha_{\Lz} s_{+} |f_{\perp}^{A}|^{2} \rho_{3}^{-} 
- 8 \alpha_{\Lz} s_{-} |f_{\perp}^{V}|^{2} \rho_{3}^{+} 
~, \\
K_{7} &=  -16 \alpha_{\Lz} \sqrt{s_{+}s_{-}} \left( 
\frac{(m_{\Lb} + m_{\Lz})}{\sqrt{q^2}} f_{0}^{V} f_{\perp}^{A} - 
\frac{(m_{\Lb} - m_{\Lz})}{\sqrt{q^2}} f_{0}^{A} f_{\perp}^{V}
\right) {\rm Re}(\rho_{4})~\\
K_{8} & = 
8 s_{+} \alpha_{\Lz} \frac{(m_{\Lb} - m_{\Lz})}{\sqrt{q^2}} f_{0}^{A} f_{\perp}^{A} \rho_{3}^{-} -  
8 s_{-} \alpha_{\Lz} \frac{(m_{\Lb} + m_{\Lz})}{\sqrt{q^2}} f_{0}^{V} f_{\perp}^{V} \rho_{3}^{+} 
~, \\
K_{10} & = 16 \alpha_{\Lz} \sqrt{s_+ s_-} \left(
\frac{(m_{\Lb} + m_{\Lz})}{\sqrt{q^2}} f_{0}^{V} f_{\perp}^{A}  + 
\frac{(m_{\Lb} - m_{\Lz})}{\sqrt{q^2}}  f_{0}^{A} f_{\perp}^{V}
\right) {\rm Im}(\rho_{4})~, \\
\end{split}
\label{eq:rhodependence1}
\end{align} 
and
\begin{align}
\begin{split}
K_{11} & = -16 P_{\Lb} \sqrt{s_+ s_-}  \left( 
f_0^A f_0^V \frac{(m_{\Lb}^2 - m_{\Lz}^2)}{q^2} - f_\perp^A f_\perp^V 
\right) {\rm Re}(\rho_4) ~,   \\
K_{12} &= 32 P_{\Lb} \sqrt{s_+ s_-} f_\perp^A f_\perp^V {\rm Re}(\rho_4) ~, \\
K_{13} &= 
 8 P_{\Lb} s_ {+} |f_{\perp}^{V}|^{2} \rho_{3}^{-} +
8 P_{\Lb} s_{-} |f_{\perp}^{A}|^{2} \rho_{3}^{+} ~,  \\
K_{14} &= -4 \alpha_{\Lz} P_{\Lb} s_- \left( |f_{\perp}^{V}|^{2} - |f_{0}^{V}|^{2} \frac{(m_{\Lb} +  m_{\Lz})^{2}}{q^2}\right) \rho_{1}^{+}  \\  
& - 4 \alpha_{\Lz} P_{\Lb}  s_+ \left( |f_{\perp}^{A}|^{2} - |f_{0}^{A}|^{2}  \frac{(m_{\Lb} -  m_{\Lz})^{2}}{q^2} \right) \rho_{1}^{-}  ~,\\ 
K_{15} &= 
-8 \alpha_{\Lz} P_{\Lb} s_{-} |f_{\perp}^V|^2 \rho_1^{+}    
-8 \alpha_{\Lz} P_{\Lb} s_{+}|f_{\perp}^A|^2  \rho_1^{-} ~ , \\
K_{16} &= -32 \alpha_{\Lz} P_{\Lb}\sqrt{s_+ s_-} f_{\perp}^A f_{\perp}^V {\rm Re}(\rho_2)~,  \\
K_{17} &= 
-8 \alpha_{\Lz} P_{\Lb}  s_{-} \frac{(m_{\Lb} + m_{\Lz})}{\sqrt{q^2}} f_{0}^{V} f_{\perp}^{V} \rho_{1}^{+} 
+8 \alpha_{\Lz} P_{\Lb} s_{+} \frac{(m_{\Lb} - m_{\Lz})}{\sqrt{q^2}} f_{0}^{A} f_{\perp}^{A} \rho_{1}^{-} ~,\\
K_{18} &= -16 \alpha_{\Lz} P_{\Lb} \sqrt{s_+ s_-} \left(
\frac{(m_{\Lb} + m_{\Lz})}{\sqrt{q^2}} f_0^V f_\perp^A  -  
\frac{(m_{\Lb} - m_{\Lz} )}{\sqrt{q^2}} f_0^A f_\perp^V 
\right) {\rm Re}(\rho_2)~,\\
K_{19} &= 16 \alpha_{\Lz} P_{\Lb} \sqrt{s_+ s_-} \left(
\frac{(m_{\Lb} + m_{\Lz})}{\sqrt{q^2}} f_0^V f_\perp^A  + 
\frac{(m_{\Lb} - m_{\Lz} )}{\sqrt{q^2}} f_0^A f_\perp^V 
\right) {\rm Im}(\rho_2) ~,\\
K_{22} &= 16 P_{\Lb} \sqrt{s_+ s_-} \left(
\frac{(m_{\Lb} + m_{\Lz})}{\sqrt{q^2}} f_{0}^{V} f_{\perp}^{A}  - 
\frac{(m_{\Lb} - m_{\Lz} )}{\sqrt{q^2}} f_{0}^{A} f_{\perp}^{V} 
\right) {\rm Im}(\rho_4) ~,\\
K_{23} &= -16 P_{\Lb} \sqrt{s_+ s_-}\left( 
\frac{(m_{\Lb} + m_{\Lz})}{\sqrt{q^2}} f_0^V f_\perp^A  + 
\frac{(m_{\Lb} - m_{\Lz} )}{\sqrt{q^2}} f_0^A f_\perp^V 
\right) {\rm Re}(\rho_4) ~,\\
K_{24} &= -8 P_{\Lb}  s_{-} \frac{(m_{\Lb} + m_{\Lz})}{\sqrt{q^2}} f_{0}^{V} f_{\perp}^{V} \rho_{3}^{+} 
- 8 P_{\Lb} s_{+} \frac{(m_{\Lb} - m_{\Lz})}{\sqrt{q^2}} f_{0}^{A} f_{\perp}^{A} \rho_{3}^{-} ~,\\
K_{25} &= -16 \alpha_{\Lz} P_{\Lb} \sqrt{s_+ s_-} \left( 
\frac{(m_{\Lb} + m_{\Lz})}{\sqrt{q^2}} f_0^V f_{\perp}^{A} -  
\frac{(m_{\Lb} - m_{\Lz})}{\sqrt{q^2}} f_0^A f_{\perp}^{V} 
\right)  {\rm Im}(\rho_2) ~,\\
K_{27} &= 
8 \alpha_{\Lz} P_{\Lb} s_{-} \frac{(m_{\Lb} + m_{\Lz})}{\sqrt{q^2}} f_0^V f_{\perp}^{V} \rho_1^+ + 
8 \alpha_{\Lz} P_{\Lb} s_{+} \frac{(m_{\Lb} - m_{\Lz})}{\sqrt{q^2}} f_0^A f_{\perp}^{A} \rho_1^- ~, \\ 
K_{28} &= 16 \alpha_{\Lz} P_{\Lb} \sqrt{s_+ s_-} \left( 
\frac{(m_{\Lb} + m_{\Lz})}{\sqrt{q^2}} f_{0}^{V} f_{\perp}^{A} + 
\frac{(m_{\Lb} - m_{\Lz})}{\sqrt{q^2}} f_{0}^{A} f_{\perp}^{V} 
\right) {\rm Re}(\rho_2)  ~, \\
K_{30} &= -16 \alpha_{\Lz} P_{\Lb} \sqrt{s_+ s_-}\frac{( m_{\Lb}^2 - m_{\Lz}^2 )}{q^2} f_0^A f_0^V {\rm Im}(\rho_2) ~,\phantom{\frac{1}{1}} \\
K_{32} &= 4 \alpha_{\Lz} P_{\Lb} s_-\frac{( m_{\Lb} + m_{\Lz} )^2}{q^2}|f_0^V|^2\rho_1^+ -
4 \alpha_{\Lz} P_{\Lb} s_+\frac{( m_{\Lb} - m_{\Lz} )^2}{q^2}|f_0^A|^2\rho_1^- ~, \\ 
K_{33} &= 4 \alpha_{\Lz} P_{\Lb}  s_- |f_{\perp}^{V}|^{2}  \rho_1^+ 
-  4 \alpha_{\Lz} P_{\Lb}s_+ |f_{\perp}^{A}|^{2}  \rho_1^- ~,\\
K_{34} &= -16 \alpha_{\Lz} P_{\Lb} \sqrt{ s_+ s_- } f_{\perp}^{A} f_{\perp}^{V} {\rm Im}(\rho_2) ~. 
\end{split} 
\label{eq:rhodependence}
\end{align}
The remaining $K_i$'s vanish in the low-recoil and zero lepton mass limits. 
In Eqs.~\ref{eq:rhodependence1} and \ref{eq:rhodependence}: 
$f_{0}^V$, $f_{0}^{A}$, $f_{\perp}^{V}$ and $f_{\perp}^{A}$ are the vector and axial-vector helicity form-factors for the $\Lb \to \Lz$ transition;  
$m_{\Lb}$ and $m_{\Lz}$ are the masses of the \Lb and \Lz baryon, respectively;
and $s_\pm = (m_{\Lb} \pm m_{\Lz} )^2 - q^{2}$. 
The four contributing tensor form-factors have been removed by exploiting Isgur-Wise relationships~\cite{Feldmann:2011xf} to relate the tensor form-factors to the vector and axial-vector form-factors.  

\clearpage

\setboolean{inbibliography}{true}
\bibliographystyle{LHCb}
\bibliography{references}

\end{document}